\documentclass{aa}
\usepackage[varg]{txfonts}
\usepackage{graphicx}   % Including figure files
\usepackage{amsmath}    % Advanced maths commands
\usepackage{amssymb}    % Extra maths symbols
\usepackage{multicol}        % Multi-column entries in tables
\usepackage{multirow}
\usepackage{pdflscape}  % Landscape pages
\usepackage[T1]{fontenc}
\usepackage{ae,aecompl}
\usepackage{mathptmx}
\usepackage{natbib}
\usepackage{color}
\usepackage{xspace}
\usepackage{hyperref}
\usepackage{rotating}
\usepackage{dcolumn,longtable,booktabs}
\usepackage{lscape}
\usepackage{longtable}
\usepackage{supertabular}
\usepackage[flushleft]{threeparttable}
\usepackage{mathtools}

\newcommand{\atlas}{ATLAS$^{\mathrm{3D}}$}
\newcommand{\hi } {{\rm H}\,{\small\rm I} \,}
\newcommand{\hiA} {{\rm H}\,{\small\rm I}}
\newcommand{\hiN} {{\rm H}\,{\scriptsize\rm I} \,}
\newcommand{\hiAN} {{\rm H}\,{\scriptsize\rm I}}

\newcommand{\hiAT} {{\rm H}\,{\tiny\rm I}}

\newcommand{\micron}{\rm$\mu$m}
\begin{document}

\bibliographystyle{aa}

\title{Cold gas and dust: Hunting spiral-like structures in early-type galaxies.
\thanks{Research supported in part by the Scientific and Technological Research Council of Turkey (TUBITAK) under the postdoc fellowship programme 1059B191701226.}
}
\author{M. K. Y{\i}ld{\i}z \inst{1,2,3,4}
       \and
       R. F. Peletier \inst{4}
       \and
       P.-A. Duc \inst{1}
       \and
       P. Serra \inst{5}
       }
\institute{Universit$\acute{e}$ de Strasbourg, CNRS, Observatoire astronomique de Strasbourg, UMR 7550, F-67000 Strasbourg, France\\
          \email{mustafa.yildiz@astro.unistra.fr, pierre-alain.duc@astro.unistra.fr}
          \and
          Astronomy and Space Sciences Department, Science Faculty, Erciyes University, Kayseri, 38039 Turkey \\
          \email{mkyildiz@erciyes.edu.tr}
          \and
          Erciyes University, Astronomy and Space Sciences Observatory Applied and Research Center (UZAYB\.{I}MER), 38039, Kayseri, Turkey
          \and
          Kapteyn Astronomical Institute, University of Groningen,
          P. O. Box 800, 9700 AV Groningen, Netherlands \\
          \email{peletier@astro.rug.nl}
          \and
          INAF - Osservatorio Astronomico di Cagliari, Via della Scienza 5, I-09047 Selargius (CA), Italy \\
          \email{paolo.serra@inaf.it}
         }
\titlerunning{Hunting spiral-like structures in ETGs}
\authorrunning{Y{\i}ld{\i}z et al.}
\date{Received \today}

\abstract
{Observations of neutral hydrogen (\hiAT) and molecular gas show that 50\% of all nearby early-type galaxies contain some cold gas. Molecular gas is always found in small gas discs in the central region of the galaxy, while neutral hydrogen is often distributed in a low-column density disc or ring typically extending well beyond the stellar body. Dust is frequently found in early-type galaxies as well.}
{The goal of our study is to understand the link between dust and cold gas in nearby early-type galaxies as a function of \hiAT \ content.}
{We analyse deep optical $g-r$ images obtained with the MegaCam camera at the Canada-France-Hawaii Telescope for a sample of 21 \hiAT-rich and 41 \hiAT-poor early-type galaxies.}
{We find that all \hiAT-rich galaxies contain dust seen as absorption. Moreover, in 57 per cent of these \hiAT-rich galaxies, the dust is distributed in a large-scale spiral pattern. Although the dust detection rate is relatively high in the \hiAT-poor galaxies ($\sim$59 per cent), most of these systems exhibit simpler dust morphologies without any evidence of spiral structures. We find that the \hiAT-rich galaxies possess more complex dust morphology extending to almost two times larger radii than \hiAT-poor objects. We measured the dust content of the galaxies from the optical colour excess and find that \hiAT-rich galaxies contain six times more dust (in mass) than \hiAT-poor ones. In order to maintain the dust structures in the galaxies, continuous gas accretion is needed, and the substantial \hiAT \ gas reservoirs in the outer regions of early-type galaxies can satisfy this need for a long time. We find that there is a good correspondence between the observed masses of the gas and dust, and it is also clear that dust is present in regions further than $3~R_{\mathrm{eff}}$.}
{Our findings indicate an essential relation between the presence of cold gas and dust in early-type galaxies and offer a way to study the interstellar medium in more detail than what is possible with \hiAT \ observations.
}

\maketitle

\section{Introduction}
\label{sec:introduction}
Integral field spectroscopy shows that many, probably most, giant ellipticals and lenticulars (hereafter ETGs) contain an inner, fast rotating component ($< 1 R_{\mathrm{eff}}$) \citep[e.g.][]{2006MNRAS.366.1151S, 2007MNRAS.379..401E,2011MNRAS.414..888E,2012MNRAS.421..872C,2012A&A...538A...8S,2015ApJ...798....7B}. In those same inner regions, ionised gas is detected frequently \citep[e.g.][]{2005A&A...433..497R,2011MNRAS.417..882D,2016A&A...588A..68G,2018A&A...611A..28G}, and young stellar populations are not uncommon \citep[e.g.][]{2000AJ....119.1645T,2007ApJS..173..619K,2010MNRAS.408...97K}. Studies that are carried out with integral field units instruments, generally focus on the central regions of galaxies due to a small field of view (FOV) \citep[e.g.][]{2011MNRAS.413..813C, 2011MNRAS.418.1452L,2012ApJS..198....2K,2013MNRAS.432.1768K,2014MNRAS.444.3340W}. However, deep optical images show that we can study ETGs at considerable distance from the centre \citep{2014MNRAS.440.1458D,2015MNRAS.446..120D,2017A&A...601A..86K}. Moreover, these deep optical studies provide good quality data for not only the outer regions but also the inner regions of galaxies.

Similarly, observations of neutral hydrogen (\hiA) and molecular gas, such as CO, show that ETGs have gas discs \citep[e.g.][]{1991A&A...243...71V, 2002AJ....124..788Y, 2006MNRAS.371..157M, 2007A&A...465..787O, 2010MNRAS.409..500O, 2011MNRAS.410.1197C}. Observations carried out as a part of the \atlas \ survey\footnote{The \atlas \ Project is a multi-wavelength survey of a complete sample of 260 ETGs within the local (42Mpc) volume. http://www-astro.physics.ox.ac.uk/atlas3d/} show that $\sim$50 per cent of all nearby ETGs contain some cold gas (CO and \hiA) \citep{2011MNRAS.414..940Y, 2012MNRAS.422.1835S}. The molecular gas is typically found in small gas discs in the central regions and is linked to small amounts of star formation (SF) \citep{2013MNRAS.432.1796A,2013MNRAS.429..534D}. In contrast, \hiA \ has been found distributed in discs or rings that are much larger than the stellar body \citep{2012MNRAS.422.1835S}. If the column density of the gas is high enough, these \hiA \ discs host star formation with an efficiency (star formation per unit gas mass) similar to that in the outer parts of spirals \citep{2015MNRAS.451..103Y, 2017MNRAS.464..329Y}.

Given their cold gas content, it is no surprise that many ETGs host some dust too. The presence of dust in these galaxies has long been known and discussed in the literature \citep[e.g.][]{1985MNRAS.214..177S, 1994MNRAS.271..833G, 1995AJ....110.2622L, 1995AJ....110.2027V, 1999A&AS..136..269F, 2001AJ....121.2928T}. These studies indicate a relation between dust and ionised gas \citep[e.g, see review by ][]{1997ASPC..121..620F}. Previous studies have revealed dust structures in the very central regions of ETGs by using the \textit{Hubble Space Telescope} (HST) \citep{2007ApJ...655..718S, 2013ApJ...766..121M}. For example, \citet{2007ApJ...655..718S} have found complex dust structures in the very central regions of ETGs, and even some spiral-shape dust structures ($<$ 25 per cent).

Dust in ETGs is not only detected via absorption at optical wavelengths but also in emission in infrared (IR) bands, particularly in the far-IR \citep{1989ApJS...70..329K,1998ApJ...499..670B,2004ApJS..154..229P,2007ApJ...660.1215T,2012ApJ...748..123S,2013A&A...552A...8D,2013ApJ...766..121M,2017A&A...602A..68O}. For example, by using \textit{Herschel} observations, \citet{2013A&A...552A...8D} find that the dust mass of ETGs in the Virgo cluster ranges from $10^{4.9}$ to $10^{7}$  M$_{\sun}$. \citet{2013ApJ...766..121M} also find a similar range of ETG dust masses, from $10^{5}$ to $10^{6.5}$ M$_{\sun,}$ based on Spitzer data.

One of the primary objectives of the above mentioned studies is to understand the origin of the dust and thus determine the following: whether it is internal (e.g. via mass loss from evolving stars), external (e.g. via galaxy minor mergers), or a combination of the two (as argued by \citealt{2013ApJ...766..121M}). \citet{2015MNRAS.449.3503D} claim that the dust in their sample of bulge-dominated galaxies with large dust lanes was accreted through minor mergers.

In this paper, we analyse ground-based deep optical images with lower angular resolution than the HST ones but with much higher radial coverage, which puts us in a position to study the origin of the dust in these galaxies. We aim to bring together the deep optical images and the \hiA, and if possible CO data to understand the relation between cold gas and dust across the entire stellar body. In particular, we try to understand whether the large \hiA \ discs play a role in the formation and evolution of the dust structures in the local ETGs. In order to do so, we investigated the distribution of dust in our sample galaxies by using deep colour maps (i.e. $g'-r'$). We classified the dust morphology and compared \hiA-rich to \hiA-poor ETGs. We modelled the dust-free colours in the galaxies and subtracted them from the colour maps to measure the colour excess. We verified whether the amount of colour excess is consistent with the known \hiA \ content. We also estimated the amount of dust in galaxies by using the total optical colour excess and compared our results to those obtained from far-IR observations.

We organise this paper in the following way. In Section \ref{sec:sample}, we describe the target and control sample selection. In Section \ref{sec:DATA_REDUC}, we describe the primary data (\hiA \ and optical) and the methodology used for our work, which includes modelling the dust-free colour, measuring the dust mass, the colour excess, and comparing it with the \hiA \ surface density in our sample galaxies, in particular. In Section \ref{sec:properties}, we present our results about the morphology of the dust structures by classifying them, their radial extent, and the dust mass. In Section \ref{sec:Discus_main}, we discuss the relation between cold gas and dust by comparing the fraction of galaxies (i.e. detection rates) in the various dust morphological classes. We then discuss whether the age of the galaxies or misalignment between gas and stars play a role in the relationship between dust and \hiA. In Section \ref{sec:conclusions}, we present our conclusions.

%\begin{sideways}
\begin{table*}
\caption{General properties of the \hiAT-rich sample.}
\begin{center}
\begin{tabular}{lccccccccccc}
\hline
\hline
No      &Name   & $D$   &$V_{\mathrm{hel}}$     &$M_{K}$        &log$_{10}$($R_{\mathrm{eff}}$) &P.A.   &$\varepsilon$  &$E(B-V)$       &log$_{10}M$(\hi)       &log$_{10}M_{\star, r}$\\
        &       &[Mpc]  &[km/s] &[mag]  &[arcsec]       &[degrees]      & [ ]     &[mag]  &[M$_{\sun}$]   &[M$_{\sun}$]\\
        & (1)   &(2)    &(3)    &(4)    &(5)            &(6)    &(7)&           (8)     &(9)    &(10)\\
\hline
1       &NGC~2594       &35.1   &2362   &-22.36 &0.82   &306    &0.122  &0.050  &8.91   &10.47\\
2       &NGC~2685       &16.7   &875    &-22.78 &1.41   &37     &0.402  &0.051  &9.33   &10.31\\
3       &NGC~2764       &39.6   &2706   &-23.19 &1.09   &173    &0.218  &0.034  &9.28   &10.64\\
4       &NGC~2859       &27.0   &1690   &-24.13 &1.43   &269    &0.272  &0.017  &8.46   &10.97\\
5       &NGC~3414       &24.5   &1470   &-23.98 &1.38   &135    &0.079  &0.021  &8.28   &11.11\\
6       &NGC~3522       &25.5   &1228   &-21.67 &1.01   &241    &0.504  &0.020  &8.47   &10.31\\
7       &NGC~3619       &26.8   &1560   &-23.57 &1.42   &72     &0.020  &0.013  &9.00   &10.91\\
8       &NGC~3626       &19.5   &1486   &-23.30 &1.41   &179    &0.454  &0.017  &8.94   &10.54\\
9       &NGC~3838       &23.5   &1308   &-22.52 &1.04   &144    &0.382  &0.009  &8.38   &10.36\\
10      &NGC~3941       &11.9   &930    &-23.06 &1.40   &189    &0.338  &0.018  &8.73   &10.34\\
11      &NGC~3945       &23.2   &1281   &-24.31 &1.45   &159    &0.435  &0.023  &8.85   &11.02\\
12      &NGC~3998       &13.7   &1048   &-23.33 &1.30   &67     &0.458  &0.013  &8.45   &10.94\\
13      &NGC~4036       &24.6   &1385   &-24.40 &1.46   &272    &0.507  &0.019  &8.41   &11.16\\
14      &NGC~4203       &14.7   &1087   &-23.44 &1.47   &206    &0.153  &0.012  &9.15   &10.60\\
15      &NGC~4278       &15.6   &620    &-23.80 &1.50   &50     &0.230  &0.023  &8.80   &11.08\\
16      &NGC~5173       &38.4   &2424   &-22.88 &1.01   &261    &0.214  &0.024  &9.33   &10.42\\
17      &NGC~5582       &27.7   &1430   &-23.28 &1.44   &30     &0.403  &0.011  &9.65   &10.86\\
18      &NGC~5631       &27.0   &1944   &-23.70 &1.32   &34     &0.198  &0.017  &8.89   &10.89\\
19      &NGC~6798       &37.5   &2360   &-23.52 &1.23   &324    &0.383  &0.114  &9.38   &10.69\\
20      &UGC~06176      &40.1   &2677   &-22.66 &1.03   &210    &0.598  &0.015  &9.02   &10.44\\
21      &UGC~09519      &27.6   &1631   &-21.98 &0.87   &198    &0.402  &0.018  &9.27   &10.06\\
\hline
\hline
\end{tabular}
\label{table:HIrich}
\end{center}
\begin{tablenotes}[para,flushleft]\footnotesize
Note.$-$ Column (1): The name is the principal designation from LEDA \citep{2003A&A...412...45P}. Column (2): distance in Mpc \citet{2011MNRAS.413..813C}. Column (3): heliocentric velocity \citet{2011MNRAS.413..813C}. Column (4): absolute magnitude derived from the apparent magnitude in K~band \citet{2011MNRAS.413..813C}. Column (5): projected half-light effective radius \citet{2011MNRAS.413..813C}. Column (6): position angle of the galaxy calculated from the \hiN image \citet{2014MNRAS.444.3388S}. Column (7): ellipticity of the galaxy calculated from the \hiN image \citet{2014MNRAS.444.3388S}. Column (8): estimates of Galactic dust extinction \citet{2011ApJ...737..103S}. Column (9): total \hiN mass calculated assuming galaxy distances given in column~(2) \citet{2012MNRAS.422.1835S}. Column (10): stellar mass calculated by using a mass-to-light ratio and a total luminosity in the r-band \citep{2013MNRAS.432.1709C}.
\end{tablenotes}
\end{table*}
%\end{sideways}

\section{Sample}
\label{sec:sample}

To study the relation between the dust distribution and the \hiA \ properties of ETGs, we selected an \hiA-rich sample and an \hiA-poor control sample. With a few exceptions, the two samples are the same as those studied in \citet{2017MNRAS.464..329Y} in which we studied ultraviolet (UV) properties in large \hiA \ discs \citep{2012MNRAS.422.1835S}. Compared to the \hiA-rich sample of \citet{2017MNRAS.464..329Y}, we excluded galaxies in the Virgo cluster because we wish to limit the analysis to a field sample. Our final \hiA-rich sample includes 21 ETGs, which we list in Table \ref{table:HIrich}. The \hiA-poor control galaxies are selected with properties as close as possible to the ones of the following \hiA-rich objects: \textbf{i)} stellar mass $M_{\star}$ \citep{2013MNRAS.432.1709C} within +/- 0.8 dex, but typically within +/- 0.5 dex; \textbf{ii)} environment density $\Sigma_{3}$ \citep{2011MNRAS.416.1680C}\footnote{Mean surface density of galaxies inside a cylinder of height $h$ = 600~km~s$^{-1}$ (i.e. $\Delta V_{hel}<$~300 km~s$^{-1}$) centred on the galaxy, which contains the three nearest neighbours.} within +/- 0.8 dex, but typically within +/- 0.5 dex \textbf{iii)} Virgo cluster non-membership \citep{2011MNRAS.413..813C}; \textbf{iv)} kinematical classification \citep[fast or slow rotator;][]{2011MNRAS.414..888E}; and \textbf{v)} distance \citep{2011MNRAS.413..813C} within +/- 15 Mpc. With these requirements, we selected a control sample of 41 \hiA-poor ETGs, which are listed in Table \ref{table:HIcontrol}. The distribution of effective radii $R_\mathrm{eff}$ for galaxies in this control sample is consistent with that for the HI-rich sample. Finally, ten (10/21) of our \hiA-rich galaxies are CO-rich, while only four out of the 41 \hiA-poor galaxies are CO-rich.

\begin{table}
\caption{Control sample galaxies for each \hiAT-rich galaxy.}
\resizebox{\columnwidth}{!}{%
\begin{tabular}{l@{\hspace{1em}}c@{\hspace{1em}}c@{\hspace{1em}}c@{\hspace{1em}}c@{}}
\hline
\hline
\hiA-rich       &Control 1      & Control 2     & Control 3     &Control 4\\
\hline
NGC~2594        &NGC~5611       &-      &-      &-\\                    
NGC~2685        &NGC~2549       &NGC~2852       &NGC~5273       &NGC~3098\\
NGC~2764        &NGC~4078       &NGC~2592       &NGC~0770       &-\\
NGC~2859        &NGC~3230       &NGC~3458       &-      &-\\
NGC~3414        &NGC~5322       &-      &-      &-\\
NGC~3522        &NGC~3796       &PGC~050395     &-      &-\\
NGC~3619        &NGC~5308       &NGC~3658       &NGC~5342       &-\\
NGC~3626        &NGC~3377       &NGC~5500       &NGC~5473       &-\\
NGC~3838        &NGC~2950       &NGC~4283       &-      &-\\
NGC~3941        &NGC~3605       &NGC~3301       &NGC~3248       &-\\
NGC~3945        &NGC~3674       &NGC~3648       &-      &-\\
NGC~3998        &NGC~3610       &-      &-      &-\\
NGC~4036        &NGC~3613       &NGC~6548       &NGC~3665       &UGC~08876\\
NGC~4203        &NGC~3245       &NGC~3757       &-      &-\\
%NGC~4262       &NGC~4503       &NGC~4340       &NGC~4267       &NGC~4377\\
NGC~4278        &NGC~5485       &-      &-      &-\\
%NGC~5103       &NGC~2679       &NGC~0770       &PGC~051753     &-\\
NGC~5173        &NGC~3400       &-      &-      &-\\
NGC~5582        &NGC~2577       &NGC~2679       &-      &-\\
NGC~5631        &NGC~0661       &-      &-      &-\\
NGC~6798        &NGC~6547       &-      &-      &-\\
%UGC~03960      &PGC~050395     &-      &-      &-\\
UGC~06176       &UGC~04551      &-      &-      &-\\
UGC~09519       &NGC~7457       &-      &-      &-\\
\hline
\hline
\end{tabular}%
}
\label{table:HIcontrol}
\end{table}

\section{Data reduction and analysis}
\label{sec:DATA_REDUC}

This work is based on a combination of radio and optical data. We use radio data from the Westerbork Synthesis Radio Telescope and make use of the \hiA \ images and $M$(\hiA) measurements (or upper limits) published by \citet{2012MNRAS.422.1835S} for all of the galaxies in our sample. That work also makes use of the data published previously by \citet{2006MNRAS.371..157M}, \citet{2009A&A...494..489J}, and \citet{2010MNRAS.409..500O} for some galaxies in our sample. In the remaining part of this section, we describe our analysis of the optical data.

\subsection{MegaCam data reduction}
\label{sec:megacam}

We make use of deep optical images obtained in $g'$ and $r'$ bands by using the MegaCam camera on the Canada-France-Hawaii Telescope. The observations were obtained as part of the MATLAS/\atlas \ programme. \citet{2015MNRAS.446..120D} describe all the details of these deep observations and of the data reduction, which makes use of the \verb'Elixir-LSB'\footnote{An observing mode optimised for the detection of extended low surface brightness features \citet{2015MNRAS.446..120D}.} pipeline. Here, we remind the reader that the typical surface brightness sensitivity of these images is 28.5 mag arcsec$^{-2}$ in the $g'$ and $r'$ bands. For this work, we applied a few additional reduction steps, including background subtraction and point-spread-function (PSF) matching in the two bands.

In order to calculate the background of each image, we masked out all pixels above a one sigma level by using \verb'SExtractor' \citep{1996A&AS..117..393B}. We then selected 20 regions, each $50 \times 50$ pixels in size, and measured the median pixel value in each of them. We took the background level as the mean of these 20 median values and calculated its error as their standard deviation. We repeated this process for all $g'$- and $r'$-band images and subtracted the resulting background level.

Since we are interested in analysing the $g'-r'$ colour maps, we needed to first ensure that the images of the two bands have the same resolution. For this purpose, we measured the seeing (full width half maximum: FWHM) in the two bands by using five unsaturated stars located in relatively empty regions of the sky. We used \verb'IRAF - imexamine' to measure the FWHM. We then corrected for the difference in seeing by smoothing the data of the band with the highest resolution to the lowest resolution band. After removing the background and matching the PSF in the two bands, we computed the calibrated colour maps. \footnote{We did not apply any Galactic foreground correction, which  would only change the scaling on the colour.}

\subsection{Modelling dust-free colours and measuring the colour excess}
\label{sec:modelling}

In this work, we use the MegaCam $g'-r'$ deep colour maps to study the spatial distribution of the dust in \hiA-rich and \hiA-poor ETGs. The signal-to-noise ratio of the colour maps decreases with increasing radius. Therefore, we derived the radial profiles along annuli with a fixed shape (ellipticity and position angle) but which logarithmically increasing in width from inside out. The ellipticity and position angle of the annuli are fixed to the values published by \citet{2011MNRAS.414.2923K}.

Before deriving the dust-free colour profiles, we masked the dusty regions in the images as much as possible. Here, we used a procedure that contains a combination of automatic and manual masking. We explain this procedure in detail in Appendix~\ref{App:AppendixB}. In summary, we \it i) \rm took the median value of the clipped annulus as a first guess for the dust-free colour; \it ii) \rm we repeated this for all the annuli and created a model; \it iii) \rm we applied a median filter to remove sharp edges between annuli; \it iv) \rm we required that the dust-free colour remains constant or becomes bluer when going outwards; and \it v) \rm we finally forced the model to be a smooth function of radius since this kind of behaviour for ETGs is expected. 

The best model representing the dust-free colour was subtracted from the original $g'-r'$ colour map to obtain a colour excess map. We limited our analysis to a maximum radius of 3~R$_{\mathrm{eff}}$. The analysis of the colour profiles at a larger radius, which is beyond the scope of this paper, would require a more careful treatment of the effects of the PSF wings. %In Fig. \ref{fig:model_profile}, we show an example of a dust-free colour model overlaid on the observed colour profile. 

There are some uncertainties as to how the colour excess is calculated, which is based on the assumption that a dust-free colour can be obtained from the image and that the stellar populations do not change quickly with the radius. Also, we assume that no young stellar populations are present in the centre. This is not the case for some objects, such as NGC~2685. For this object, the measured colour excess is a lower limit. However, this problem is alleviated by the fact that young stars most likely occupy the same regions as the dust. 

Having obtained the colour excess maps, we can calculate an average colour excess in a given annulus or a region. For this, we used the following equation:

\begin{equation}
\begin{aligned}
\label{eq:eqW}
\langle E(g'-r')\rangle~=~-2.5\times log_{10}(\frac {\Sigma(I_{r} \times R^{2} \times q \times 2.512^{-E(g'-r')})}{\Sigma (I_{r} \times R^{2} \times q)})
\end{aligned}
,\end{equation}

\begin{figure*}
\begin{center}
\includegraphics[scale=1.0]{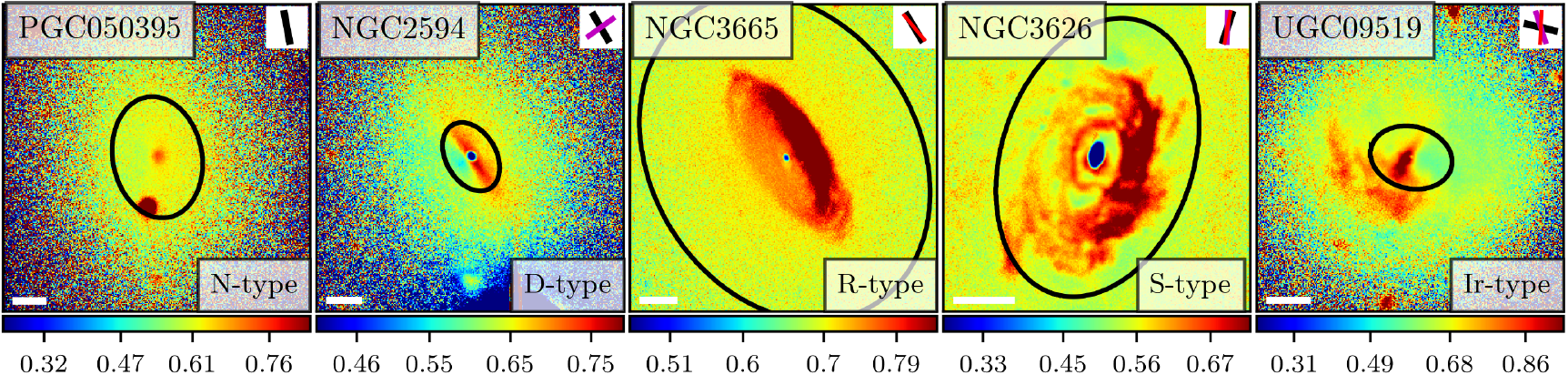}
\end{center}
\caption{Example $g'-r'$ colour maps for our dust classification. The tick labels along the colour-bar indicate median$-4\sigma$, median$-2\sigma$, median, and median$+2\sigma$, respectively. The colour bars are limited between median$-5\sigma$ and median$+3\sigma$. The median and standard deviation values of  the $g'-r'$ colour are calculated in the region of 0-1~R$_{\mathrm{eff}}$, which is shown with the black ellipse. The white scale bar at the bottom corner indicates 1 kpc. On the top-right, we show an illustration of the position angles of the various galaxy components: in black the optical-disc from \citet{2011MNRAS.414.2923K}; in purple the \hiAN-disc from \citet{2014MNRAS.444.3388S}; and in red the CO-disc from \citet{2013MNRAS.432.1796A}. Similar $g'-r'$ colour maps of all of the galaxies studied here are shown in Appendix \ref{App:AppendixC}.}
\label{fig:five_gals}
\end{figure*}

\noindent where $I_{r}$ is the relative surface brightness (in intensity units) in the $r'$-band, $R$ is the radius of the annulus, and $q$ is the flattening value\footnote{A way to describe the ellipticity of a circle or sphere: $\epsilon$ = 1 - b/a, where b/a is the ratio of minor axis to major axis and equals to the flattening value, $q$.} of the annulus which is assumed to be constant throughout the galaxy. In Sec. \ref{App:AppendixA}, we explain how we used these maps to estimate the amount of \hiA \ from the extinction caused by the dust.

\subsection{Dust mass calculation and other parameters}
\label{sec:dust_masses}

There are several ways to estimate the dust mass of a galaxy. First, we estimated the dust mass by using the total extinction obtained from the colour excess maps (Sec. \ref{sec:modelling}). For example, for this we followed the method used in the literature by \citet{1994MNRAS.271..833G},\citet{2007A&A...461..103P}, and \citet{2012MNRAS.422.1384F} (see Sec. \ref{App:AppendixA} for the equations and details for the optical method). As a second estimate, we adopted the dust mass values of \citet{2019A&A...622A..87K}, which are based on the spectral energy distribution (SED) fits of \citet{2017A&A...605A..74K} for \atlas \ galaxies. Finally, the IR dust estimate based on the  graphical method of \citet{2007ApJ...657..810D} could not be applied because the available mid- and far-IR data are of insufficient sensitivity.\footnote{Most of the galaxies in our sample have IR flux values below the detection limit or very large errors \citep[see,][]{2017A&A...605A..74K}.}

\begin{figure*}
\begin{center}
\includegraphics[scale=1.0]{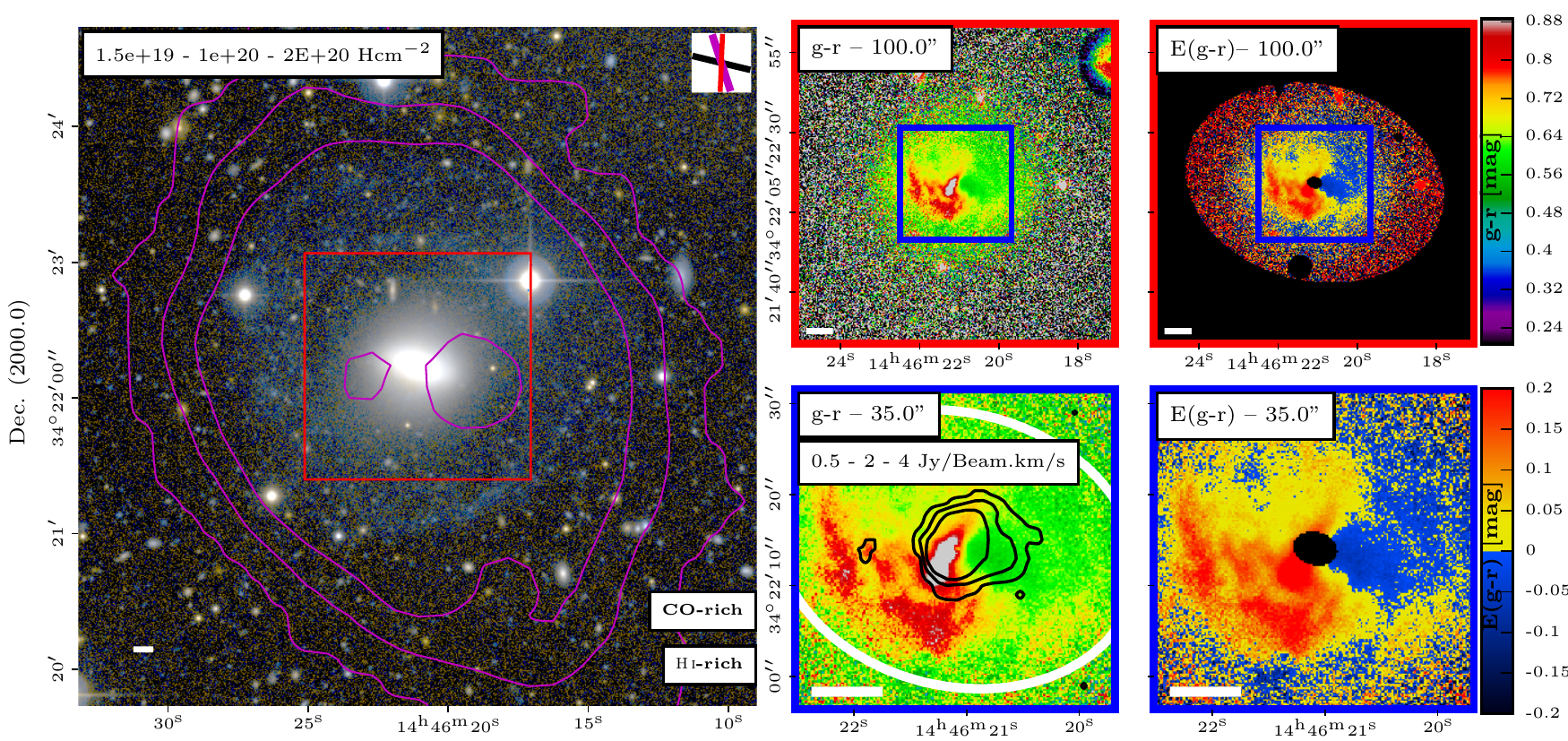}
\caption{\textit{Left}: True colour image of UGC~09519 with overlaid \hiAT \ column density contours (magenta; see top-left box for the contour levels). The size of the true colour image is 300$\times$300 arcsec$^{2}$. \textit{Top-middle}: $g'-r'$ image covering 100$\times$100 arcsec$^{2}$ area indicated in the left panel with a red box. \textit{Top-right}: $E(g'-r')$ colour excess map (100$\times$100 arcsec$^{2}$). \textit{Bottom-middle}: $g'-r'$ image covering the 35$\times$35 arcsec$^{2}$ area indicated in the top-middle panel with a blue box. Since CO is generally detected in the inner regions, we show the CO intensity contours with black lines in this panel \citep[see][]{2013MNRAS.432.1796A}. The white ellipse indicates the dust radius determined through visual inspection (see Table \ref{table:dust_morp2}). \textit{Bottom-right}: $E(g'-r')$ colour excess map (35$\times$35 arcsec$^{2}$). The contour levels are indicated in the relevant panels. The white bar in the panels represents 1 kpc. On the top-right corner of the left panel, we also present the position angles as in Fig. \ref{fig:five_gals}. See Appendix \ref{App:AppendixC} for the rest of the sample.}
\label{fig:example3}
\end{center}
\end{figure*}

We make use of the data of \citet{2011MNRAS.414..940Y} and \citet{2013MNRAS.432.1796A} who provide the total H$_{2}$ mass values and the distribution of CO emission (from the zero-moment maps), respectively. \citet{2011MNRAS.414..940Y} used the Institut de Radioastronomie Millimetrique (IRAM) 30-metre single-dish radio telescope to look at the molecular gas content for all the \atlas \ galaxies. \citet{2013MNRAS.432.1796A} observed some of these CO-detected ETGs with the \textit{Combined Array for Research in Millimeter Astronomy} (CARMA) interferometer. We retrieved the CO maps of \citet{2013MNRAS.432.1796A} to compare the dust and the CO distribution. We cannot compare \hiA \ and the other component of interstellar matter (ISM) since the resolution of the WSRT beam is much lower than those of the CARMA and the CHFT.

\section{Results}
\label{sec:properties}

\subsection{Dust morphological classes}
\label{sec:dust_morp}

The first step of our analysis is a visual inspection of the MegaCam $g'-r'$ colour maps to establish which \hiA-rich and \hiA-poor galaxies harbour some dust. Almost all galaxies in our sample show redder colours in the inner regions. However, we label a galaxy as `dusty': (i) if the reddened region is distinctly different from the underlying stellar body of the galaxy (either with a different morphology or a different position angle); and (ii) in case of a similar morphology and position angle, if the red colour is in excess of the model colour (see below). Otherwise, we label galaxies as no-dust objects (N). Our method for selecting dusty galaxies is somewhat conservative and, therefore, we may have missed some of them.

We classify the dust-detected ETGs as follows: dusty disc (D), irregular dust structure (Ir), dusty ring (R), and spiral shape (S). We present an example galaxy for each of these morphological types in Fig. \ref{fig:five_gals}. We show all the galaxies in Appendix \ref{App:AppendixD}. Below, we elaborate on the characteristics of these classes.

When the position angle of these reddened regions is aligned with the stellar disc, it is hard to separate the dust from the old stellar populations. Either a large amount of dust or old and red stellar populations might cause the redder colours in these systems. Thus, for these cases, we calculated a theoretical $g'-r'$ colour with the following equation given by \citet{2005AJ....129...61B}:
\begin{equation}
g'-r' = \left\{ \begin{array}{l}0.7 ~ + ~ 0.25(T-9.5) ~ + ~ 0.375~Z \qquad \text{ if } T>9.5 \\
0.7 ~ + ~ 0.50(T-9.5) ~ + ~ 0.375~Z \qquad  \text{ if } T<9.5 \end{array}\right.
,\end{equation}

\noindent where T is the age in units of Gyr and $Z$ is the metallicity. This equation estimates a $g'-r'$ colour for stellar populations older than 1 Gyr based on the single-burst models of \citet{2003MNRAS.344.1000B}. We used the stellar ages and metallicity values from \citet{2015MNRAS.448.3484M}\footnote{Here, we used the single-burst-stellar-population equivalent (SSP-equivalent) ages. One should be careful since these SSP-equivalent ages are strongly biased towards the young ages \citep{2007MNRAS.374..769S}.} and we classify a galaxy as a D object, that is, a galaxy with a dusty disc, if the observed g'-r' colour\footnote{Here, we removed the Galactic extinction by using $E(g'-r') = E(B-V) \times 0.9796$.} is redder than the theoretical one given the high surface brightness in the inner regions. We can trust a small difference between the observed and theoretical colours (e.g. 0.01 mag). It is important to note that we only used this test if a galaxy has a red disc-type region that is well aligned with the stellar component (six galaxies in our sample).

Many of the galaxies in our sample show a ring-like structure in the very central regions ($\sim < 0.5$ kpc). However, since a number of galaxies are saturated in the centre\footnote{Short exposure observations taken with the MegaCam camera are available, and we will make use of them in the future projects.}, we did not take these very inner regions into account. Therefore, only if a galaxy exhibits a relatively sizeable ($> 0.5$ kpc) dusty ring do we label it as an R object (see Fig. \ref{fig:five_gals}).

Several galaxies in our sample exhibit dust that is distributed in a spiral pattern. In these cases, we label them as S galaxies. These spiral-shaped arms could be confined to a relatively inner region, but in several cases, they reach a large radius (see Fig. \ref{fig:five_gals}). Some of the galaxies show very complex dust structures, and we label them as irregular (Ir) dusty galaxies. 

We list our classification together with those of \citet{2011MNRAS.414.2923K} for all the sample galaxies in Table \ref{table:dust_morp2}. We quantify the colour excess and the amount of dust in the next sections. We detected dust in a total of 44 galaxies, which is 71 per cent of our sample of 62 ETGs. Out of these, 12, 14, 6, and 12 are S-, IR-, R-, and D-type, respectively.

\begin{table*}
\caption{Dust morphology, extension, and mass for the \hiAT-rich and \hiAT-poor galaxies.}
\begin{center}
\begin{tabular}{l@{\hspace{0.8em}}c@{\hspace{0.8em}}c@{\hspace{0.8em}}c@{\hspace{0.8em}}c@{\hspace{0.8em}}|c@{\hspace{0.8em}}c@{\hspace{0.8em}}c@{\hspace{0.8em}}|c@{\hspace{0.8em}}c@{\hspace{0.8em}}c@{\hspace{0.8em}}}
\hline
\hline
Gal     &Dist.  &log$_{10}R_{\mathrm{eff}}$     &q      &H\text{\scriptsize{I}}-Type$^{c}$      &\multicolumn{3}{c|}{Dust morphology}     &\multicolumn{3}{c}{Dust Radius}\\
        &[Mpc]  &[arcsec]       &-      &-      &SL$_{07}$-T$_{01}$     &K$_{11}$       &This work    &[arcsec]       &[$R_{\mathrm{eff}}$]   &[Kpc]\\
(1)     &(2)    &(3)    &(4)    &(5)    &(6)    &(7)    &(8)    &(9)    &(10)   &(11)\\
\hline
NGC~0661        &30.6   &1.12   &0.69   &Poor   &-      &N      &N      &0      &0.0    &0.0\\
NGC~0770        &36.7   &0.94   &0.71   &Poor   &-      &N      &D      &5      &0.6    &0.9\\
NGC~2549        &12.3   &1.28   &0.31   &Poor   &-N     &N      &N      &0      &0.0    &0.0\\
NGC~2577        &30.8   &1.15   &0.59   &Poor   &-      &N      &Ir     &14     &1.0    &2.1\\
NGC~2592        &25     &1.09   &0.79   &Poor   &-D     &N      &N      &0      &0.0    &0.0\\
NGC~2594        &35.1   &0.82   &0.68   &Rich   &-      &N      &D      &10     &1.5    &1.7\\
NGC~2679        &31.1   &1.35   &0.93   &Poor   &-      &N      &N      &0      &0.0    &0.0\\
NGC~2685        &16.7   &1.41   &0.6    &Rich$^{c}$     &-      &F      &S      &100    &3.9    &8.1\\
NGC~2764        &39.6   &1.09   &0.51   &Rich$^{c}$     &-      &FB     &S      &35     &2.8    &6.7\\
NGC~2852        &28.5   &0.85   &0.86   &Poor   &-      &N      &R      &12     &1.7    &1.7\\
NGC~2859        &27     &1.43   &0.85   &Rich   &-      &N      &S      &22     &0.8    &2.9\\
NGC~2950        &14.5   &1.19   &0.59   &Poor   &N-N    &N      &N      &0      &0.0    &0.0\\
NGC~3098        &23     &1.12   &0.23   &Poor   &-      &N      &D$^{+}$        &50     &3.8    &5.6\\
NGC~3230        &40.8   &1.26   &0.39   &Poor   &-      &N      &N      &0      &0.0    &0.0\\
NGC~3245        &20.3   &1.40   &0.54   &Poor$^{c}$     &-      &N      &Ir     &15     &0.6    &1.5\\
NGC~3248        &24.6   &1.20   &0.6    &Poor   &-      &N      &D      &5      &0.3    &0.6\\
NGC~3301        &22.8   &1.30   &0.31   &Poor   &-      &N      &Ir     &11     &0.6    &1.2\\
NGC~3377        &10.9   &1.55   &0.67   &Poor   &l,0.26-F1      &N      &Ir     &19     &0.5    &1.0\\
NGC~3400        &24.7   &1.23   &0.56   &Poor   &-      &N      &N      &0      &0.0    &0.0\\
NGC~3414        &24.5   &1.38   &0.78   &Rich   &S,0.15-F1      &N      &S      &30     &1.3    &3.6\\
NGC~3458        &30.9   &1.06   &0.71   &Poor   &-      &N      &N      &0      &0.0    &0.0\\
NGC~3522        &25.5   &1.01   &0.52   &Rich   &-      &N      &Ir     &8      &0.8    &1.0\\
NGC~3605        &20.1   &1.23   &0.6    &Poor   &-      &N      &R      &7      &0.4    &0.7\\
NGC~3610        &20.8   &1.20   &0.81   &Poor   &N      &N      &N      &10     &0.6    &1.0\\
NGC~3613        &28.3   &1.42   &0.54   &Poor   &-N     &N      &N      &0      &0.0    &0.0\\
NGC~3619        &26.8   &1.42   &0.91   &Rich$^{c}$     &-      &FBR    &S      &40     &1.5    &5.2\\
NGC~3626        &19.5   &1.41   &0.67   &Rich$^{c}$     &-      &D      &S      &30     &1.2    &2.8\\
NGC~3648        &31.9   &1.12   &0.56   &Poor   &-      &N      &N      &0      &0.0    &0.0\\
NGC~3658        &32.7   &1.28   &0.84   &Poor   &-      &N      &N      &0      &0.0    &0.0\\
NGC~3665        &33.1   &1.49   &0.78   &Poor$^{c}$     &-      &D      &R      &18     &0.6    &2.9\\
NGC~3674        &33.4   &1.05   &0.36   &Poor   &-      &N      &N      &0      &0.0    &0.0\\
NGC~3757        &22.6   &0.95   &0.85   &Poor   &-      &N      &N      &0      &0.0    &0.0\\
NGC~3796        &22.7   &1.06   &0.6    &Poor   &-      &N      &Ir     &7      &0.6    &0.8\\
NGC~3838        &23.5   &1.04   &0.44   &Rich   &-      &N      &Ir     &9      &0.8    &1.0\\
NGC~3941        &11.9   &1.40   &0.75   &Rich   &-      &N      &Ir     &26     &1.0    &1.5\\
NGC~3945        &23.2   &1.45   &0.65   &Rich   &i,x,   &FBR    &S      &60     &2.1    &6.8\\
NGC~3998        &13.7   &1.30   &0.78   &Rich   &il,0.43        &N      &Ir     &35     &1.8    &2.3\\
NGC~4036        &24.6   &1.46   &0.4    &Rich$^{c}$     &il,x   &F      &S      &30     &1.0    &3.6\\
NGC~4078        &38.1   &0.92   &0.44   &Poor   &-      &N      &D      &8      &1.0    &1.5\\
NGC~4203        &14.7   &1.47   &0.89   &Rich$^{c}$     &sl,0.23        &N      &S      &50     &1.7    &3.6\\
NGC~4278        &15.6   &1.50   &0.91   &Rich   &i,x    &N      &Ir     &30     &0.9    &2.3\\
NGC~4283        &15.3   &1.09   &0.96   &Poor$^{c}$     &-      &N      &D      &8      &0.7    &0.6\\
NGC~5173        &38.4   &1.01   &0.87   &Rich$^{c}$     &-F3    &B      &Ir     &23     &2.2    &4.3\\
NGC~5273        &16.1   &1.57   &0.84   &Poor$^{c}$     &sl,0.14        &N      &Ir     &18     &0.5    &1.4\\
NGC~5308        &31.5   &1.25   &0.2    &Poor   &N-N    &N      &D$^{+}$        &50     &2.8    &7.6\\
NGC~5322        &30.3   &1.60   &0.64   &Poor   &-      &N      &N      &0      &0.0    &0.0\\
NGC~5342        &35.5   &0.97   &0.46   &Poor   &-      &N      &D      &18     &1.9    &3.1\\
NGC~5473        &33.2   &1.32   &0.79   &Poor   &-      &N      &N      &0      &0.0    &0.0\\
NGC~5485        &25.2   &1.45   &0.74   &Poor   &-      &D      &R      &25     &0.9    &3.1\\
NGC~5500        &31.7   &1.18   &0.8    &Poor   &-      &N      &N      &0      &0.0    &0.0\\
NGC~5582        &27.7   &1.44   &0.65   &Rich   &-      &N      &D      &8      &0.3    &1.1\\
NGC~5611        &24.5   &1.00   &0.45   &Poor   &-      &N      &D      &6      &0.6    &0.7\\
NGC~5631        &27     &1.32   &0.93   &Rich   &-      &D      &S      &35     &1.7    &4.6\\
NGC~6547        &40.8   &1.06   &0.33   &Poor   &-      &N      &D$^{+}$        &25     &2.2    &4.9\\
NGC~6548        &22.4   &1.35   &0.89   &Poor   &-      &N      &R      &12     &0.5    &1.3\\
NGC~6798        &37.5   &1.23   &0.53   &Rich$^{c}$     &-      &N      &S      &30     &1.8    &5.5\\
NGC~7457        &12.9   &1.56   &0.53   &Poor   &N      &N      &D      &15     &0.4    &0.9\\

\hline
\hline
\end{tabular}
\label{table:dust_morp2}
\end{center}
\end{table*}
\begin{table*}
\addtocounter{table}{-1}
\caption{\textit{Continued}}
\begin{center}
\begin{tabular}{l@{\hspace{0.8em}}c@{\hspace{0.8em}}c@{\hspace{0.8em}}c@{\hspace{0.8em}}c@{\hspace{0.8em}}|c@{\hspace{0.8em}}c@{\hspace{0.8em}}c@{\hspace{0.8em}}|c@{\hspace{0.8em}}c@{\hspace{0.8em}}c@{\hspace{0.8em}}}
\hline
\hline
Gal     &Dist.  &log$_{10}R_{\mathrm{eff}}$     &q      &H\text{\scriptsize{I}}-Type$^{c}$      &\multicolumn{3}{c|}{Dust morphology}     &\multicolumn{3}{c}{Dust Radius}\\
        &[Mpc]  &[arcsec]       &-      &-      &SL$_{07}$-T$_{01}$     &K$_{11}$       &This work    &[arcsec]       &[$R_{\mathrm{eff}}$]   &[Kpc]\\
(1)     &(2)    &(3)    &(4)    &(5)    &(6)    &(7)    &(8)    &(9)    &(10)   &(11)\\
\hline
PGC~050395      &37.2   &1.04   &0.73   &Poor   &-      &N      &N      &0      &0.0    &0.0\\
UGC~04551       &28     &1.03   &0.39   &Poor   &-N     &N      &Ir     &10     &0.9    &1.4\\
UGC~06176       &40.1   &1.03   &0.51   &Rich$^{c}$     &-      &F      &S      &31     &2.9    &6.0\\
UGC~08876       &33.9   &0.93   &0.37   &Poor   &-      &N      &R      &12     &1.4    &2.0\\
UGC~09519       &27.6   &0.87   &0.75   &Rich$^{c}$     &-      &F      &Ir     &20     &2.7    &2.7\\
\hline
\hline
\end{tabular}
\end{center}
\begin{tablenotes}[para,flushleft]\footnotesize
Note.$-$ Column (1): The name is the principal designation from LEDA \citep{2003A&A...412...45P}. Column (2): distance in Mpc \citet{2011MNRAS.413..813C}. Column (3): projected half-light effective radius \citet{2011MNRAS.413..813C}. Column (4): the ratio of the minor axis to the major axis obtained from the ellipticity value \citet{2011MNRAS.414.2923K}. Column (5): \hiN content of the galaxy \citep{2012MNRAS.422.1835S}. The $^{c}$ sign indicates CO detection \citep{2011MNRAS.414..940Y}. Column ($6-7$): Labels used to define the dust morphology in the previous studies: \citet{2007ApJ...655..718S} and \citet{2001AJ....121.2928T} (6); \citet{2011MNRAS.414.2923K} (7). If the size of the dust structure is given in the previous studies, we give this value next to the label. Column (8): labels used to define the dust morphologies in this study. Column ($9-11$): Dust radius measured in this study in the units of arcsec, effective radius, and Kpc, respectively. Column ($12$): Estimated dust masses from the total optical extinction ($A(g')$). Column ($13$): Dust masses derived from SED fits by using far-IR data \citep{2019A&A...622A..87K}.\\
Comments.$-$ Labels of the dust morphological classes, D: dusty disc; Ir: irregular dust; R: dusty ring; S: spiral-shaped structure; N: no-dust structure; D$^{+}$: almost edge-on dusty disc. Labels used in the previous studies are different than our classes: F1: filamentary low; F3: filamentary high; l: dust lane or filament.
\end{tablenotes}
\end{table*}

\subsection{Extended dust structures and relation with the cold gas properties}
\label{sec:large_dust}

A significant result of our study is the distribution of the dust features in ETGs. For the first time, our results show the detailed distribution of the dust in ETGs up to a large radius. In Table \ref{table:dust_morp2}, we list the dust radii estimated visually based on our MegaCam images and, for comparison, those available in the literature. As an example, Fig. \ref{fig:example3} shows the $g'-r'$ and $E(g'-r')$ images of UGC~9519 together with an ellipse indicating the dust radius\footnote{This is defined as the distance between the outermost region where we detected extinction and the centre of the galaxy.}. The white ellipse in Fig. \ref{fig:example3} has a semi-major axis that is equal to the dust radius, as well as the position angle and ellipticity from \citet{2011MNRAS.414.2923K}.

The average dust radius across the sample is $\sim$ 2.9 $\pm$ 0.3 kpc (1.3 $\pm$ 0.1 R$_{\mathrm{eff}}$). 
There are three nearly edge-on galaxies in our sample (inclination $>$ 80 deg, labelled as D$^{+}$ in Table \ref{table:dust_morp2}), and since the sensitivity to dust absorption increases with increasing inclination, it is easier to detect dust in these galaxies. If we do not take into account these galaxies, the average dust radius becomes $\sim$ 2.6 $\pm$ 0.3 kpc, which is not very different.

Here we discuss the dust radius as a function of cold gas content. As can be seen in Table \ref{table:dust_morp2}, in all the S-type galaxies, which are also \hiA-rich, the dust extends further than 3.5~kpc. We find the average dust radius for the \hiA-rich and \hiA-poor ETGs to be $\sim$~3.7 $\pm$ 0.4 kpc and $\sim$~2.1 $\pm$ 0.3 kpc, respectively. However, if we do not take into account the nearly edge-on systems, the average value for \hiA-poor ETGs becomes $\sim$~1.4 $\pm$ 0.2 kpc, while the average value does not change for the \hiA-rich ETGs. Therefore, we can conclude that \hiA-rich ETGs contain about a two times larger dust structure than \hiA-poor ETGs. If we define gas-rich galaxies as those detected in either \hiN or CO, then we find that the dust structures extend to $\sim$~3.3 $\pm$ 0.4 and $\sim$~1.6 $\pm$ 0.3 kpc in gas-rich and gas-poor galaxies, respectively. The average dust radius for the CO-rich (\hiA-poor) galaxies is the same as the gas-poor galaxies. The S-type dust structures (present in only gas-rich ETGs) can reach out to $\sim$~5 $\pm$ 0.5 kpc from the galaxy centre. Irregular dust structures (S+Ir) in gas-rich systems extend up to $\sim$~4 $\pm$ 0.3 kpc. Therefore, we conclude that if there is cold gas in a galaxy (with \hiA), the dust structures extend to two times larger radii (on average) than in a gas-poor galaxy.

\begin{figure}
\includegraphics[scale=1.0]{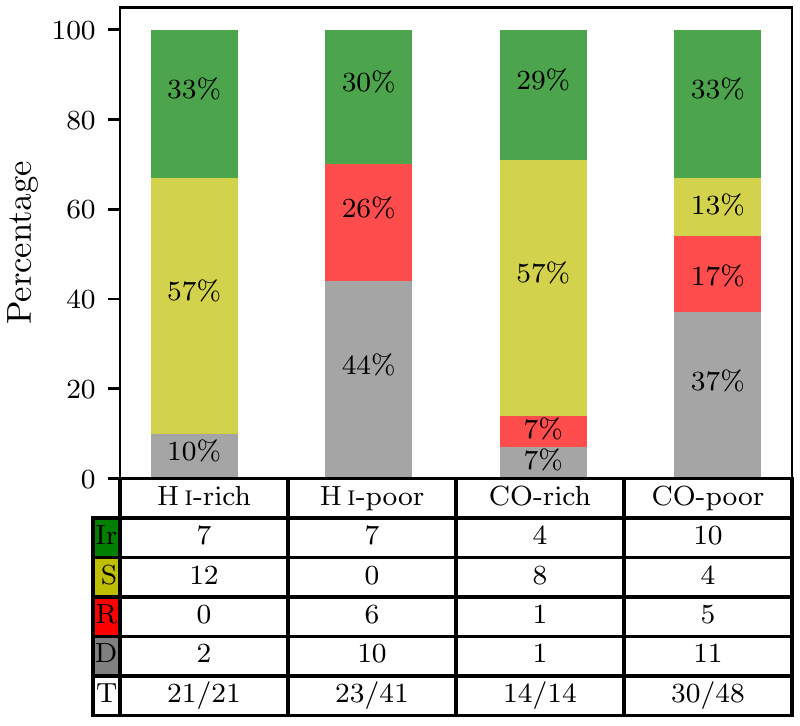}
\caption{Number of galaxies containing dust according to their dust-type and cold gas content. The colour code is based on dust morphology defined in this paper. The percentage of the groups are written on the related bars, and the quantities are given in the lower table. The bottom row presents the total number of dusty galaxies for each gas content class.}
\label{fig:number_dust}
\end{figure}

The principal goal of this work is to understand the relation between dust and cold gas, mainly \hiA, in ETGs. In this respect, we find an unambiguous result: all \hiA-rich ETGs in our sample host dust (mostly undetected in \cite{2011MNRAS.414.2923K}), while the dust detection rate of \hiA-poor galaxies is just 56 per cent. The presence of \hiA ,\ anywhere in the galaxy, dramatically increases the dust detection rate of ETGs in the inner 3 R$_{\mathrm{eff}}$. It is important to note that we can identify dust structures extending out to more than 3 R$_{\mathrm{eff}}$, but we do not study these dust structures in this paper. We show the number of galaxies in each class based on their cold gas content in Fig. \ref{fig:number_dust}. As can be seen from the figure, all the S-type ETGs contain large \hiA-discs.

We conclude that the presence of \hiA \ out to large radii always implies the presence of dust in the stellar body, whereas the opposite is not true. This is also the case for the CO-rich ETGs based on the results of \citet{2011MNRAS.414..940Y}, but it is important to note that these are not all the same objects as our \hiA-rich ones. In general, more nearby ETGs have been detected in \hiA \ than in CO (especially outside the Virgo cluster). ETGs that host a small \hiA \ disc, which we do not study in this work, also fit in this picture: nearly all of them host CO and dust \citep{2012MNRAS.422.1835S}.

\begin{figure}
\begin{center}
\includegraphics[scale=1.0]{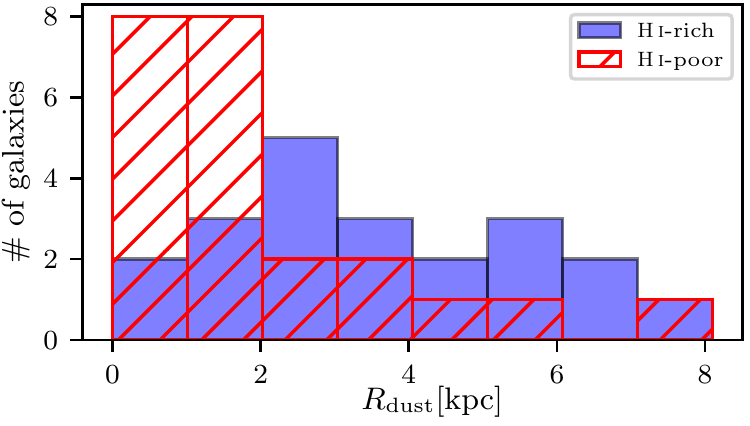}
\caption{Histogram showing the number of galaxies as a function of dust radius. Small (i.e. $R_{\mathrm{dust}} \leq$ 1~kpc) and large (i.e. $R_{\mathrm{dust}}>$ 1~kpc) dust structures tend to be in \hiAT-rich (blue bars) and \hiAT-poor (red hatched bars) ETGs, respectively.}
\label{fig:histogram2}
\end{center}
\end{figure}

Another notable result of this study is that we only detected spiral-shaped dust distributions in \hiA-rich galaxies (8/12 are also CO-rich; see Fig. \ref{fig:number_dust}). Moreover, 90 per cent of the dusty \hiA-rich galaxies (19 out of 21) shows complex dust structures that are either irregular or spiral in shape. In other words, the difference between gas-rich and gas-poor galaxies is even more dramatic if we consider the spatial distribution of the dust. We present a histogram in Fig. \ref{fig:histogram2} showing the number of galaxies with and without \hiA \ for the large and small dust structures. As can be seen in the histogram, large dust structures tend to be in \hiA-rich galaxies while the small size dust structures reside in the \hiA-poor ETGs.

\subsection{Colour excess and \hiA \ radial profiles}
\label{sec:rad_pro}

We propagated the errors on the colour profiles and calculated the average colour excess in each annulus by using Eq. \ref{eq:eqW}. We also estimated the \hiA \ surface density of each annulus based on the colour excess as explained in Appendix \ref{App:AppendixA}, and compared it with observed \hiA \ and CO profiles. This allowed us to test whether the radial distribution of the colour excess and that of the cold gas are related. We provide all the radial profiles of the sample galaxies in Appendix \ref{App:AppendixD}.

\begin{figure}
\begin{center}
\includegraphics[scale=1.0]{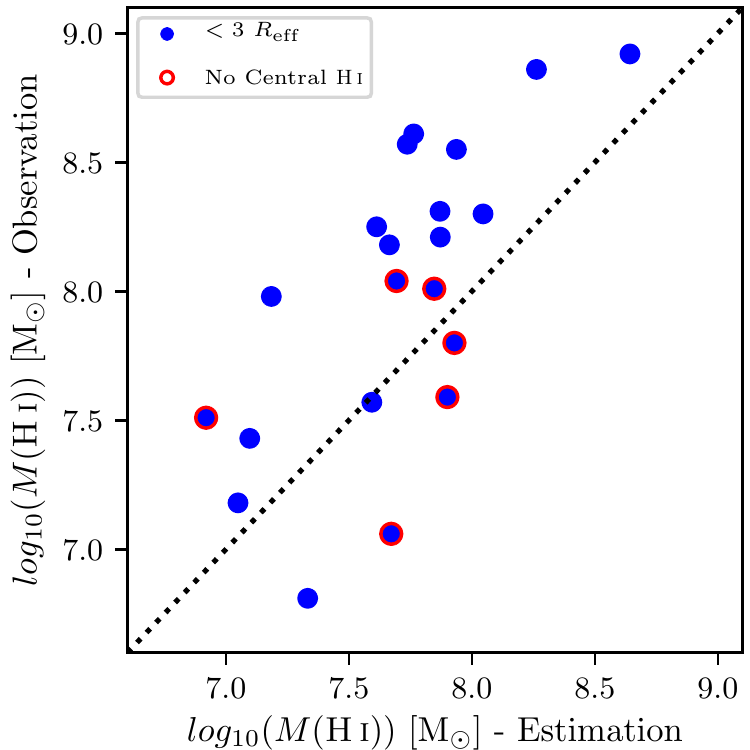}
\caption{Observed \hiN mass vs. the \hiN mass estimated based on the colour excess (see text). The observed \hiN mass was calculated based on the \hiN images in  \citet{2012MNRAS.422.1835S} within the same region where we detected a colour excess. The dotted line shows the one-to-one relation. Red outer circles indicate galaxies with no central \hiAN \ ($<$ 1~R$_{\mathrm{eff}}$).}
\label{fig:derivedH}
\end{center}
\end{figure}

In addition to the estimation of the \hiA \ surface density in each annulus, we also estimated the total \hiA \ mass in the region between 0-3 R$_{\mathrm{eff}}$ (see Appendix \ref{App:AppendixA}). We plotted the estimated and observed values in this aperture in Fig. \ref{fig:derivedH}.
As can be seen from the figure, the galaxies follow the one-to-one relation with a scatter. We note that six \hiA-rich galaxies (NGC~2859, NGC~3941, NGC~3945, NGC~4036, NGC~4278, and NGC~5582) in our sample have enormous \hiA \ ring or disc structures, while they have no \hiA \ in their very central regions (red circles in Fig. \ref{fig:derivedH}). While these galaxies show very red colours in the inner regions due to dust, they show blue colours in the outer regions where the highest \hiA \ column density rings are located \citep{2017MNRAS.464..329Y}. We would like to point out that these galaxies are also CO-poor with the exception of NGC~4278. Therefore, we conclude that the presence of a large \hiA \ disc or ring in an ETG indicates a dust structure in the centre, even if there is no \hiA \ in central regions. It would be interesting to continue to study these objects in future studies. Considering that our \hiA \ mass estimate in Fig. \ref{fig:derivedH} is based on the colour excess-\hiA \ relation of the Milky Way (Appendix~\ref{App:AppendixB}), Fig. \ref{fig:derivedH} suggests that \hiA-rich ETGs follow a different relation than our Galaxy.

\section{Discussion}
\label{sec:Discus_main}

\subsection{Dust detection rate and extension}
\label{sec:Discuss_detection}
The wide-field, deep optical images analysed here allow for a more complete characterisation of the distribution of dust in nearby, non-cluster ETGs than in previous works. As a result, we find that more than half of these galaxies (both gas-rich and -poor) contain dust structures (71 $\pm$ 6 per cent). 

Our dust detection rate is much higher than that reported by \citet{2011MNRAS.414.2923K} for the full \atlas \ sample based on SDSS images, which is 19 per cent (this rate also holds for the 62 galaxies that are in common with our study alone). Therefore, the dust detection rate increases by a factor of approximately four when going from an SDSS-like depth and resolution to that of our MegaCam images. Images obtained with the HST, therefore, are more sensitive than the SDSS ones in the centre, and they give an ETG dust detection rate of $\sim$~50 per cent, which is closer to those reported here. For example, \citet{2005AJ....129.2138L} detected dust in 47 per cent of all galaxies in their sample of 77 ETGs.

In previous work based on the HST data, our D objects are typically not labelled as dusty, but this is primarily due to definition criteria (see Sec. \ref{sec:dust_morp}). For example, \citet{2005AJ....129.2138L} had stricter requirements for the position-angle misalignment. They did not classify a galaxy as dusty if the dust and stars are aligned (same position-angle). If we ignore the 12 D objects in our sample, the dust detection rate becomes entirely consistent with that of previous HST studies (32/62 = 52 $\pm$ 6 per cent). Concerning the remaining dusty galaxies, 10/12 S, 6/14 Ir, and 2/6 R were already known to host dust, although the dust distribution could not be studied to the same large radius as in this work.

Using HST images of 68 galaxies, \citet{2007ApJ...655..718S} found that 63 per cent of their sample contains dust structures and all the observed dust is in the central kiloparsec region. However, we find that most of the ETGs ($\sim$ 77 per cent) have extended dust structures more than 1 kiloparsec (see Fig. \ref{fig:histogram2}). Therefore, our study complements the previous investigations carried out using the HST data. For example, while the circumnuclear dust structure in NGC~4203 as seen by the HST extends up to just 0.23 kpc \citep{2007ApJ...655..718S}, our study shows that the spiral dust structures extend up to 2 kpc \citep[see also][]{2015MNRAS.451..103Y}. One should consider that while we use deep optical $g'-r'$ colours to define the dust structures in ETGs, these authors use unsharp-masked, single-band HST images. 

Finally, \citet{1997ApJ...481..710C} showed that NGC~5322 has a clear dust structure; however, we do not label this galaxy as dusty. \citet{2005AJ....129.2138L} claimed that virtually all the dust is located in the central 4~arcsec$^{2}$ region. In the deep MegaCam colour map, this galaxy has a saturated central part, and, therefore, we cannot identify the dust structure.

\begin{figure}
\includegraphics[scale=1.0]{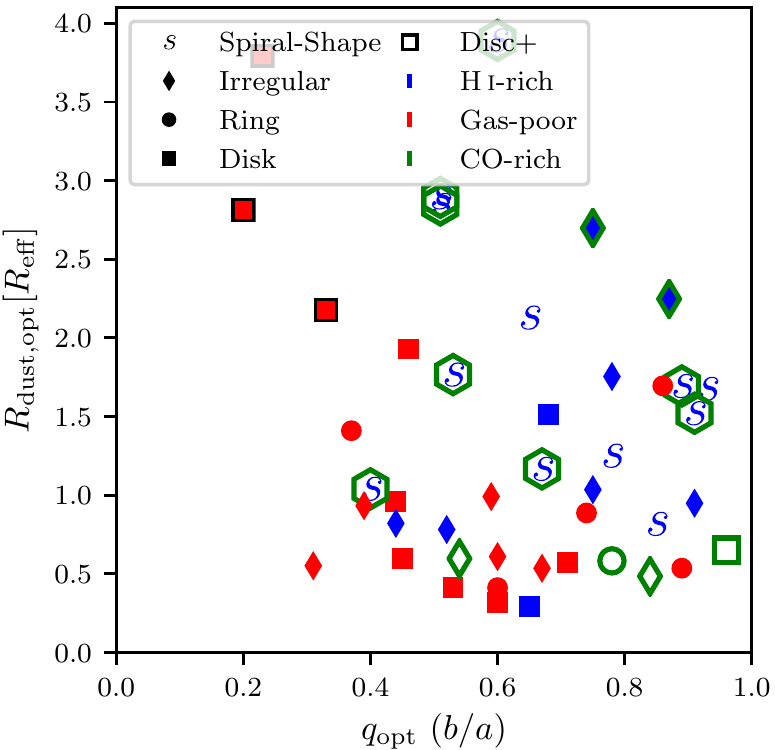}
\caption{Dust radius as a function of ellipticity. The colours of the markers indicate gas content of the galaxies. The green hexagons indicate galaxies detected in CO. The 'Disc+' legend (black marker edge colour) indicates galaxies with high inclination i.e. $\gtrsim$ 80 degrees.}
\label{fig:Rdust_inc}
\end{figure}

Dust detection is easier when the inclination of a galaxy is higher. Therefore, we checked whether the dust radius is related to the ellipticity (a proxy for inclination), and we present our result in Fig. \ref{fig:Rdust_inc}. We conclude that gas-rich ETGs contain more extended dust than gas-poor ETGs regardless of the inclination. We do find a significant correlation between the dust radius and axis ratio for gas poor galaxies\footnote{The correlation coefficient and p-value are $-0.52$ and $0.001$}, which could be expected for galaxies with optically thin dust structures.

Our detection rates based on dust absorption are similar to those reported in studies using infrared emission data. For example, \citet{2012ApJ...748..123S} report that the detection rate is 24 per cent and 62 per cent for elliptical and lenticular galaxies, respectively. The overall detection rate for their ETGs is $\sim$~48 per cent. In a similar study, but only for Virgo ETGs, \citet{2013A&A...552A...8D} detect dust in 41 per cent of the lenticulars, and 17 per cent of the ellipticals (32 per cent for all ETGs).

Although the overall detection rate of \citet{2013A&A...552A...8D} is close to our results, they find dust in half of their \hiA \ detected galaxies (8/16). Yet, we find that all large \hiA \ disc galaxies, which are all outside of Virgo, contain dust. Their lower detection rate might be partly due to the fact that they searched for dust at the position of the optical galaxy centre, while in reality dust can be offset from that.

\citet{2012MNRAS.422.1835S} report 43 $\pm$ 9 per cent CO and dust detection rate in all \hiA-rich ETGs by using the study of \citet{2011MNRAS.414.2923K}. If we revisit their study, a similar rate holds for our gas-rich ETGs (44 $\pm$ 10). However, by using deep-optical imaging, we find that all the gas-rich galaxies contain dust structures.

\subsection{Dust mass}
\label{sec:Discus_coldgas}

As discussed in the previous section, we find a different dust morphology in the \hiA-rich and \hiA-poor ETGs thanks to deep optical imaging. Yet, the question is begged as to whether we can properly measure the mass of dust in ETGs. In order to answer this question, we used the total extinction and calculated the dust mass for our sample galaxies, and then we compared our results with those derived from the SED fits including far-IR data \citep{2019A&A...622A..87K}. We tabulated the resulting dust masses in Table \ref{table:dust_mass_app}.

\begin{figure}
\includegraphics[scale=1.0]{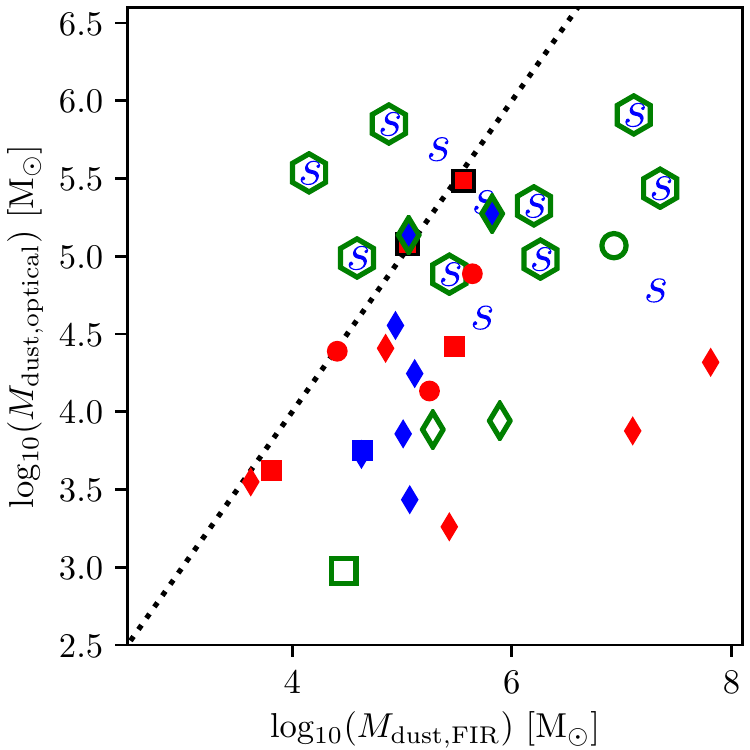}
\caption{Dust masses estimated from optical absorption in this study vs. infrared emission \citep{2019A&A...622A..87K}. The black dotted line shows the one-to-one relation. The symbols are the same as in Fig. \ref{fig:Rdust_inc}.}
\label{fig:MdvsMd}
\end{figure}

Not surprisingly, by using optical images, we underestimated the dust mass for most of the sample ETGs, which is similar to what was found in previous studies \citep[e.g.][]{2007A&A...461..103P,2012MNRAS.422.1384F,2012MNRAS.423...49K,2015A&A...579A.103V}. Fig. \ref{fig:MdvsMd} shows the comparison between our results and those obtained from far-IR emission. However, our result that the dust mass obtained from the total extinction value is six times higher in gas-rich than in gas-poor galaxies is consistent with previous studies (see comparison in Table \ref{table:aver_c}).We also find that the correlation coefficients and p-values of the relation shown in Fig. \ref{fig:MdvsMd} are $0.41 - 0.05$ and $0.18 - 0.51$ for the gas-rich and gas-poor galaxies, respectively. Although the significance of the correlation is low, it does increase for gas-rich ETGs.

\begin{table}
\caption{Comparison of the average dust masses for gas-rich (GR) and gas-poor (GP) galaxies.}
\begin{tabular}{@{\extracolsep{\fill}}ccccc}
%\begin{tabular}{l@{\hspace{0.65em}}c@{\hspace{0.65em}}c@{\hspace{0.65em}}c@{\hspace{0.65em}}c@{\hspace{0.65em}}}
\hline
\hline
$M_{\mathrm{dust,GR}}$  &$M_{\mathrm{dust,GP}}$ &GR/GP  &$M_{\star}$ range&     Ref.\\ %[2.0pt]
$\mathrm{M}_{\sun}$     &$\mathrm{M}_{\sun}$    & -     &$\mathrm{M}_{\sun}$    \\
\hline
5.2     &4.4    &6.3    &$\sim$10.0~-~11.5&     1\\
6.5     &6.4 (5.8)\tablefootmark{a}     &5.0    &$\sim$10.0~-~11.5&     2\tablefootmark{b}\\
6.5     &5.9    &4.0    &$\sim$9.0~-~11.5&      3\tablefootmark{c}\\
6.5     &5.9    &4.0    &$\sim$8.5~-~11.0&      4\tablefootmark{d}\\
\hline
\hline
\end{tabular}
\label{table:aver_c}
\begin{tablenotes}[para,left]\footnotesize
\tablebib{(1) This work. (2) \citet{2019A&A...622A..87K} (3) \citet{2013A&A...552A...8D} (4) \citet{2012ApJ...748..123S}.}\\
\tablefoot {\tablefoottext{a}{ When we removed the two outlier ETGs from the statistics (see text).} \tablefoottext{b}{Calculated by using SED fitting including 140 \micron \ data. We calculated the average value for the same sample galaxies studied in this work.} \tablefoottext{c}{Calculated by using \textit{Herschel}/SPIRE observations for Virgo ETGs.}\tablefoottext{d}{Calculated by using \textit{Herschel}/SPIRE observations for 62 ETGs (39 S0+S0a, and 23 E).}}
\end{tablenotes}
\end{table}

In Table \ref{table:aver_c}, the reason for the small difference between gas-rich and gas-poor galaxies in \citet{2019A&A...622A..87K} is that two galaxies are contaminated or miss-classified: NGC~2577 and NGC~3301. As can be seen in Fig. \ref{fig:MdvsMd}, these galaxies have the highest estimated dust mass. However, the data of NGC~2577 comes from \textit{IRAS} observations, and there is a nearby large spiral galaxy $\sim$~6 arcmin north-west. Moreover, these two galaxies are located in a similar direction with the \textit{IRAS} scanning track. Therefore, the high dust mass of this galaxy is likely due to the contamination from the nearby source. For NGC~3301, single-dish observations detected \hiA \ in this galaxy \citep{2005ApJS..160..149S}. However, we consider it as gas-poor since it was not detected in \citet{2012MNRAS.422.1835S}. If we remove these two galaxies, the difference between the logarithmic dust mass of gas-rich and poor galaxies becomes 0.7 dex, which is similar to our result obtained based on the optical data.

It is clear that the ratio of dust mass between the gas-rich and gas-poor ETGs ($\frac{M_{d,GR}}{M_{d,GP}} = \sim 5 $) is similar regardless of whether we use optical, 140 \micron \ or 500 \micron \ data. Moreover, it is the same for the field and cluster ETGs. These results are most likely due to a different origin of the gas and dust in these galaxies. Although far-IR data provides a better way to measure dust mass, the spatial resolution is very low at the long wavelengths. On the other hand, as we discussed in Sec. \ref{sec:Discuss_detection}, deep optical images provide excellent data to study the spatial distribution of the dust in ETGs.

\subsection{Origin of the dust}

The origin of the dust in ETGs has been long discussed in the literature. These discussions follow two main lines: internal dust production and external dust accretion. There are also studies, such as \citet{2013ApJ...766..121M}, which favour a mixed origin instead of a single one. There are uncertainties regarding the way in which the presence of cold gas in ETGs is linked to the origin of the dust.

If the stellar evolution is responsible for the dust produced in galaxies, we should see a correlation between the dust mass and the stellar mass of galaxies, especially if they are similar in terms of morphology, star formation history, and environment. \citet{2019A&A...622A..87K} find no such correlation and conclude that stellar mass loss is not a significant source of dust in ETGs.

The spiral shapes, irregular and clumpy dust distributions, and physical extent of the dust in \hiA-rich ETGs indicate that the dust was most likely accreted through a gas-rich minor merger. Our conclusion is supported by the fact that extended \hiA \ in several of these galaxies often comes from the external sources,
as implied by the lack of a correlation between \hiA \ mass and stellar mass in ETGs \citep{2012MNRAS.422.1835S} and by the frequent kinematical misalignment between \hiA \ and stars \citep{2014MNRAS.444.3388S}. Our results are also consistent with previous studies, such as \citet{2015MNRAS.449.3503D}, who find that the origin of the cold gas is most likely due to the minor mergers in their sample ETGs \citep[see also][]{2016MNRAS.462..382G}. \citet{2012MNRAS.423...49K} also stress that dust contribution from stellar mass loss is not enough to create the observed dust alone. Moreover, it is possible that dust is coupled with the \hiA \ in the very outer regions, and, therefore, could survive the interactions with the hot gas when accreting towards the centre \citep[see also][]{2012MNRAS.422.1384F}.

In \hiA-poor ETGs, dust is mostly detected in the central regions. This is different from \hiA-rich galaxies and suggests that the dust origin is likely different. It is possible that dust in \hiA-poor ETGs mostly has an internal origin. Indeed, another scenario proposed by \citet{2007ApJ...660.1215T} might be valid for gas-poor ETGs: cold dust is buoyantly transported from the core of the galaxies, where old stellar populations are located and dust is produced via stellar evolution. They claimed that active galactic nucleus energy outbursts might drive the buoyant outflow and cause the widespread dust distribution observed in optical colour maps.

As mentioned in the Introduction, \citet{2013ApJ...766..121M} discuss a hybrid model that includes both external accretion and internal dust production, which is further supported by recent studies \citep[e.g.][]{2017MNRAS.470.1991B}. For this hybrid model, they require low metallicity gas, and dusty ETGs should experience more mergers than dust-free galaxies. In this model, the external accretion of cold gas can be responsible for the dust distributions observed in ETGs. Our \hiA-rich ETGs have a vast gas reservoir, and the depletion time of the \hiA \ gas varies between 10 Gyr and 100 Gyr when moving from the inner regions to the outer regions \citep[see][]{2017MNRAS.464..329Y}. The lifetime of \hiA \ and dust is most likely similar in the field, and, therefore, there is enough gas to accrete and prevent the dust structures from being depleted.

As a final note, we remind the reader that our sample of nearby ETGs includes only galaxies outside massive clusters. Our results may not hold for galaxies in clusters, where the environment could significantly affect the relation between gas at a large radius and dust in the stellar body. In the case of field galaxies, though, we conclude that the primary factor for the presence of dust in gas-rich ETGs is the external sources.

\section{Conclusions and future prospects}
\label{sec:conclusions}

We used deep optical images to study the distribution of dust in nearby ETGs. Thanks to the excellent resolution of the MegaCam images, we can classify the dust morphology in these ETGs. We compared \hiA-rich and \hiA-poor ETGs to understand whether the presence of dust and its spatial distribution are related to the presence of a large, extended reservoir of \hiA \ in these galaxies.\\

1- We find that all of the \hiA-rich galaxies show dusty structures, while 56 per cent of the \hiA-poor galaxies show dust.

2- In 76 per cent of the \hiA-rich galaxies, the dust distribution has a spiral-like morphology.

3- The spatially extended dust structures tend to be in \hiA-rich galaxies, while the small size dust structures are generally found in the \hiA-poor ETGs.

4- All of the galaxies with \hiA \ in the outskirts, but not in the centre, host dust in the relatively inner regions. Therefore, the presence of dust is linked to the presence of \hiA \ even when the \hiA \ is found at a much larger radius. These dusty inner regions may also indicate that cold gas used to be present in the centre of these galaxies.

5- Gas-rich ETGs contain less regular dust systems, which extend to a larger radius than gas-poor objects.

Given the extended and complex morphology of dust in \hiA-rich ETGs, and the fact that the \hiA \ is mostly accreted externally in these galaxies, we conclude that external accretion appears to be the dominant source of dust in these objects. As seen above, it is clear that there is a relation between the dust and cold gas (\hiA \ and CO) presence in ETGs.

\section{Acknowledgements}

MKY acknowledges the financial support from the Scientific and Technological Research Council of Turkey (TUBITAK) under the postdoc fellowship programme 1059B191701226. We also thank Tom Jarrett and Martin Bureau for valuable discussions on the infrared emission and dust mass calculations. 

\bibliography{35090_MKYILDIZ.bib}

\appendix

\section{Model calculation}
\label{App:AppendixB}

In this appendix we give the details of the procedure used to calculate the dust-free colour from the $g'-r'$ images. In order to calculate the dust-free colours, we first needed to mask out all regions with dust, and to do this we started by automatically finding a masking threshold. To do so, we did the following: (1) We took a 100$\times$100 pixel$^{2}$ central area from the $g'-r'$ images and calculated its mean and standard deviation ($\sigma$) value. (2) We masked out all pixels above $mean+4\sigma$ and below $mean-4\sigma$. (3) We calculated the median and standard deviation from the remaining pixels. (4) We masked all pixels above and below $median \pm 3\sigma$ in the original image.

Having inspected the images, additional regions are manually masked if it is necessary. This masking procedure removes most of the dusty regions in the inner regions and also the noisy pixels in the outer regions. Having applied the masking procedure, we derived the profiles (as explained in Sec. \ref{sec:modelling}). There is a possibility that a small part of the galaxy might contain a star-forming region. In this case, the dust-free colour is bluer than it should be, and the colour excess is overestimated. Therefore, we did not directly average the pixels in an annulus. Instead, we used four quadrants and a second masking process as follows:\\
\begin{figure}
\includegraphics[scale=1.0]{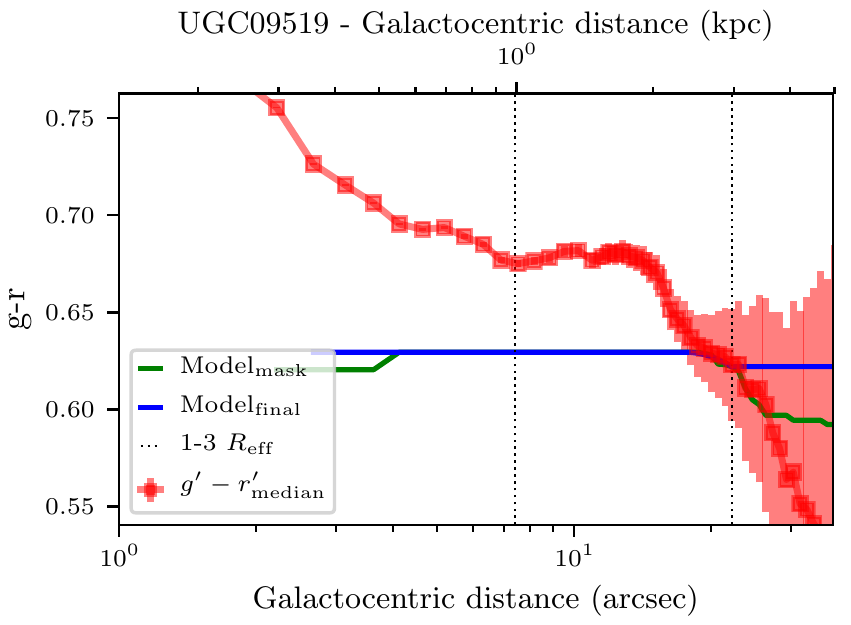}
\caption{Radial profiles for the $g'-r'$ images and the two models obtained in our dust-free colour calculation process. The red profile shows the median colour value of the original annuli. The green line shows the model, which follows the requirement that the colour cannot get redder with increasing radius (see text). The blue profile shows the final model, which is flat before 0.5 $R_{\mathrm{eff}}$, and after 3~$R_{\mathrm{eff}}$. The dotted vertical lines indicate 1~$R_{\mathrm{eff}}$ and 3~$R_{\mathrm{eff}}$, respectively.}
\label{fig:mask_methodRP}
\end{figure}
(a) We divided each annulus into four quadrants, and masked the pixels above median$+1.5\sigma$ and below median$-3\sigma$ in each quadrant. Here, we estimated the standard deviation by using the median absolute deviation. Using this asymmetric masking removes remaining red pixels in the annulus and keeps the blue pixels. (b) We calculated the median value from the remaining pixels in each quadrant for each annulus. This results in four median values representing the dust-free colours in the quadrants. (c) To get a better dust-free colour value, we used the second minimum of these four median values mentioned above as the dust-free colour\footnote{There is uncertainty in determining which model (i.e. represented by the median value) is the best. We tested whether the dust-free colour would change when taking more sectors or when taking the median or the mean, and found that it would change by 0.01 mag at most. We therefore ignored the uncertainty in our models.}. (d) Having derived the profiles, we applied a seven-pixel size median filter to remove sharp changes between the annuli.\\

\noindent The two-step process given above creates the first model for the dust-free colour. We manipulated this first model as follows: (i) We set the colour within $R_{\mathrm{eff}}$/2 to the value at $R_{\mathrm{eff}}$/2. If an effective radius is larger than 15 arcsec, we used the colour at 7.5$''$\footnote{Since the deep optical images are generally saturated in the centre, we assume that the colour of the innermost region is constant.}. (ii) Since early-type galaxies without a significant cold ISM are known to have smooth colour profiles, which become slowly bluer with an increasing radius beyond $R_{\mathrm{eff}}$/2 \citep[e.g.][]{1989AJ.....98..538F,1990AJ....100.1091P,1999MNRAS.310..703P,2012MNRAS.419.2031P}, we did not allow our model to become redder towards a larger radius. Therefore, the model can either become bluer or stay constant when going outwards. (iii) We smoothed the model to remove the sharp edges. (iv) We assumed that the model colour does not change after 3~$R_{\mathrm{eff}}$ and we obtained our final model to represent a dust-free colour. (v) We calculated the errors on the colour (based on the errors in $g'$ and $r'$ bands, see Sec. \ref{sec:megacam} for details), and we truncated the colour radial profile at the first annulus where the error is higher than 0.2 mag, or at 3~$R_{\mathrm{eff}}$ at most. We show the original and masked $g'-r'$ median profiles together with the three model radial profiles in Fig. \ref{fig:mask_methodRP}. In Fig. \ref{fig:mask_methodRP}, the models are always below the $g'-r'$ colour profile because of the masking procedure explained above.% (see also Fig. \ref{fig:mask_method}).\\ 

\section{\hiA \ and dust mass estimation}
\label{App:AppendixA}

Our sample selection is based on \hiA \ detectability, and since we measured the colour excess in our sample galaxies, we can apply a simple test to check whether our colour excess measurements are reasonable. \citet{1978ApJ...224..132B} presented the following relation between the \hiA \ column density and the colour excess $E(B-V)$ for the Milk Way:\\
\begin{equation}
\begin{aligned}
\label{eq:eqHI}
N(H~\text{\footnotesize{I}}) = 4.8 \times10^{21}~\mathrm{cm}^{-2} \times E(B-V)~\mathrm{mag}^{-1}
\end{aligned}
,\end{equation}

\noindent where the proportionality constant holds for \hiA \ only, and not for \hiA \ + H$_{2}$. Since we can get the $E(g'-r')$ values from the colour excess maps (Eq. \ref{eq:eqW}) and convert them to $E(B-V)$ values, we can estimate the amount of hydrogen expected to reside in the regions where we measured extinction. In this study, we adopted the MilkWay value ($R_{\mathrm{v}}=A_{\mathrm{v}} / E(B-V)$ = 3.1), which is appropriate for an average interstellar dust grain size. 

One should always consider that the value of the constant in Eq. \ref{eq:eqHI} can change for different $R_{\mathrm{v}}$ values, as \citet{1978ApJ...224..132B} claimed that larger grain sizes can significantly increase the constant given in Eq. \ref{eq:eqHI}. For example, they measured this value to be 3.2 times higher (i.e. $15.4\times10^{21}\mathrm{cm}^{-2}$) for a dark cloud star, $\rho$ Oph A.

We calculated $E(g'-r')$ by assuming $A_{\mathrm{g'}}$=3.689~$\times~E$(B-V) and $A_{\mathrm{r'}}$=2.709~$\times~E$(B-V) \citep{2014ApJS..210....4M}. Therefore Eq. \ref{eq:eqHI} becomes:
\begin{equation}
\begin{aligned}
\label{eq:eqHI2}
\frac {\langle N(H~\text{\footnotesize{I}})\rangle} {\langle E(g'-r') \rangle} = 4.9 \times10^{21} \mathrm{cm}^{-2} \mathrm{mag}^{-1}
\end{aligned}
.\end{equation}

By using the conversion from column density to surface density for \hiA \ (1~M$_{\sun}$ pc$^{-2}$ =   $8 \times10^{19} \mathrm{cm}^{-2}$) we obtained,
\begin{equation}
\begin{aligned}
\label{eq:eqHI3}
M_{\mathrm{H\text{\scriptsize{I}}}} = 36.8  \times \langle E(g'-r')\rangle \times \mathrm{A} ~~~~\mathrm{M}_{\sun}
\end{aligned}
,\end{equation}
\noindent where $M_{\mathrm{H\text{\scriptsize{I}}}}$ is the \hiA \ mass in $\mathrm{M}_{\sun}$ measured in the area A in pc$^{2}$, and$\langle E(g'-r')\rangle$ is the average colour excess in mag in this area.

To calculate the dust mass, we used optical extinction. There are different geometrical models to calculate the total extinction, such as `the sandwich model' \citep[e.g.][]{1989MNRAS.239..939D} or a simple screen model. Here, we used a simple screen model to obtain a lower limit on the dust mass.

We used colour excess maps to calculate the extinction in the $g'$ band ($A_{g'}$). To do so, we adopted the mean $R_{V}$ value of \citet{1994MNRAS.271..833G} for the ETGs with large scale dust lanes. 

The total extinction due to dust depending on the wavelengths is given by:

\begin{equation}
\begin{aligned}
\label{eq:TotalExt}
A_{\lambda}=1.086\,\, C_{ext}({\lambda})\times l_{d}
\end{aligned}
,\end{equation}
where $l_{d}$ is the dust column length along the line of sight, and $C_{ext}({\lambda})$ is the extinction cross-section at wavelength $\lambda$. The geometry is one of the main determinants for estimating the dust mass.

We can calculate the extinction cross-section by using a dust grain size distribution function ($n(a)$) as shown in the following equation.
\begin{equation}
\begin{aligned}
C_{ext}(\lambda)=\int_{a_-}^{a_+}Q_{ext}(a,\lambda)\, \pi\, a^{2}\,n(a)\,da
\end{aligned}
,\end{equation}

\noindent where $Q_{ext}(a,\lambda)$ is the extiction efficiency factor, and $a_-$ and $a_+$ are the lower and upper cutoffs of the grain size distribution, respectively. To obtain the extiction efficiency, we adopted the following values for graphite and silicate grains,

\begin{displaymath}
Q_{ext,silicate} = \left\{ \begin{array}{ll} 
 0.8\, a/{a_{silicate}} & \textrm{for $a < a_{silicate}$},\\
0.8 & \textrm{for $a \geq a_{silicate}$}
\end{array} \right.
\end{displaymath}

\begin{displaymath}
Q_{ext,graphite}=\left\{ \begin{array}{ll} 
2.0\, a/{a_{graphite}} & \textrm{for $a < a_{graphite}$},\\ 
2.0 & \textrm{for $a \geq a_{graphite.}$} \end{array} \right.
\end{displaymath}

We adopted the lower cutoff size, 0.005 \micron, and the upper cutoff size $a_{+} = <a> / a_{\rm Gal} \times 0.22$ \micron. We note that $a_{\rm Gal}$ is the mean grain size value of the ETGs reported in \citet{1994MNRAS.271..833G}.
We also used the same size distribution function as in \citet{1994MNRAS.271..833G}:

\begin{displaymath}
n(a)=n_0\,\,a^{-3.5} \hspace{10mm}     (a_- \leq a \leq a_+)
\end{displaymath}

With these assumptions, we estimated the extinction cross-section by further assuming that the distribution function does not change across the dusty regions. From the total extinction and extinction cross-section, we estimated the dust column length. The dust mass surface density is then:

\begin{equation}
{\Sigma}_d= \int_{a_-}^{a_+}\frac{4}{3}\,\pi\, a^3\, \rho_d\, n(a)\,da\times l_d
,\end{equation}

where $\rho_{d}$ is the specific grain mass density, which is taken to be $\sim$ 3 g cm$^{-3}$ for graphite and silicate grains (Draine \& Lee 1984). To obtain the total dust mass, we simply multiplied the dust surface density with the total dusty area,  $\rm M_d=\Sigma_{d} \, \times$ Area, expressed in solar mass units.

\begin{table*}
\caption{Estimated dust masses from the optical extinction and infrared emission.}
\begin{center}
\begin{tabular}{ccccccccccc}
\hline
\hline
Galaxy Name     &\multicolumn{2}{c}{Dust Mass}  &Galaxy Name    &\multicolumn{2}{c}{Dust Mass}   &Galaxy Name    &\multicolumn{2}{c}{Dust Mass}\\
        &\multicolumn{2}{c}{log$_{10} M_{\mathrm{dust}}~ [\mathrm{M}_{\sun}]$}  &       &\multicolumn{2}{c}{log$_{10} M_{\mathrm{dust}}~ [\mathrm{M}_{\sun}]$}        &       &\multicolumn{2}{c}{log$_{10} M_{\mathrm{dust}}~ [\mathrm{M}_{\sun}]$}\\
(1)     &(2)    &(3)    &(4)    &(5)    &(6)    &(7)    &(8)    &(9)\\
\hline
NGC~0661        &-      &5.5    &NGC~3522       &3.9    &5      &NGC~5173       &5.3    &5.8\\
NGC~0770        &3.4    &-      &NGC~3605       &3.6    &-      &NGC~5273       &3.9    &5.3\\
NGC~2549        &-      &4.6    &NGC~3610       &-      &3.5    &NGC~5308       &5.1    &5.1\\
NGC~2577        &4.3    &7.8    &NGC~3613       &-      &5.3    &NGC~5322       &-      &5.1\\
NGC~2592        &-      &5.3    &NGC~3619       &5.9    &4.9    &NGC~5342       &5.2    &-\\
NGC~2594        &4.4    &-      &NGC~3626       &5      &6.3    &NGC~5473       &-      &5.2\\
NGC~2679        &-      &5.4    &NGC~3648       &-      &5.3    &NGC~5485       &4.9    &5.6\\
NGC~2685        &5.4    &7.4    &NGC~3658       &-      &5.3    &NGC~5500       &-      &5\\
NGC~2764        &5.9    &7.1    &NGC~3665       &5.1    &6.9    &NGC~5582       &3.8    &4.6\\
NGC~2852        &4.4    &4.4    &NGC~3674       &-      &5.2    &NGC~5611       &2.9    &-\\
NGC~2859        &4.6    &5.7    &NGC~3757       &-      &4.9    &NGC~5631       &5.3    &5.7\\
NGC~2950        &-      &4.3    &NGC~3796       &3.3    &5.4    &NGC~6547       &5.5    &5.6\\
NGC~3098        &4.5    &5.1    &NGC~3838       &3.4    &5.1    &NGC~6548       &3.8    &5.1\\
NGC~3230        &-      &5.5    &NGC~3941       &3.7    &4.6    &NGC~6798       &5.5    &4.2\\
NGC~3245        &3.9    &5.9    &NGC~3945       &5.7    &5.3    &NGC~7457       &3.6    &3.8\\
NGC~3248        &2.8    &5.1    &NGC~3998       &4.6    &4.9    &PGC~050395     &-      &4.7\\
NGC~3301        &3.9    &7.1    &NGC~4036       &5      &4.6    &UGC~04551      &4.4    &4.9\\
NGC~3377        &3.5    &3.6    &NGC~4078       &4.4    &5.5    &UGC~06176      &5.3    &6.2\\
NGC~3400        &-      &4.4    &NGC~4203       &4.9    &5.4    &UGC~08876      &4.1    &5.3\\
NGC~3414        &4.8    &7.3    &NGC~4278       &4.2    &5.1    &UGC~09519      &5.1    &5.1\\
NGC~3458        &-      &-      &NGC~4283       &3      &4.5    &               &       &\\     
\hline
\hline
\end{tabular}
\label{table:dust_mass_app}
\end{center}
\begin{tablenotes}[para,flushleft]\footnotesize
Note.$-$ Column (1, 4, 7): The name is the principal designation from LEDA \citep{2003A&A...412...45P}. Column ($2-3,~5-6,~8-9$): Estimated dust masses from the total optical extinction ($A(g')$) and from SED fits by using far-IR data \citep{2019A&A...622A..87K}.
\end{tablenotes}
\end{table*}

\section{Figures of classes}
\label{App:AppendixC}
In this appendix, we show the $g'-r'$ colour maps of the galaxies according to their classes. Five figures in the appendix demonstrate five dust classes determined in this work (see, \ref{sec:dust_morp}).

\begin{figure*}
\includegraphics[scale=0.25]{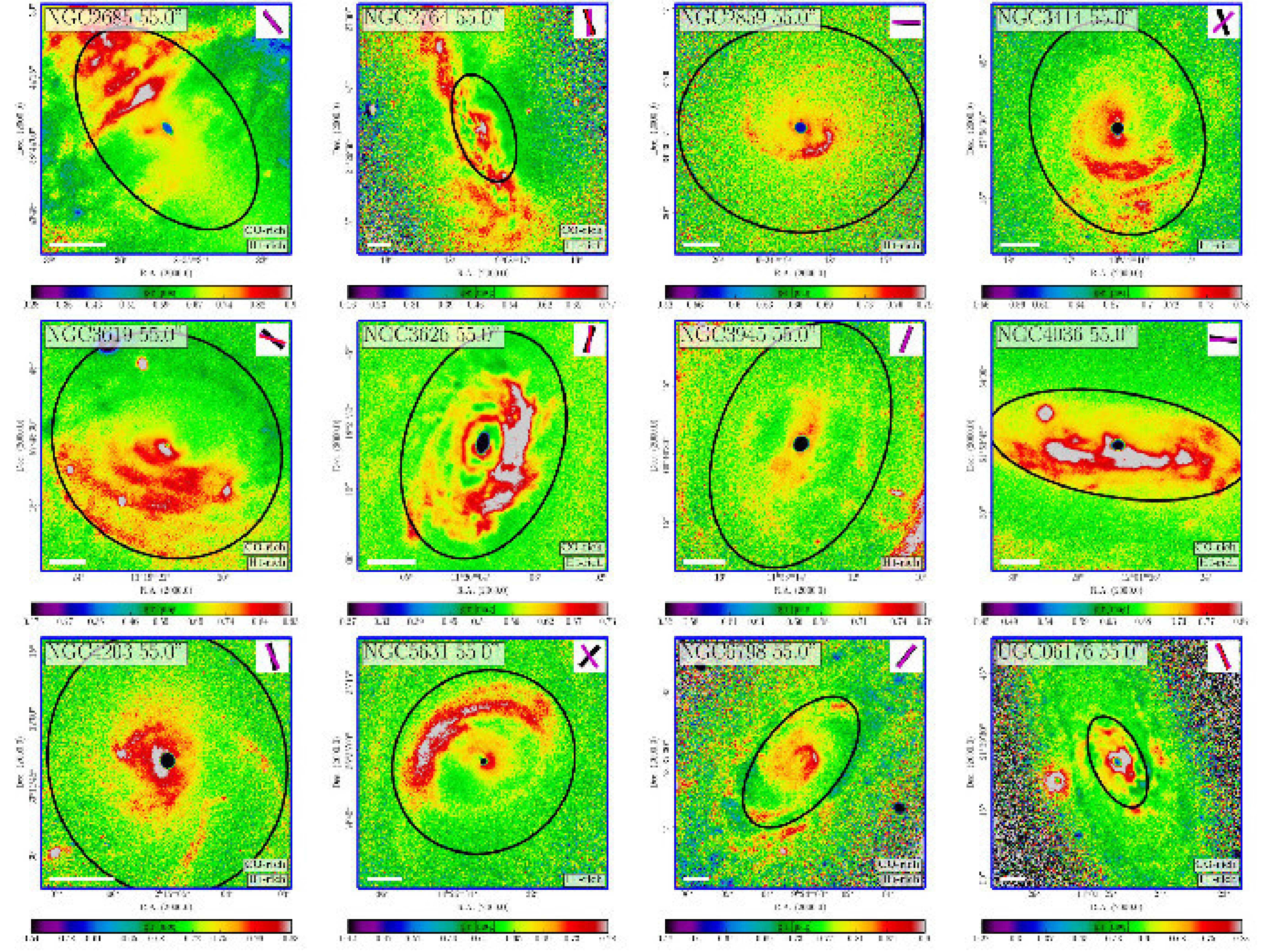}
\caption{Spiral-like structure (S)-type dusty galaxies. Each tick on the colour-bar indicates median$-5\sigma$, median$-4\sigma$, median$-3\sigma$, median$-2\sigma$, median$-\sigma$, median, median$+\sigma$, median$+2\sigma$, and median$+3\sigma$, respectively. The median and standard deviation values are calculated in the region of 0-1~R$_{\mathrm{eff}}$, which is shown with the black ellipse. The white scale bar at the bottom corner indicates 1 kpc. On the top-right, we show an illustration of the position angles of the various galaxy components: in purple the \hiAN-disc from \citet{2014MNRAS.444.3388S}; in black the optical-disc from \citet{2011MNRAS.414.2923K}; and in red the CO-disc from \citet{2013MNRAS.432.1796A}.}
\label{fig:class_SS}
\end{figure*}

\begin{figure*}
\includegraphics[scale=0.25]{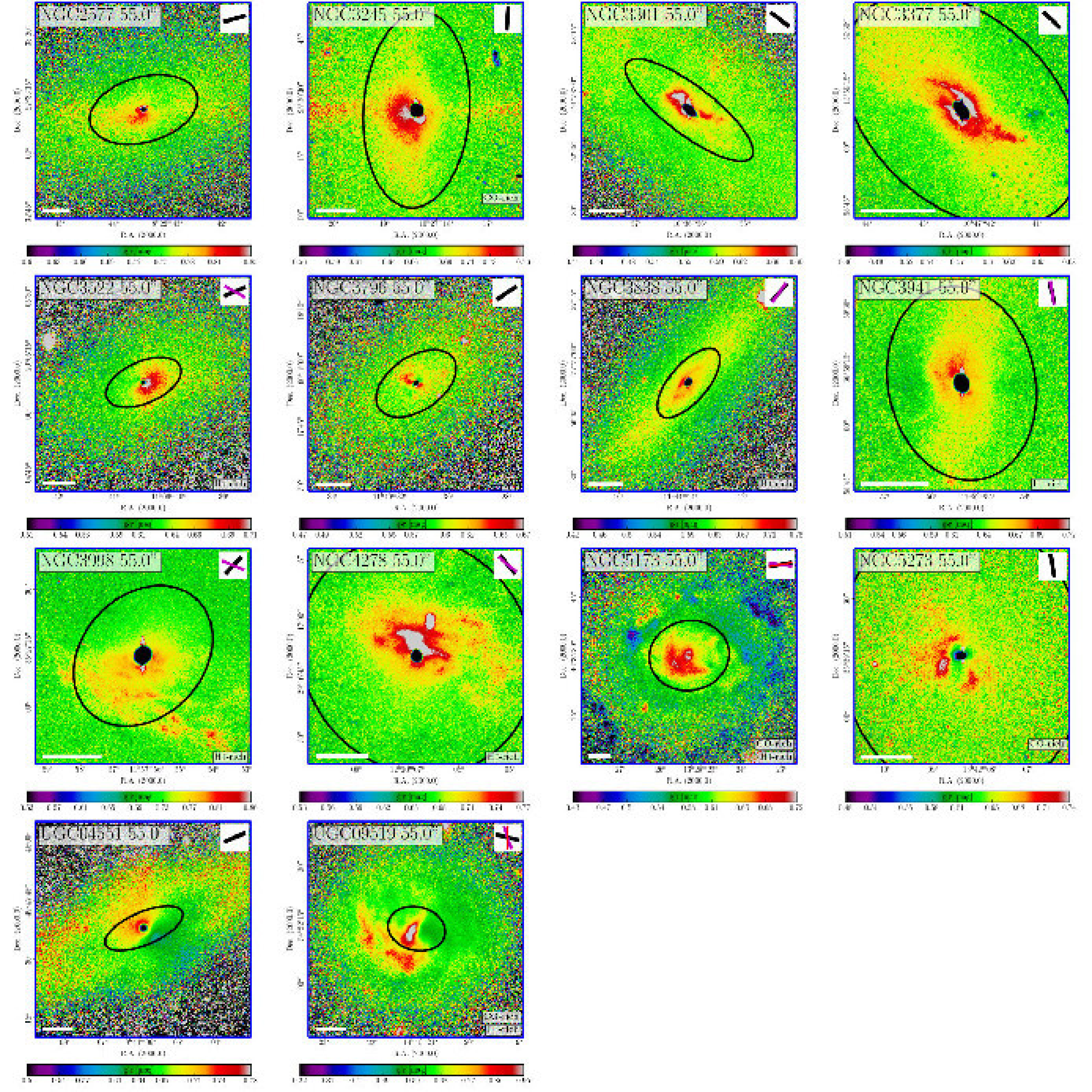}
\caption{Irregular (Ir)-type dusty galaxies. The colour-bar and other components are similar to that of Fig. \ref{fig:class_SS}}
\label{fig:class_IR}
\end{figure*}

\begin{figure*}
\begin{center}
\includegraphics[scale=0.32]{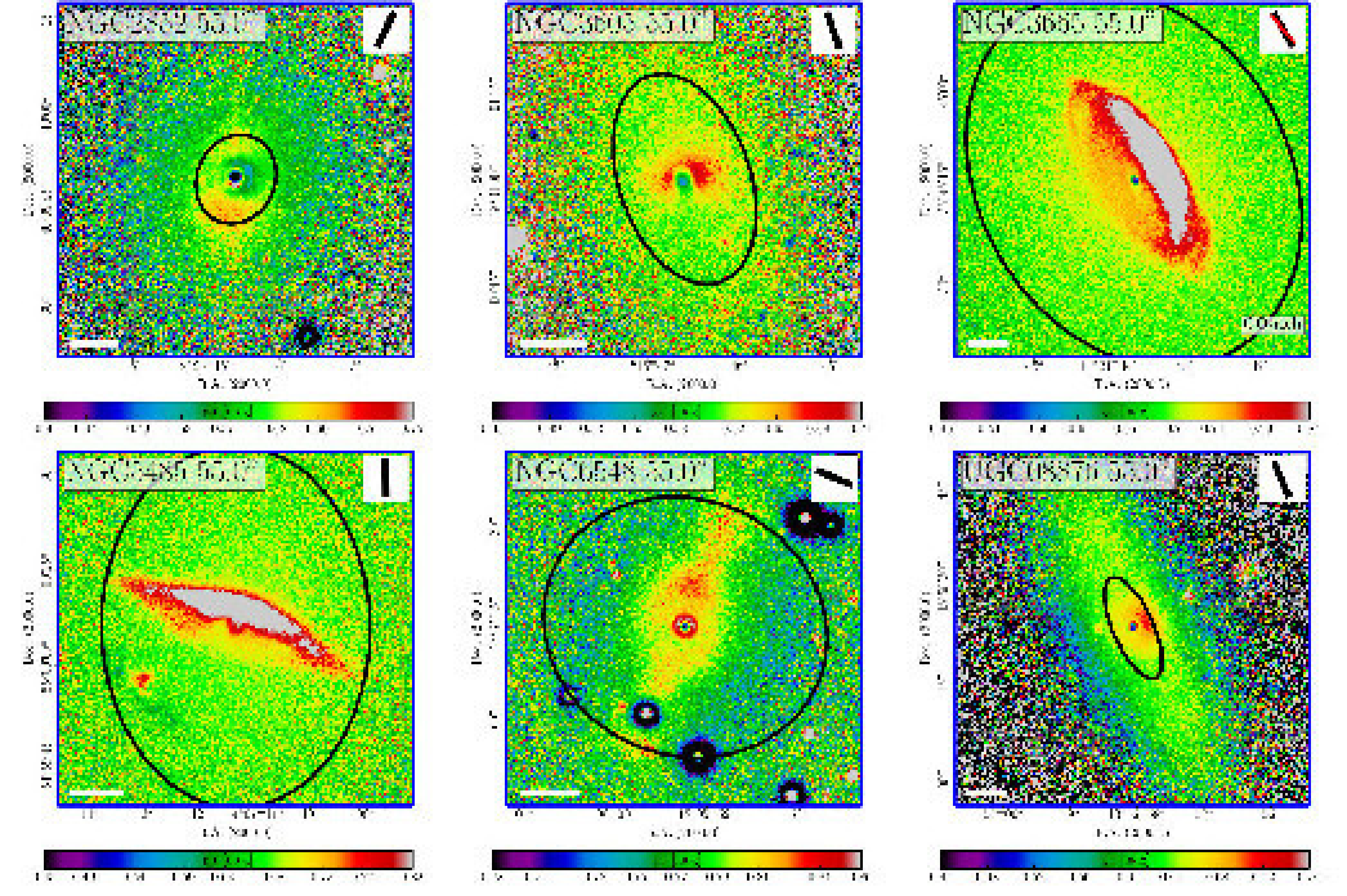}
\caption{Ring (R)-type dusty galaxies. The colour-bar and other components are similar to that of Fig. \ref{fig:class_SS}}
\label{fig:class_R}
\end{center}
\end{figure*}

\begin{figure*}
\includegraphics[scale=0.25]{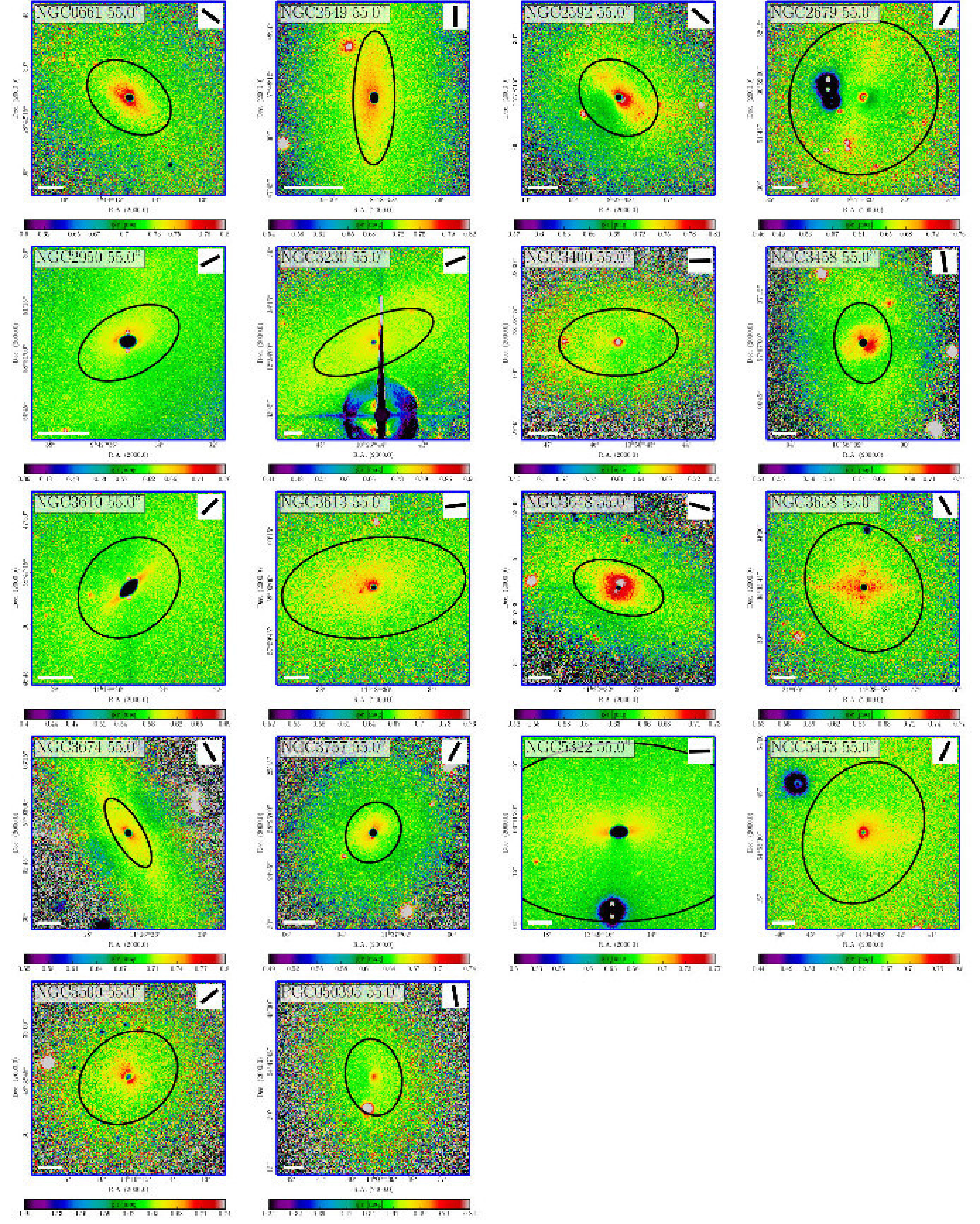}
\caption{No (N)-type dusty galaxies. The colour-bar and other components are similar to that of Fig. \ref{fig:class_SS}}
\label{fig:class_N}
\end{figure*}

\begin{figure*}[htb!]
\begin{center}
\includegraphics[scale=0.25]{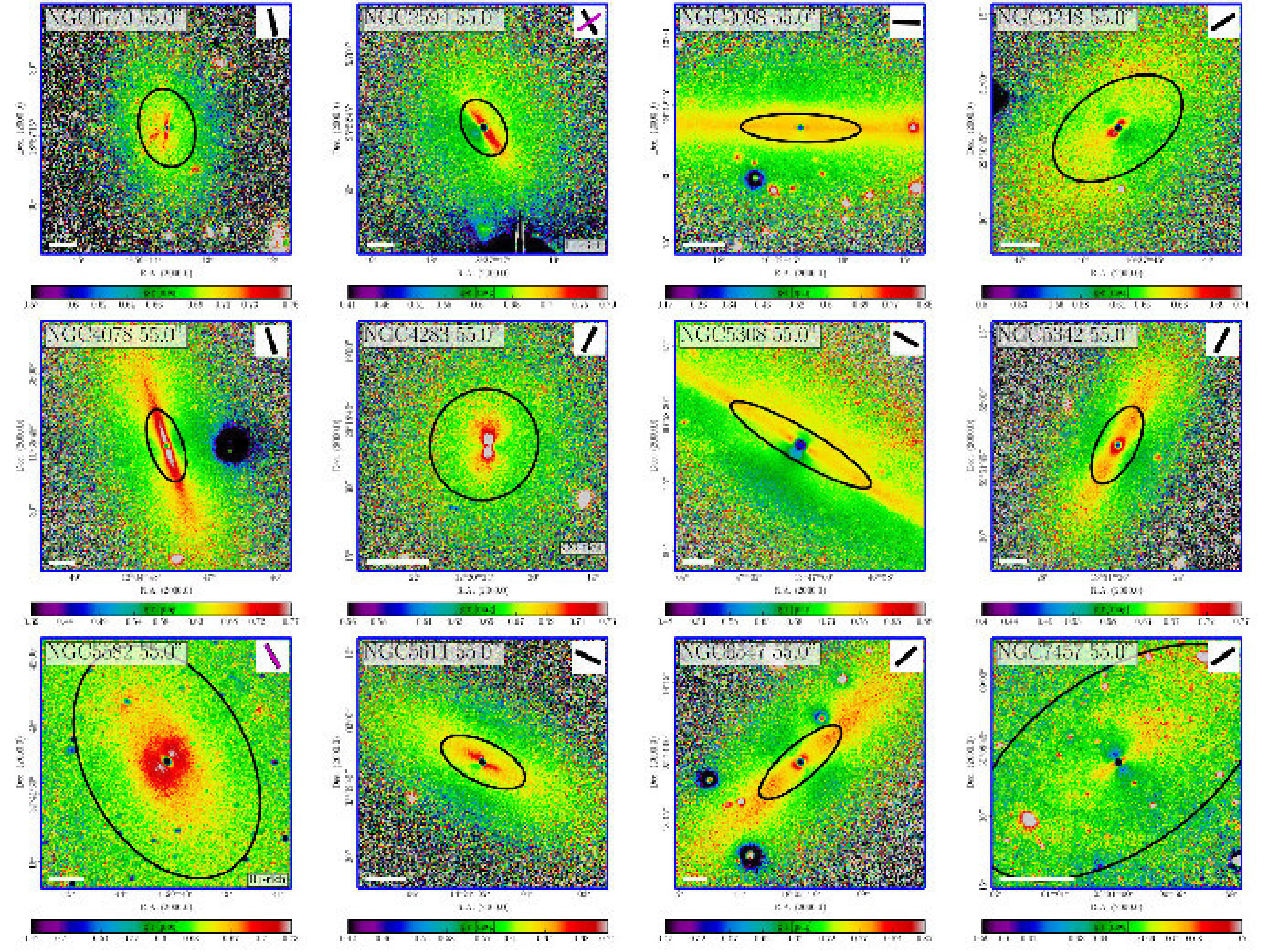}
\caption{Disc (D)-type dusty galaxies. The colour-bar and other components are similar to that of Fig. \ref{fig:class_SS}}
\label{fig:class_D}
\end{center}
\end{figure*}

\section{Figures of individual galaxies}
\label{App:AppendixD}

Here, we give the true colour images\footnote{We obtained the images by using the \textit{STIFF} program \citep{2012ASPC..461..263B} and we used the following configurations that are different from the default: Saturation level=2; Gamma=3.5; and Gamma correction factor=1.5, grey level=0.00001.} together with $g'-r'$ colour and $E(g'-r')$ colour excess maps of the sample galaxies. The images are the same as Fig. \ref{fig:example3}. The size of the true colour images is 300$\times$300 arcsec$^{2}$. The colour and colour excess maps in the top panel show the central area with a size of 100$\times$100 arcsec$^{2}$. The maps in the bottom panel are more zoomed-in and show an area with a size of 35$\times$35 arcsec$^{2}$. 

If \hiA \ is observed in a galaxy, we show column density contours with magenta lines. The \hiA \ column density contour levels are given on the true colour images. We note that the \hiA \ extends beyond the area shown. 

If CO is observed in a galaxy, we show the CO intensity contours with black lines levels in the bottom zoomed-in panel since CO is mostly detected in the inner regions. The CO intensity contour levels are indicated in the same panel. 

If a galaxy is classified as dusty, we indicate the dust radius with a white ellipse, which is determined by visual inspection (see Table \ref{table:dust_morp2}). The white bar in the panels represents 1 kpc spatial size. We also show an illustration of the position angle as in Fig. \ref{fig:five_gals}.

Additionally, we show two radial profiles for the sample galaxies. \textit{Top}: The radial profiles of the $g'-r'$ colour maps, and the final model obtained in the dust-free colour calculation process (see Appendix \ref{App:AppendixB}). The red profile with squares and error bars show the median value of the annuli. The solid blue profile shows the final model, which is flattened before 0.5 R$_{\mathrm{eff}}$ and after 3~R$_{\mathrm{eff}}$. \textit{Bottom}: Radial profiles of the colour excess maps (blue). The colour excess values are given on the right-y axis, and the corresponding average \hiAT \ mass density values are given on the left-y axis. If \hiA \ or CO is observed in a galaxy, we also show a mass density profile wtih a green or yellow line, respectively. 

%\pagestyle{plain}

%Page1
\centering
\begin{figure*}
\makebox[\textwidth][c]{\begin{minipage}[b][10.3cm]{.85\textwidth}
  \vspace*{\fill}
      \includegraphics[scale=0.85]{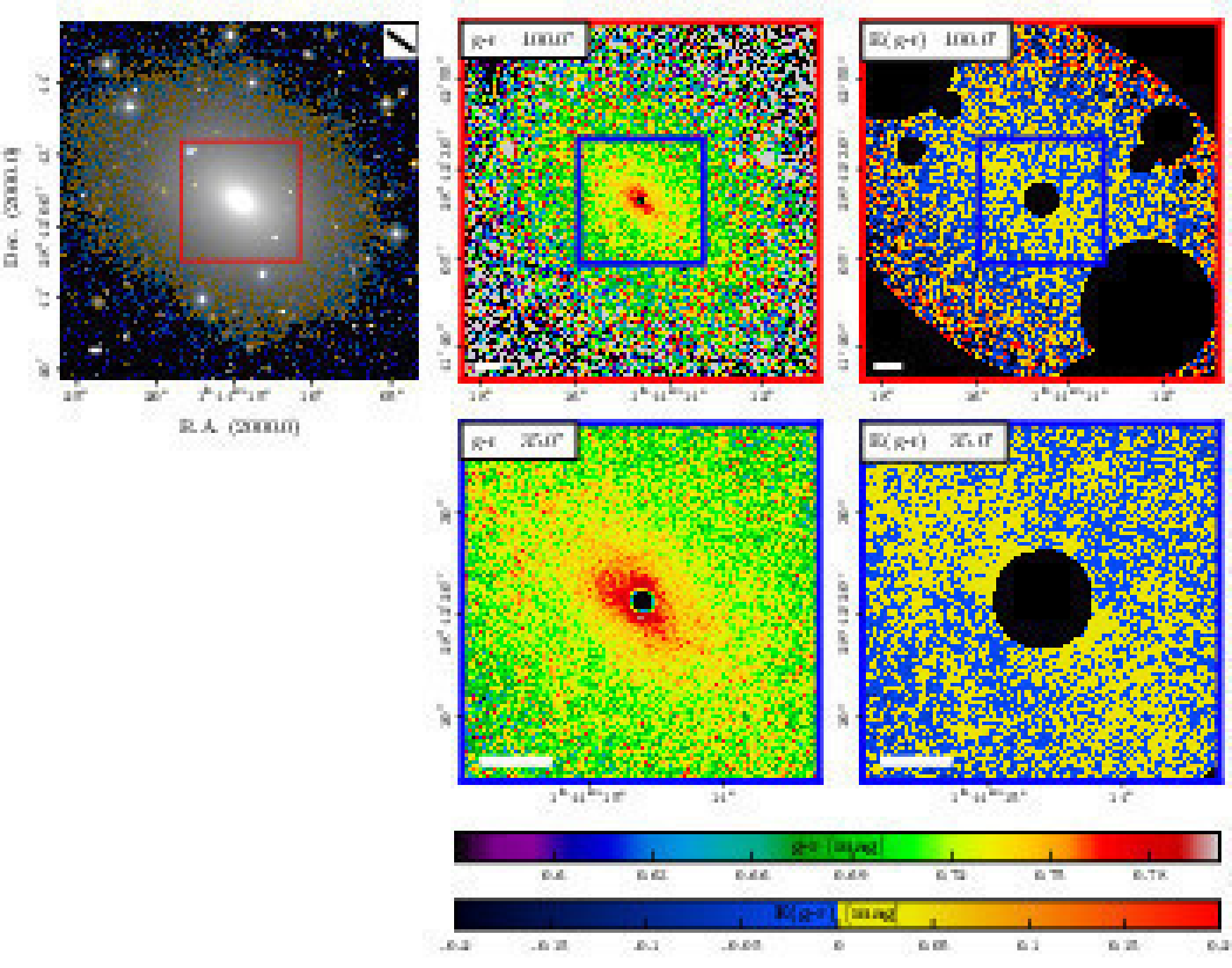}
\end{minipage}}
\makebox[\textwidth][c]{\begin{minipage}[l][-0.7cm][b]{.85\linewidth}
      \includegraphics[scale=0.55, trim={0 0.7cm 0 0},clip]{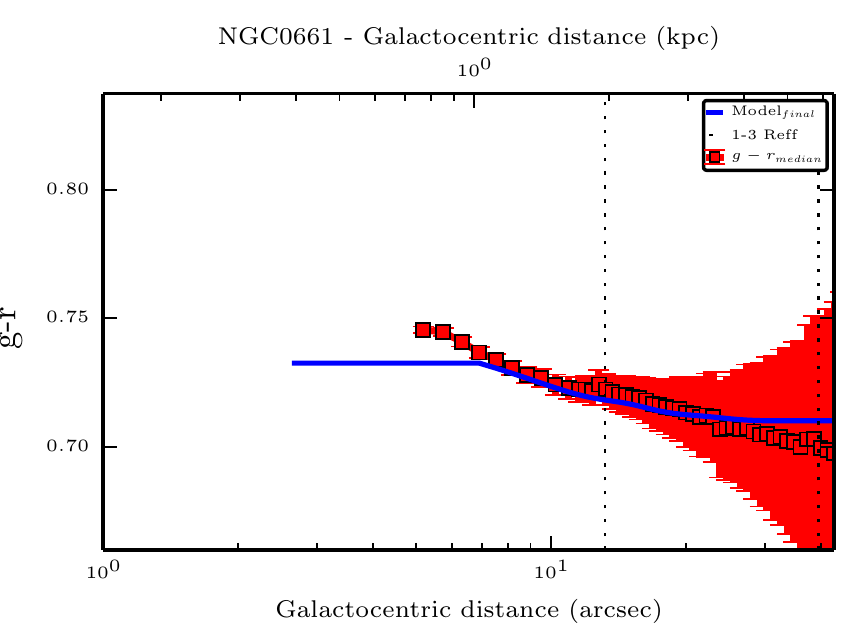}
      \includegraphics[scale=0.55]{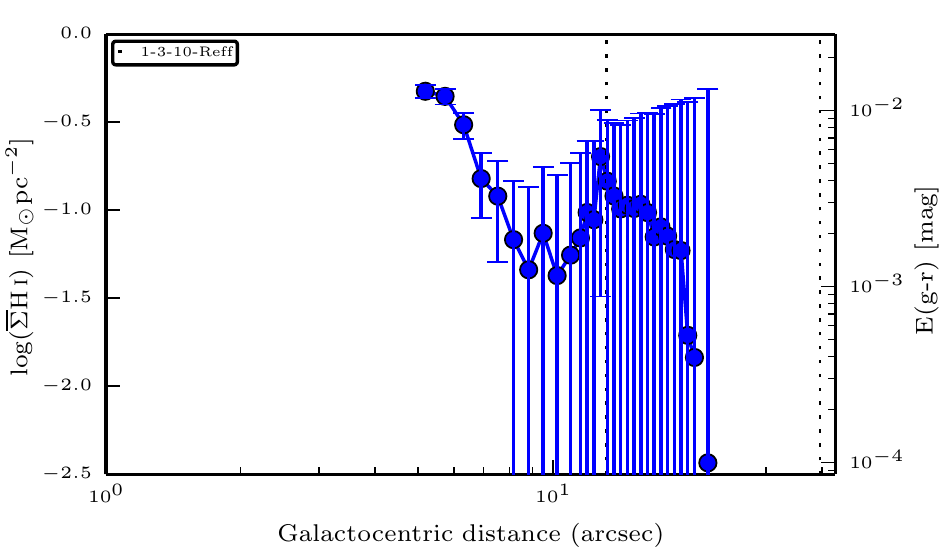}
\end{minipage}}
\caption{True colour image of NGC~0661 together with $g'-r'$ colour and $E(g'-r')$ colour excess maps. All the parameters are the same as in Fig. \ref{fig:example3}. \textit{Bottom-left:} The radial profiles of NGC~0661 are similar as in \ref{fig:mask_methodRP}.}
\label{fig:0661}
\end{figure*}

\begin{figure*}
\makebox[\textwidth][c]{\begin{minipage}[b][11.6cm]{.85\textwidth}
  \vspace*{\fill}
      \includegraphics[scale=0.85]{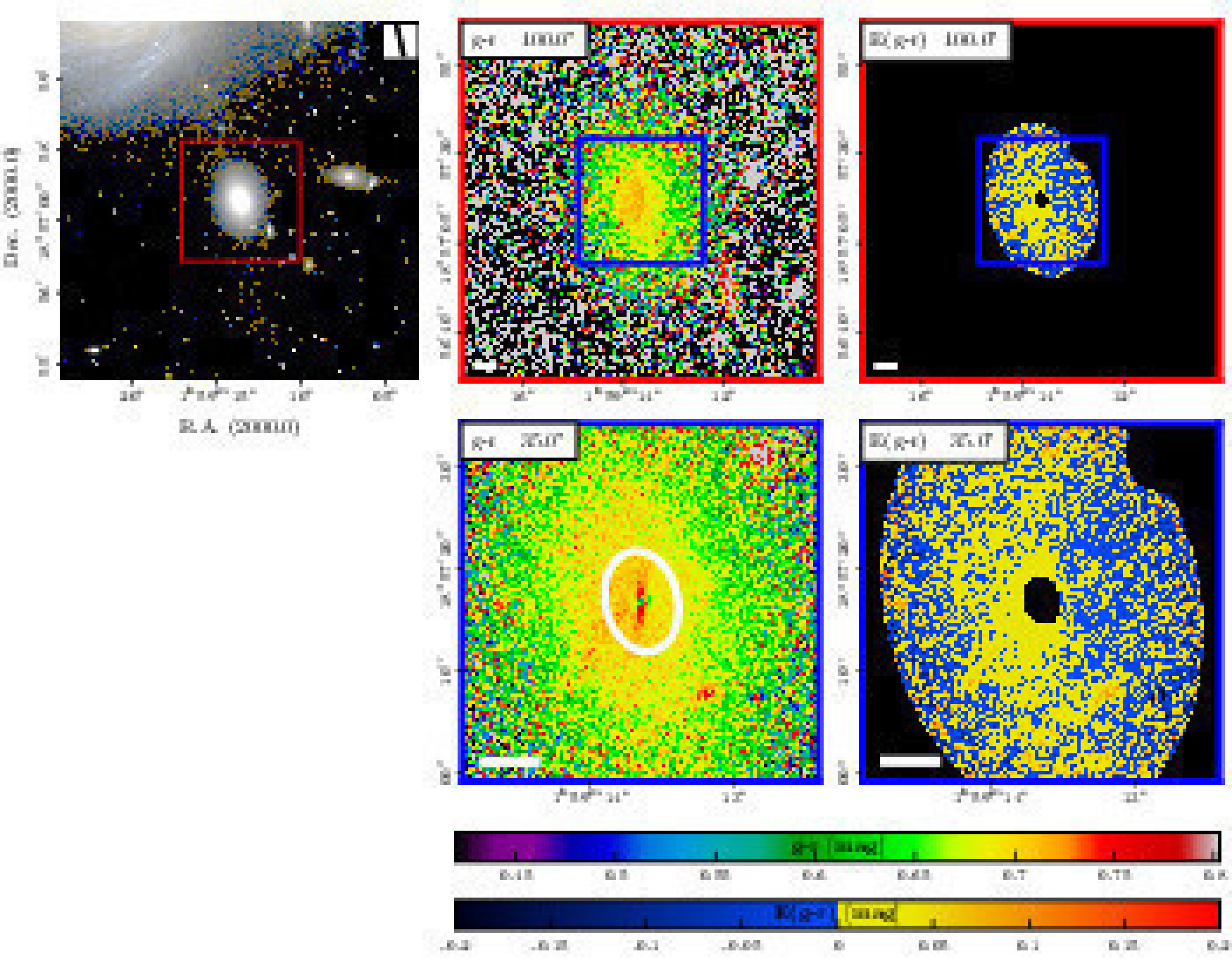}
\end{minipage}}
\makebox[\textwidth][c]{\begin{minipage}[l][-0.7cm][b]{.85\linewidth}
      \includegraphics[scale=0.55, trim={0 0.7cm 0 0},clip]{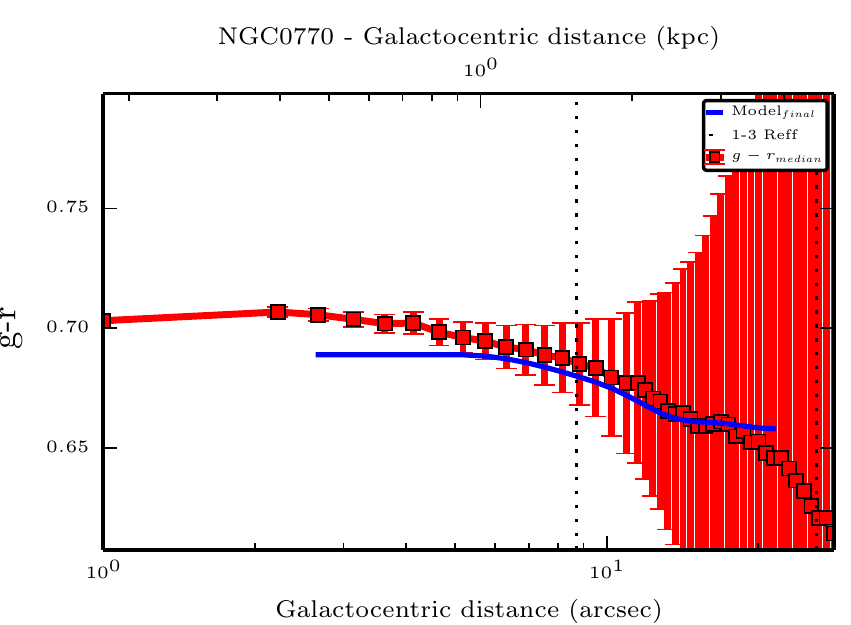}
      \includegraphics[scale=0.55]{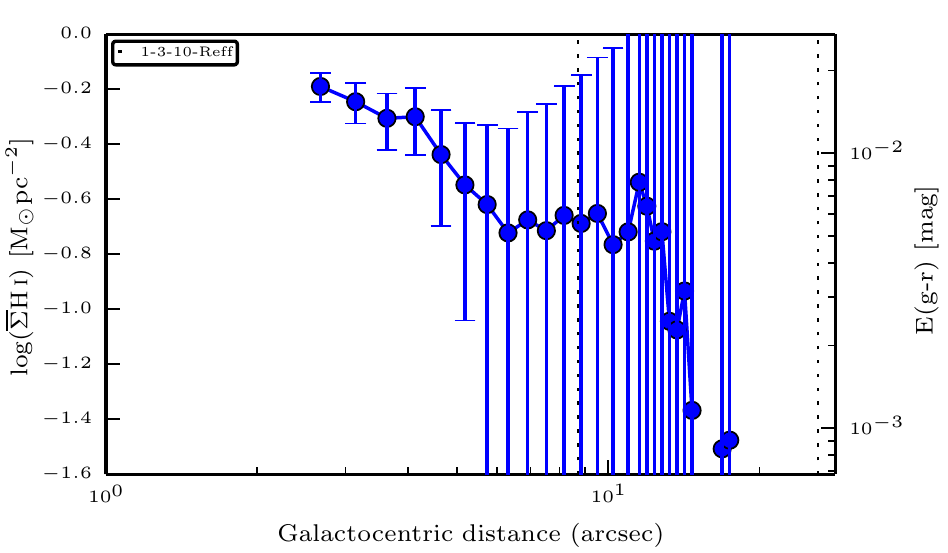}
\end{minipage}}
\caption{True image, colour map, colour excess map, and the radial profiles of NGC~0770.}
\label{fig:0770}
\end{figure*}

%Page2
\clearpage
\begin{figure*}
\makebox[\textwidth][c]{\begin{minipage}[b][10.5cm]{.85\textwidth}
  \vspace*{\fill}
      \includegraphics[scale=0.85]{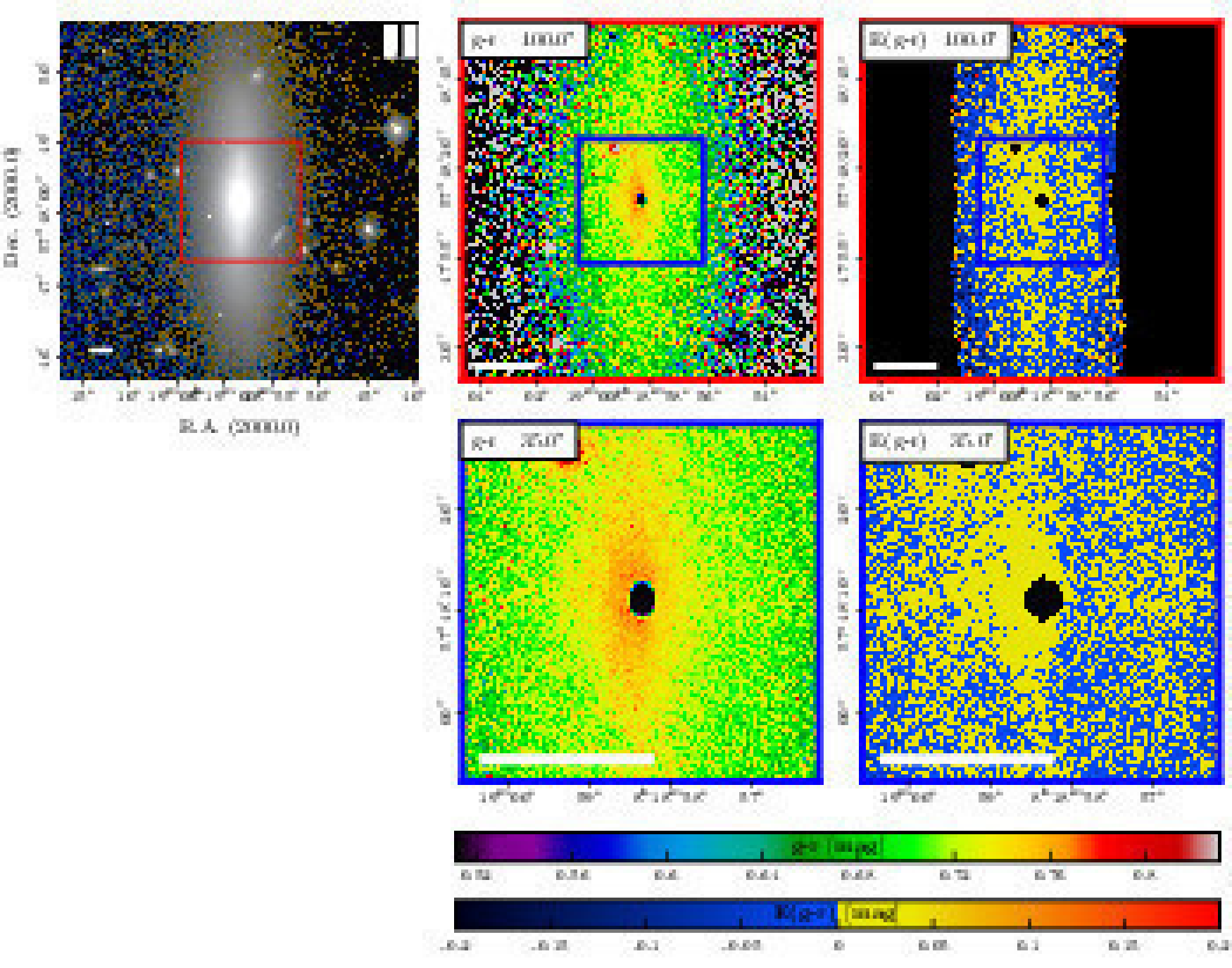}
\end{minipage}}
\makebox[\textwidth][c]{\begin{minipage}[l][-0.7cm][b]{.85\linewidth}
      \includegraphics[scale=0.55, trim={0 0.7cm 0 0},clip]{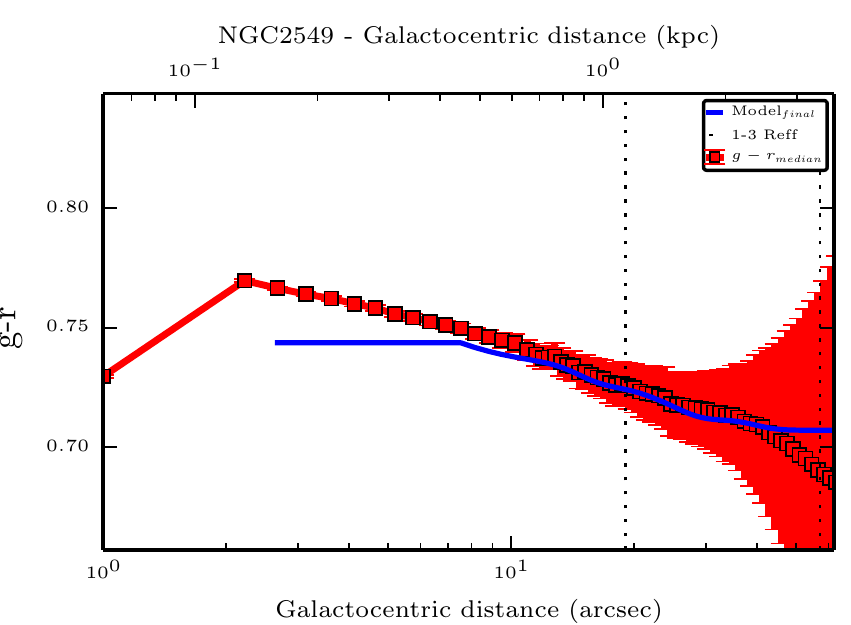}
      \includegraphics[scale=0.55]{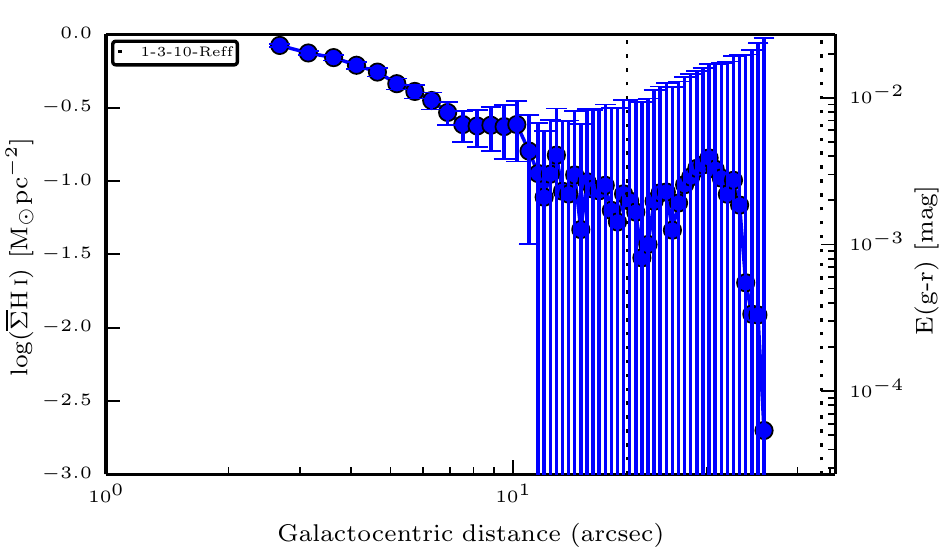}
\end{minipage}}
\caption{True image, colour map, colour excess map, and the radial profiles of NGC~2549.}
\label{fig:2549}
\end{figure*}

\begin{figure*}
\makebox[\textwidth][c]{\begin{minipage}[b][11.6cm]{.85\textwidth}
  \vspace*{\fill}
      \includegraphics[scale=0.85]{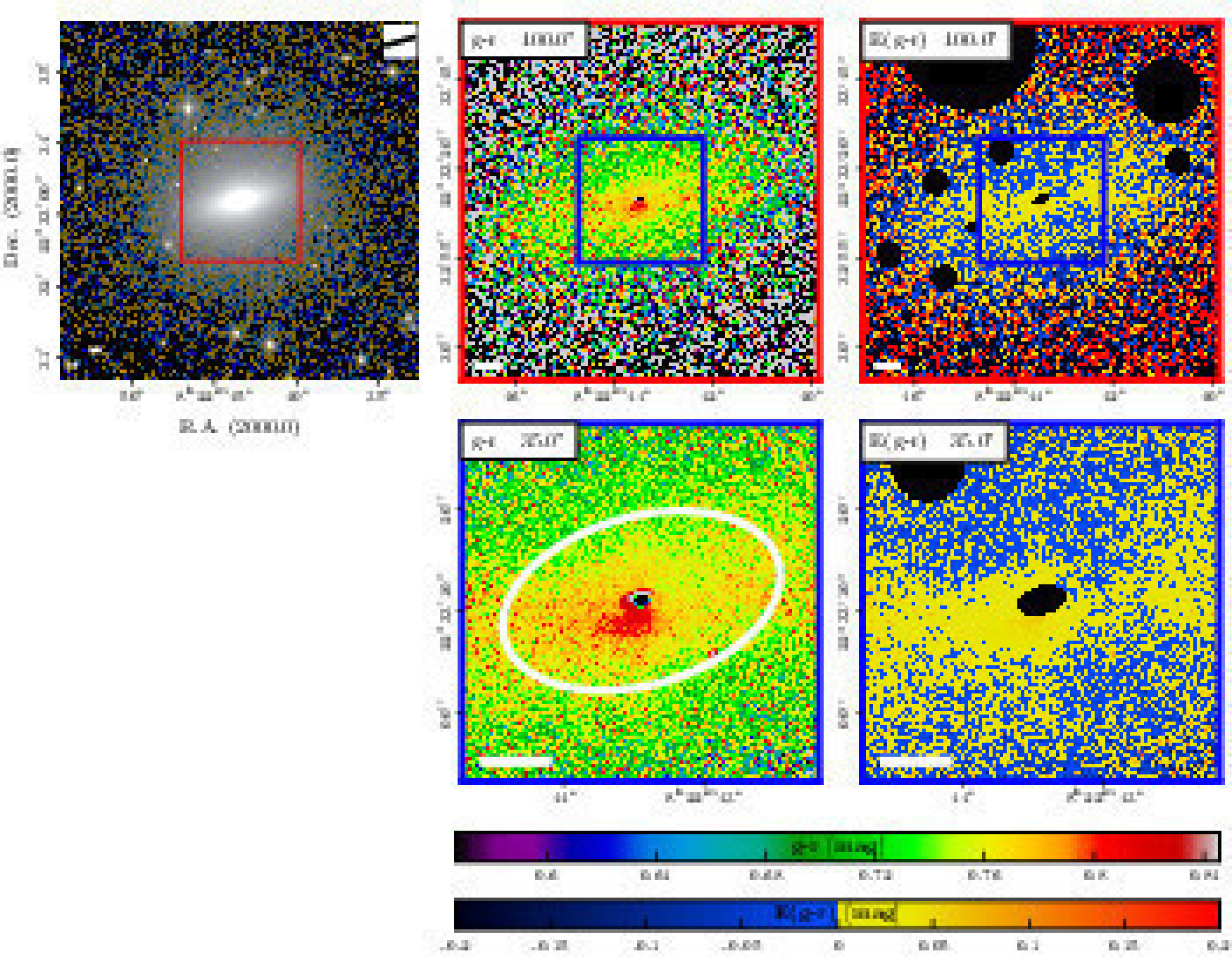}
\end{minipage}}
\makebox[\textwidth][c]{\begin{minipage}[l][-0.7cm][b]{.85\linewidth}
      \includegraphics[scale=0.55, trim={0 0.7cm 0 0},clip]{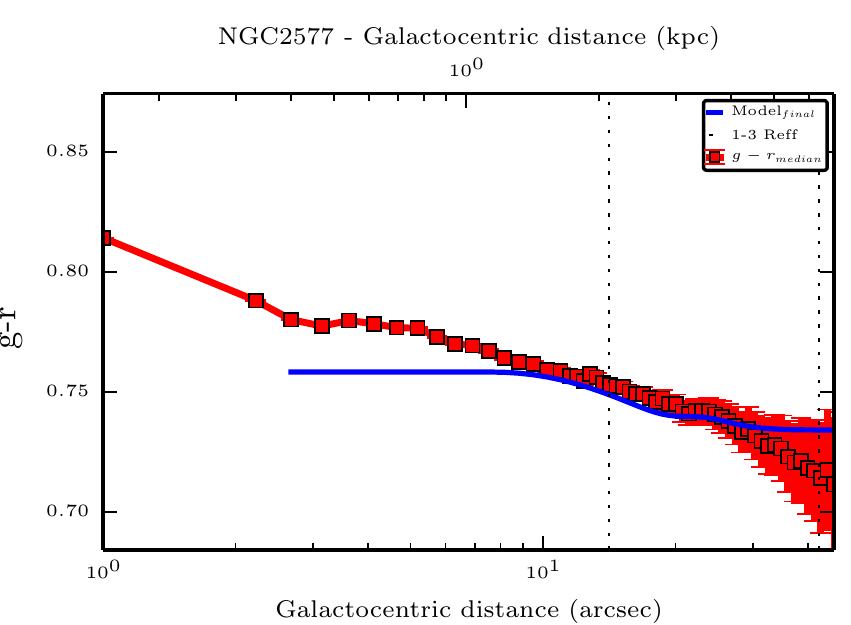}
      \includegraphics[scale=0.55]{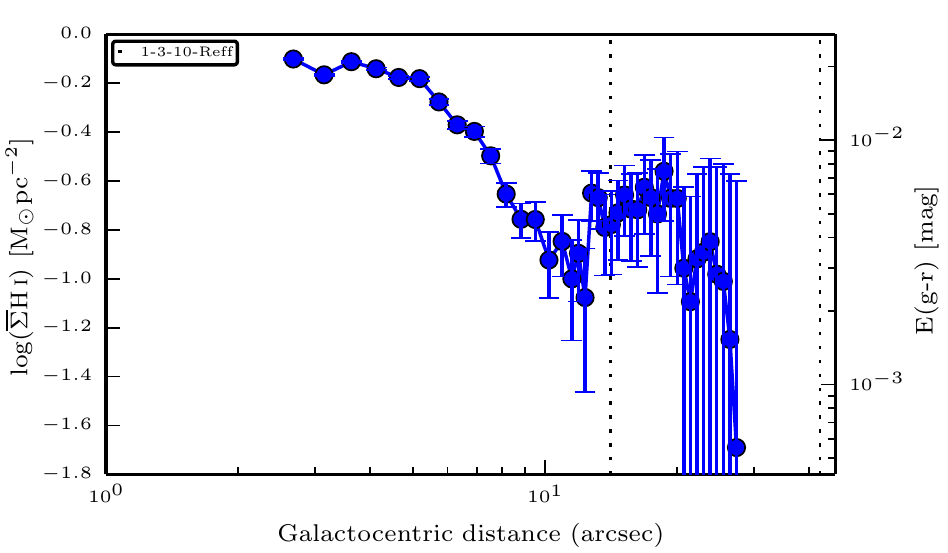}
\end{minipage}}
\caption{True image, colour map, colour excess map, and the radial profiles of NGC~2577.}
\label{fig:2577}
\end{figure*}

%Page3
\clearpage
\begin{figure*}
\makebox[\textwidth][c]{\begin{minipage}[b][10.5cm]{.85\textwidth}
  \vspace*{\fill}
      \includegraphics[scale=0.85]{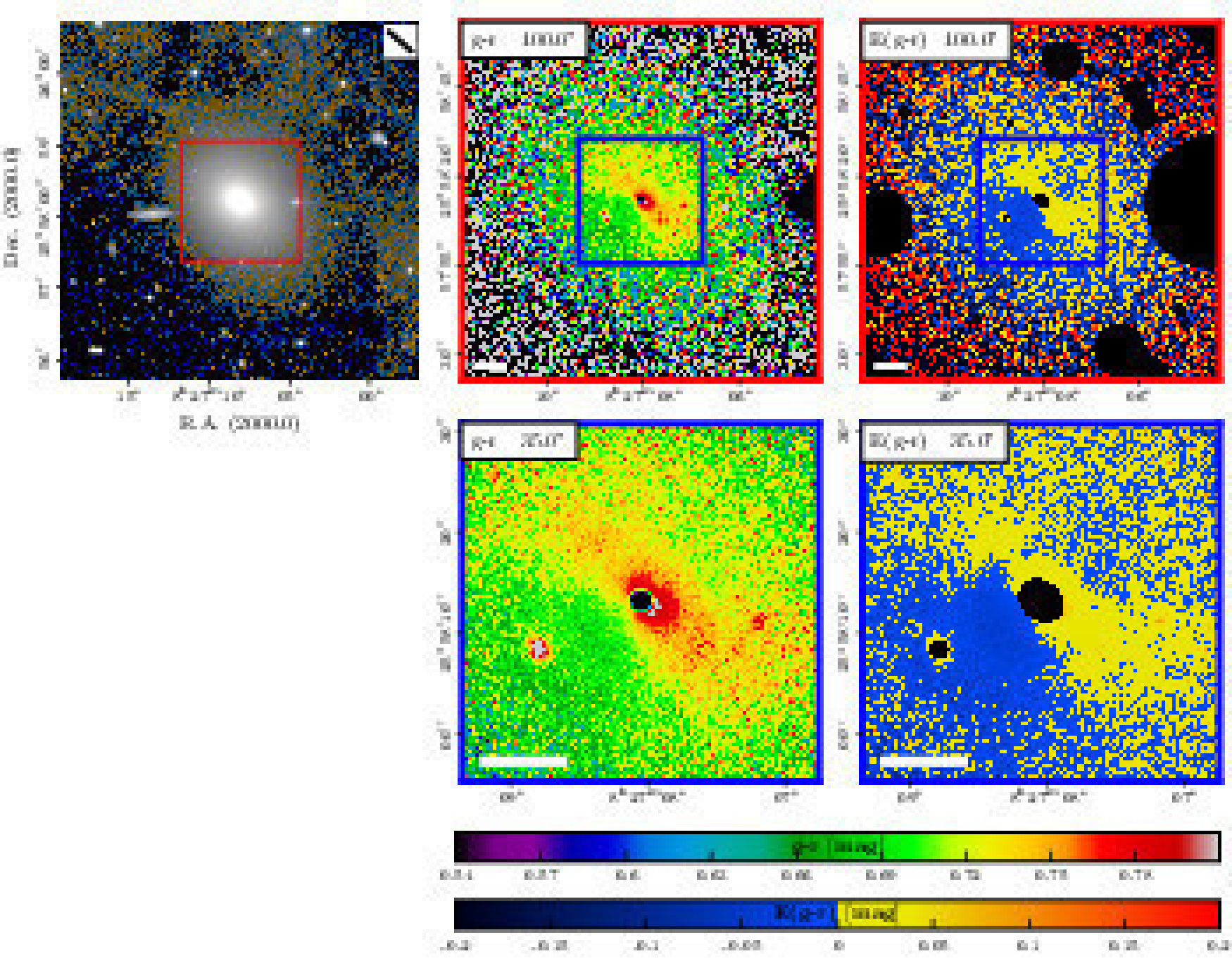}
\end{minipage}}
\makebox[\textwidth][c]{\begin{minipage}[l][-0.7cm][b]{.85\linewidth}
      \includegraphics[scale=0.55, trim={0 0.7cm 0 0},clip]{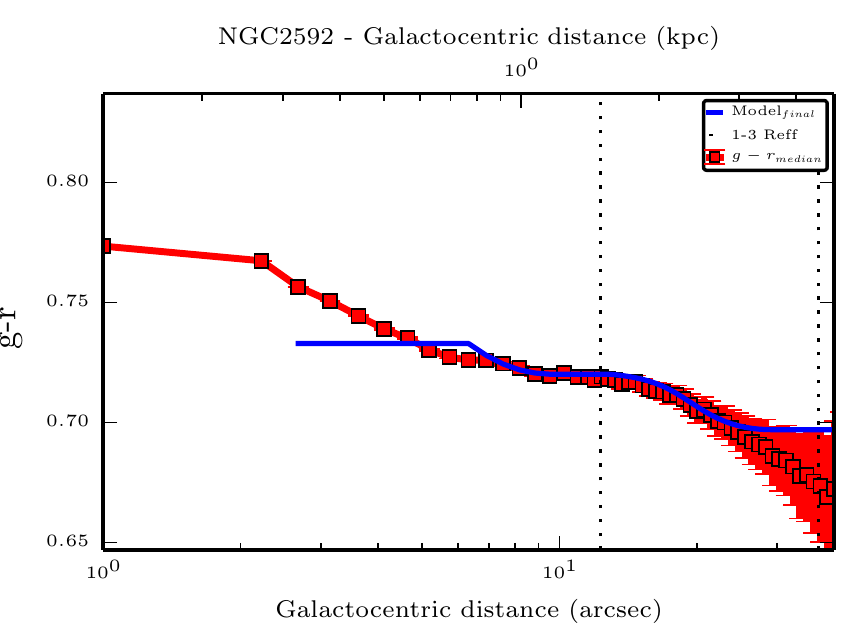}
      \includegraphics[scale=0.55]{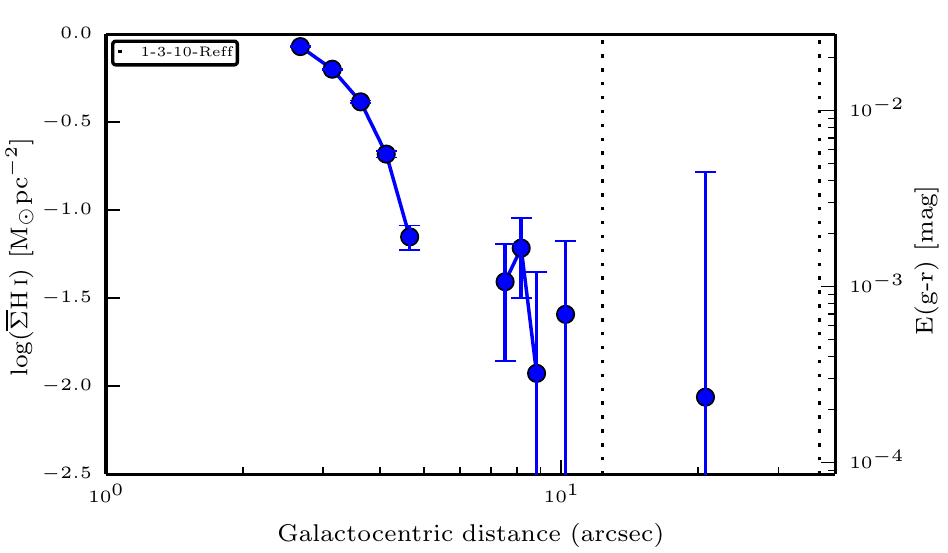}
\end{minipage}}
\caption{True image, colour map, colour excess map, and the radial profiles of NGC~2592.}
\label{fig:2592}
\end{figure*}

\begin{figure*}
\makebox[\textwidth][c]{\begin{minipage}[b][11.6cm]{.85\textwidth}
  \vspace*{\fill}
      \includegraphics[scale=0.85]{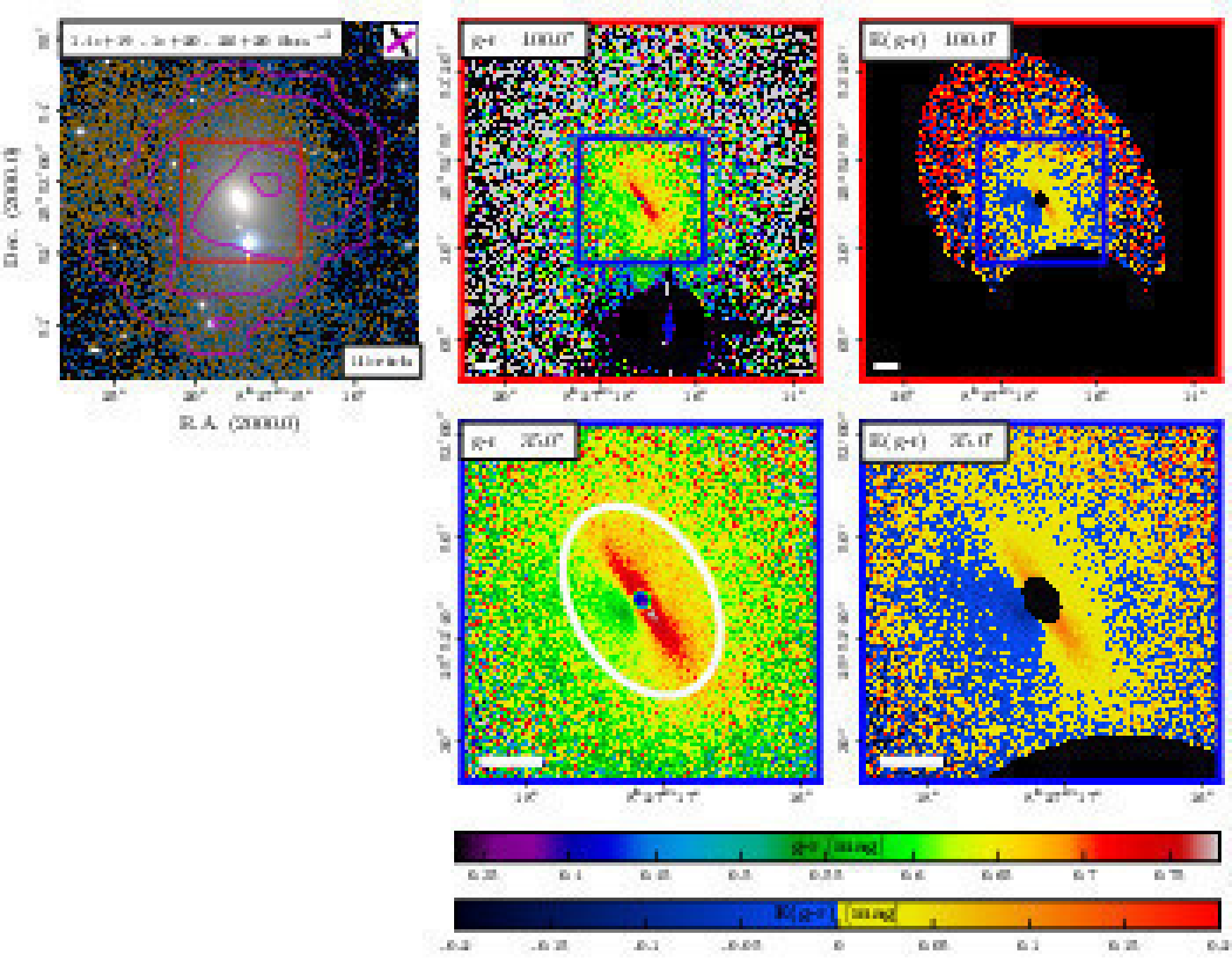}
\end{minipage}}
\makebox[\textwidth][c]{\begin{minipage}[l][-0.7cm][b]{.85\linewidth}
      \includegraphics[scale=0.55, trim={0 0.7cm 0 0},clip]{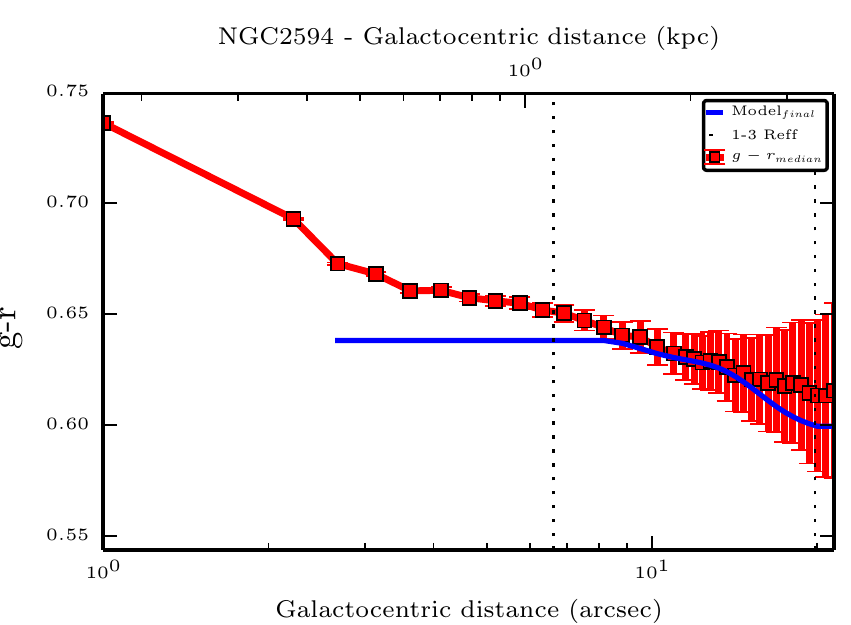}
      \includegraphics[scale=0.55]{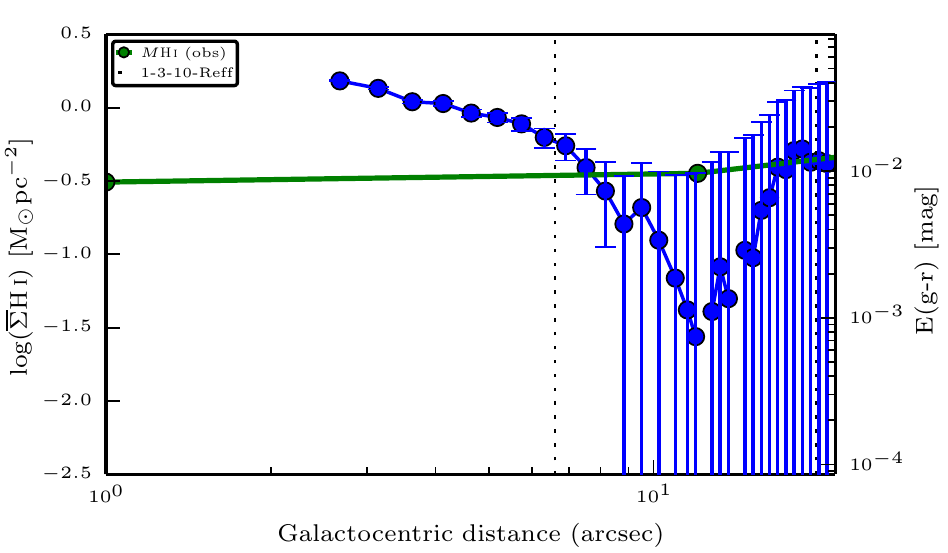}
\end{minipage}}
\caption{True image, colour map, colour excess map, and the radial profiles of NGC~2594.}
\label{fig:2594}
\end{figure*}

%Page4
\clearpage
\begin{figure*}
\makebox[\textwidth][c]{\begin{minipage}[b][10.5cm]{.85\textwidth}
  \vspace*{\fill}
      \includegraphics[scale=0.85]{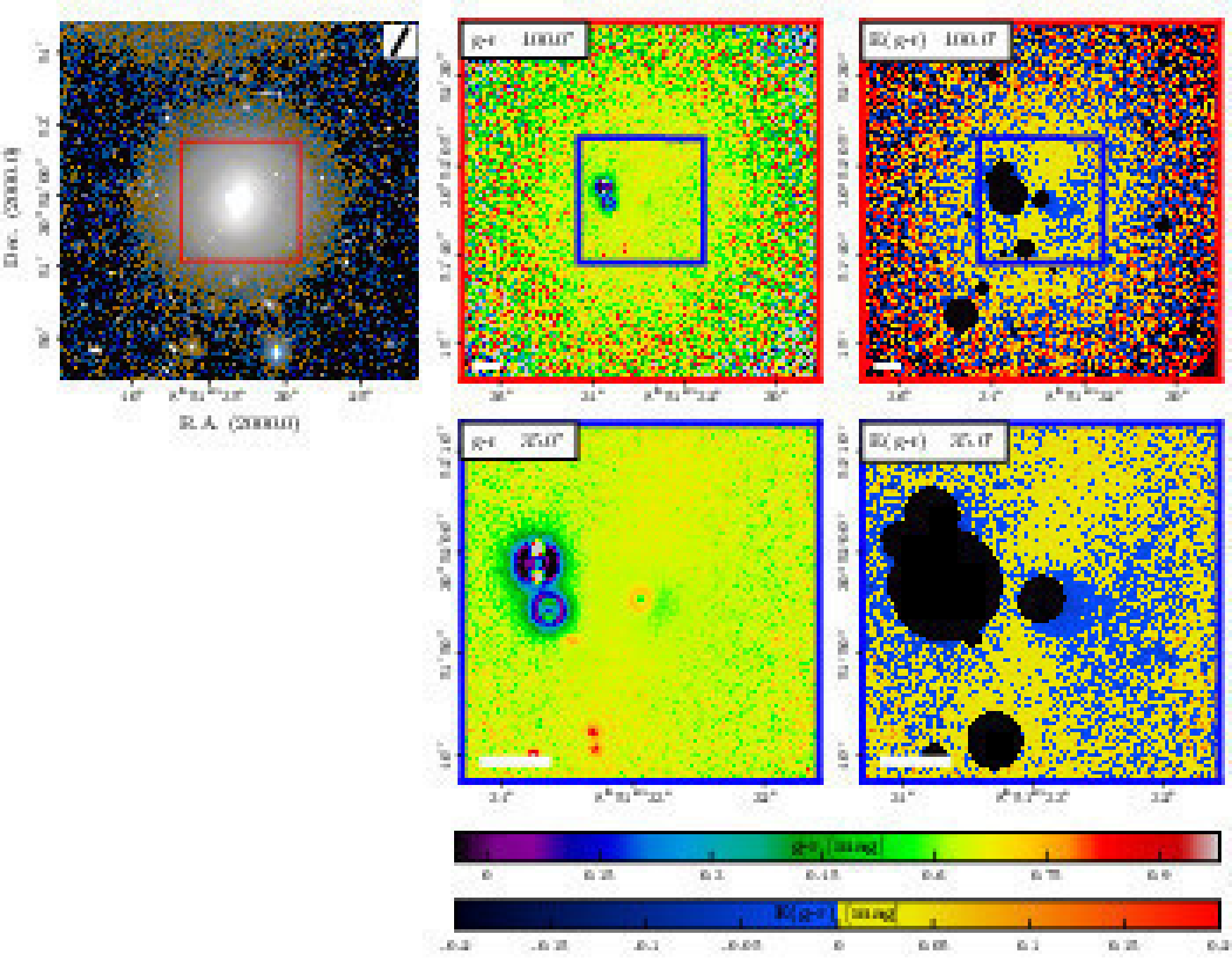}
\end{minipage}}
\makebox[\textwidth][c]{\begin{minipage}[l][-0.7cm][b]{.85\linewidth}
      \includegraphics[scale=0.55, trim={0 0.7cm 0 0},clip]{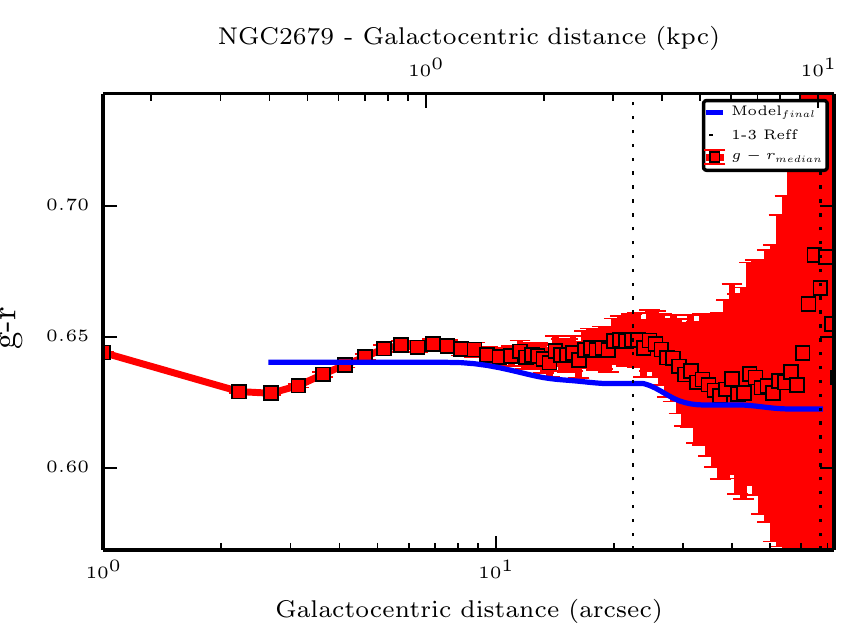}
      \includegraphics[scale=0.55]{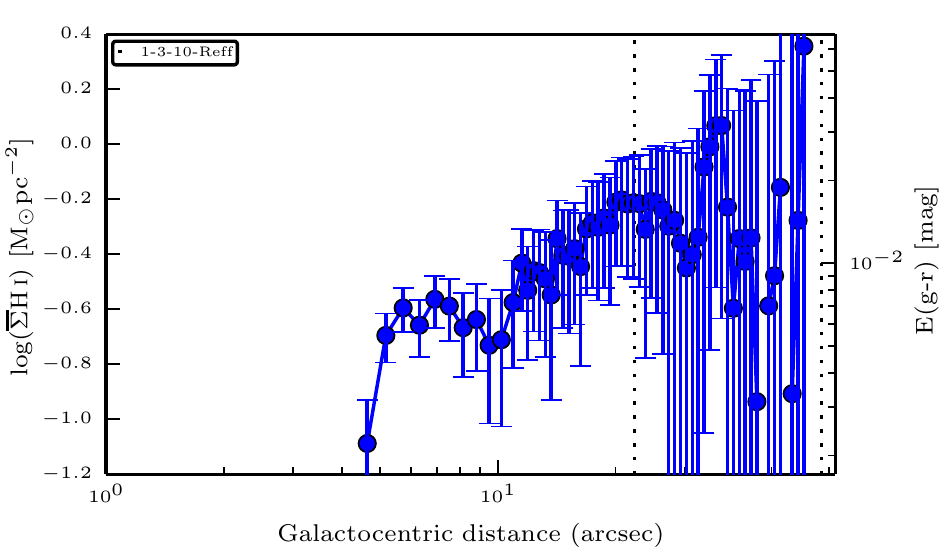}
\end{minipage}}
\caption{True image, colour map, colour excess map, and the radial profiles of NGC~2679.}
\label{fig:2679}
\end{figure*}

\begin{figure*}
\makebox[\textwidth][c]{\begin{minipage}[b][11.6cm]{.85\textwidth}
  \vspace*{\fill}
      \includegraphics[scale=0.85]{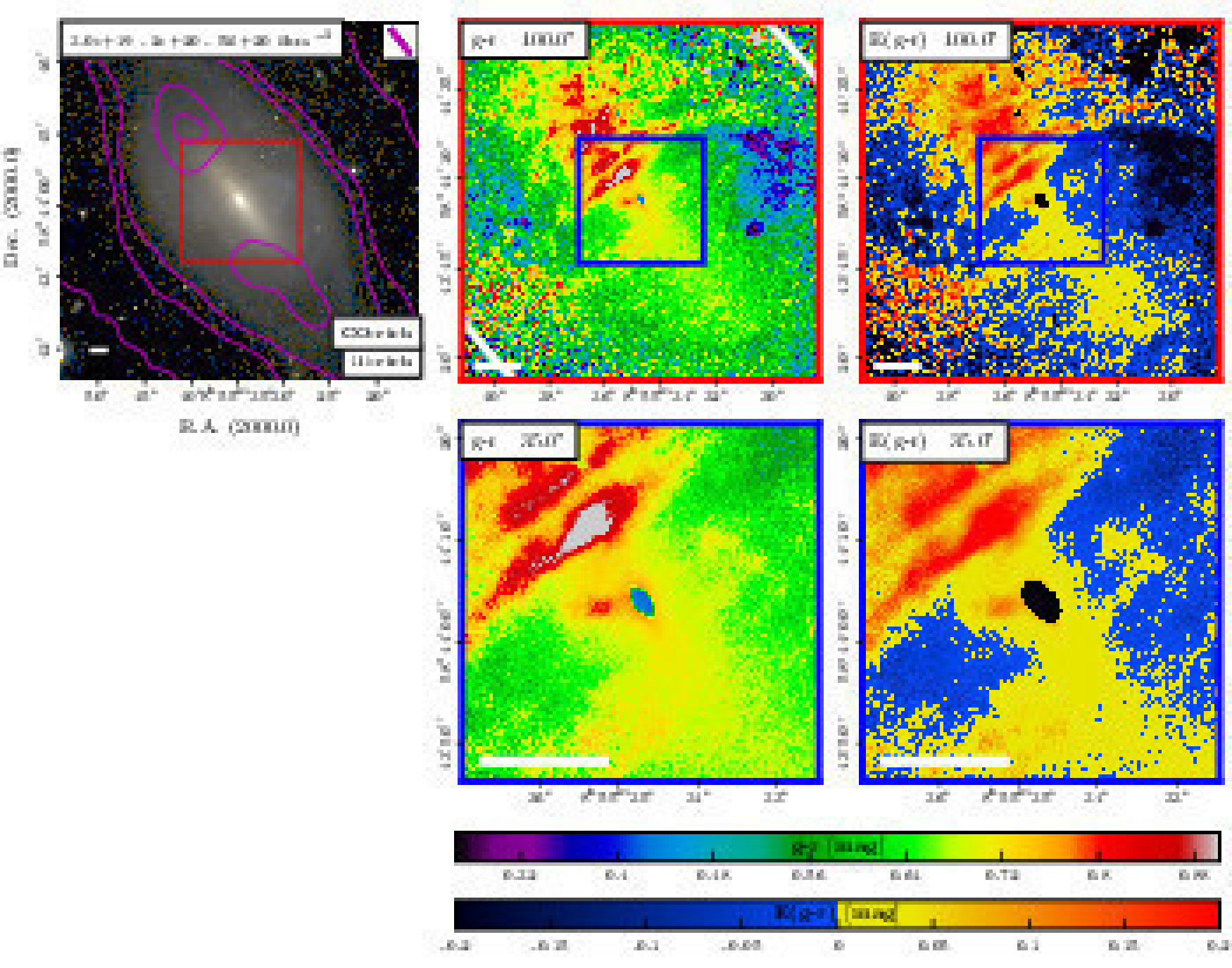}
\end{minipage}}
\makebox[\textwidth][c]{\begin{minipage}[l][-0.7cm][b]{.85\linewidth}
      \includegraphics[scale=0.55, trim={0 0.7cm 0 0},clip]{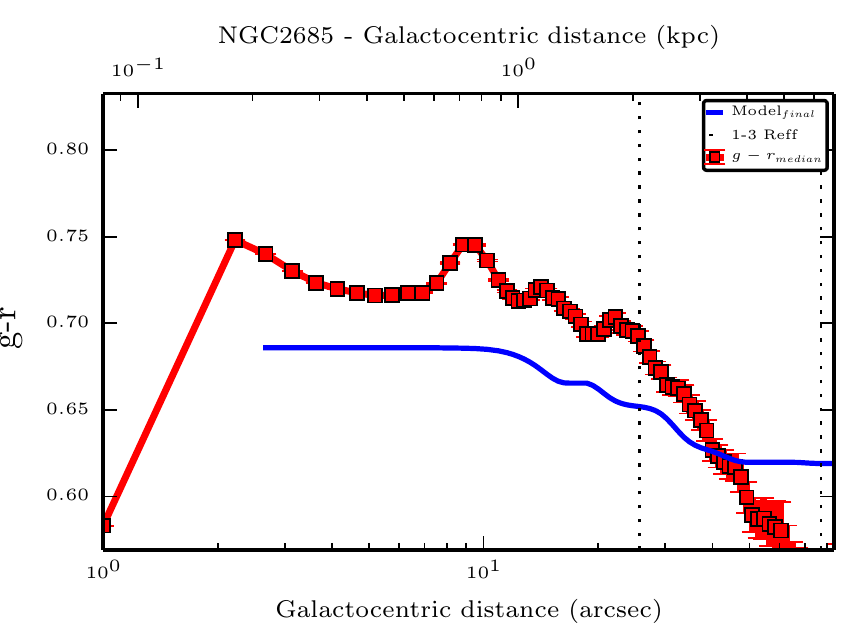}
      \includegraphics[scale=0.55]{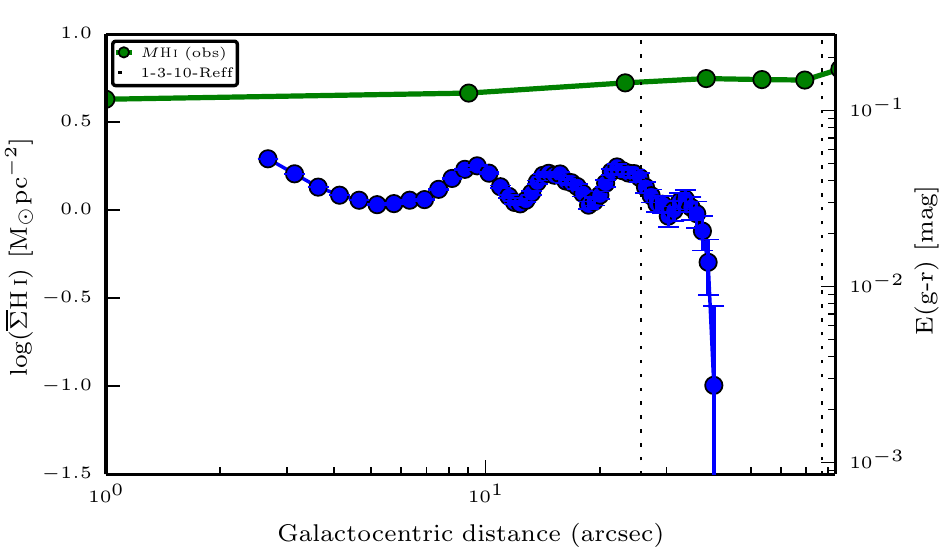}
\end{minipage}}
\caption{True image, colour map, colour excess map, and the radial profiles of NGC~2685.}
\label{fig:2685}
\end{figure*}

%Page5
\clearpage
\begin{figure*}
\makebox[\textwidth][c]{\begin{minipage}[b][10.5cm]{.85\textwidth}
  \vspace*{\fill}
      \includegraphics[scale=0.85]{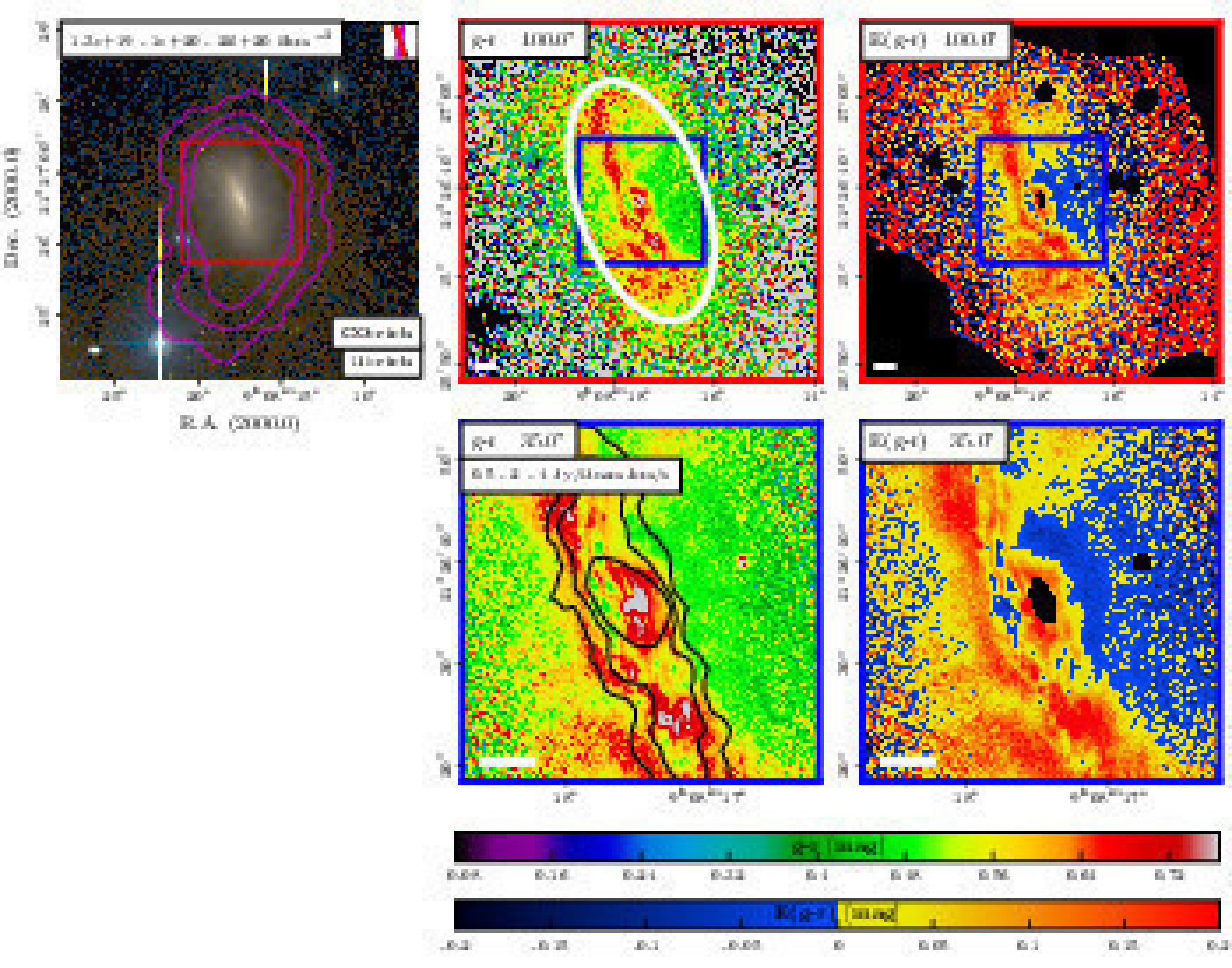}
\end{minipage}}
\makebox[\textwidth][c]{\begin{minipage}[l][-0.7cm][b]{.85\linewidth}
      \includegraphics[scale=0.55, trim={0 0.7cm 0 0},clip]{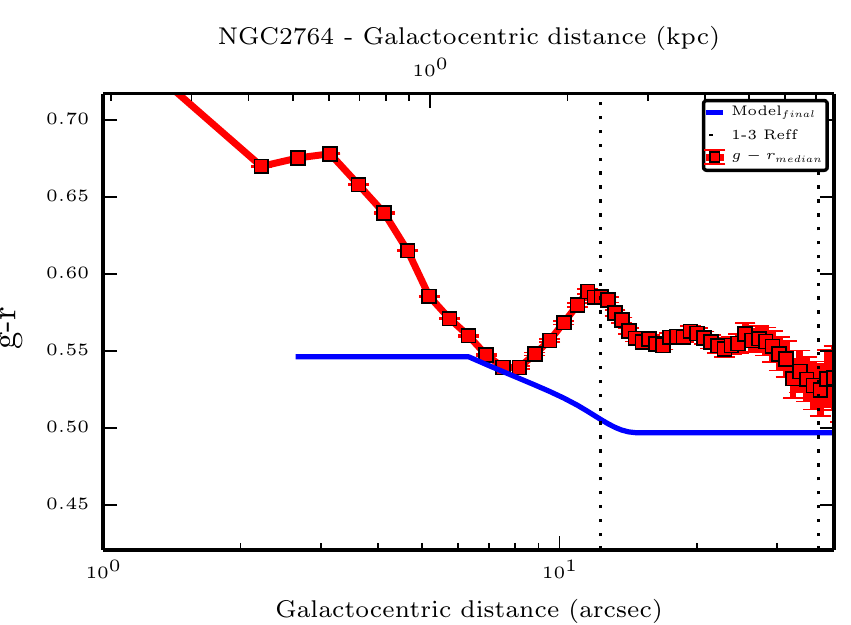}
      \includegraphics[scale=0.55]{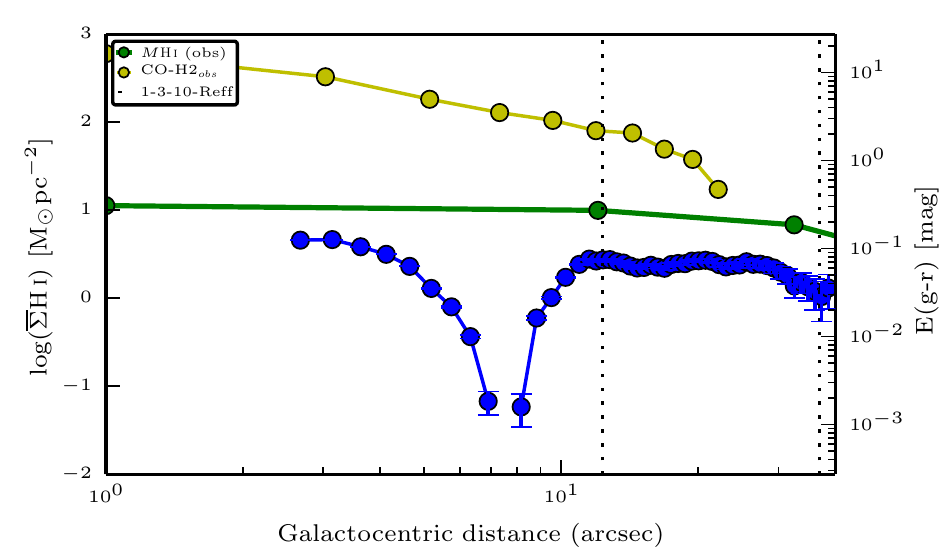}
\end{minipage}}
\caption{True image, colour map, colour excess map, and the radial profiles of NGC~2764.}
\label{fig:2764}
\end{figure*}

\begin{figure*}
\makebox[\textwidth][c]{\begin{minipage}[b][11.6cm]{.85\textwidth}
  \vspace*{\fill}
      \includegraphics[scale=0.85]{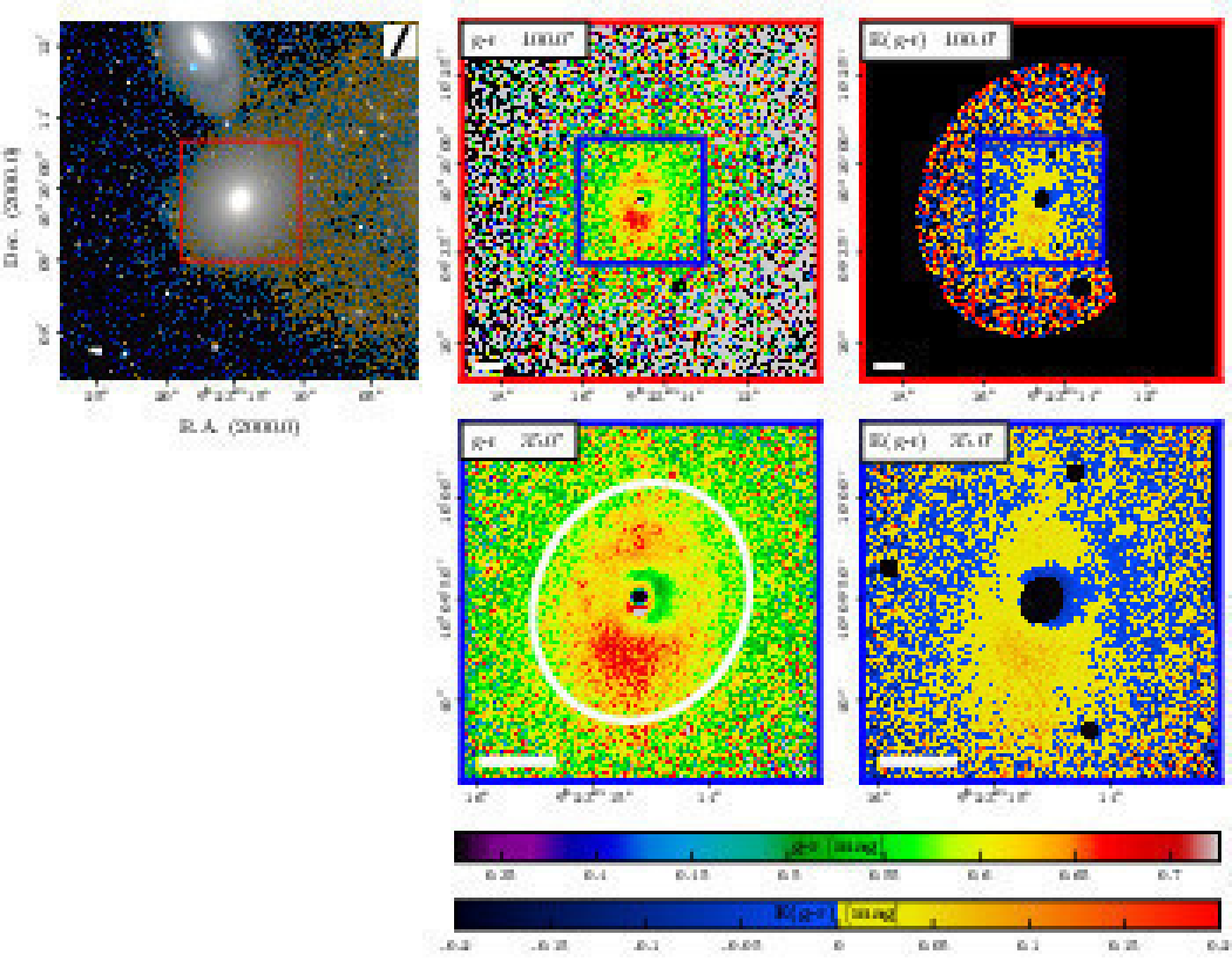}
\end{minipage}}
\makebox[\textwidth][c]{\begin{minipage}[l][-0.7cm][b]{.85\linewidth}
      \includegraphics[scale=0.55, trim={0 0.7cm 0 0},clip]{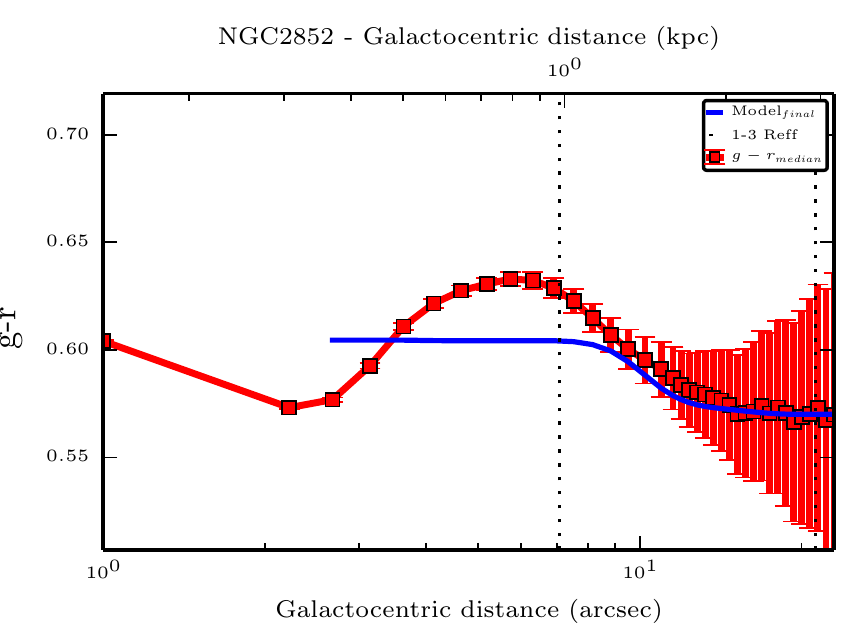}
      \includegraphics[scale=0.55]{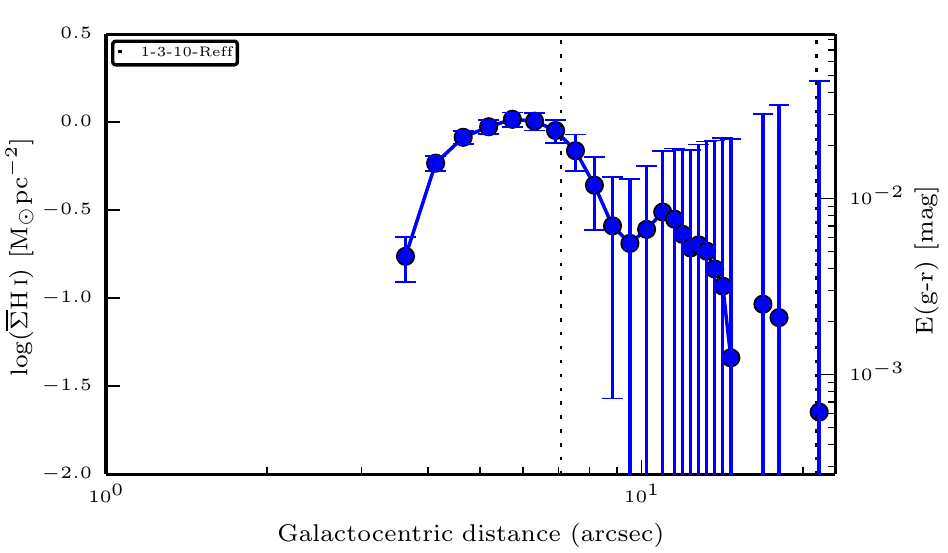}
\end{minipage}}
\caption{True image, colour map, colour excess map, and the radial profiles of NGC~2852.}
\label{fig:2852}
\end{figure*}

%Page6
\clearpage
\begin{figure*}
\makebox[\textwidth][c]{\begin{minipage}[b][10.5cm]{.85\textwidth}
  \vspace*{\fill}
      \includegraphics[scale=0.85]{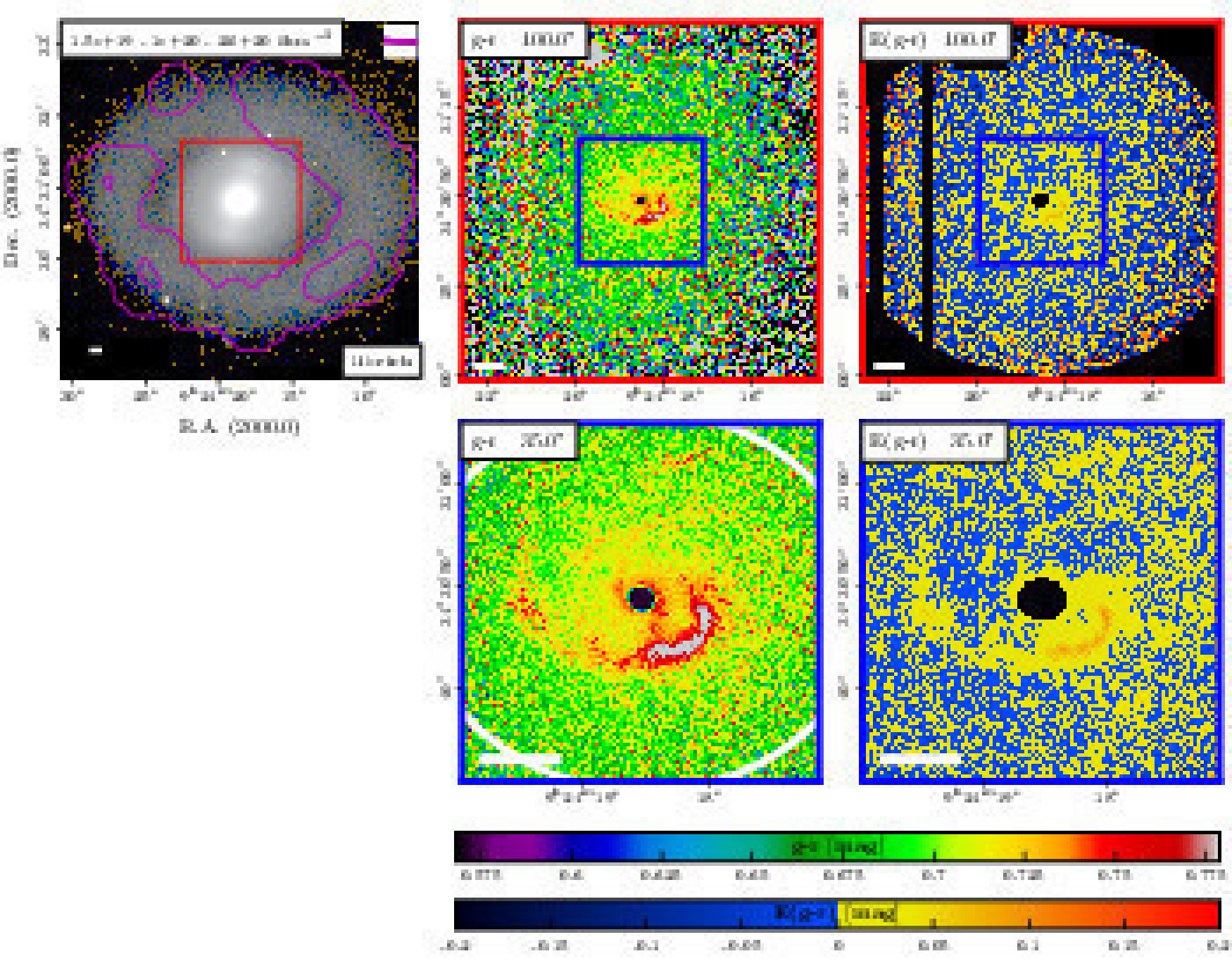}
\end{minipage}}
\makebox[\textwidth][c]{\begin{minipage}[l][-0.7cm][b]{.85\linewidth}
      \includegraphics[scale=0.55, trim={0 0.7cm 0 0},clip]{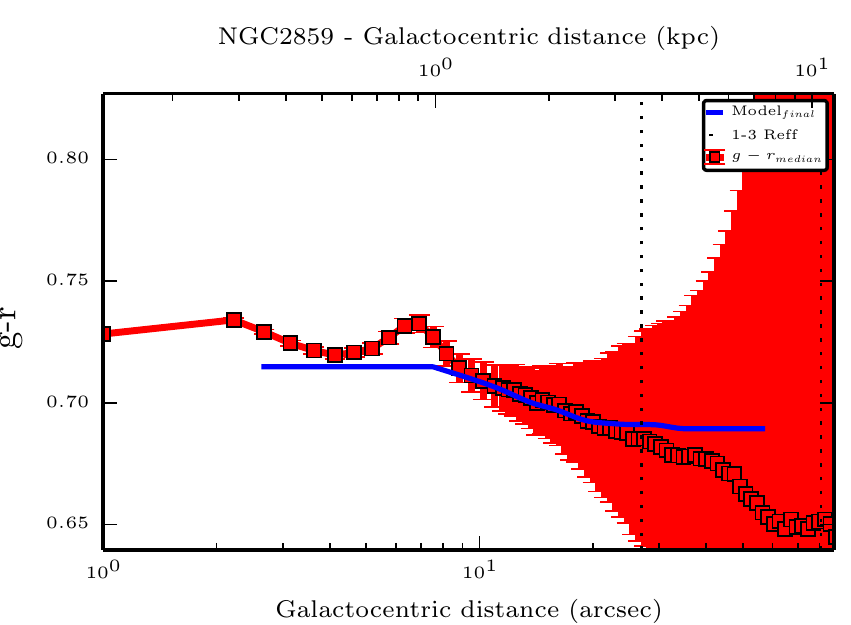}
      \includegraphics[scale=0.55]{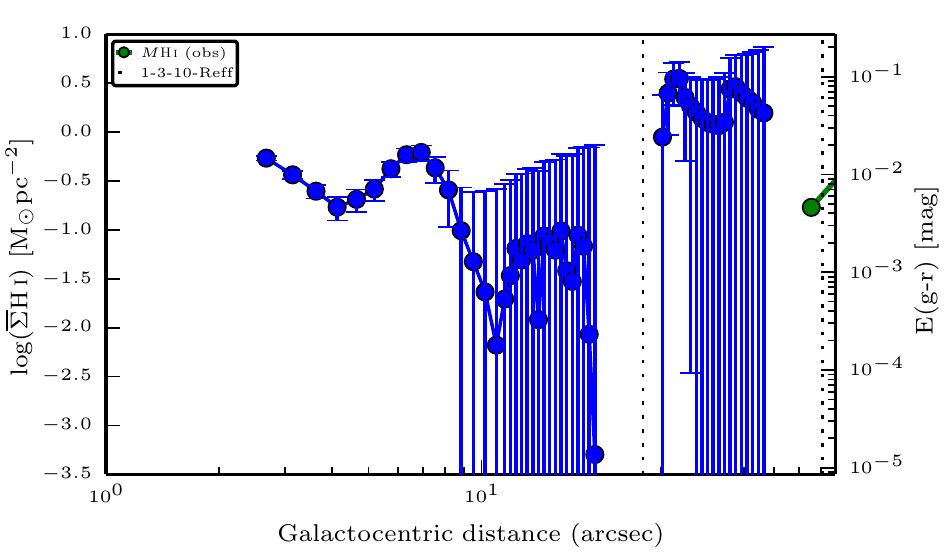}
\end{minipage}}
\caption{True image, colour map, colour excess map, and the radial profiles of NGC~2859.}
\label{fig:2859}
\end{figure*}

\begin{figure*}
\makebox[\textwidth][c]{\begin{minipage}[b][11.6cm]{.85\textwidth}
  \vspace*{\fill}
      \includegraphics[scale=0.85]{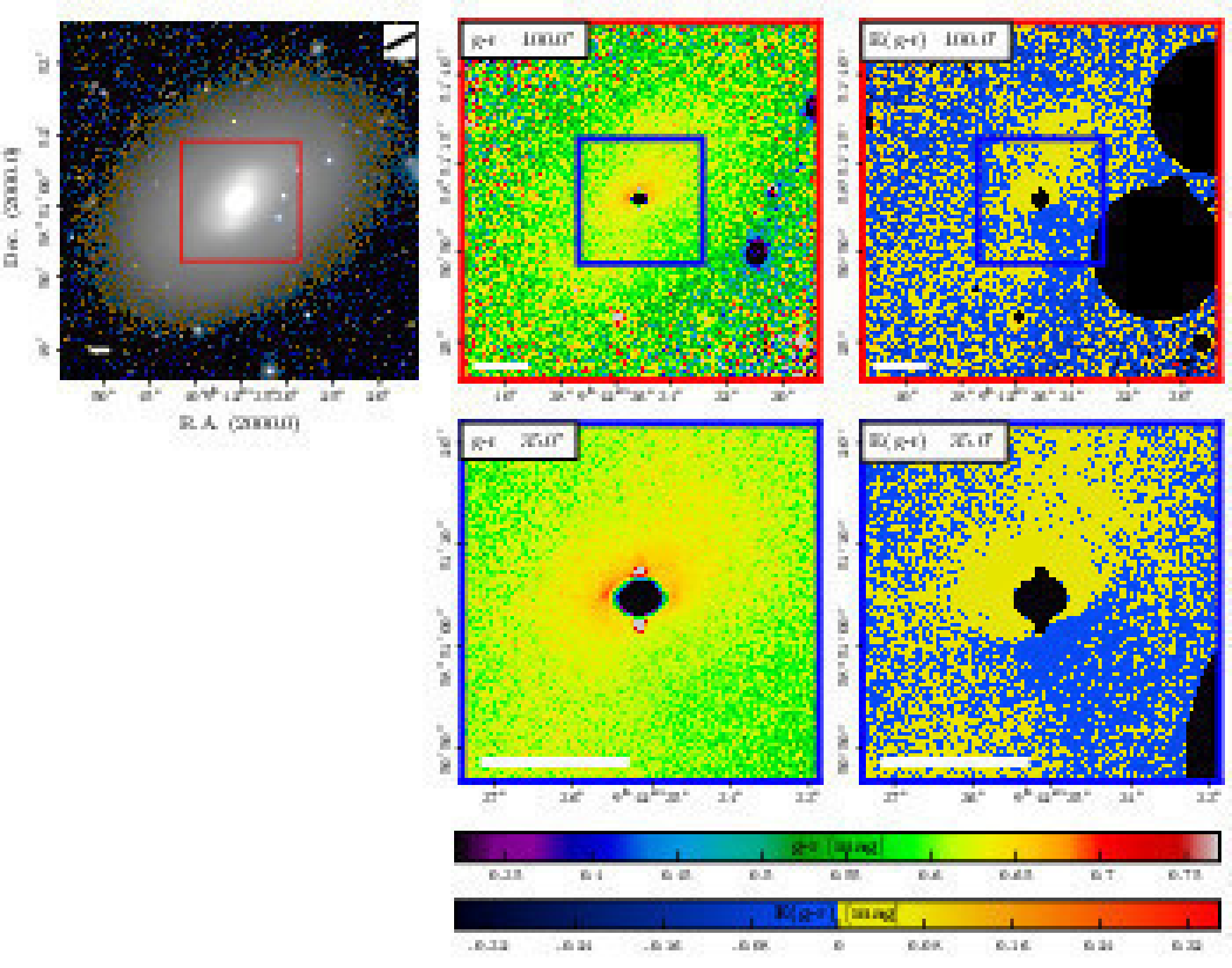}
\end{minipage}}
\makebox[\textwidth][c]{\begin{minipage}[l][-0.7cm][b]{.85\linewidth}
      \includegraphics[scale=0.55, trim={0 0.7cm 0 0},clip]{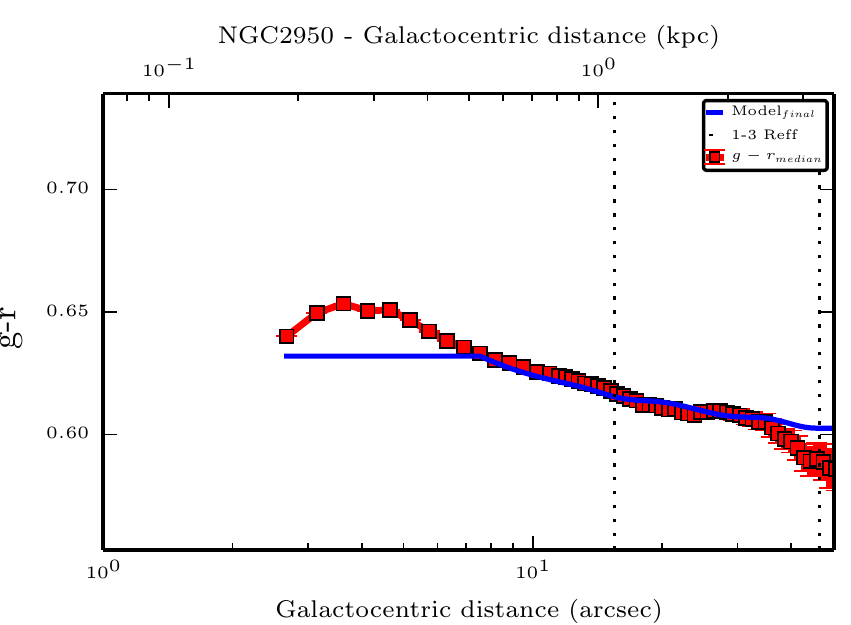}
      \includegraphics[scale=0.55]{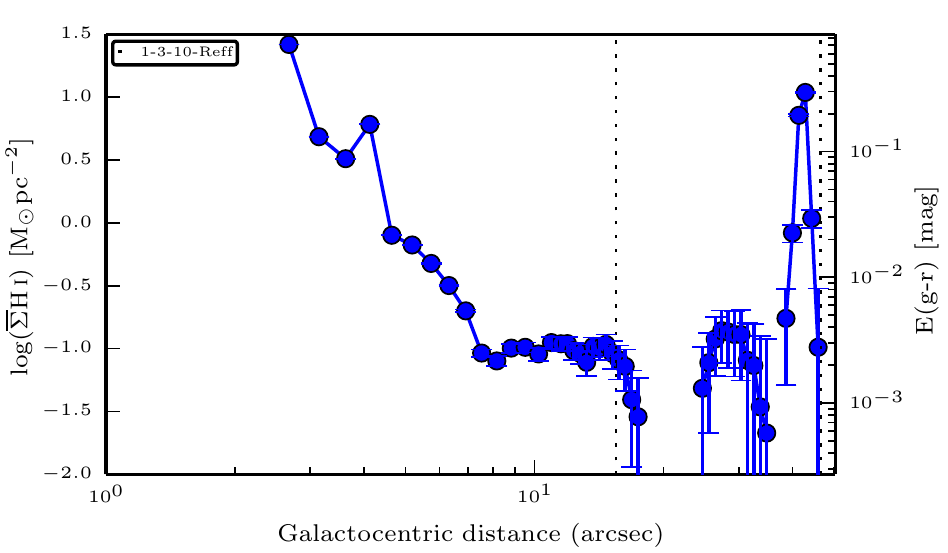}
\end{minipage}}
\caption{True image, colour map, colour excess map, and the radial profiles of NGC~2950.}
\label{fig:2950}
\end{figure*}

%Page7
\clearpage
\begin{figure*}
\makebox[\textwidth][c]{\begin{minipage}[b][10.5cm]{.85\textwidth}
  \vspace*{\fill}
      \includegraphics[scale=0.85]{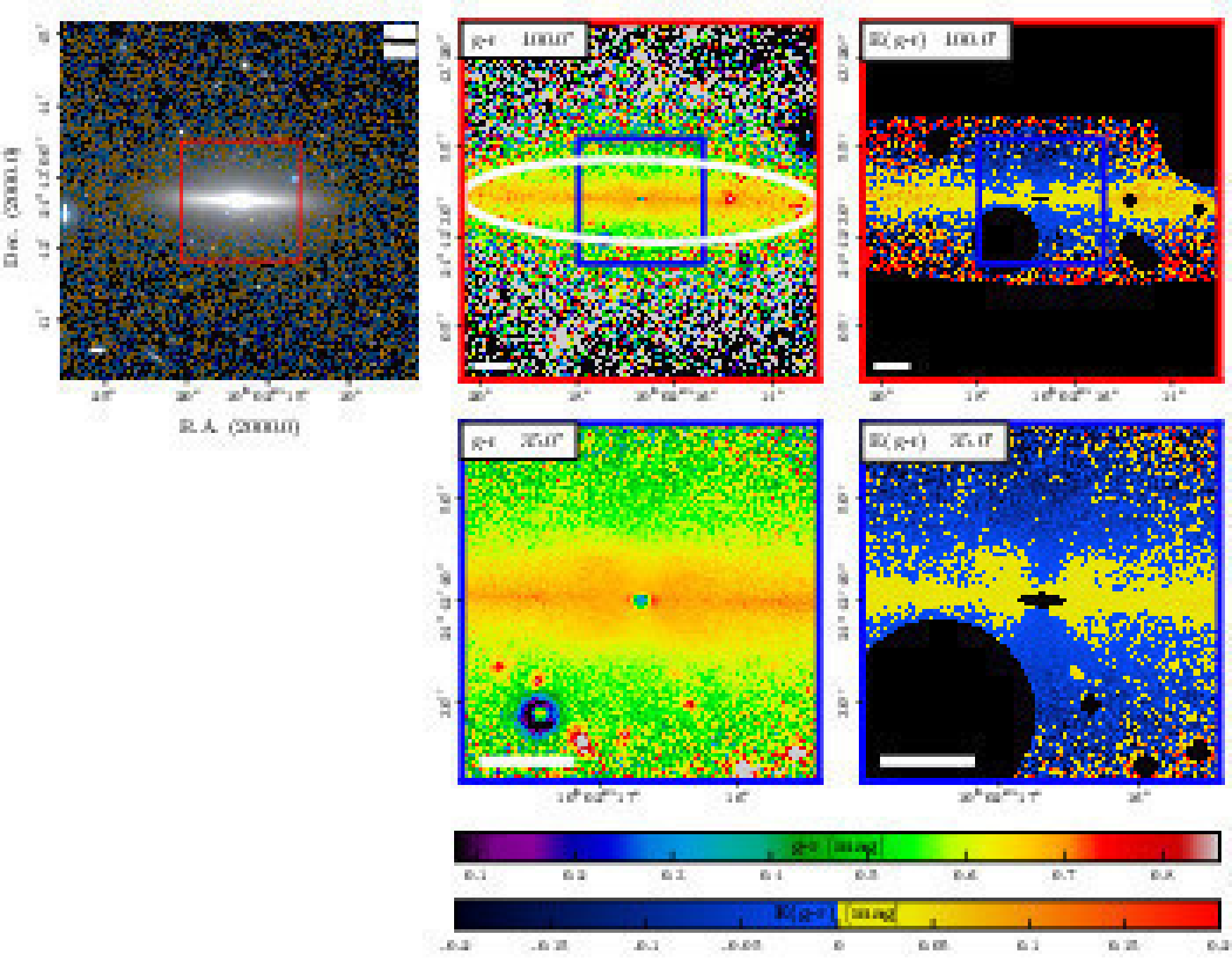}
\end{minipage}}
\makebox[\textwidth][c]{\begin{minipage}[l][-0.7cm][b]{.85\linewidth}
      \includegraphics[scale=0.55, trim={0 0.7cm 0 0},clip]{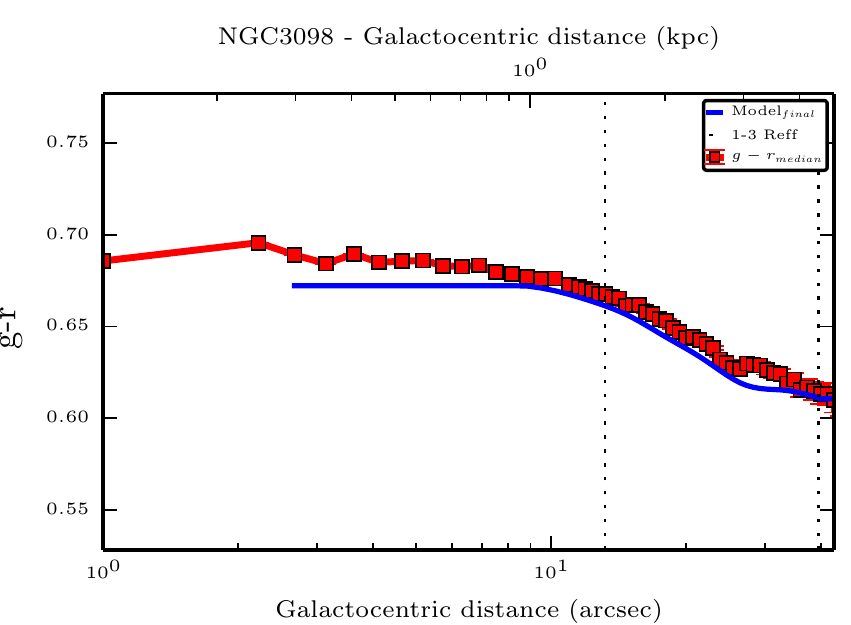}
      \includegraphics[scale=0.55]{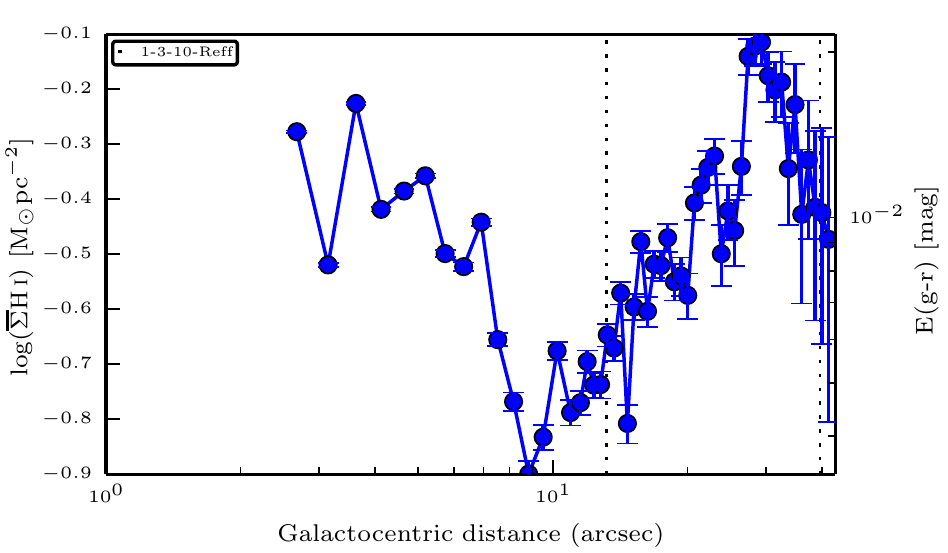}
\end{minipage}}
\caption{True image, colour map, colour excess map, and the radial profiles of NGC~3098.}
\label{fig:3098}
\end{figure*}

\begin{figure*}
\makebox[\textwidth][c]{\begin{minipage}[b][11.6cm]{.85\textwidth}
  \vspace*{\fill}
      \includegraphics[scale=0.85]{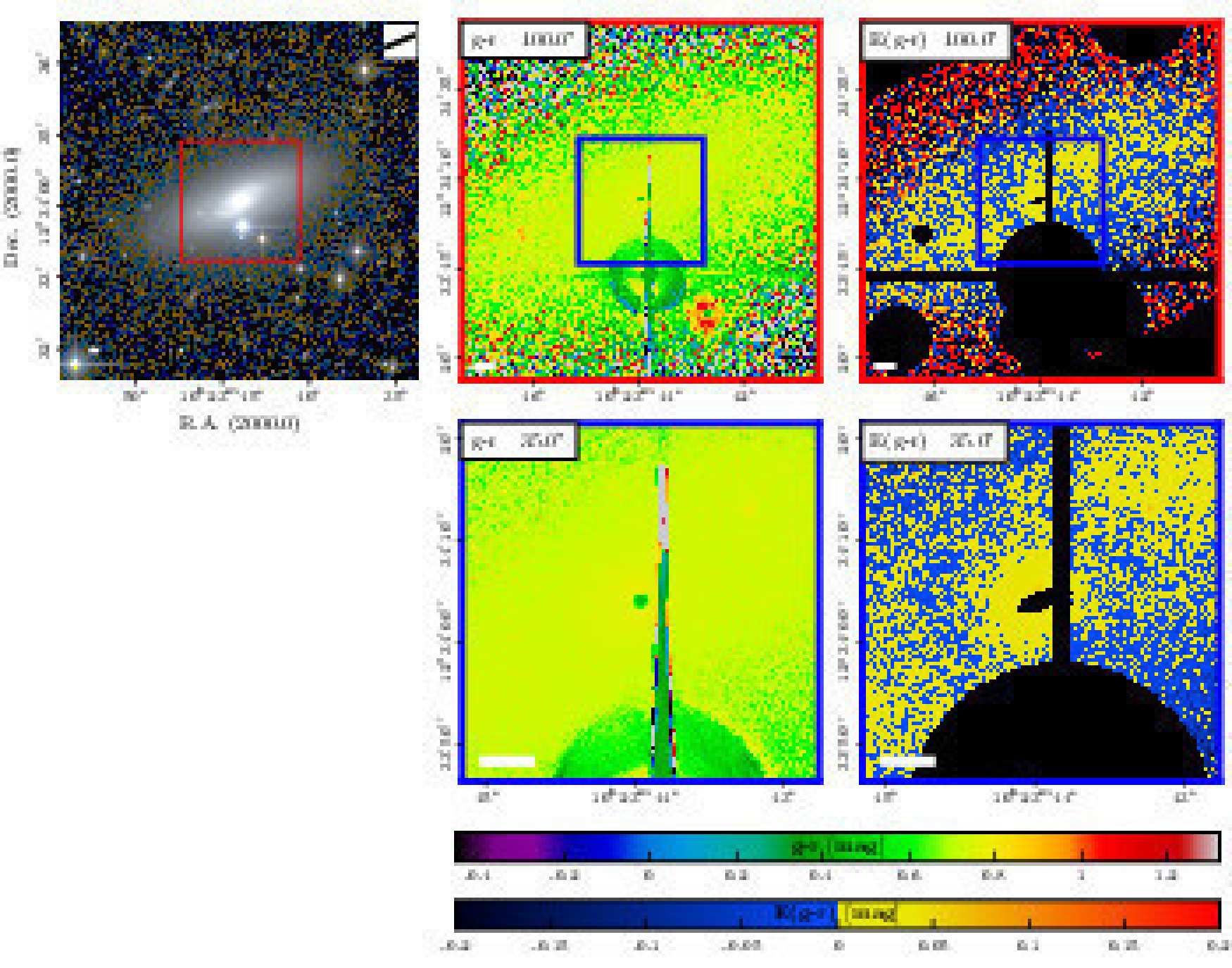}
\end{minipage}}
\makebox[\textwidth][c]{\begin{minipage}[l][-0.7cm][b]{.85\linewidth}
      \includegraphics[scale=0.55, trim={0 0.7cm 0 0},clip]{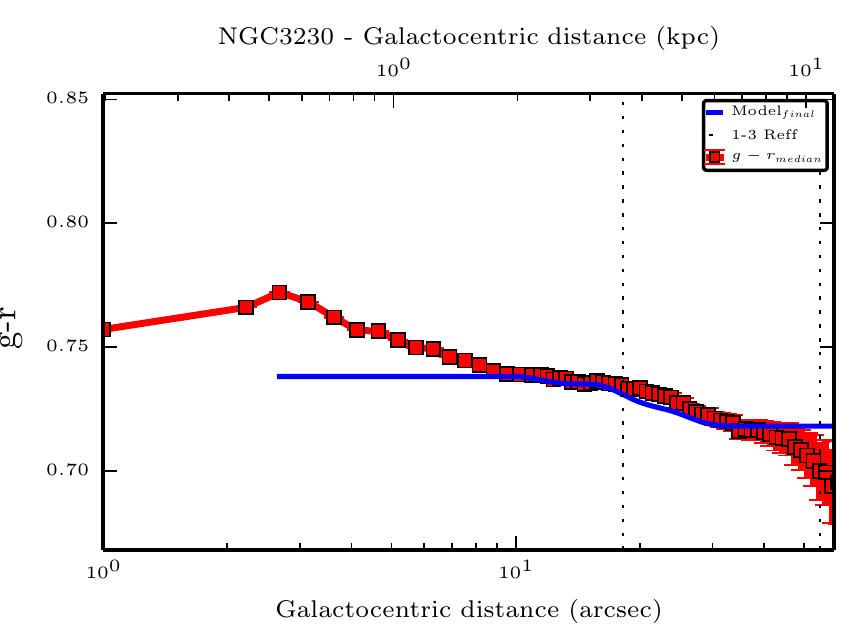}
      \includegraphics[scale=0.55]{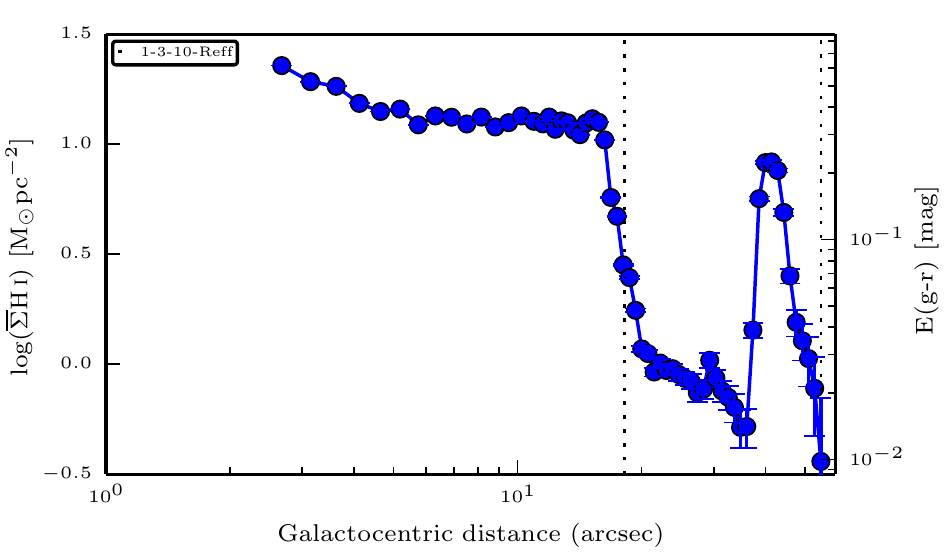}
\end{minipage}}
\caption{True image, colour map, colour excess map, and the radial profiles of NGC~3230.}
\label{fig:3230}
\end{figure*}

%Page8
\clearpage
\begin{figure*}
\makebox[\textwidth][c]{\begin{minipage}[b][10.5cm]{.85\textwidth}
  \vspace*{\fill}
      \includegraphics[scale=0.85]{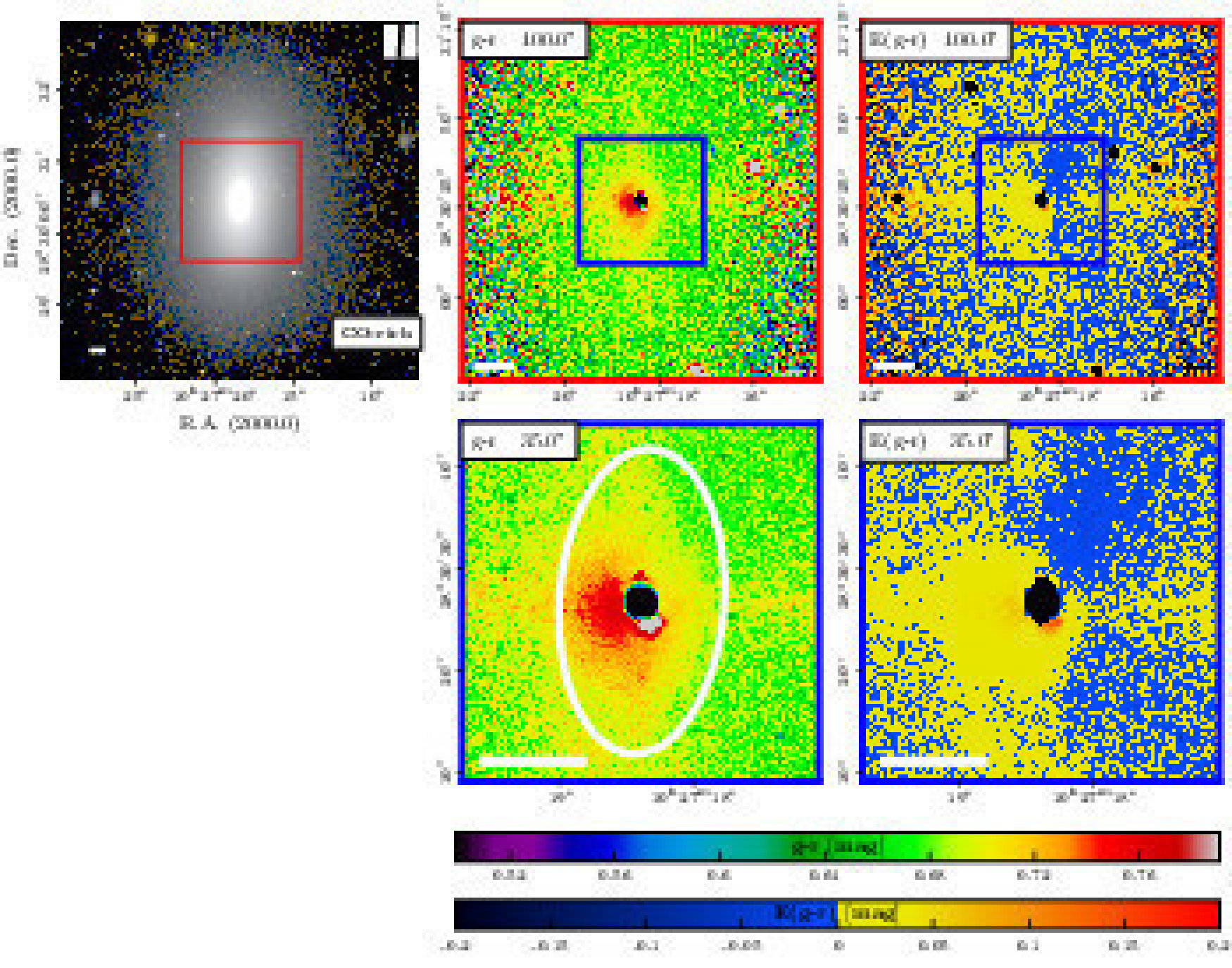}
\end{minipage}}
\makebox[\textwidth][c]{\begin{minipage}[l][-0.7cm][b]{.85\linewidth}
      \includegraphics[scale=0.55, trim={0 0.7cm 0 0},clip]{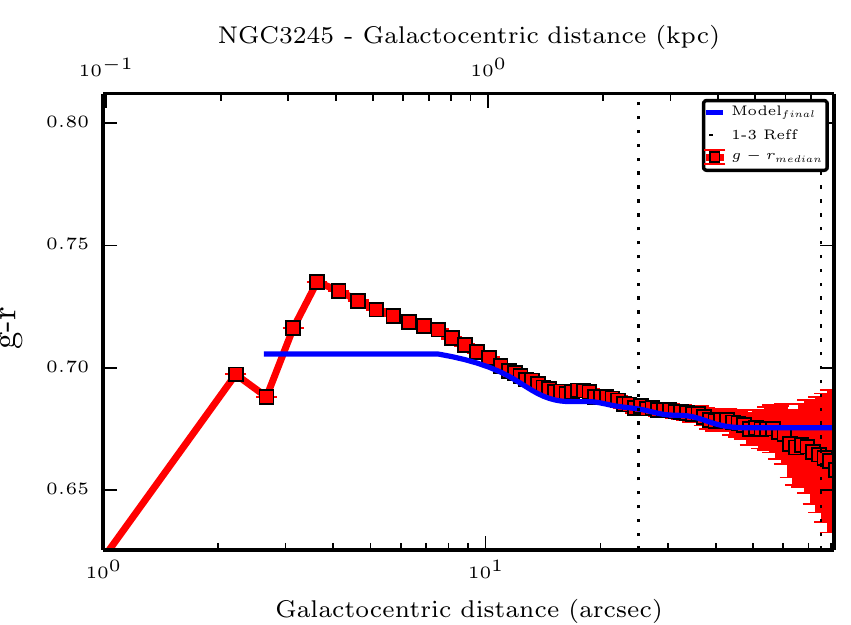}
      \includegraphics[scale=0.55]{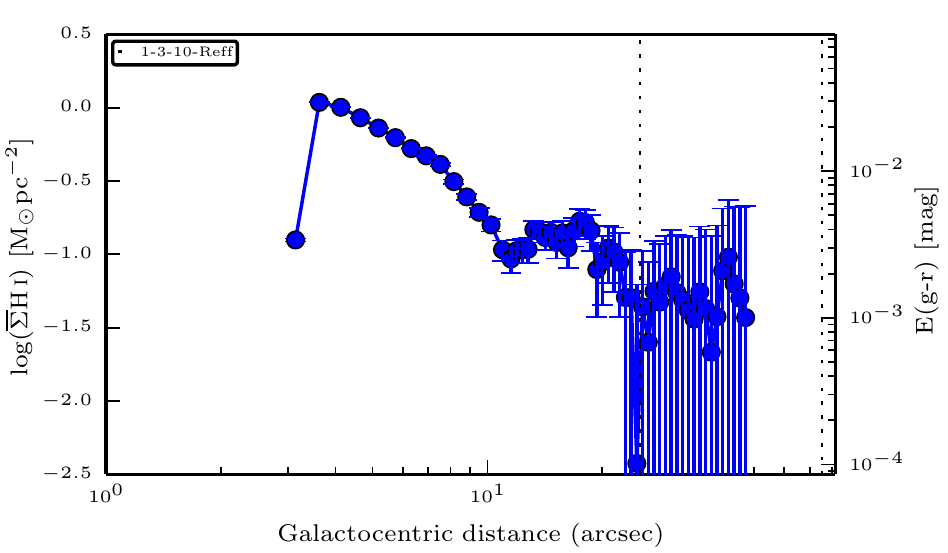}
\end{minipage}}
\caption{True image, colour map, colour excess map, and the radial profiles of NGC~3245.}
\label{fig:3245}
\end{figure*}

\begin{figure*}
\makebox[\textwidth][c]{\begin{minipage}[b][11.6cm]{.85\textwidth}
  \vspace*{\fill}
      \includegraphics[scale=0.85]{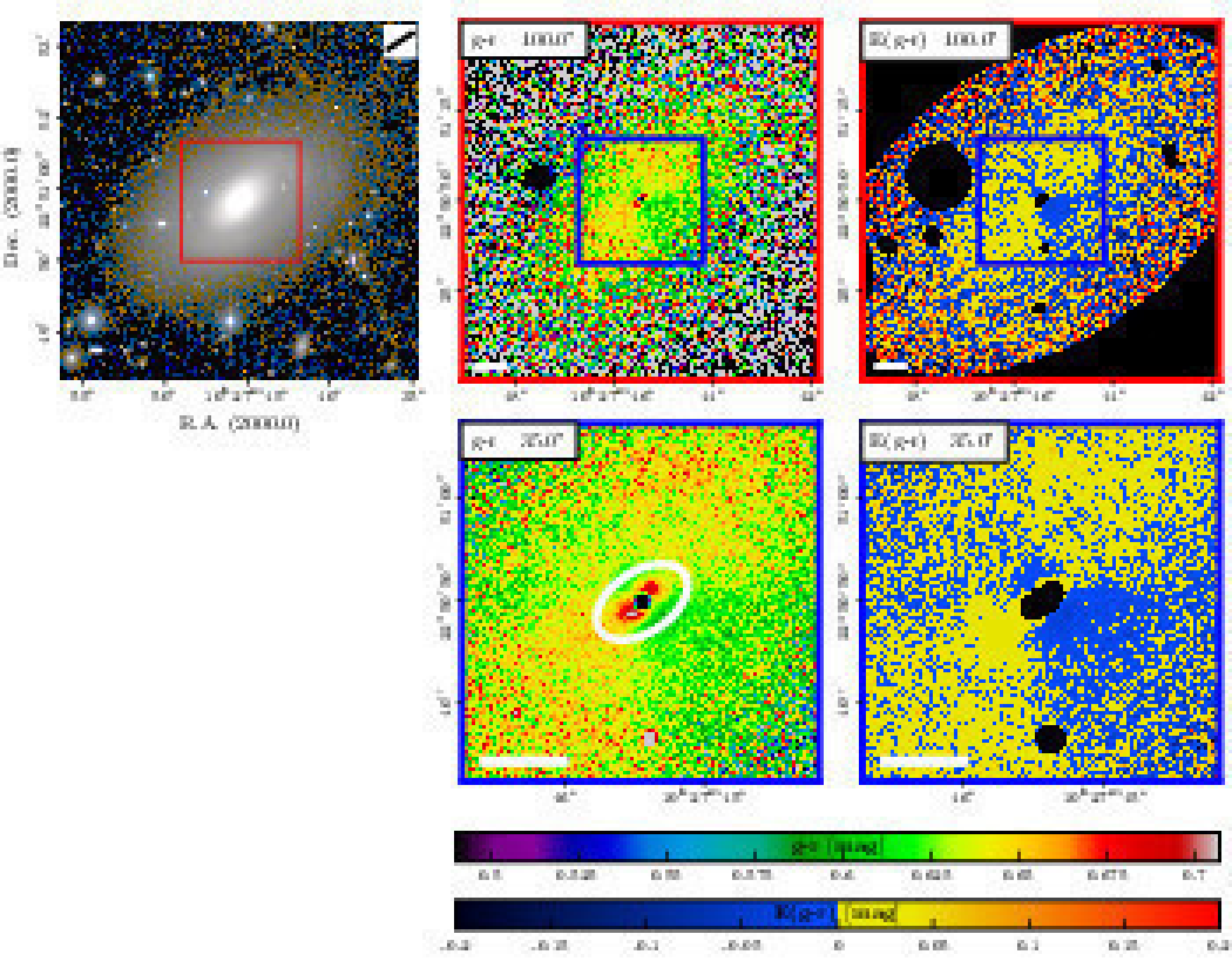}
\end{minipage}}
\makebox[\textwidth][c]{\begin{minipage}[l][-0.7cm][b]{.85\linewidth}
      \includegraphics[scale=0.55, trim={0 0.7cm 0 0},clip]{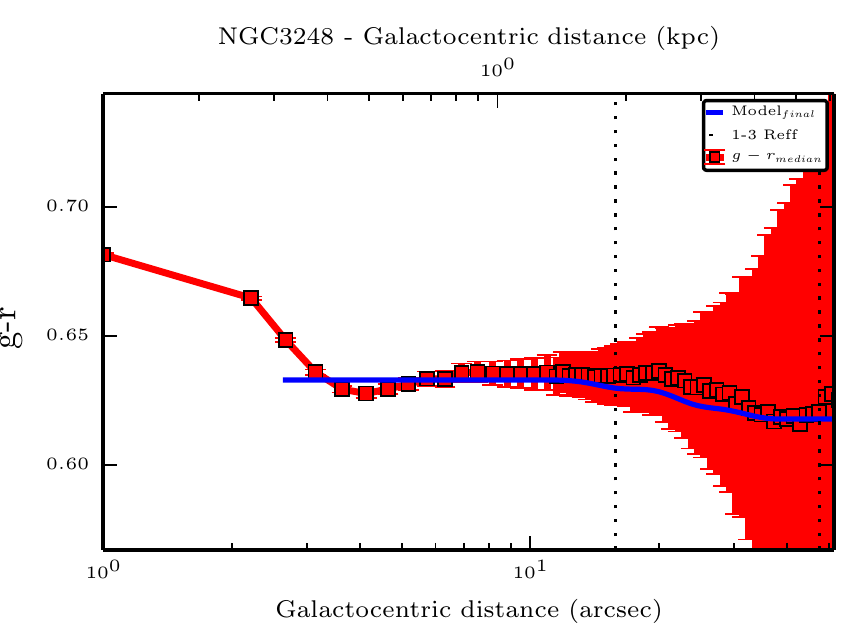}
      \includegraphics[scale=0.55]{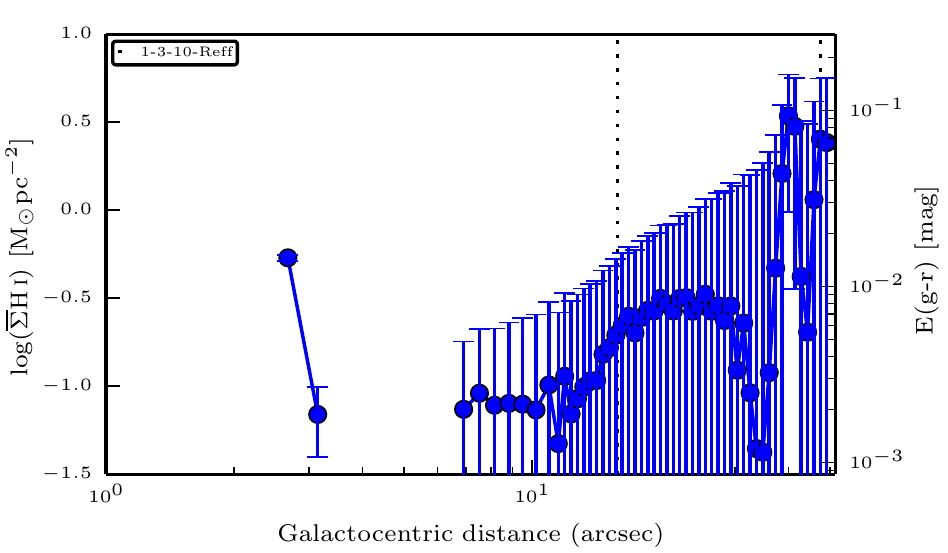}
\end{minipage}}
\caption{True image, colour map, colour excess map, and the radial profiles of NGC~3248.}
\label{fig:3248}
\end{figure*}

%Page9
\clearpage
\begin{figure*}
\makebox[\textwidth][c]{\begin{minipage}[b][10.5cm]{.85\textwidth}
  \vspace*{\fill}
      \includegraphics[scale=0.85]{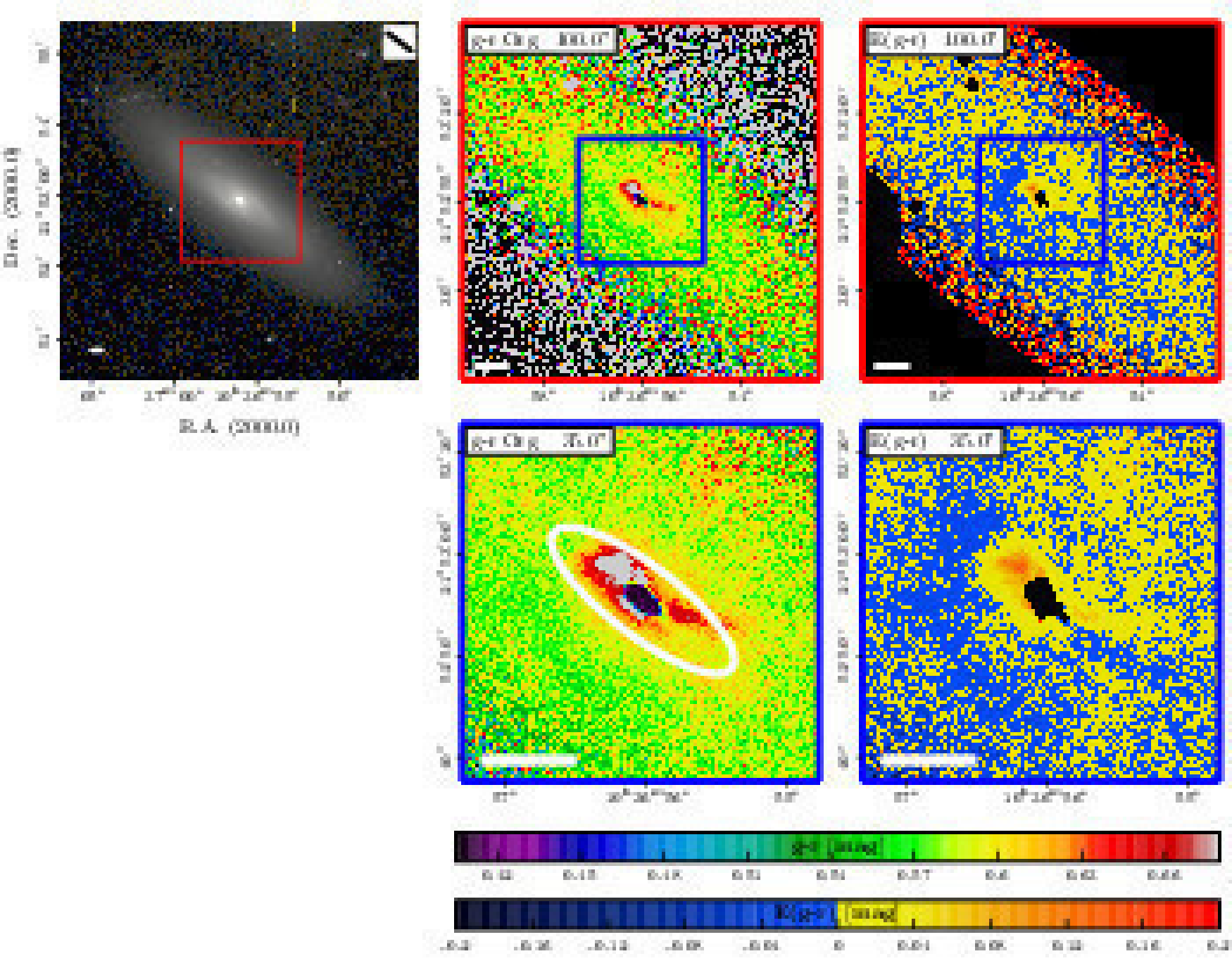}
\end{minipage}}
\makebox[\textwidth][c]{\begin{minipage}[l][-0.7cm][b]{.85\linewidth}
      \includegraphics[scale=0.55, trim={0 0.7cm 0 0},clip]{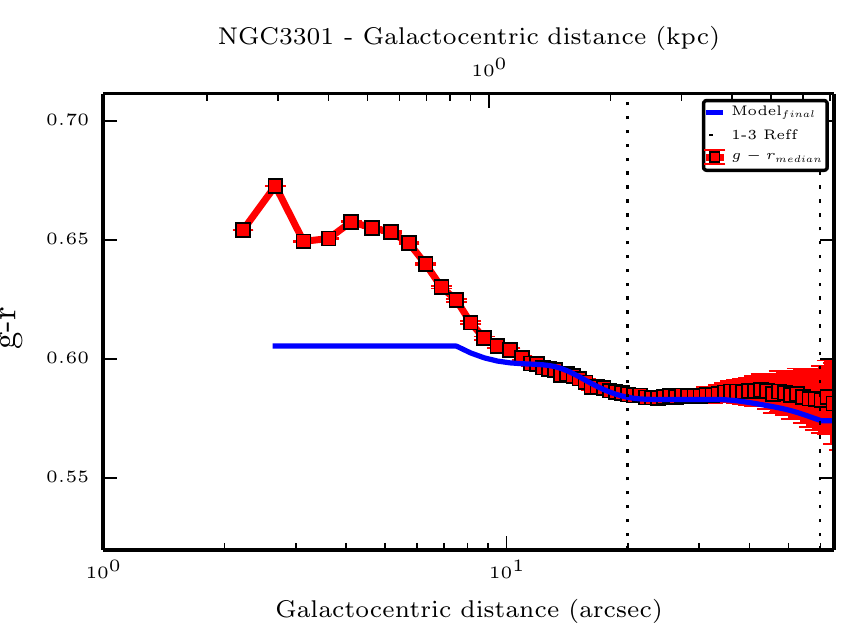}
      \includegraphics[scale=0.55]{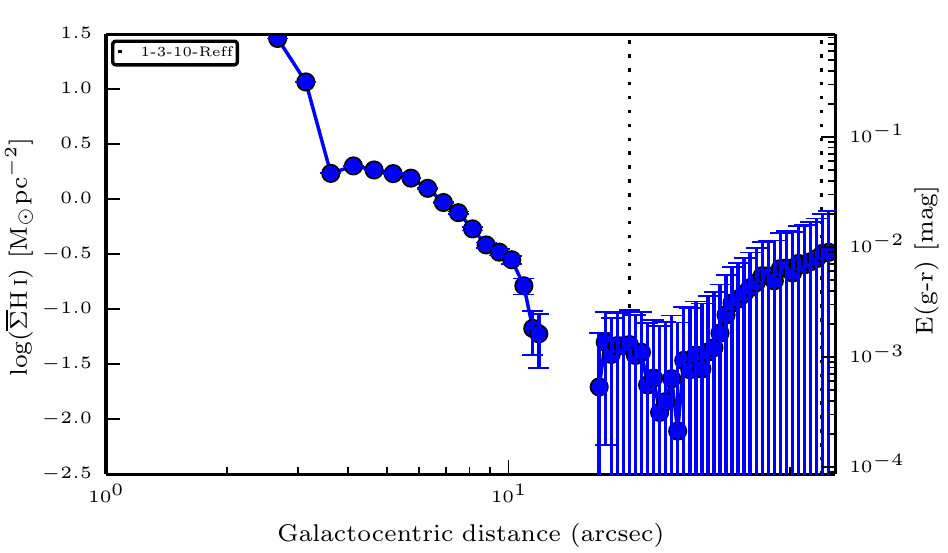}
\end{minipage}}
\caption{True image, colour map, colour excess map, and the radial profiles of NGC~3301.}
\label{fig:3301}
\end{figure*}

\begin{figure*}
\makebox[\textwidth][c]{\begin{minipage}[b][11.6cm]{.85\textwidth}
  \vspace*{\fill}
      \includegraphics[scale=0.85]{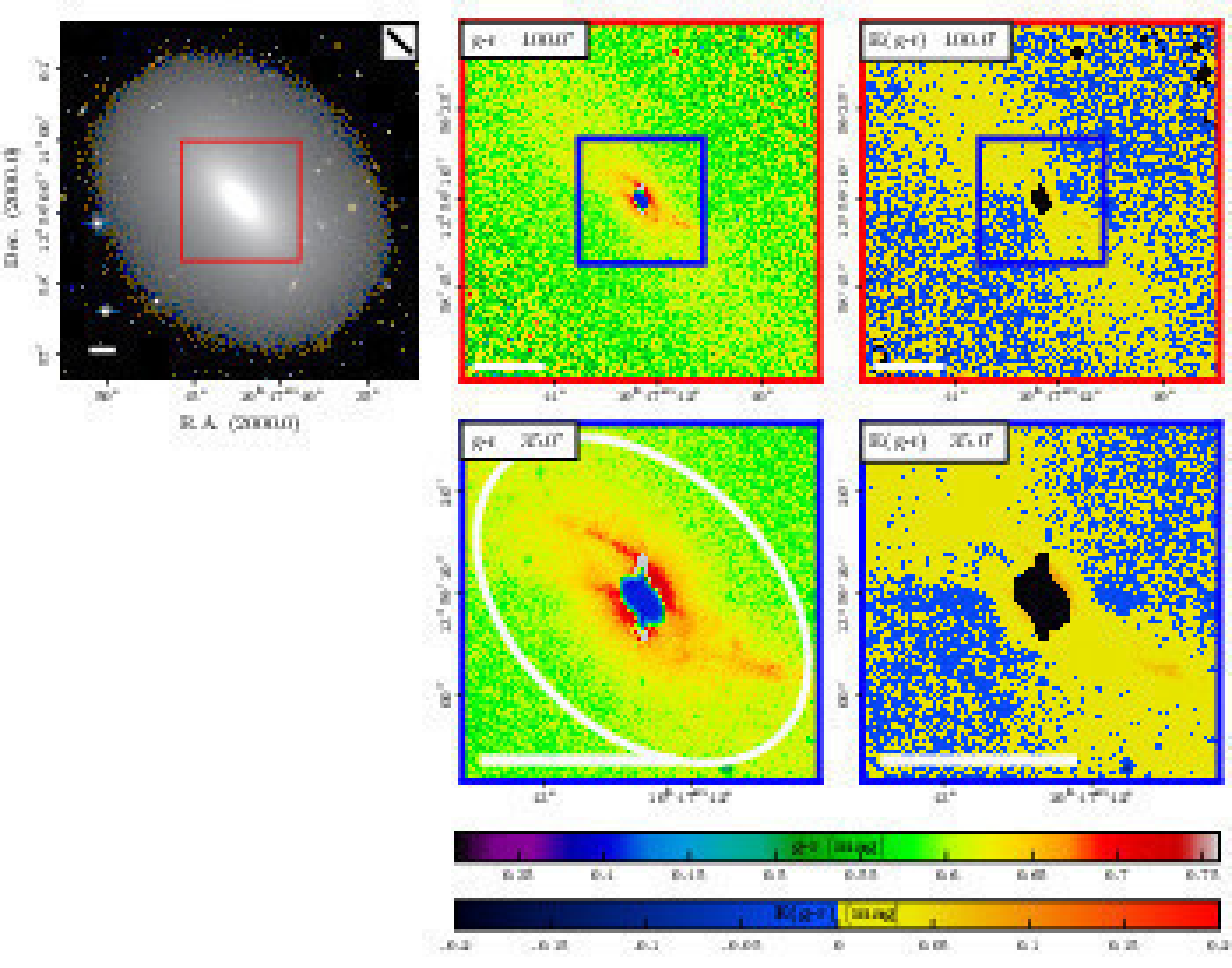}
\end{minipage}}
\makebox[\textwidth][c]{\begin{minipage}[l][-0.7cm][b]{.85\linewidth}
      \includegraphics[scale=0.55, trim={0 0.7cm 0 0},clip]{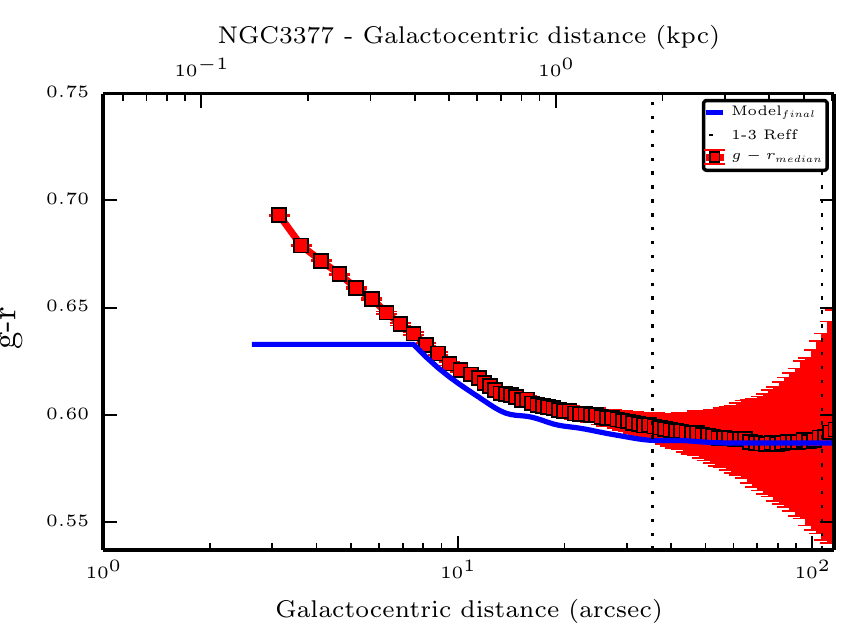}
      \includegraphics[scale=0.55]{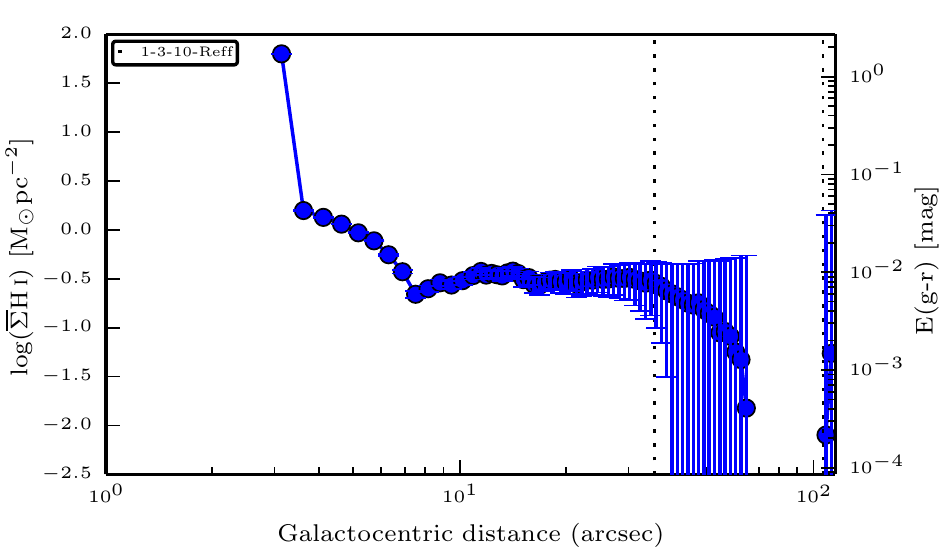}
\end{minipage}}
\caption{True image, colour map, colour excess map, and the radial profiles of NGC~3377.}
\label{fig:3377}
\end{figure*}

%Page10
\clearpage
\begin{figure*}
\makebox[\textwidth][c]{\begin{minipage}[b][10.5cm]{.85\textwidth}
  \vspace*{\fill}
      \includegraphics[scale=0.85]{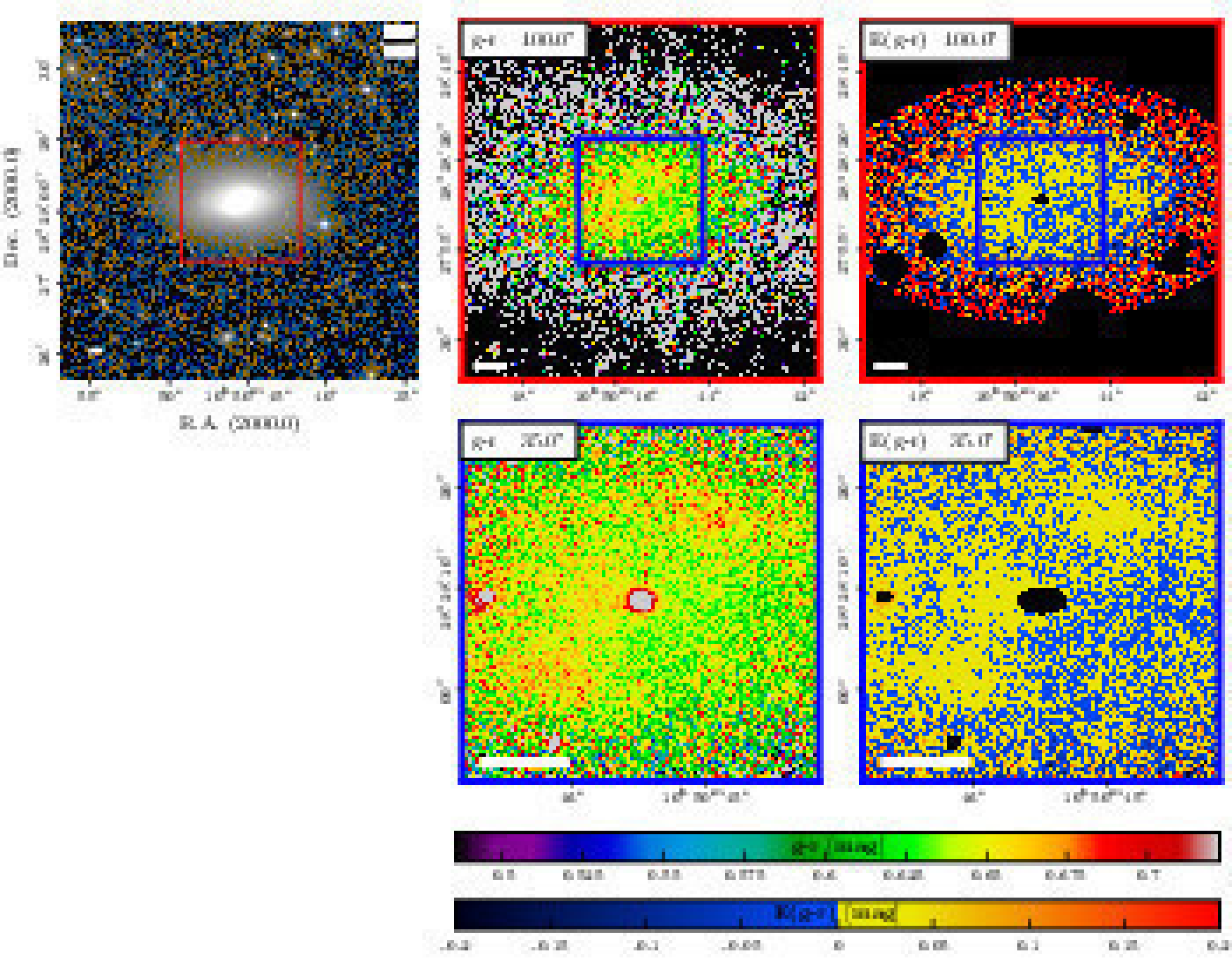}
\end{minipage}}
\makebox[\textwidth][c]{\begin{minipage}[l][-0.7cm][b]{.85\linewidth}
      \includegraphics[scale=0.55, trim={0 0.7cm 0 0},clip]{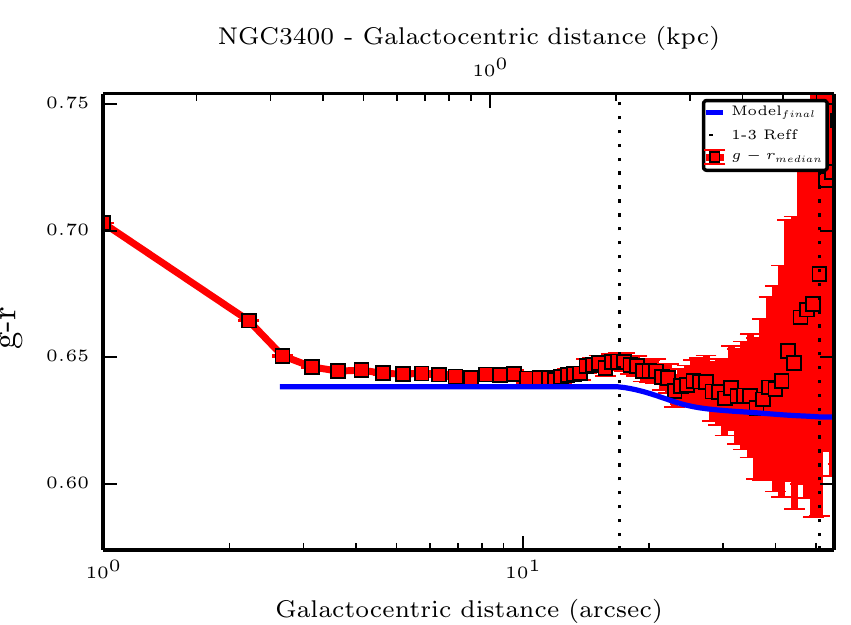}
      \includegraphics[scale=0.55]{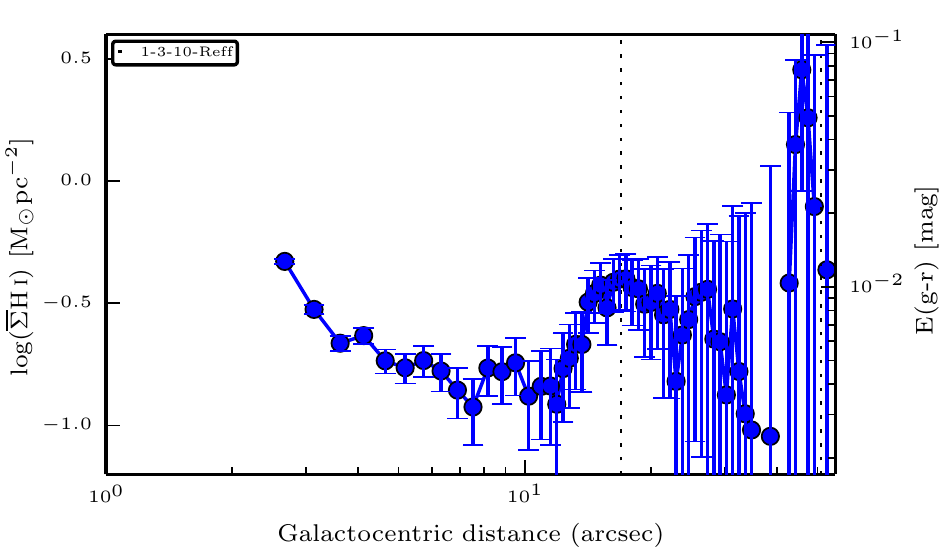}
\end{minipage}}
\caption{True image, colour map, colour excess map, and the radial profiles of NGC~3400.}
\label{fig:3400}
\end{figure*}

\begin{figure*}
\makebox[\textwidth][c]{\begin{minipage}[b][11.6cm]{.85\textwidth}
  \vspace*{\fill}
      \includegraphics[scale=0.85]{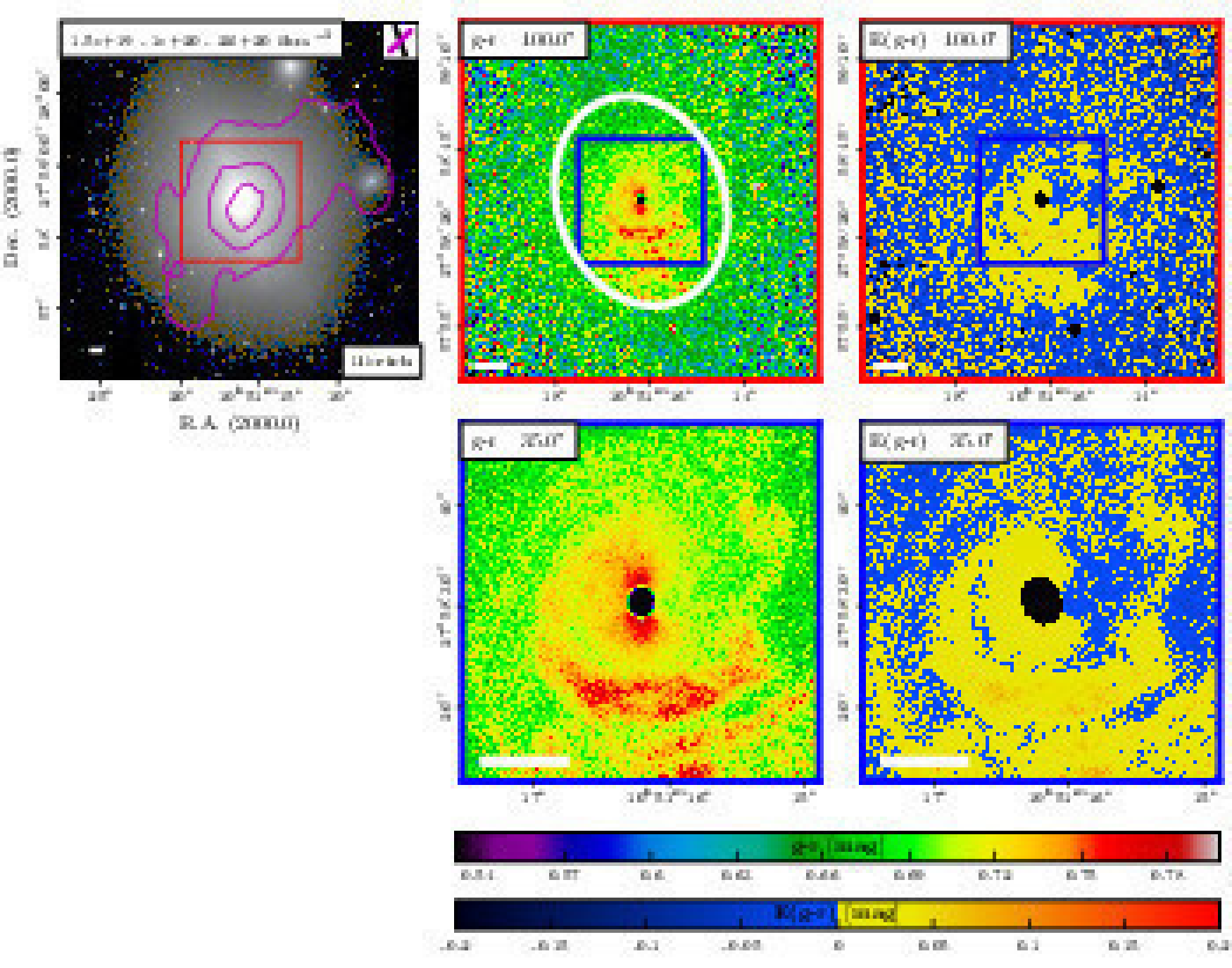}
\end{minipage}}
\makebox[\textwidth][c]{\begin{minipage}[l][-0.7cm][b]{.85\linewidth}
      \includegraphics[scale=0.55, trim={0 0.7cm 0 0},clip]{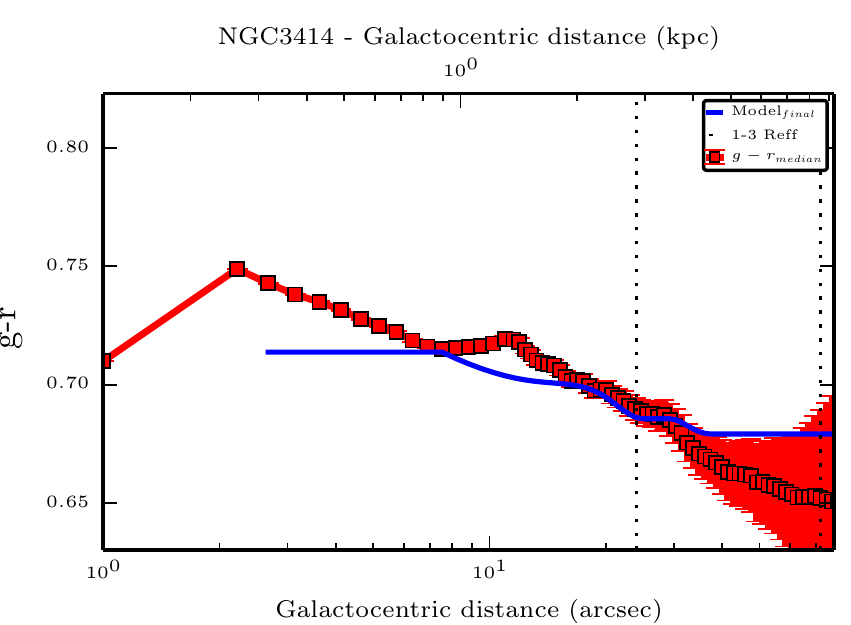}
      \includegraphics[scale=0.55]{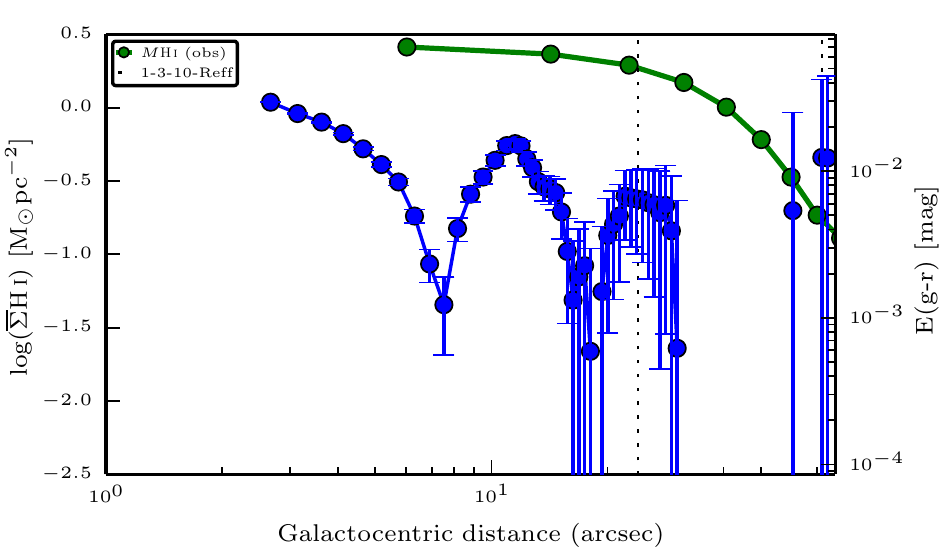}
\end{minipage}}
\caption{True image, colour map, colour excess map, and the radial profiles of NGC~3414.}
\label{fig:3414}
\end{figure*}

%Page11
\clearpage
\begin{figure*}
\makebox[\textwidth][c]{\begin{minipage}[b][10.5cm]{.85\textwidth}
  \vspace*{\fill}
      \includegraphics[scale=0.85]{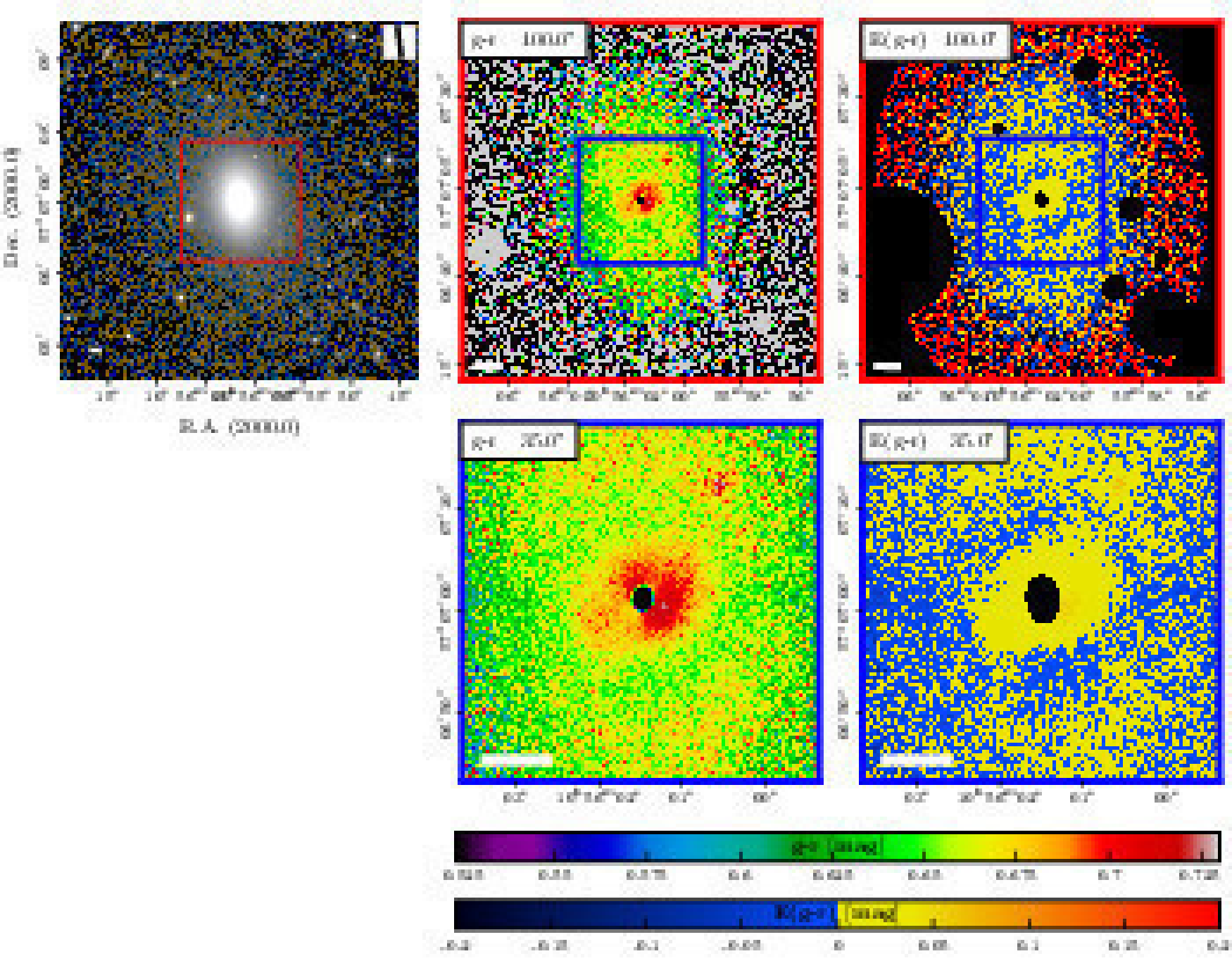}
\end{minipage}}
\makebox[\textwidth][c]{\begin{minipage}[l][-0.7cm][b]{.85\linewidth}
      \includegraphics[scale=0.55, trim={0 0.7cm 0 0},clip]{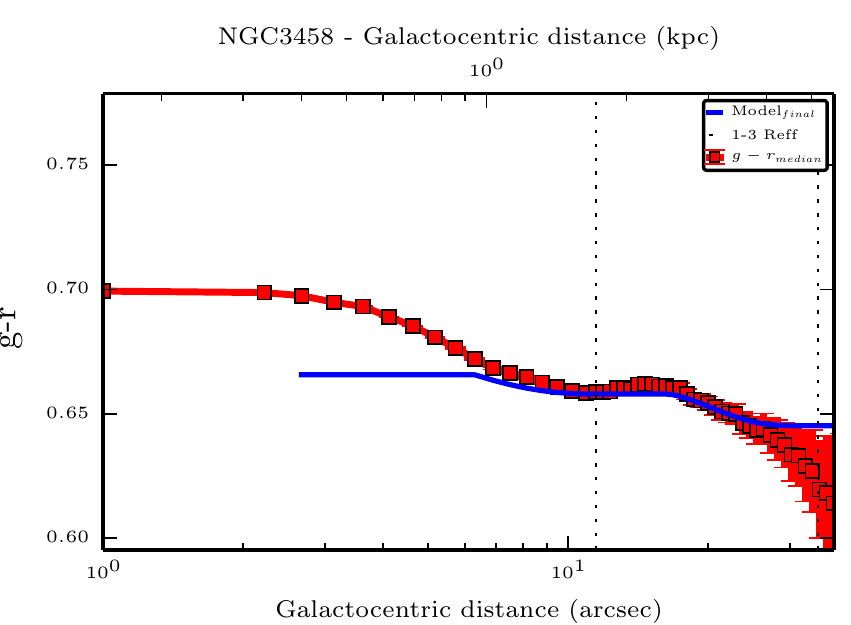}
      \includegraphics[scale=0.55]{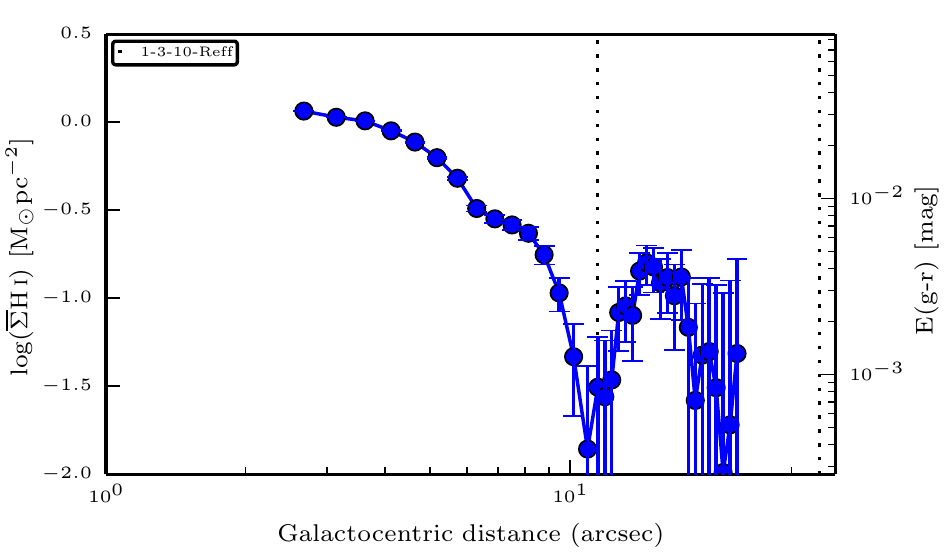}
\end{minipage}}
\caption{True image, colour map, colour excess map, and the radial profiles of NGC~3458.}
\label{fig:3458}
\end{figure*}

\begin{figure*}
\makebox[\textwidth][c]{\begin{minipage}[b][11.6cm]{.85\textwidth}
  \vspace*{\fill}
      \includegraphics[scale=0.85]{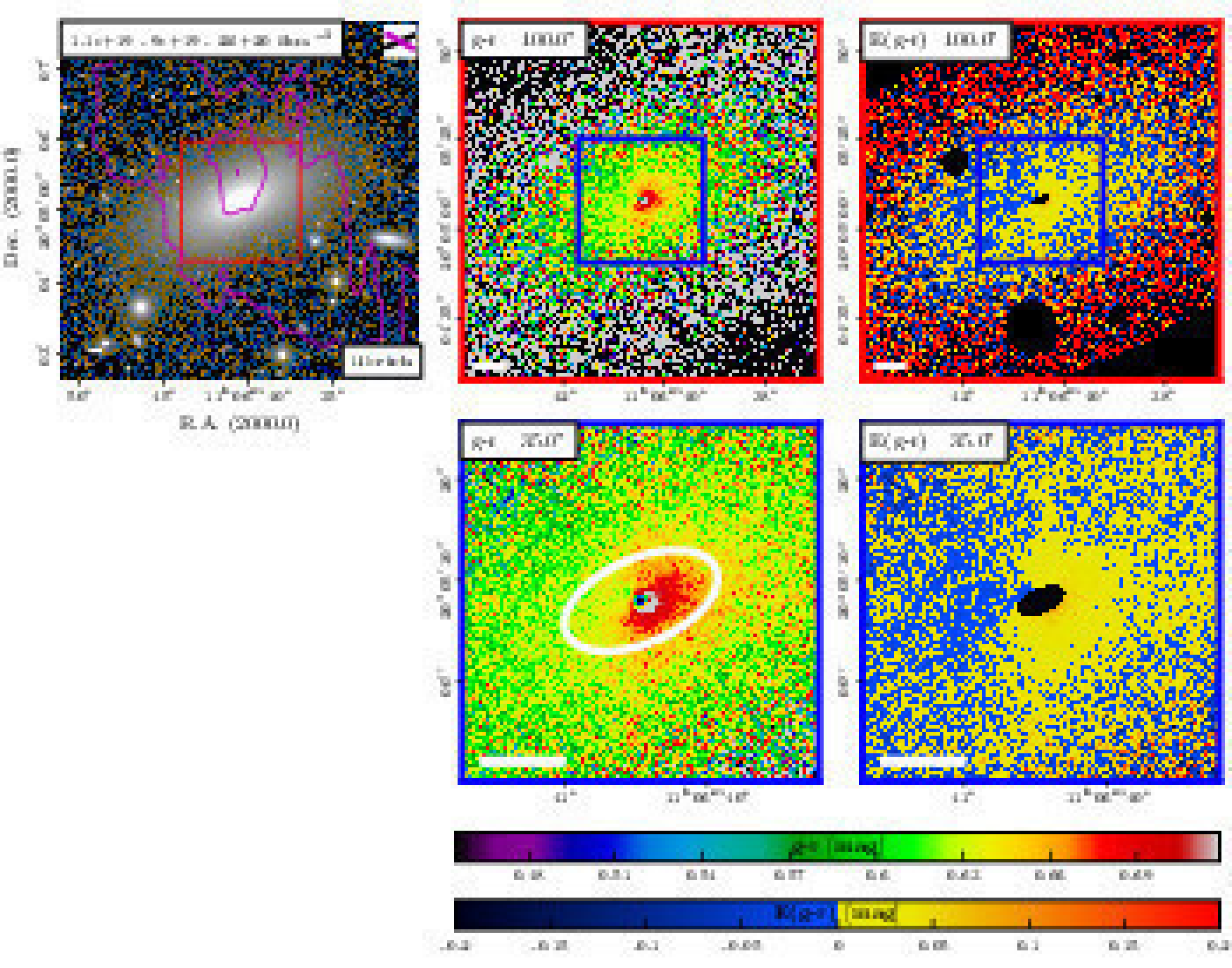}
\end{minipage}}
\makebox[\textwidth][c]{\begin{minipage}[l][-0.7cm][b]{.85\linewidth}
      \includegraphics[scale=0.55, trim={0 0.7cm 0 0},clip]{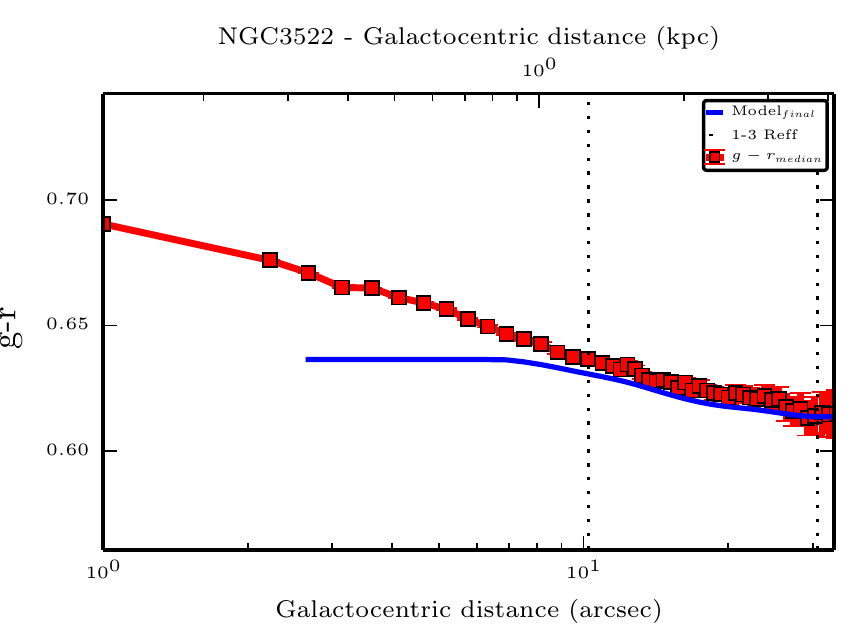}
      \includegraphics[scale=0.55]{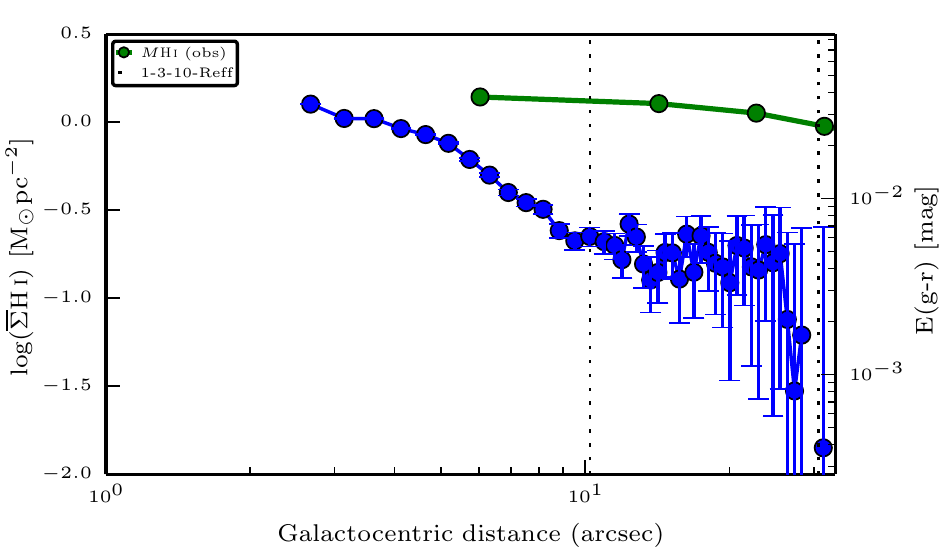}
\end{minipage}}
\caption{True image, colour map, colour excess map, and the radial profiles of NGC~3522.}
\label{fig:3522}
\end{figure*}

%Page12
\clearpage
\begin{figure*}
\makebox[\textwidth][c]{\begin{minipage}[b][10.5cm]{.85\textwidth}
  \vspace*{\fill}
      \includegraphics[scale=0.85]{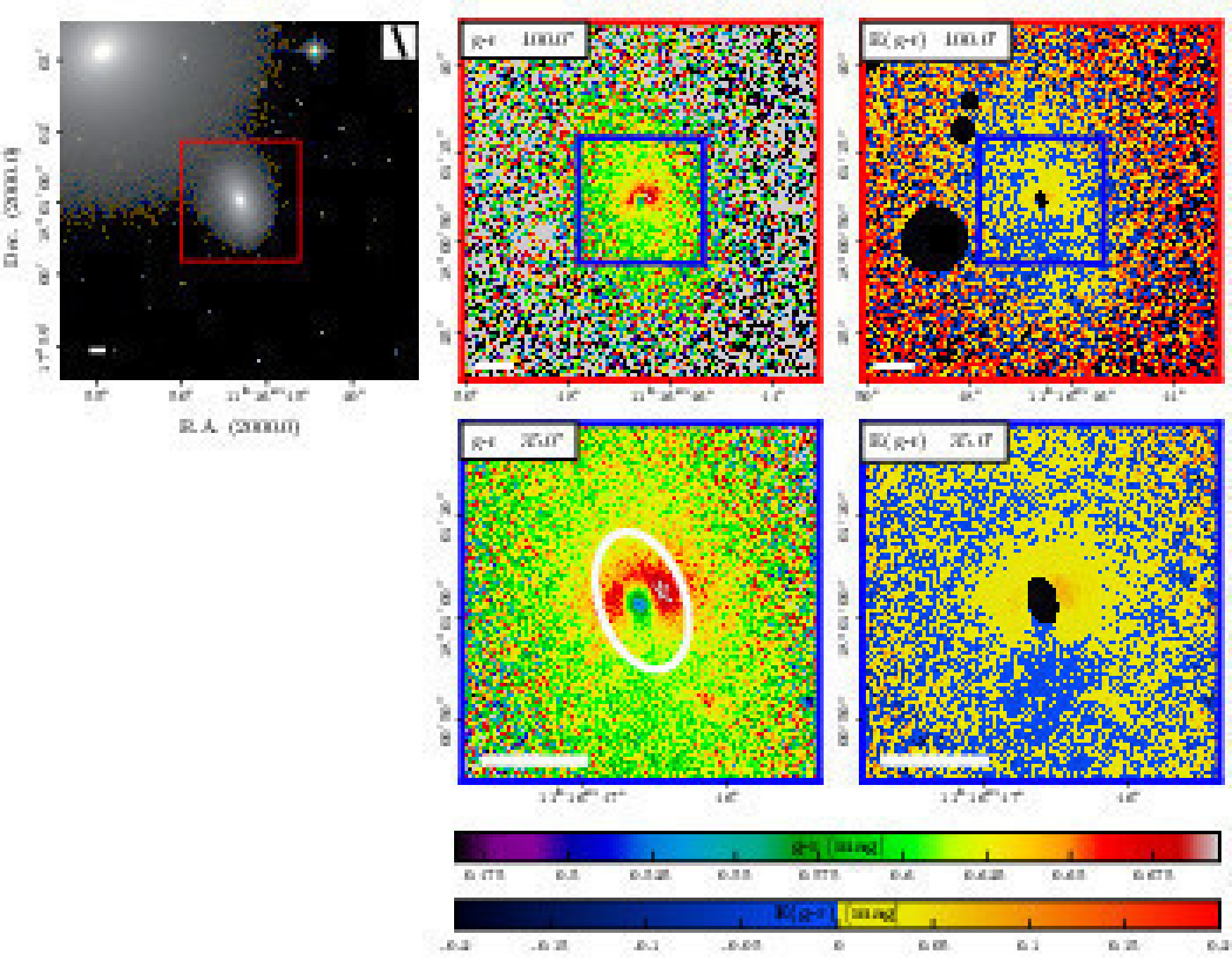}
\end{minipage}}
\makebox[\textwidth][c]{\begin{minipage}[l][-0.7cm][b]{.85\linewidth}
      \includegraphics[scale=0.55, trim={0 0.7cm 0 0},clip]{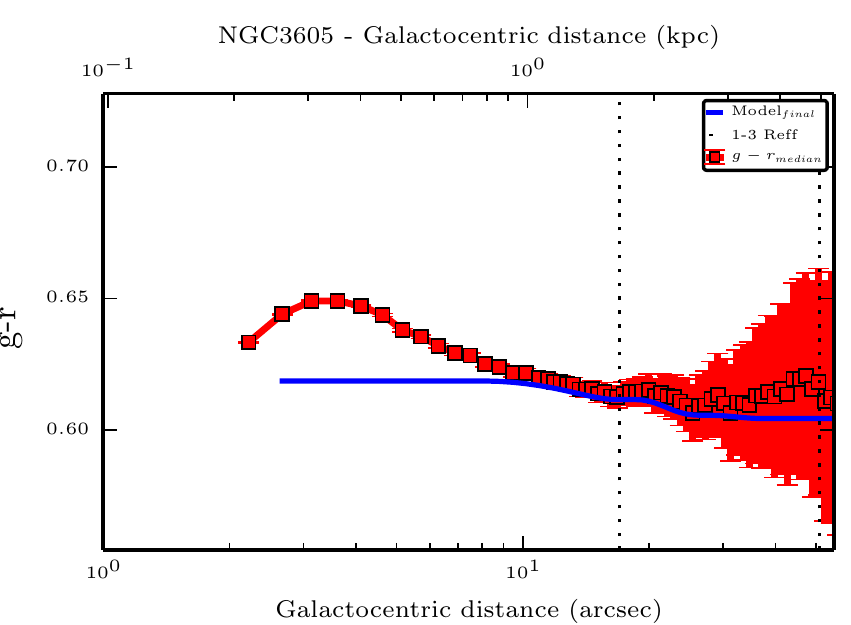}
      \includegraphics[scale=0.55]{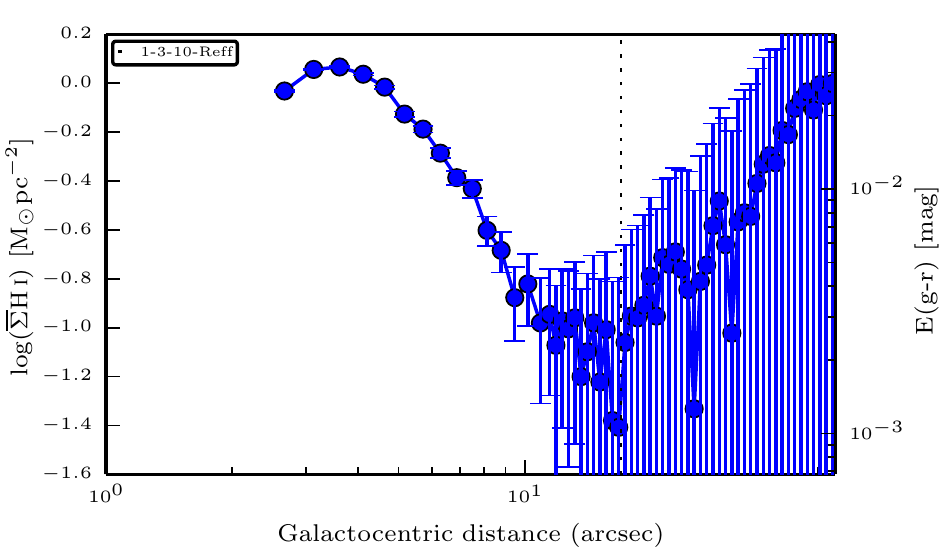}
\end{minipage}}
\caption{True image, colour map, colour excess map, and the radial profiles of NGC~3605.}
\label{fig:3605}
\end{figure*}

\begin{figure*}
\makebox[\textwidth][c]{\begin{minipage}[b][11.6cm]{.85\textwidth}
  \vspace*{\fill}
      \includegraphics[scale=0.85]{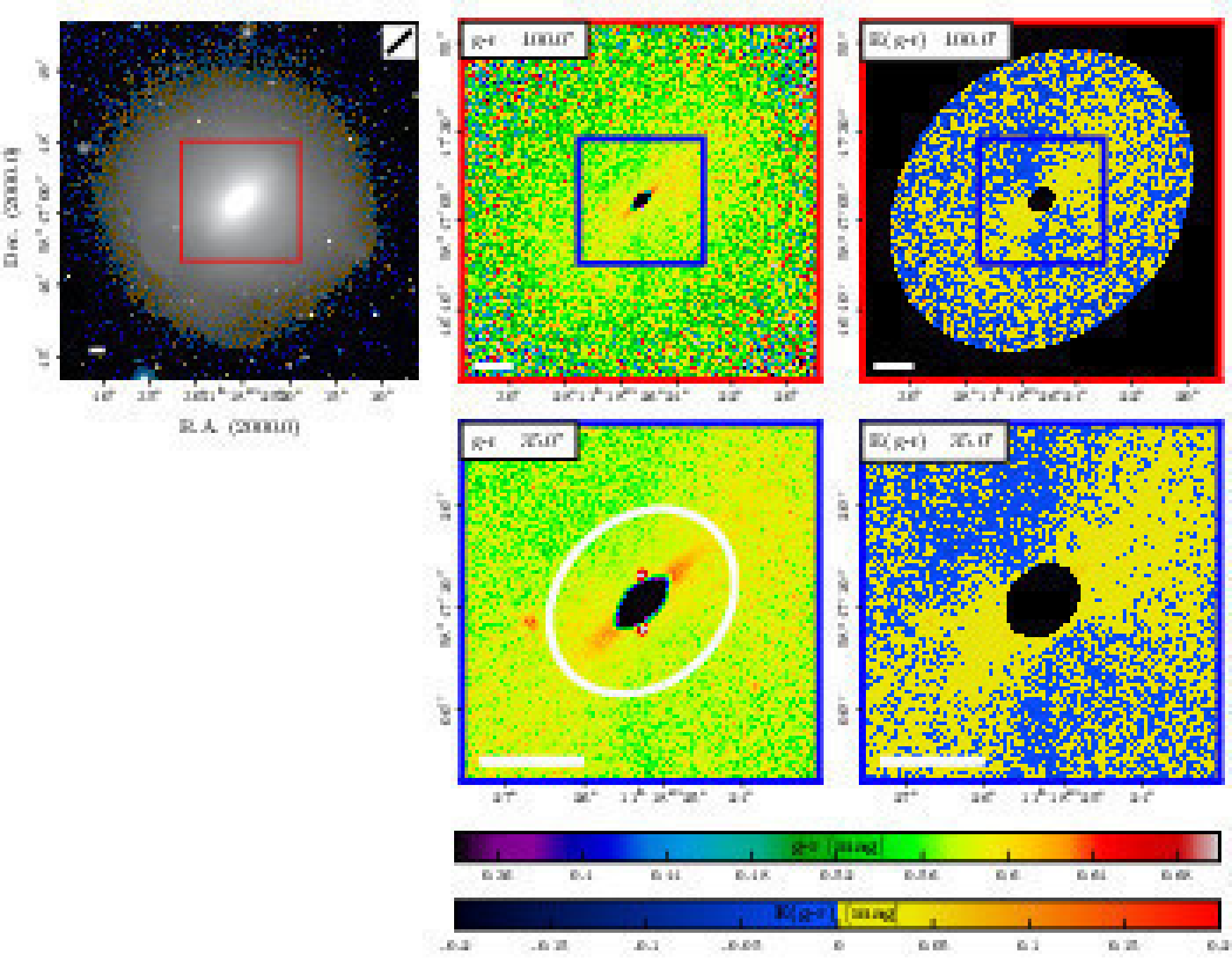}
\end{minipage}}
\makebox[\textwidth][c]{\begin{minipage}[l][-0.7cm][b]{.85\linewidth}
      \includegraphics[scale=0.55, trim={0 0.7cm 0 0},clip]{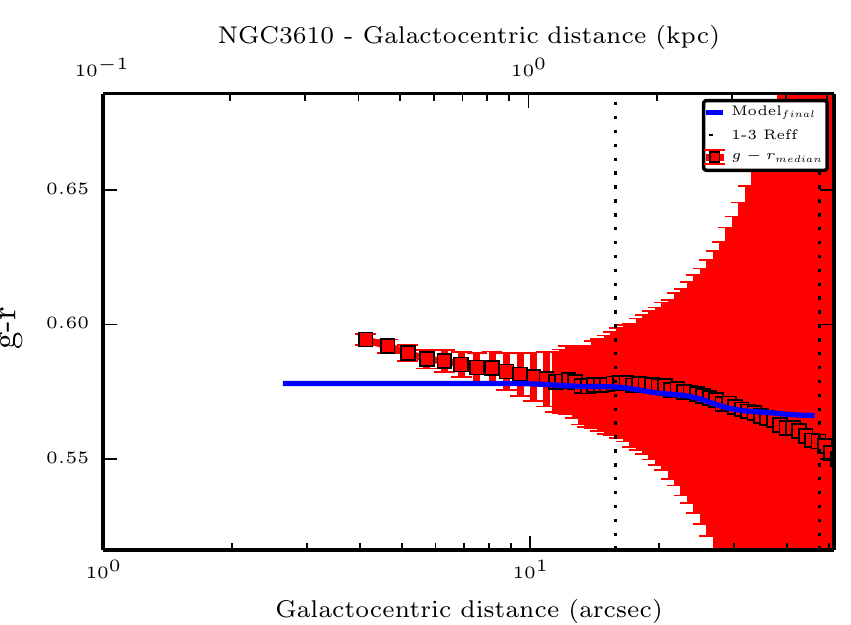}
      \includegraphics[scale=0.55]{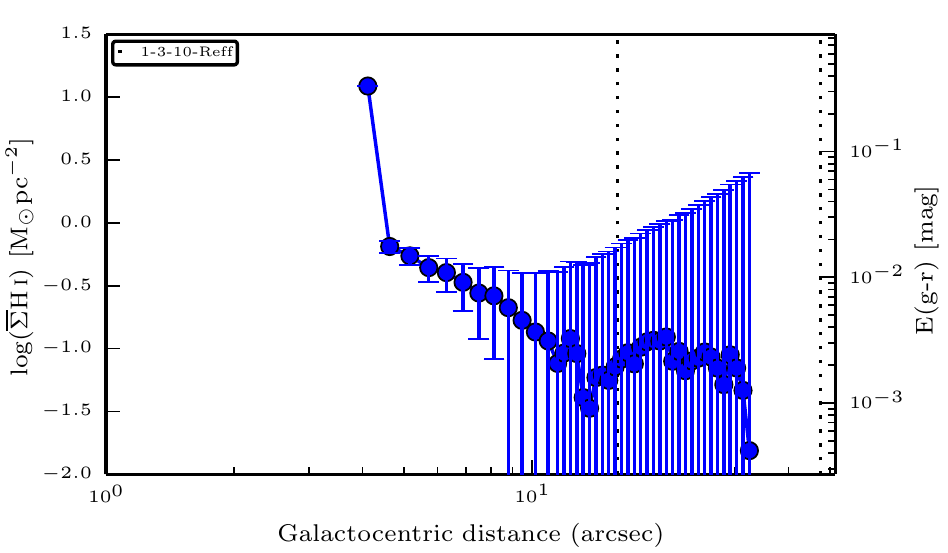}
\end{minipage}}
\caption{True image, colour map, colour excess map, and the radial profiles of NGC~3610.}
\label{fig:3610}
\end{figure*}

%Page13
\clearpage
\begin{figure*}
\makebox[\textwidth][c]{\begin{minipage}[b][10.5cm]{.85\textwidth}
  \vspace*{\fill}
      \includegraphics[scale=0.85]{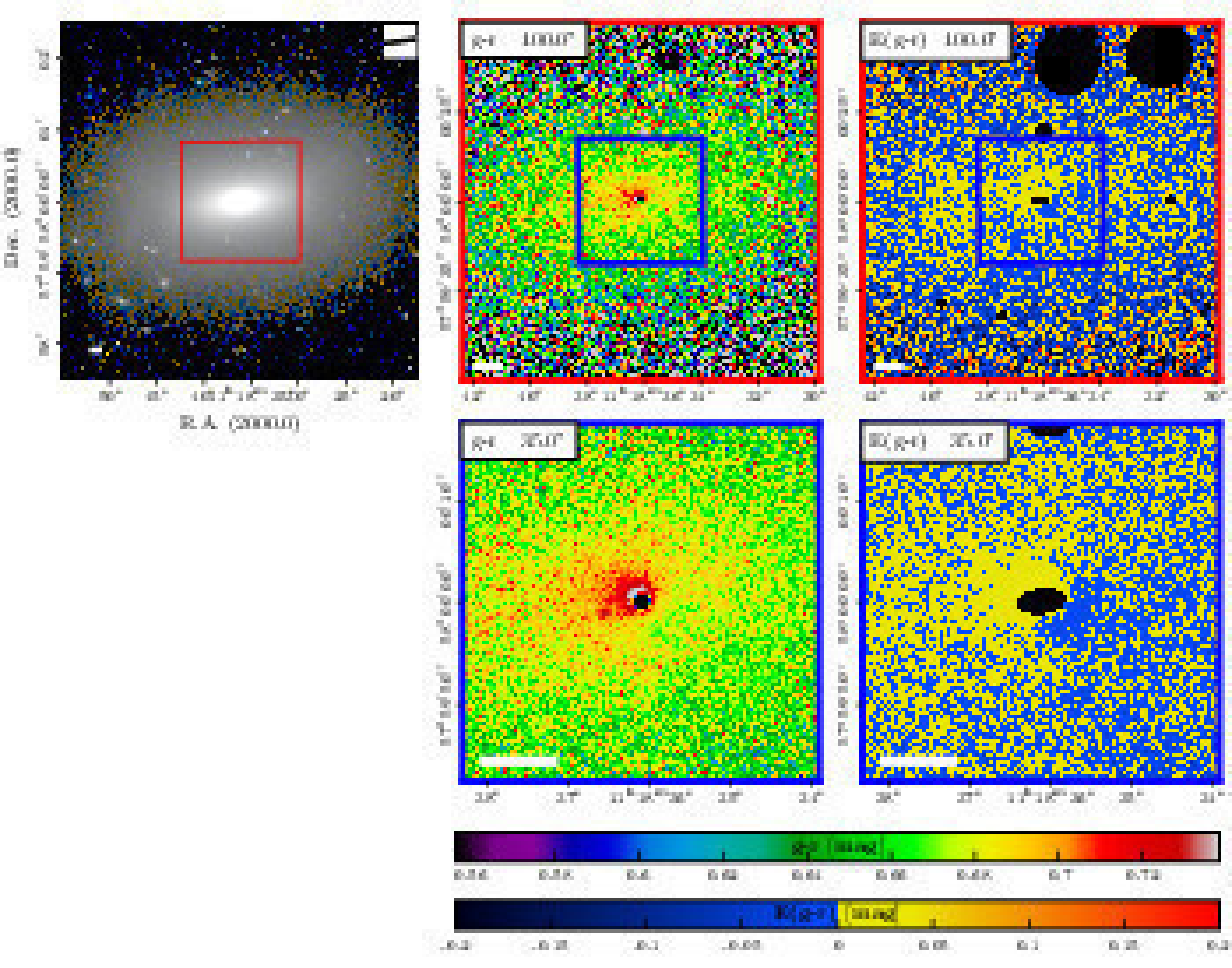}
\end{minipage}}
\makebox[\textwidth][c]{\begin{minipage}[l][-0.7cm][b]{.85\linewidth}
      \includegraphics[scale=0.55, trim={0 0.7cm 0 0},clip]{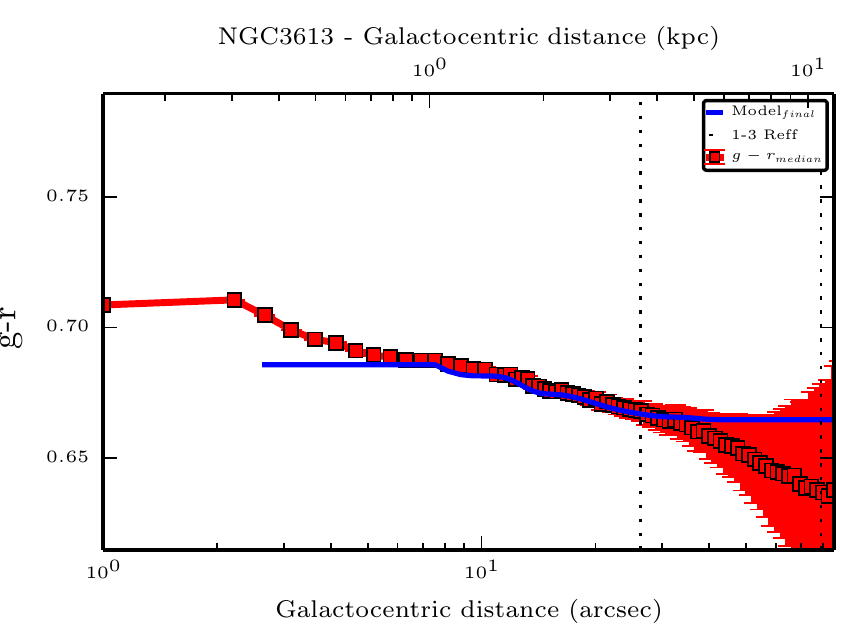}
      \includegraphics[scale=0.55]{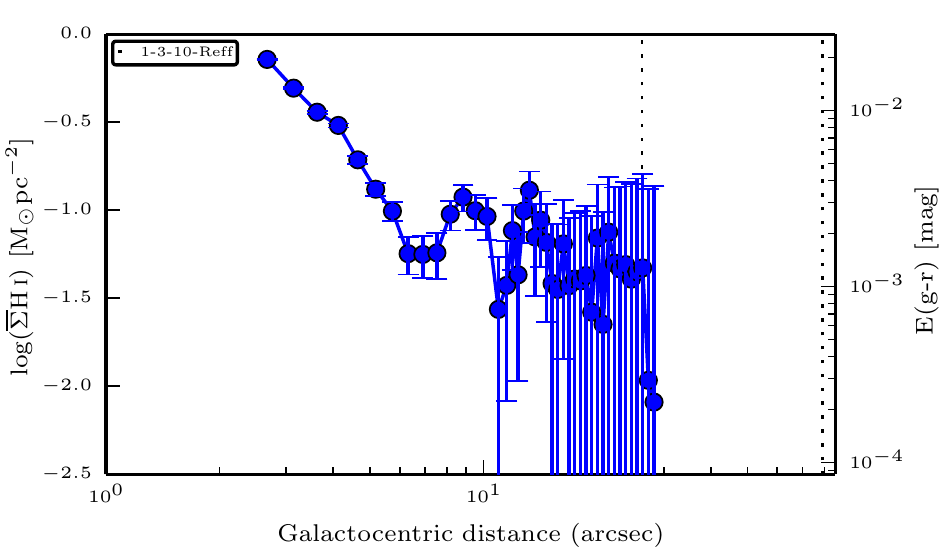}
\end{minipage}}
\caption{True image, colour map, colour excess map, and the radial profiles of NGC~3613.}
\label{fig:3613}
\end{figure*}

\begin{figure*}
\makebox[\textwidth][c]{\begin{minipage}[b][11.6cm]{.85\textwidth}
  \vspace*{\fill}
      \includegraphics[scale=0.85]{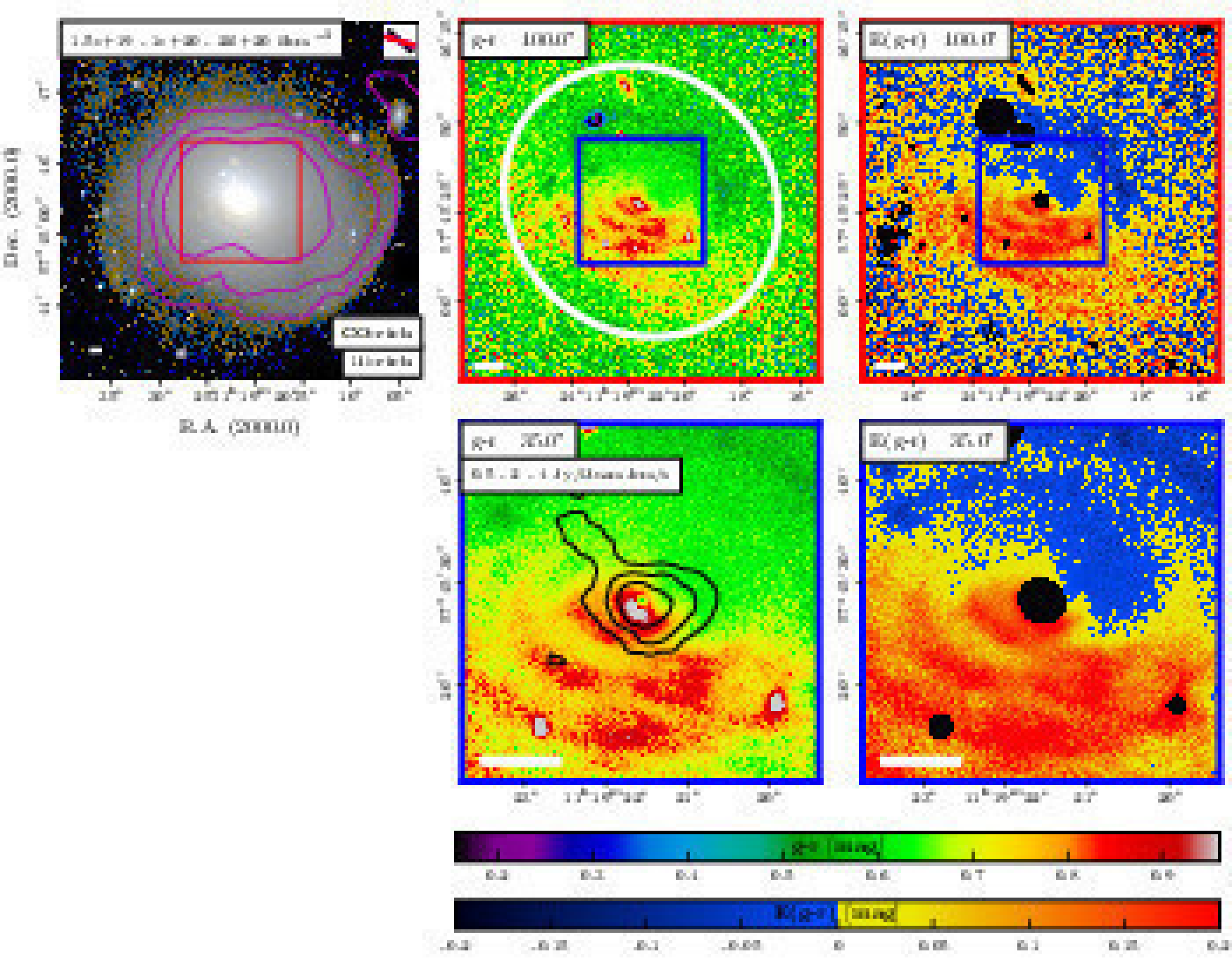}
\end{minipage}}
\makebox[\textwidth][c]{\begin{minipage}[l][-0.7cm][b]{.85\linewidth}
      \includegraphics[scale=0.55, trim={0 0.7cm 0 0},clip]{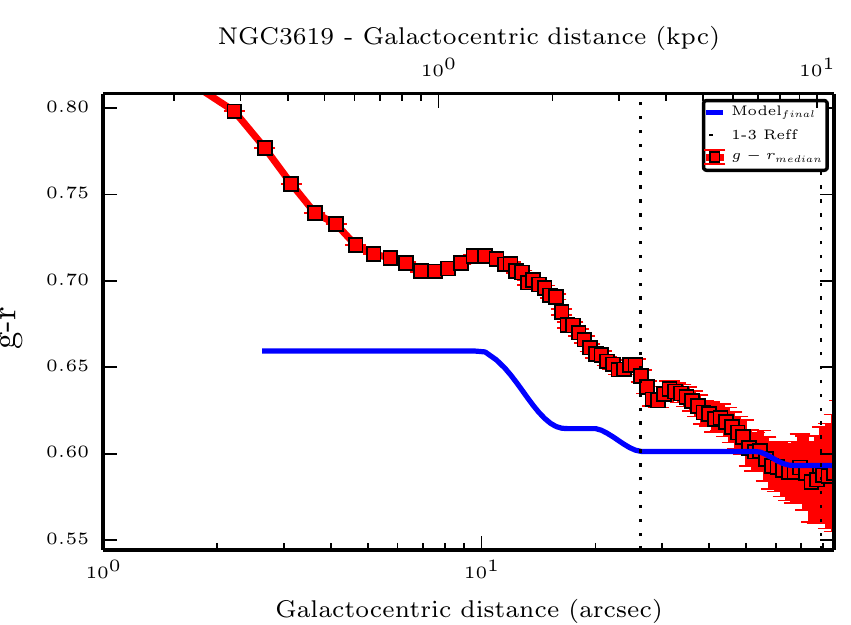}
      \includegraphics[scale=0.55]{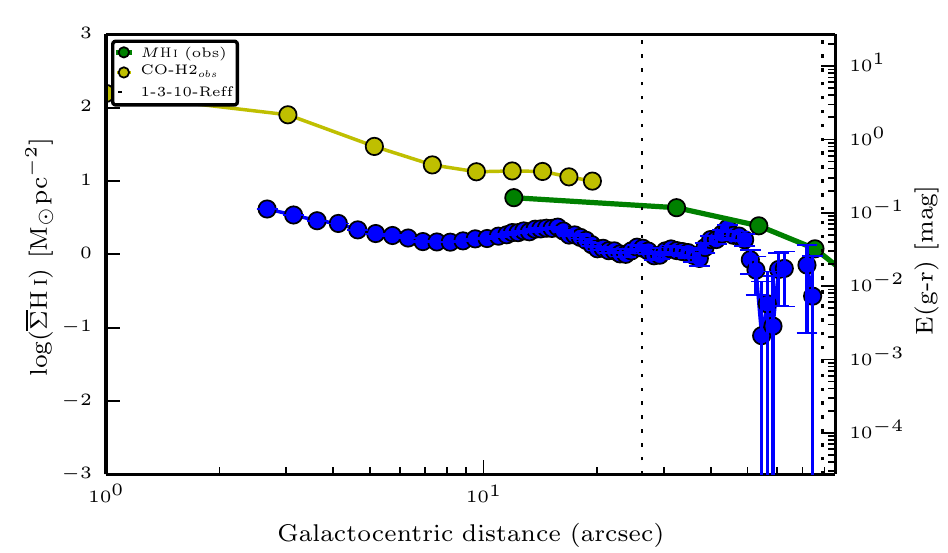}
\end{minipage}}
\caption{True image, colour map, colour excess map, and the radial profiles of NGC~3619.}
\label{fig:app_profiles}
\end{figure*}

%Page14
\clearpage
\begin{figure*}
\makebox[\textwidth][c]{\begin{minipage}[b][10.5cm]{.85\textwidth}
  \vspace*{\fill}
      \includegraphics[scale=0.85]{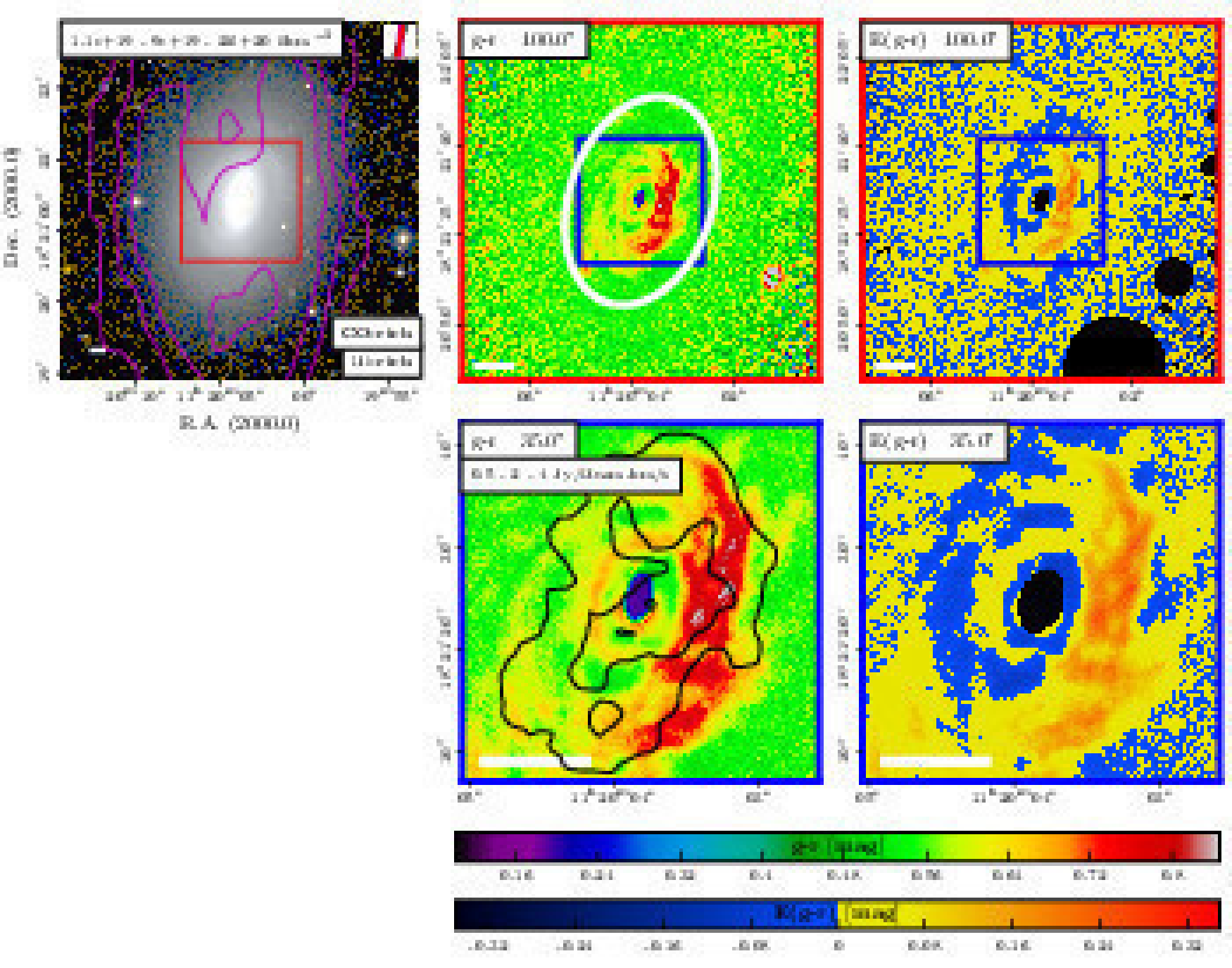}
\end{minipage}}
\makebox[\textwidth][c]{\begin{minipage}[l][-0.7cm][b]{.85\linewidth}
      \includegraphics[scale=0.55, trim={0 0.7cm 0 0},clip]{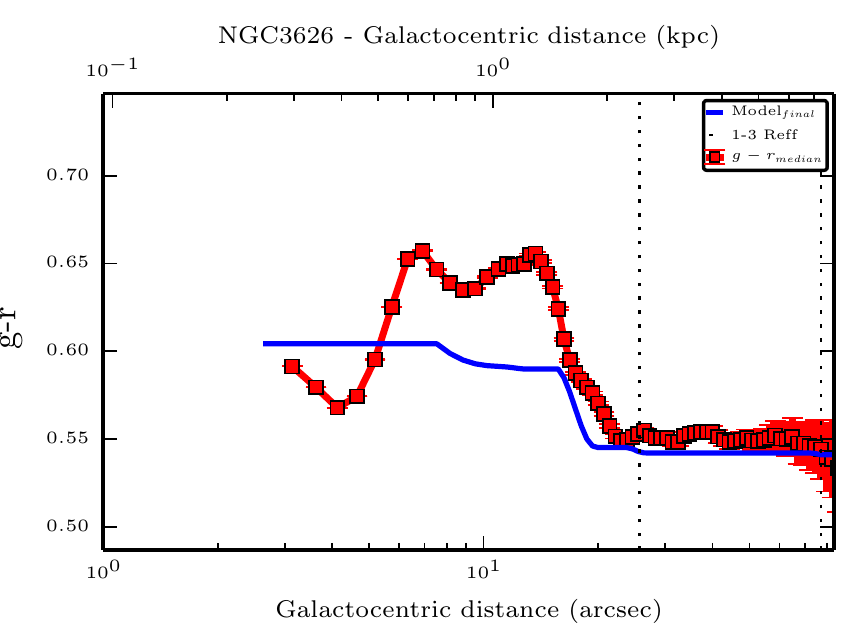}
      \includegraphics[scale=0.55]{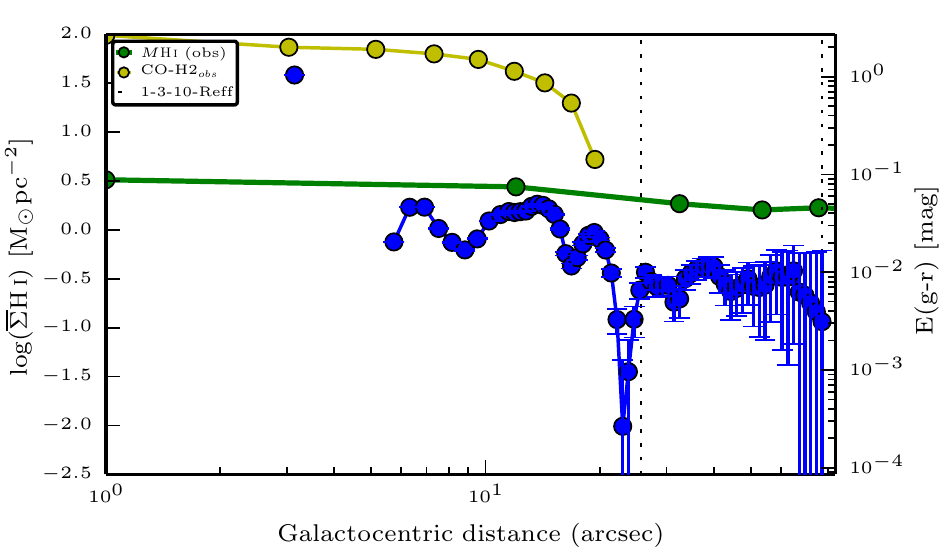}
\end{minipage}}
\caption{True image, colour map, colour excess map, and the radial profiles of NGC~3626.}
\label{fig:app_profiles}
\end{figure*}

\begin{figure*}
\makebox[\textwidth][c]{\begin{minipage}[b][11.6cm]{.85\textwidth}
  \vspace*{\fill}
      \includegraphics[scale=0.85]{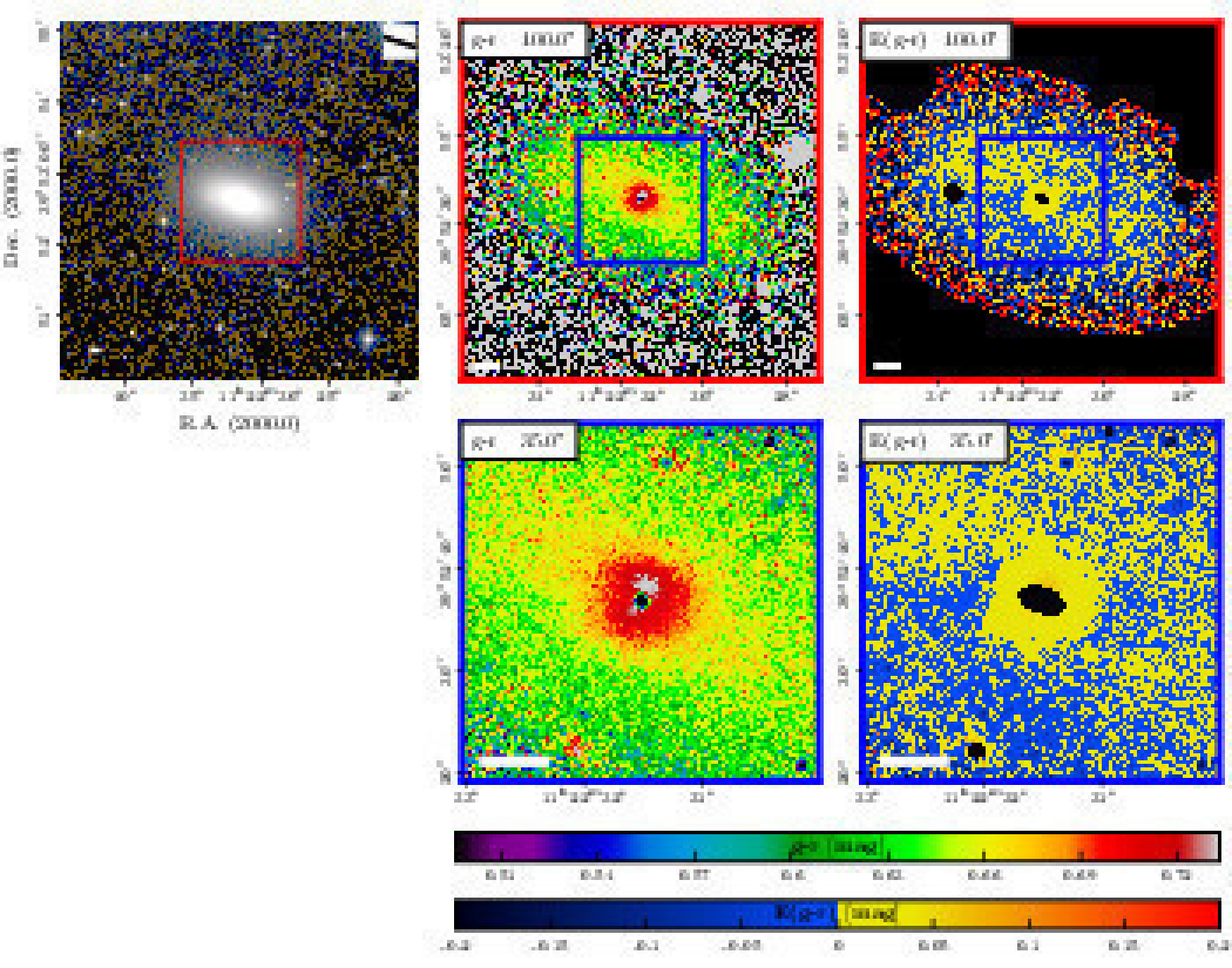}
\end{minipage}}
\makebox[\textwidth][c]{\begin{minipage}[l][-0.7cm][b]{.85\linewidth}
      \includegraphics[scale=0.55, trim={0 0.7cm 0 0},clip]{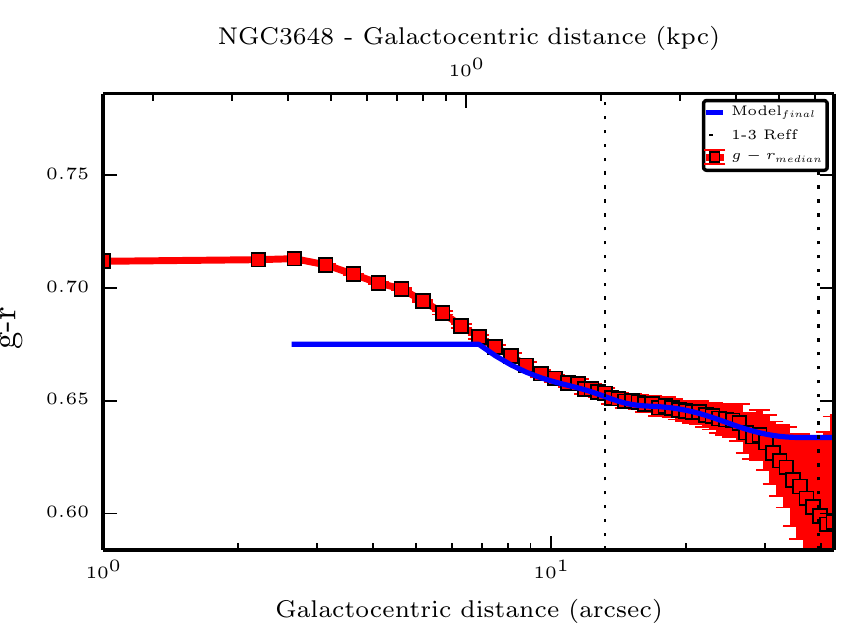}
      \includegraphics[scale=0.55]{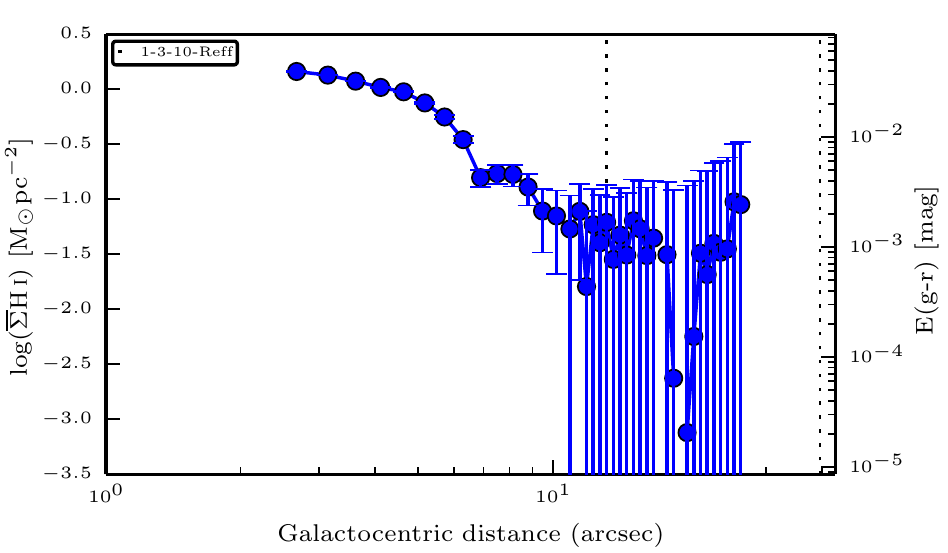}
\end{minipage}}
\caption{True image, colour map, colour excess map, and the radial profiles of NGC~3648.}
\label{fig:app_profiles}
\end{figure*}

%Page15
\clearpage
\begin{figure*}
\makebox[\textwidth][c]{\begin{minipage}[b][10.5cm]{.85\textwidth}
  \vspace*{\fill}
      \includegraphics[scale=0.85]{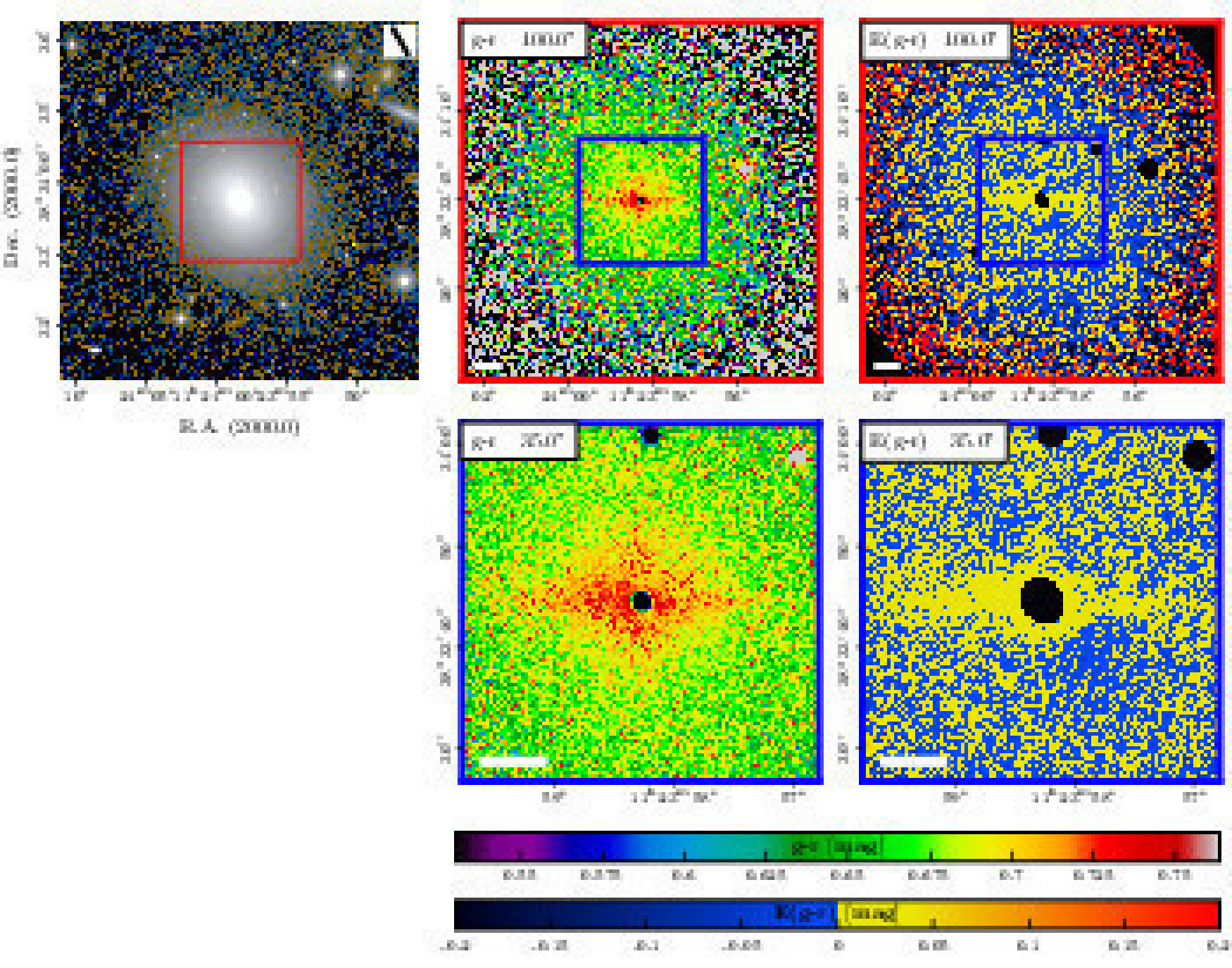}
\end{minipage}}
\makebox[\textwidth][c]{\begin{minipage}[l][-0.7cm][b]{.85\linewidth}
      \includegraphics[scale=0.55, trim={0 0.7cm 0 0},clip]{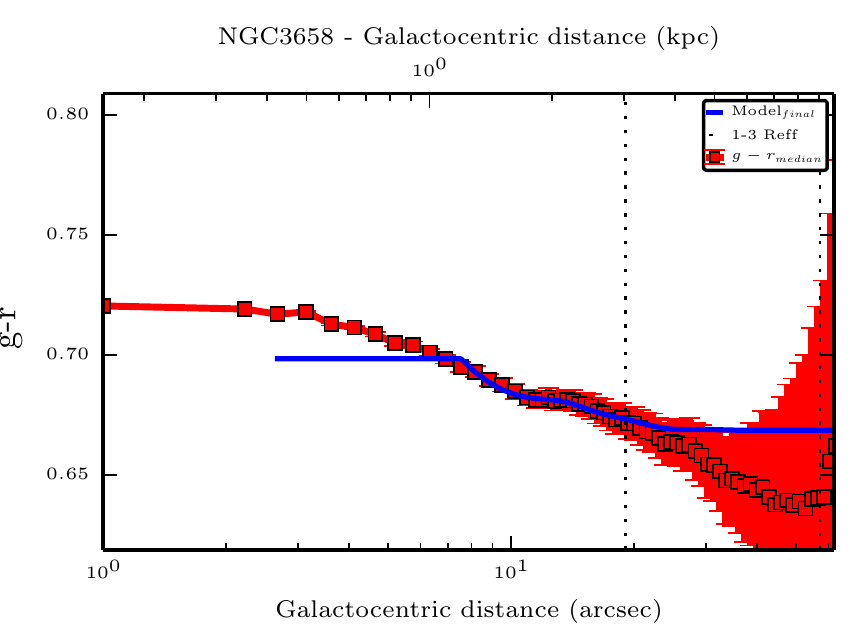}
      \includegraphics[scale=0.55]{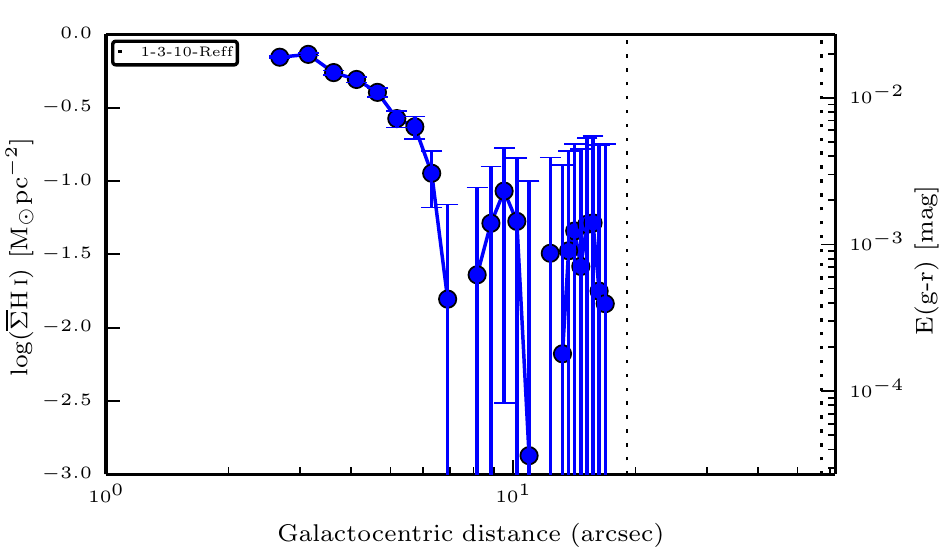}
\end{minipage}}
\caption{True image, colour map, colour excess map, and the radial profiles of NGC~3658.}
\label{fig:app_profiles}
\end{figure*}

\begin{figure*}
\makebox[\textwidth][c]{\begin{minipage}[b][11.6cm]{.85\textwidth}
  \vspace*{\fill}
      \includegraphics[scale=0.85]{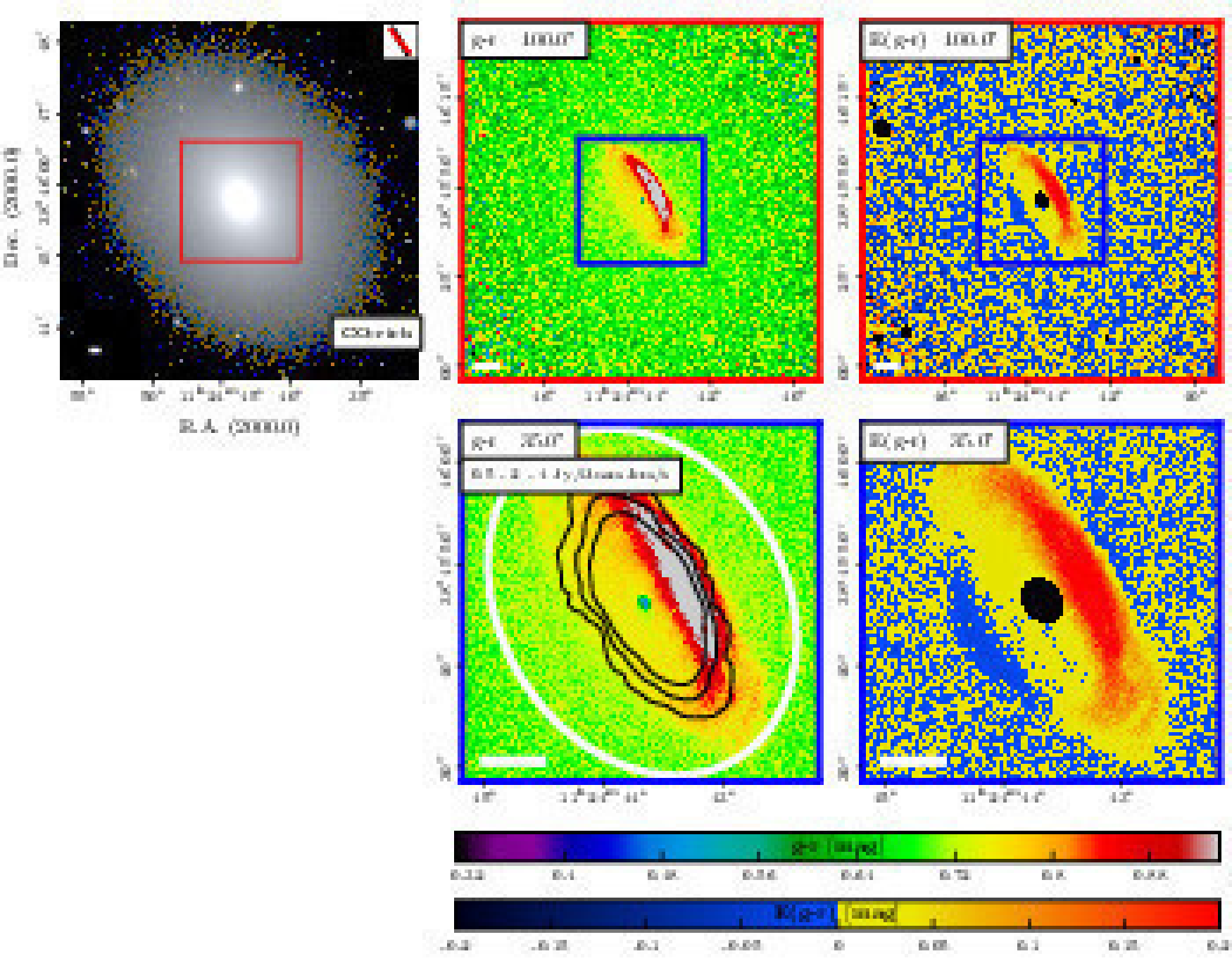}
\end{minipage}}
\makebox[\textwidth][c]{\begin{minipage}[l][-0.7cm][b]{.85\linewidth}
      \includegraphics[scale=0.55, trim={0 0.7cm 0 0},clip]{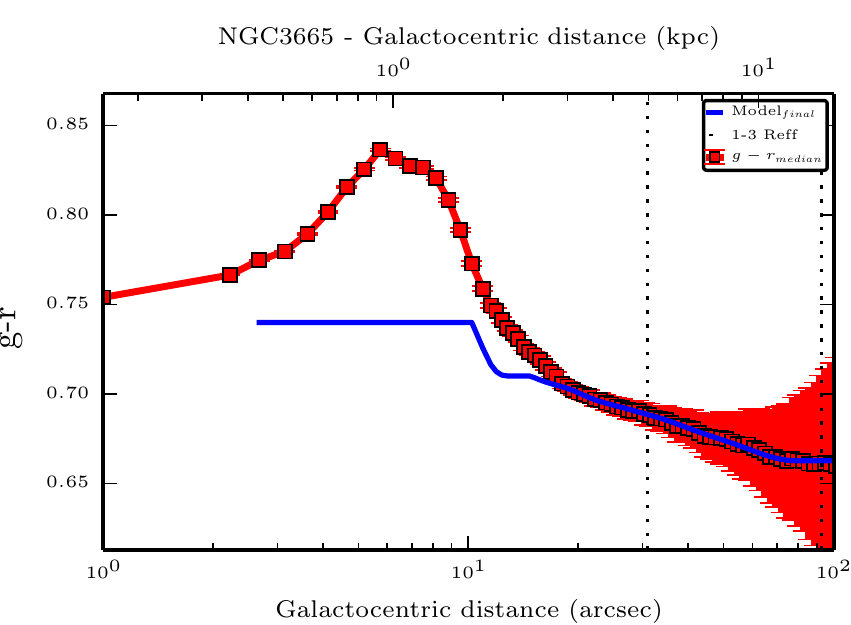}
      \includegraphics[scale=0.55]{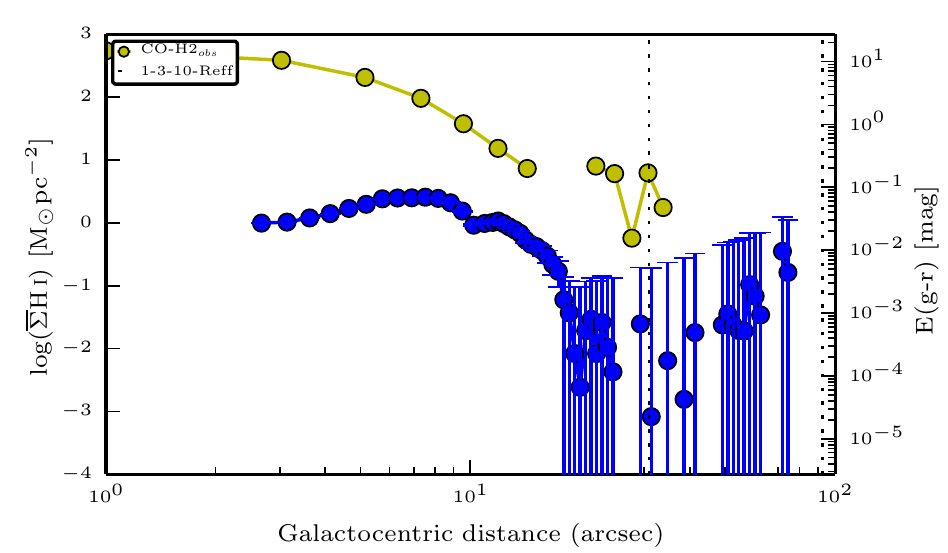}
\end{minipage}}
\caption{True image, colour map, colour excess map, and the radial profiles of NGC~3665.}
\label{fig:app_profiles}
\end{figure*}

%Page16
\clearpage
\begin{figure*}
\makebox[\textwidth][c]{\begin{minipage}[b][10.5cm]{.85\textwidth}
  \vspace*{\fill}
      \includegraphics[scale=0.85]{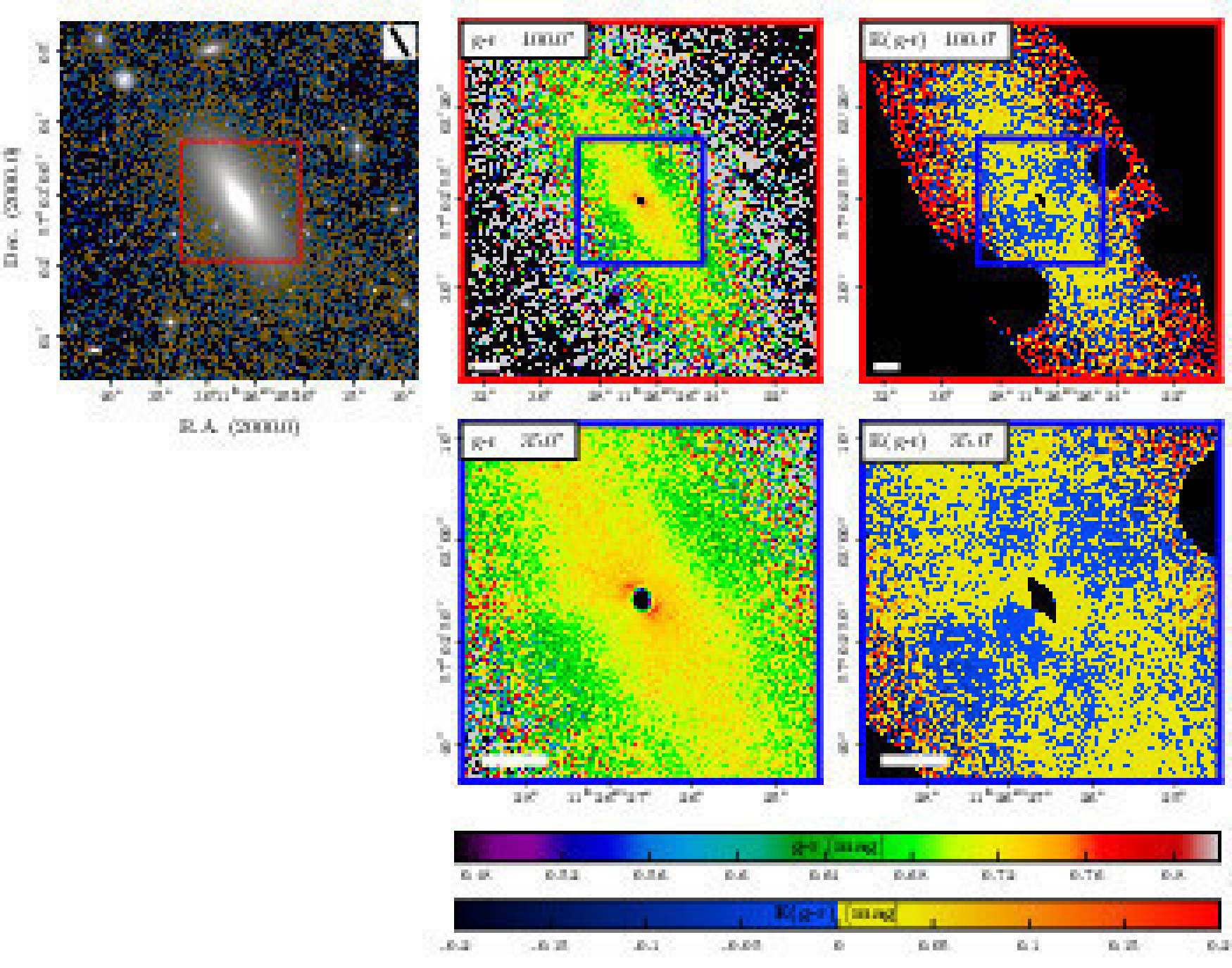}
\end{minipage}}
\makebox[\textwidth][c]{\begin{minipage}[l][-0.7cm][b]{.85\linewidth}
      \includegraphics[scale=0.55, trim={0 0.7cm 0 0},clip]{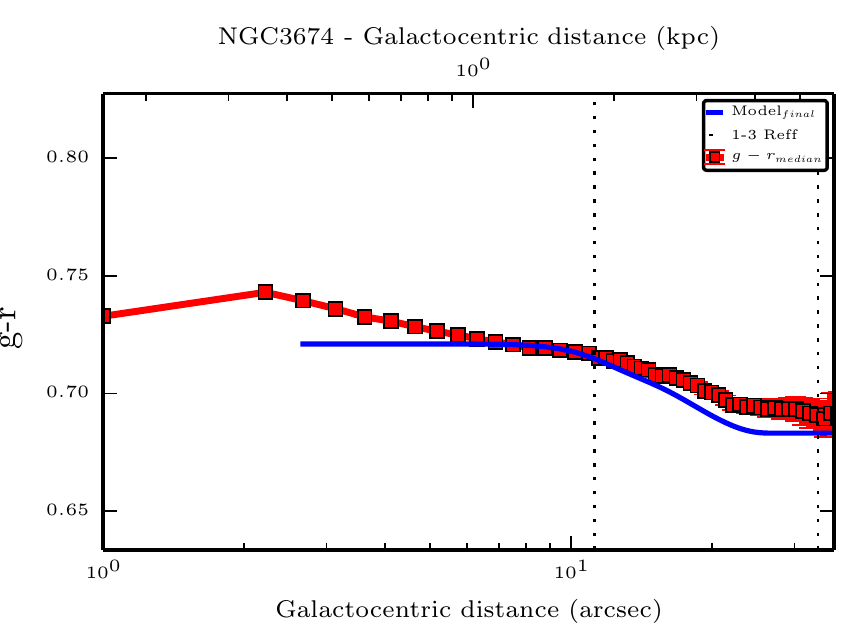}
      \includegraphics[scale=0.55]{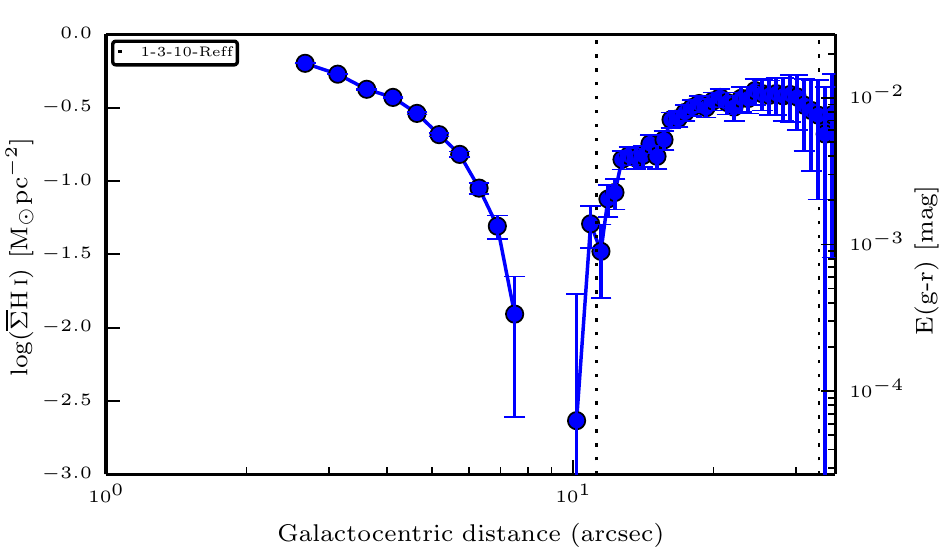}
\end{minipage}}
\caption{True image, colour map, colour excess map, and the radial profiles of NGC~3674.}
\label{fig:app_profiles}
\end{figure*}

\begin{figure*}
\makebox[\textwidth][c]{\begin{minipage}[b][11.6cm]{.85\textwidth}
  \vspace*{\fill}
      \includegraphics[scale=0.85]{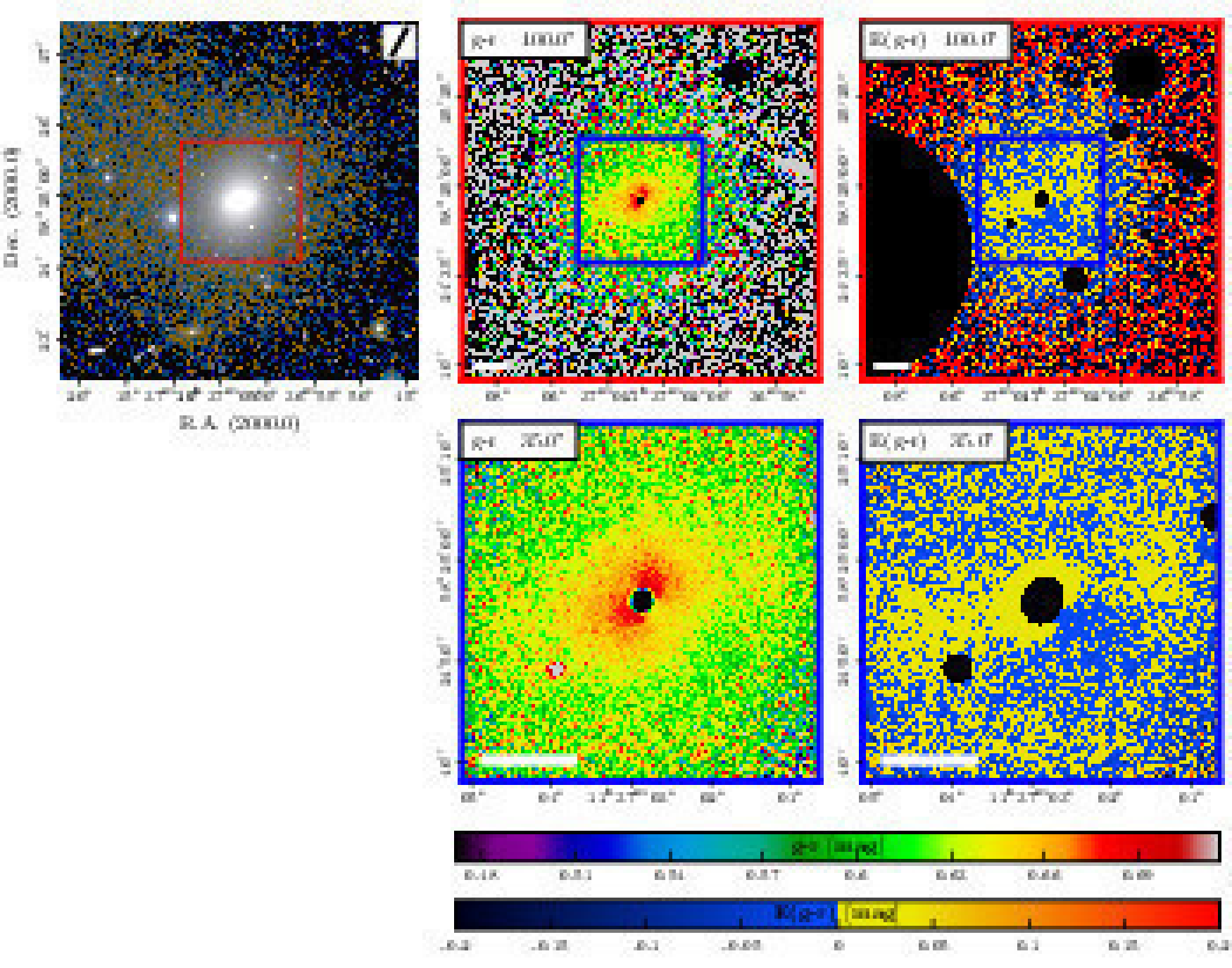}
\end{minipage}}
\makebox[\textwidth][c]{\begin{minipage}[l][-0.7cm][b]{.85\linewidth}
      \includegraphics[scale=0.55, trim={0 0.7cm 0 0},clip]{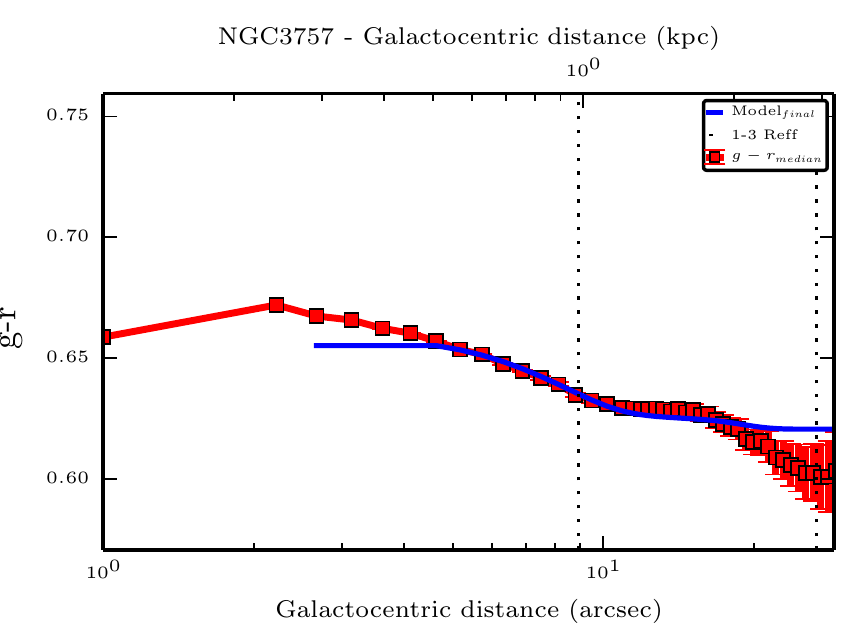}
      \includegraphics[scale=0.55]{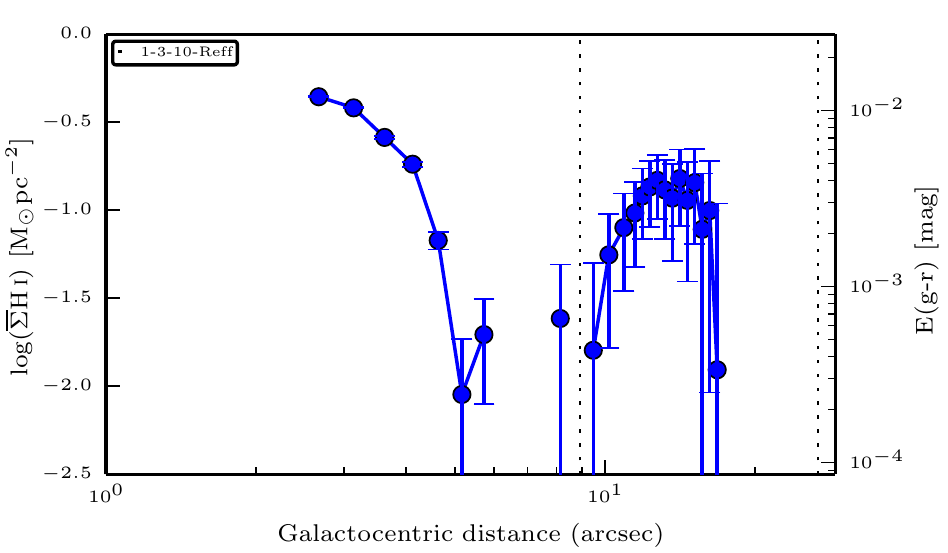}
\end{minipage}}
\caption{True image, colour map, colour excess map, and the radial profiles of NGC~3757.}
\label{fig:app_profiles}
\end{figure*}

%Page17
\clearpage
\begin{figure*}
\makebox[\textwidth][c]{\begin{minipage}[b][10.5cm]{.85\textwidth}
  \vspace*{\fill}
      \includegraphics[scale=0.85]{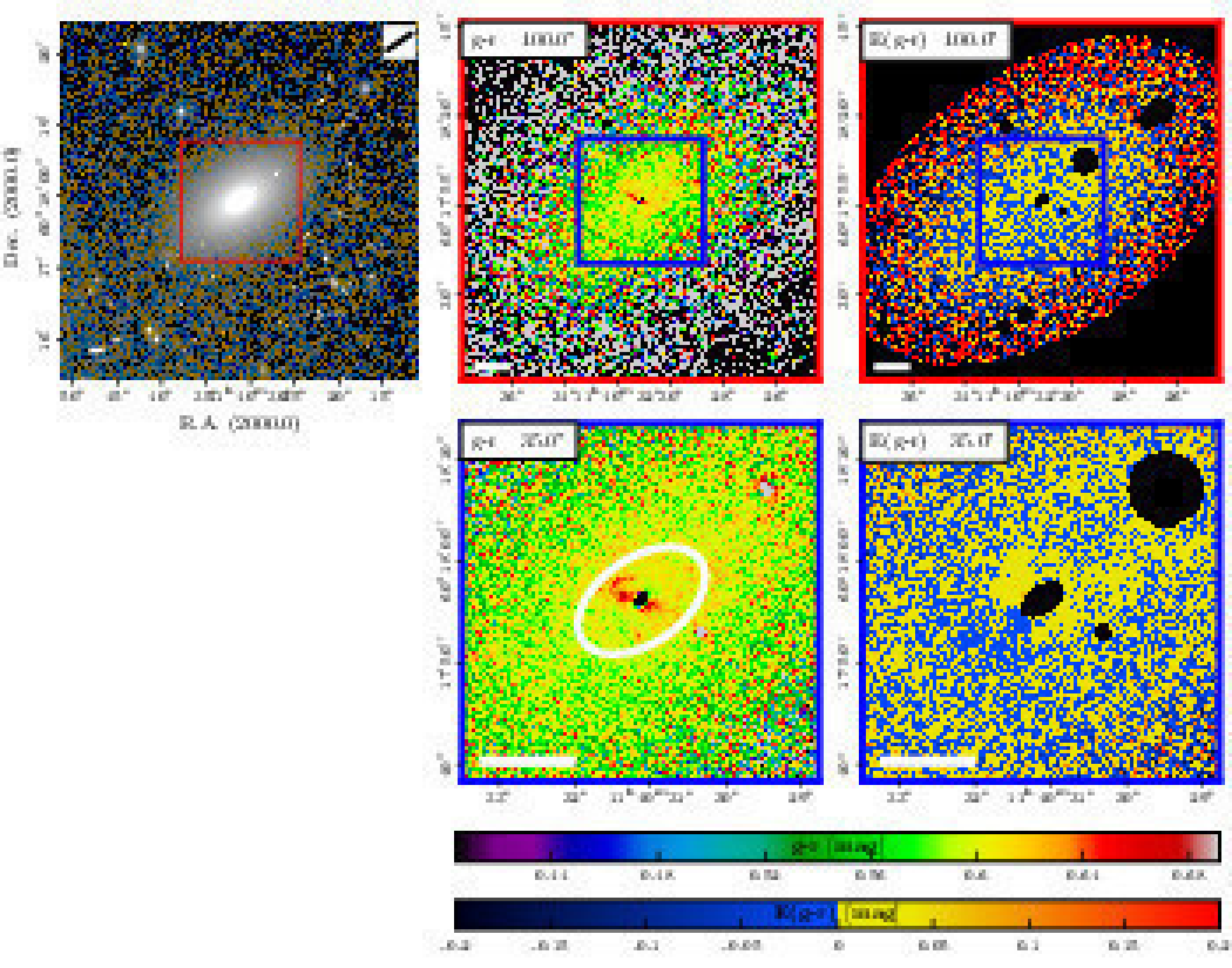}
\end{minipage}}
\makebox[\textwidth][c]{\begin{minipage}[l][-0.7cm][b]{.85\linewidth}
      \includegraphics[scale=0.55, trim={0 0.7cm 0 0},clip]{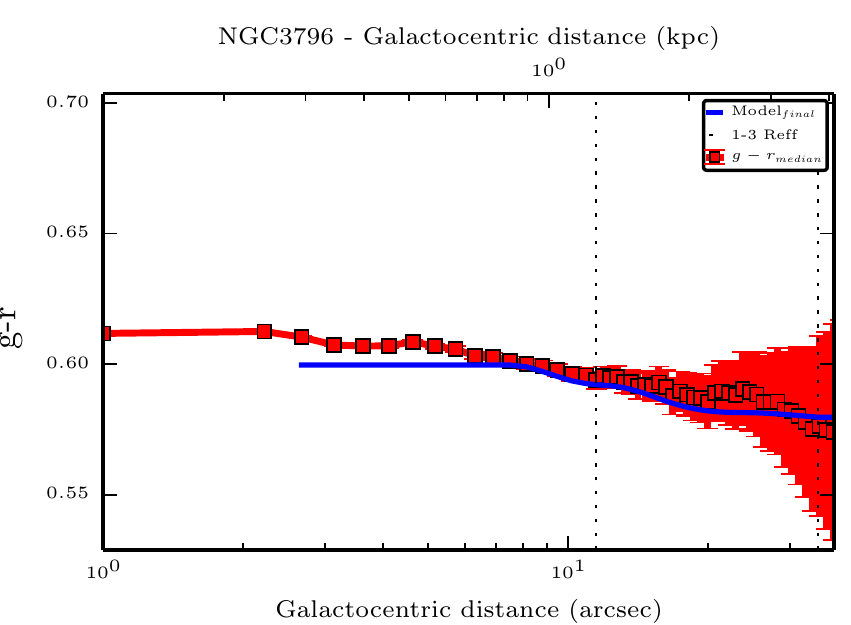}
      \includegraphics[scale=0.55]{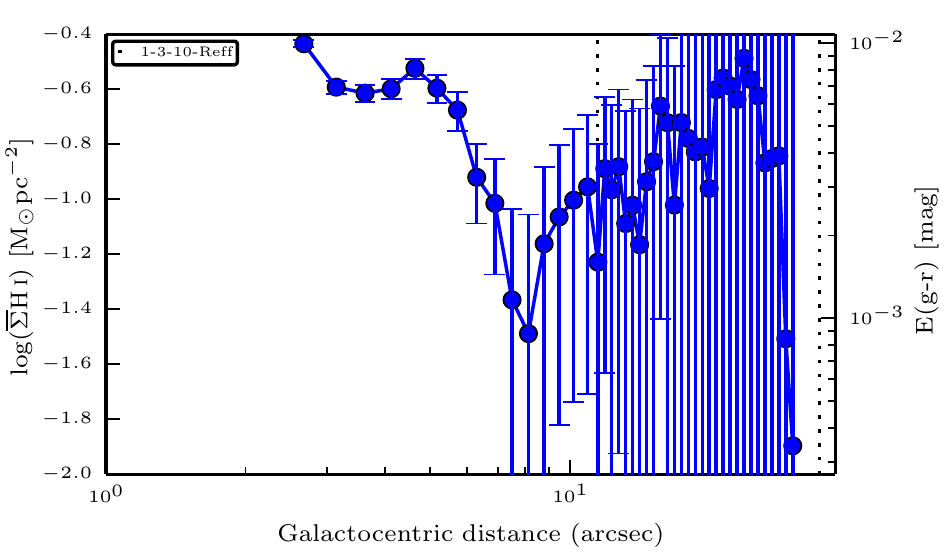}
\end{minipage}}
\caption{True image, colour map, colour excess map, and the radial profiles of NGC~3796.}
\label{fig:app_profiles}
\end{figure*}

\begin{figure*}
\makebox[\textwidth][c]{\begin{minipage}[b][11.6cm]{.85\textwidth}
  \vspace*{\fill}
      \includegraphics[scale=0.85]{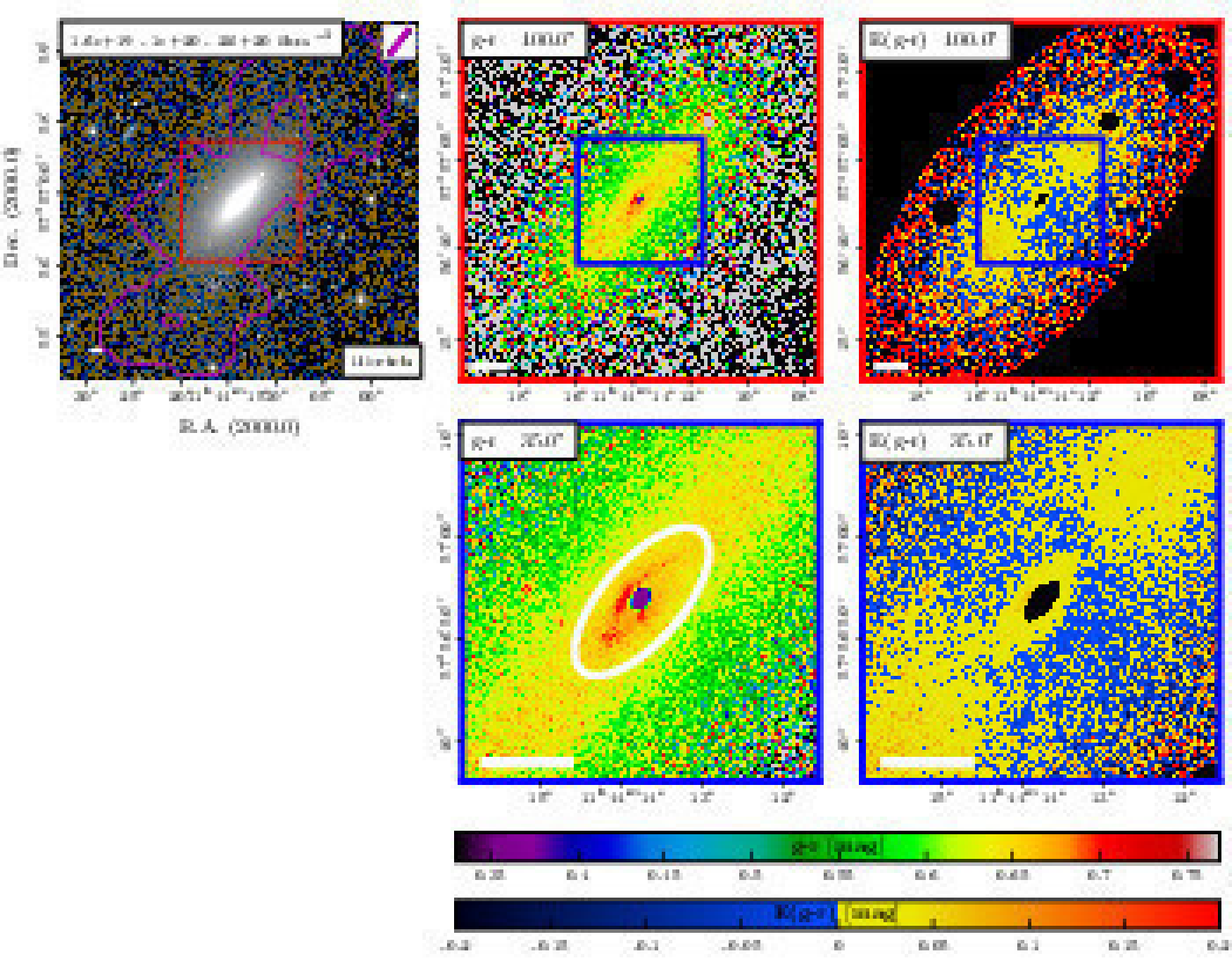}
\end{minipage}}
\makebox[\textwidth][c]{\begin{minipage}[l][-0.7cm][b]{.85\linewidth}
      \includegraphics[scale=0.55, trim={0 0.7cm 0 0},clip]{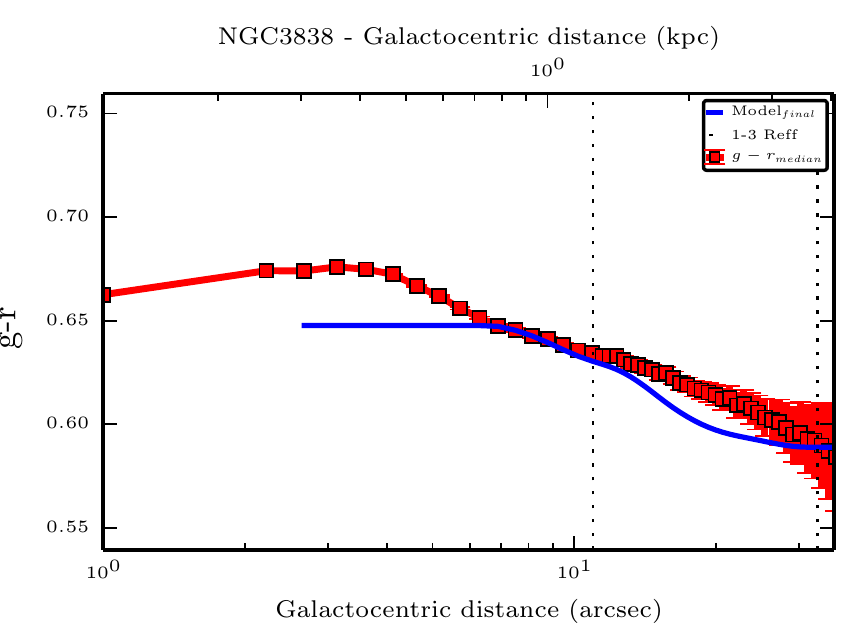}
      \includegraphics[scale=0.55]{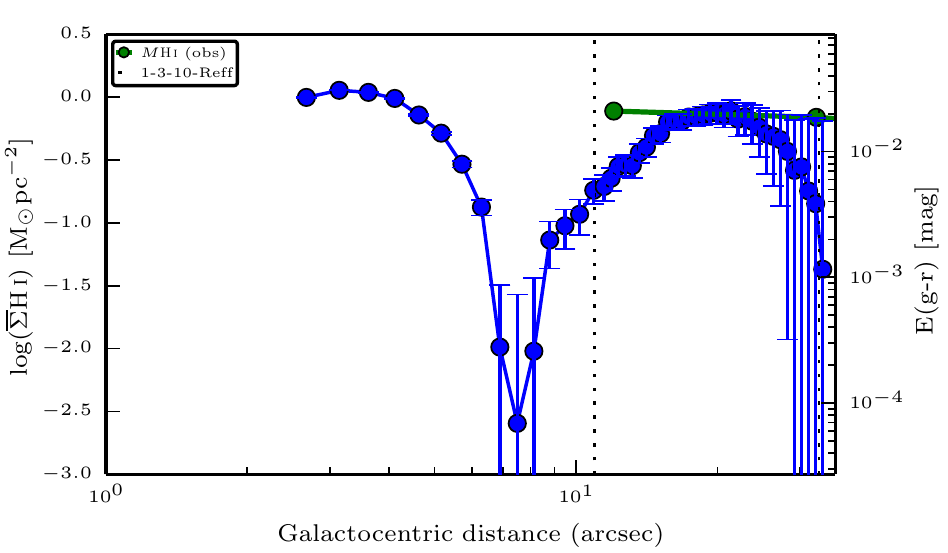}
\end{minipage}}
\caption{True image, colour map, colour excess map, and the radial profiles of NGC~3838.}
\label{fig:app_profiles}
\end{figure*}

%Page18
\clearpage
\begin{figure*}
\makebox[\textwidth][c]{\begin{minipage}[b][10.5cm]{.85\textwidth}
  \vspace*{\fill}
      \includegraphics[scale=0.85]{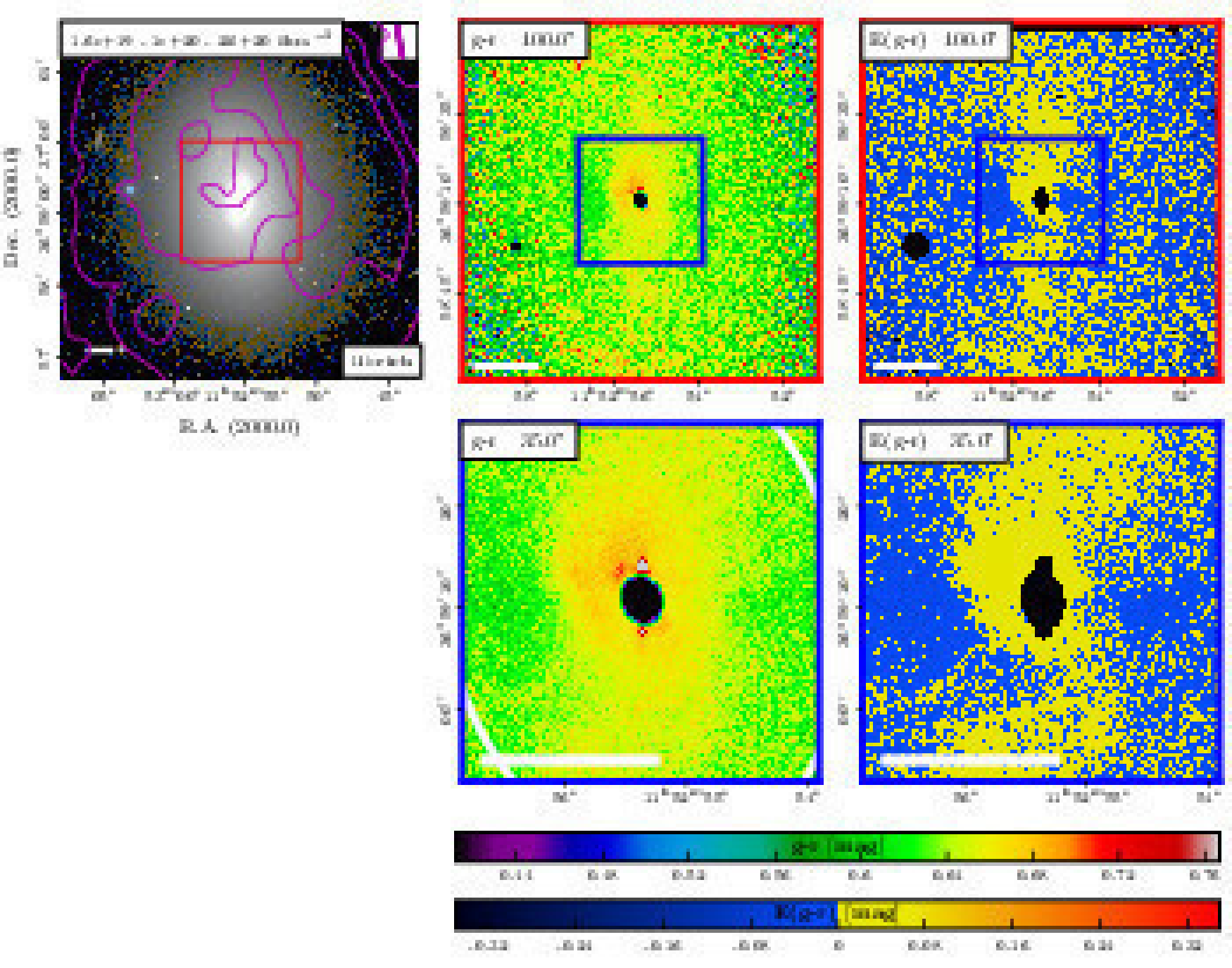}
\end{minipage}}
\makebox[\textwidth][c]{\begin{minipage}[l][-0.7cm][b]{.85\linewidth}
      \includegraphics[scale=0.55, trim={0 0.7cm 0 0},clip]{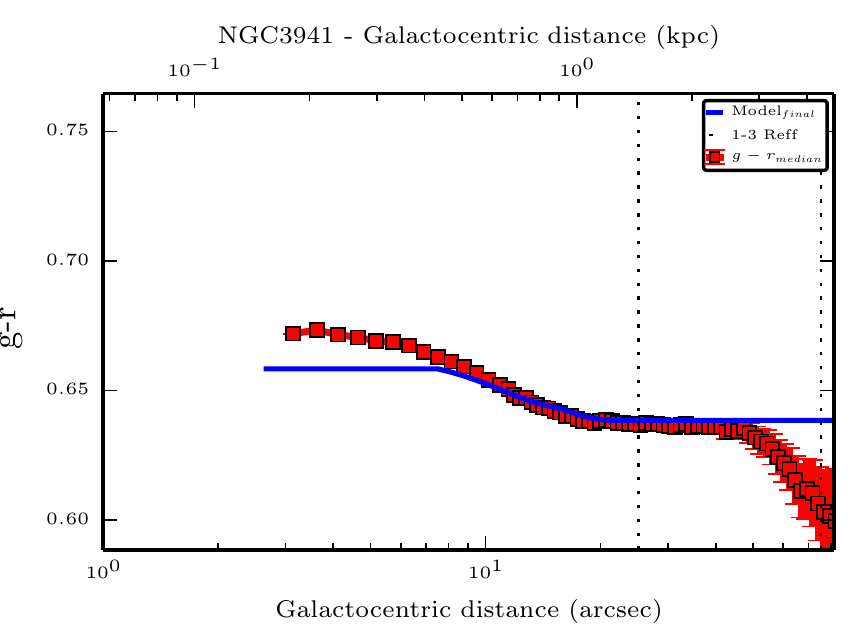}
      \includegraphics[scale=0.55]{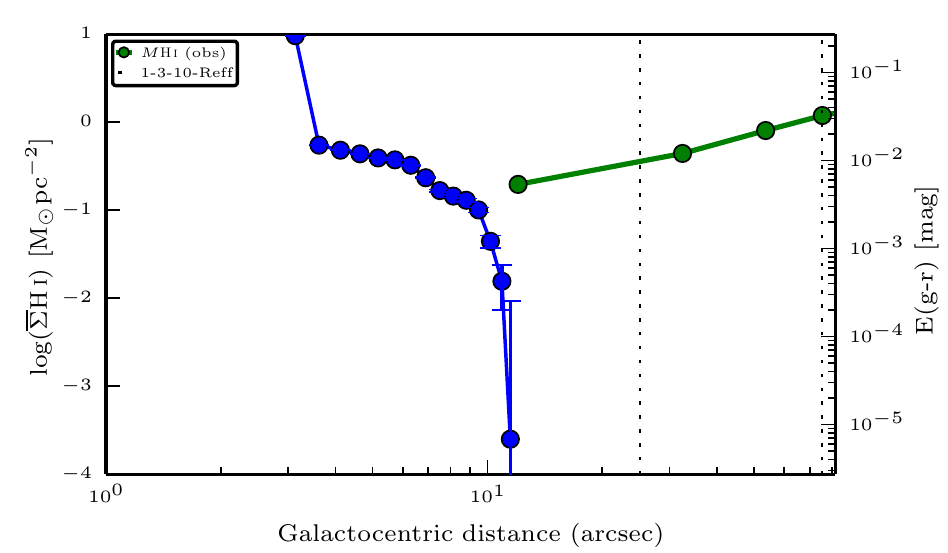}
\end{minipage}}
\caption{True image, colour map, colour excess map, and the radial profiles of NGC~3941.}
\label{fig:app_profiles}
\end{figure*}

\begin{figure*}
\makebox[\textwidth][c]{\begin{minipage}[b][11.6cm]{.85\textwidth}
  \vspace*{\fill}
      \includegraphics[scale=0.85]{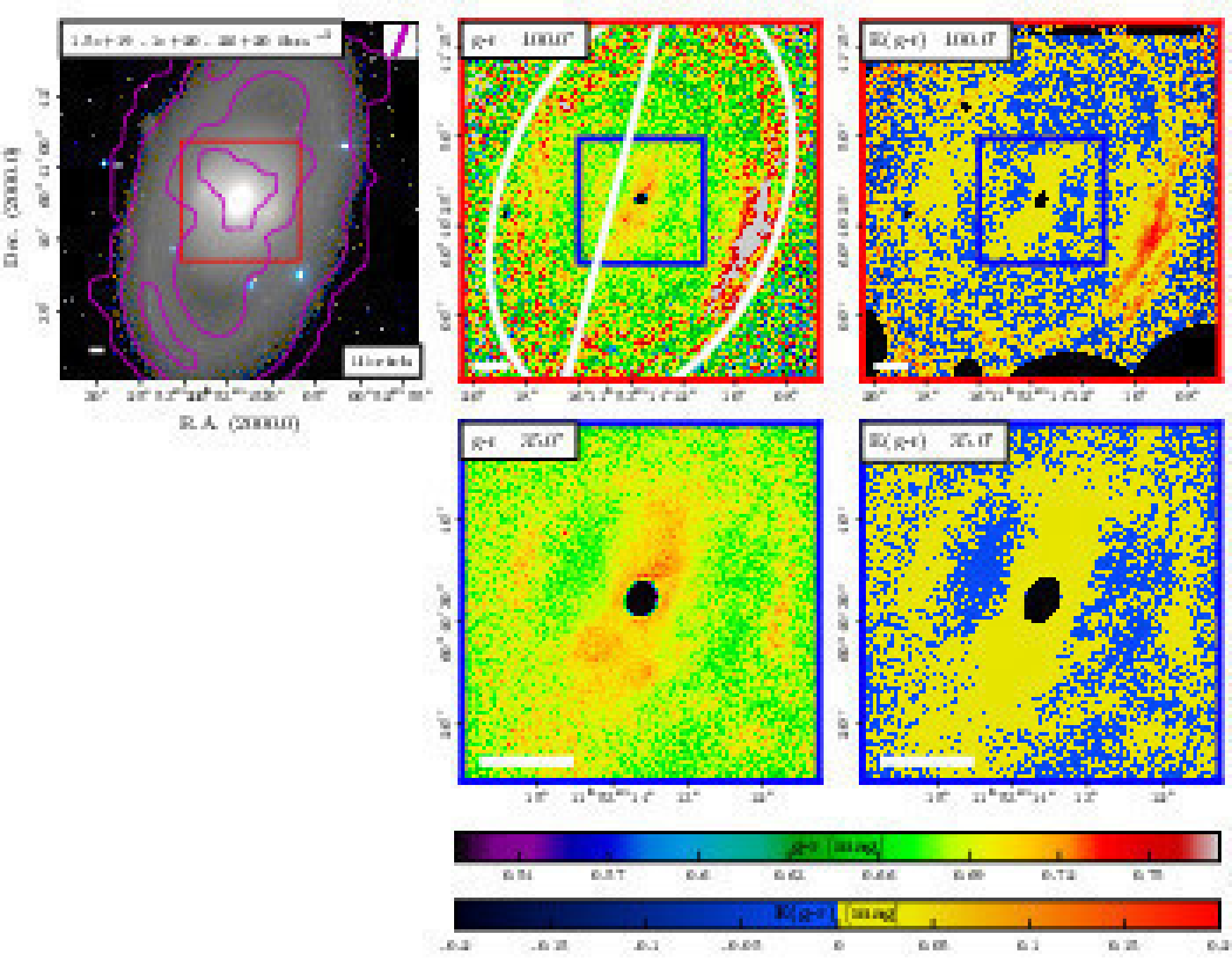}
\end{minipage}}
\makebox[\textwidth][c]{\begin{minipage}[l][-0.7cm][b]{.85\linewidth}
      \includegraphics[scale=0.55, trim={0 0.7cm 0 0},clip]{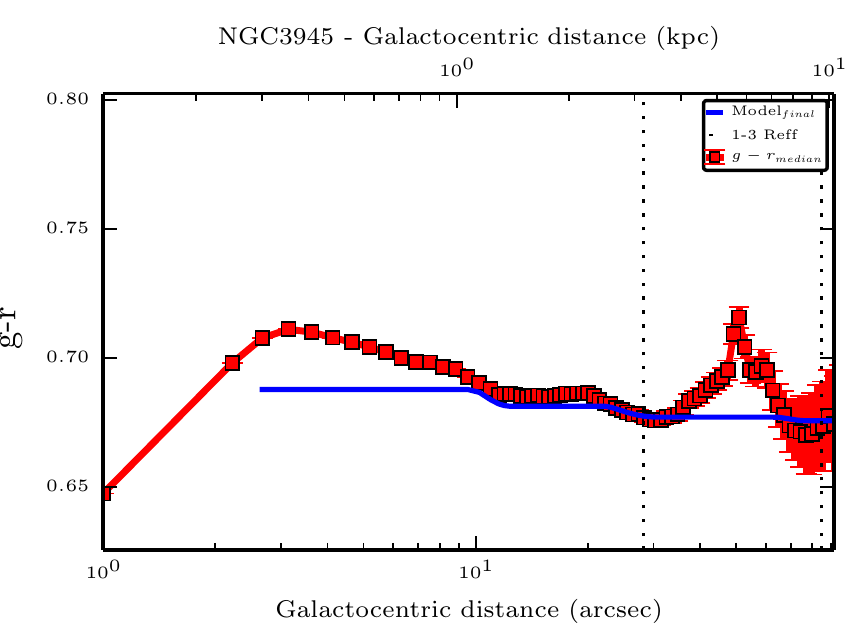}
      \includegraphics[scale=0.55]{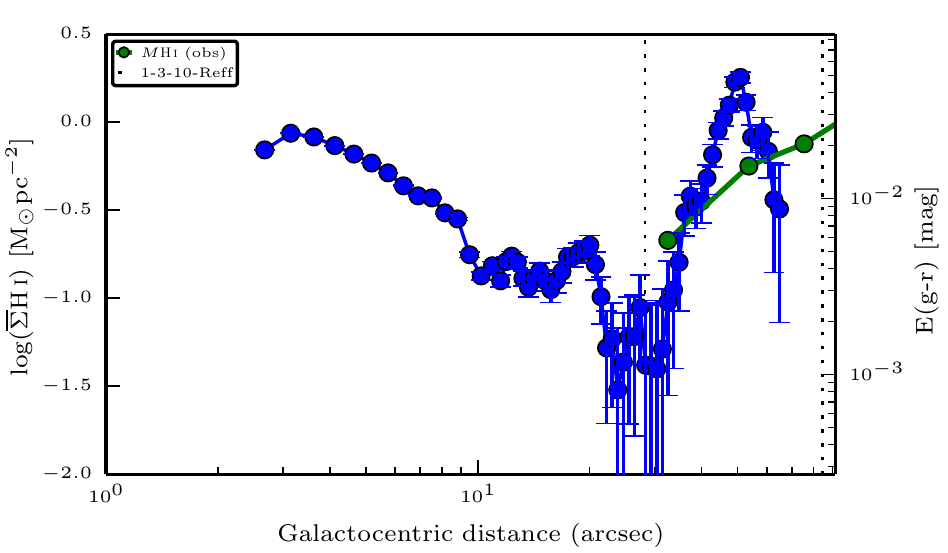}
\end{minipage}}
\caption{True image, colour map, colour excess map, and the radial profiles of NGC~3945.}
\label{fig:app_profiles}
\end{figure*}

%Page19
\clearpage
\begin{figure*}
\makebox[\textwidth][c]{\begin{minipage}[b][10.5cm]{.85\textwidth}
  \vspace*{\fill}
      \includegraphics[scale=0.85]{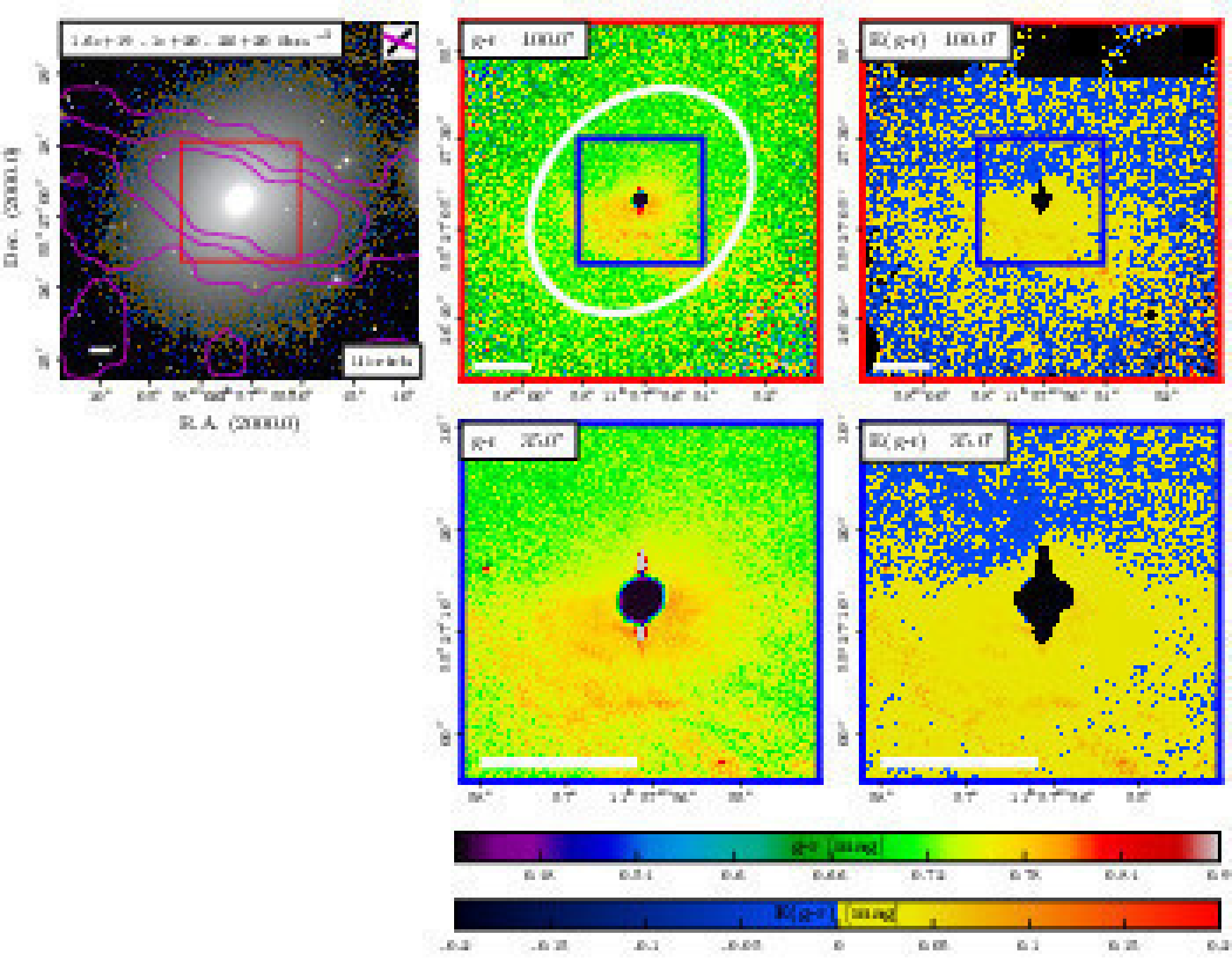}
\end{minipage}}
\makebox[\textwidth][c]{\begin{minipage}[l][-0.7cm][b]{.85\linewidth}
      \includegraphics[scale=0.55, trim={0 0.7cm 0 0},clip]{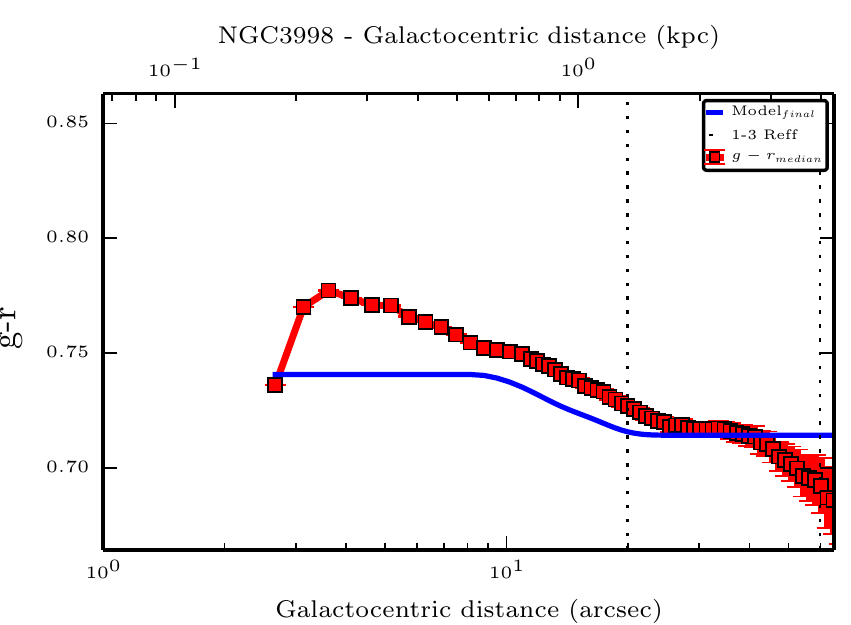}
      \includegraphics[scale=0.55]{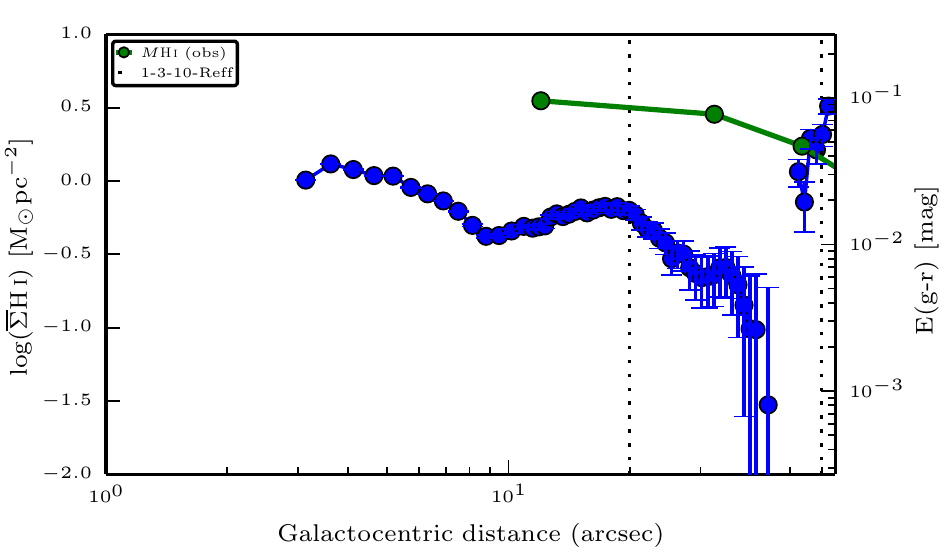}
\end{minipage}}
\caption{True image, colour map, colour excess map, and the radial profiles of NGC~3998.}
\label{fig:app_profiles}
\end{figure*}

\begin{figure*}
\makebox[\textwidth][c]{\begin{minipage}[b][11.6cm]{.85\textwidth}
  \vspace*{\fill}
      \includegraphics[scale=0.85]{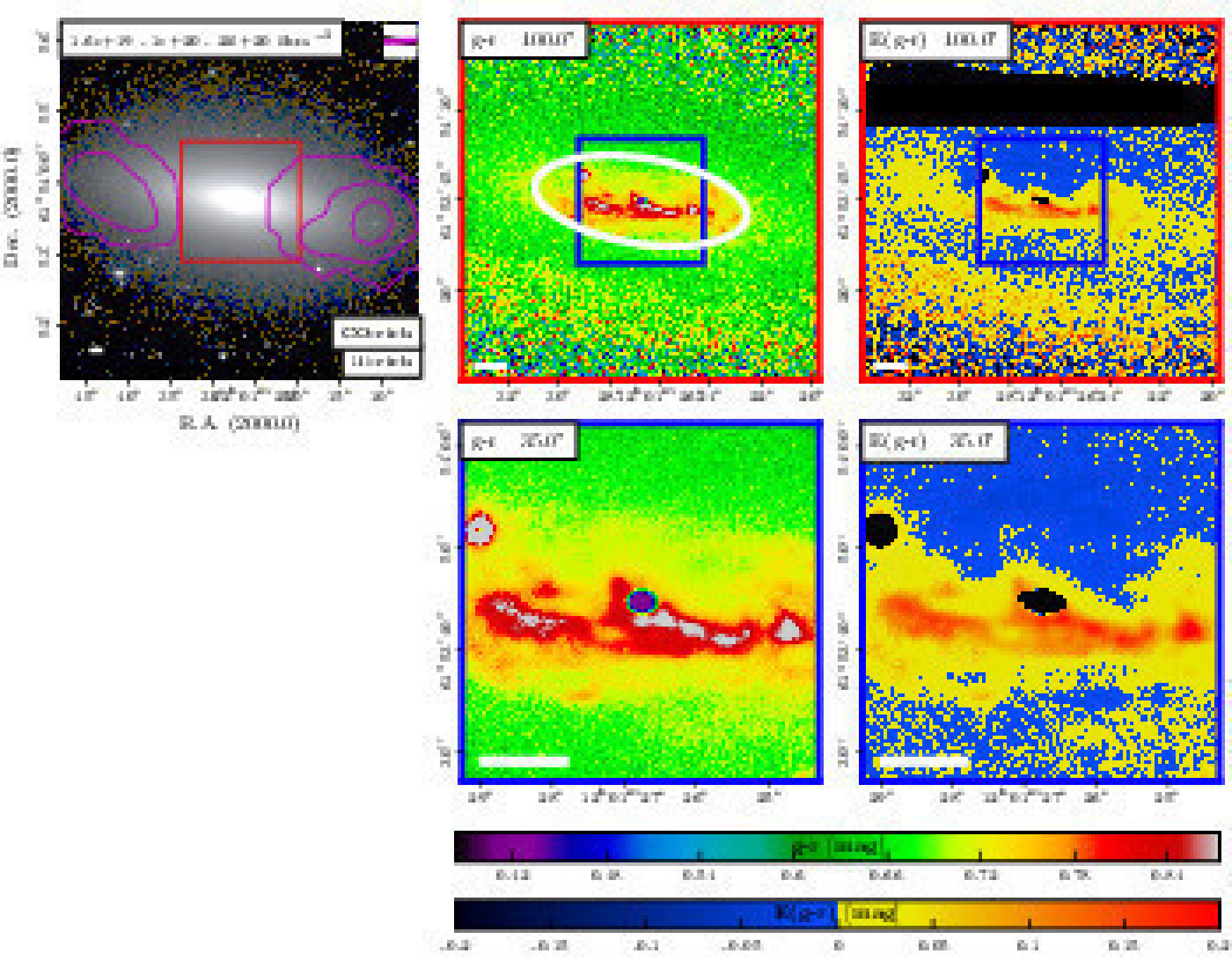}
\end{minipage}}
\makebox[\textwidth][c]{\begin{minipage}[l][-0.7cm][b]{.85\linewidth}
      \includegraphics[scale=0.55, trim={0 0.7cm 0 0},clip]{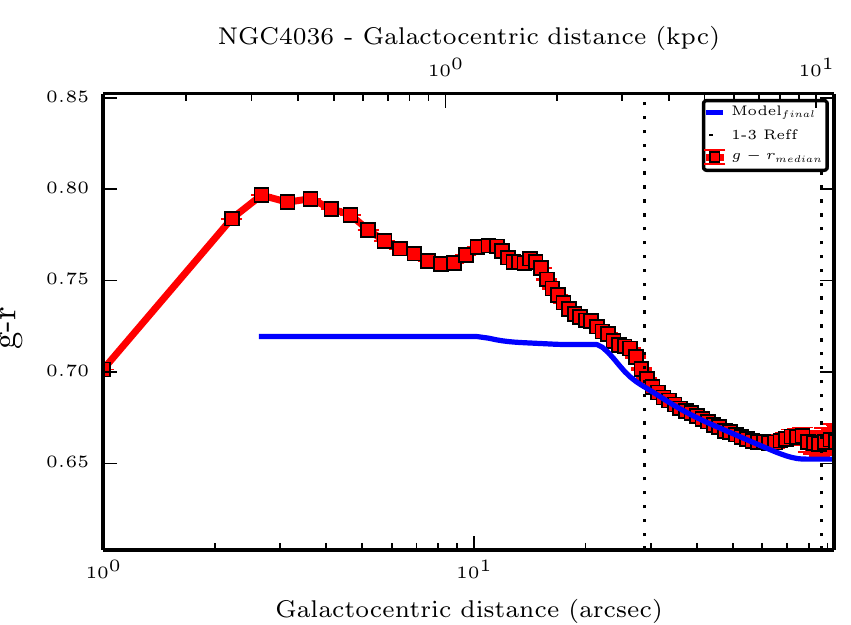}
      \includegraphics[scale=0.55]{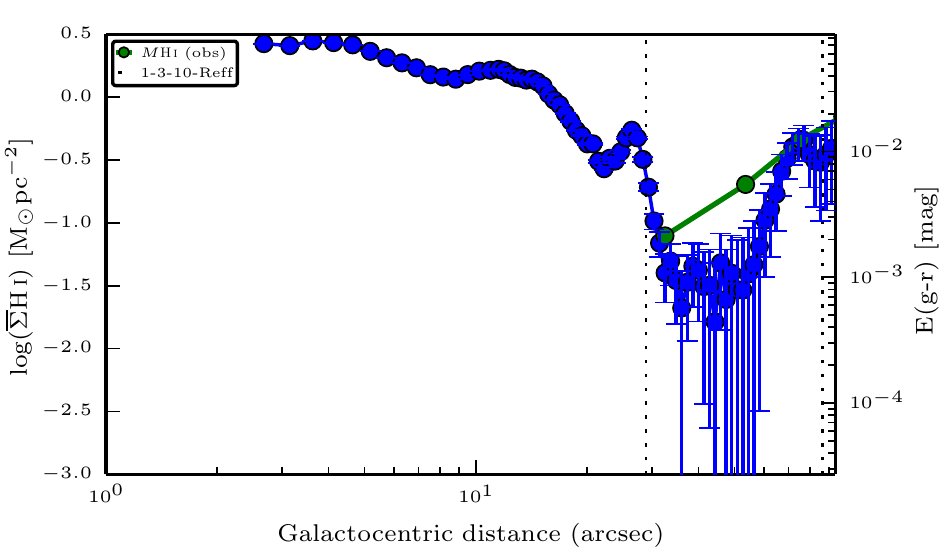}
\end{minipage}}
\caption{True image, colour map, colour excess map, and the radial profiles of NGC~4036.}
\label{fig:app_profiles}
\end{figure*}

%Page20
\clearpage
\begin{figure*}
\makebox[\textwidth][c]{\begin{minipage}[b][10.5cm]{.85\textwidth}
  \vspace*{\fill}
      \includegraphics[scale=0.85]{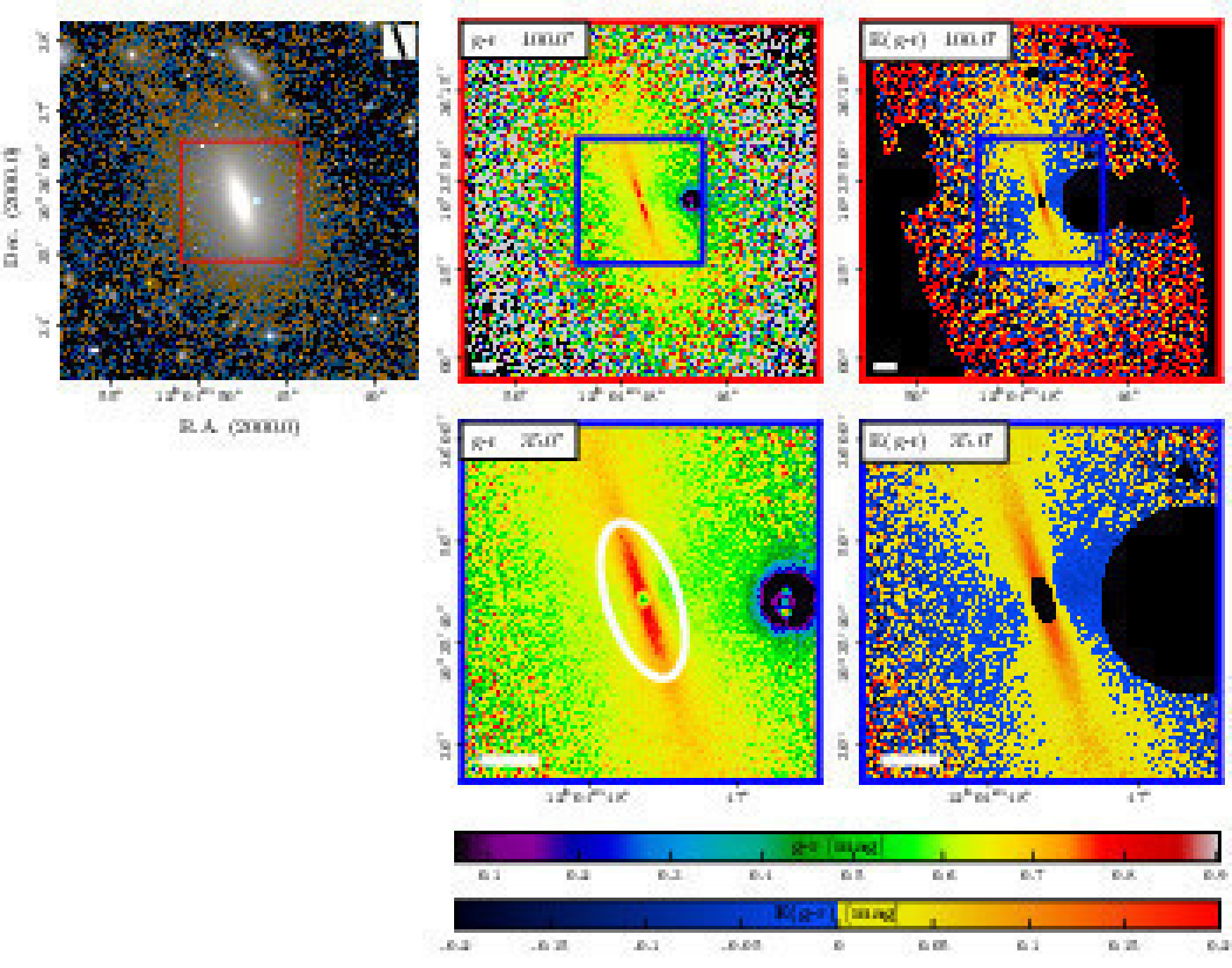}
\end{minipage}}
\makebox[\textwidth][c]{\begin{minipage}[l][-0.7cm][b]{.85\linewidth}
      \includegraphics[scale=0.55, trim={0 0.7cm 0 0},clip]{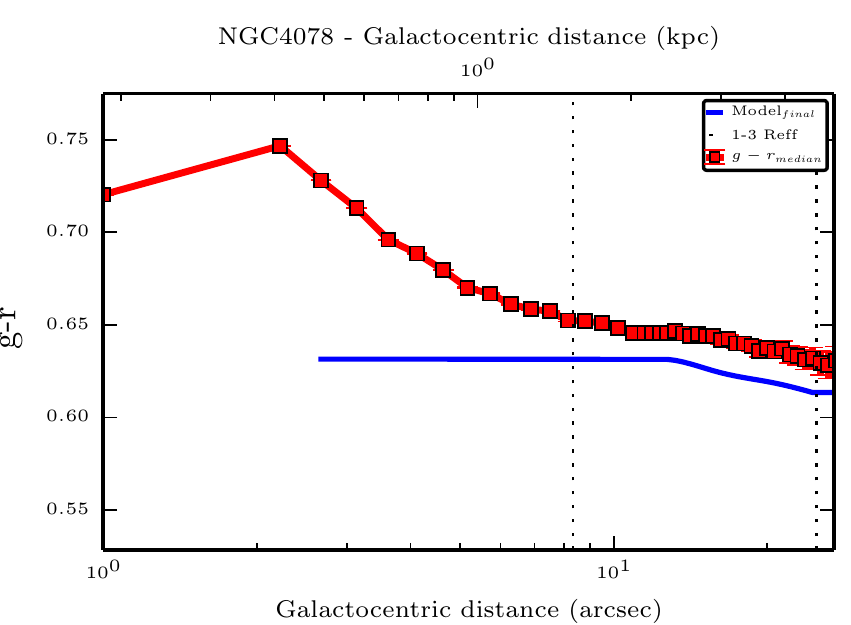}
      \includegraphics[scale=0.55]{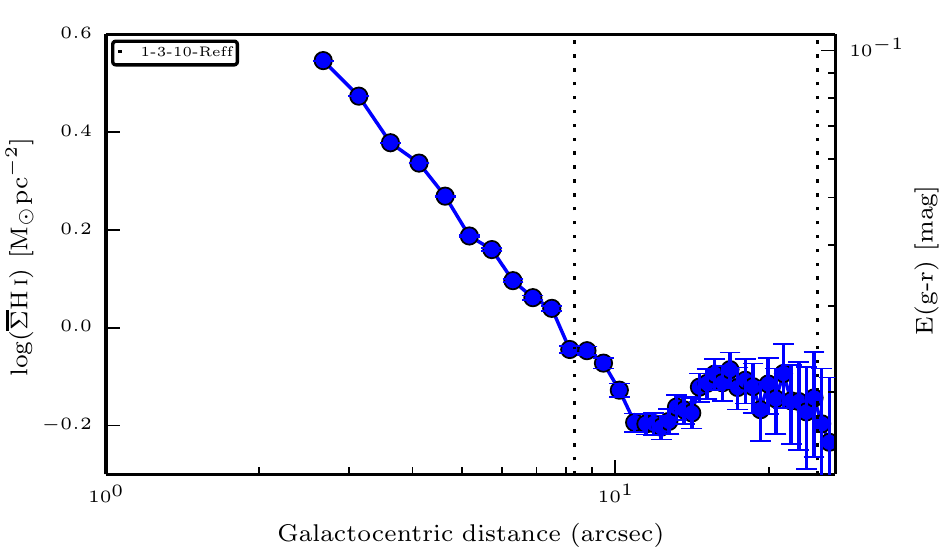}
\end{minipage}}
\caption{True image, colour map, colour excess map, and the radial profiles of NGC~4078.}
\label{fig:app_profiles}
\end{figure*}

\begin{figure*}
\makebox[\textwidth][c]{\begin{minipage}[b][11.6cm]{.85\textwidth}
  \vspace*{\fill}
      \includegraphics[scale=0.85]{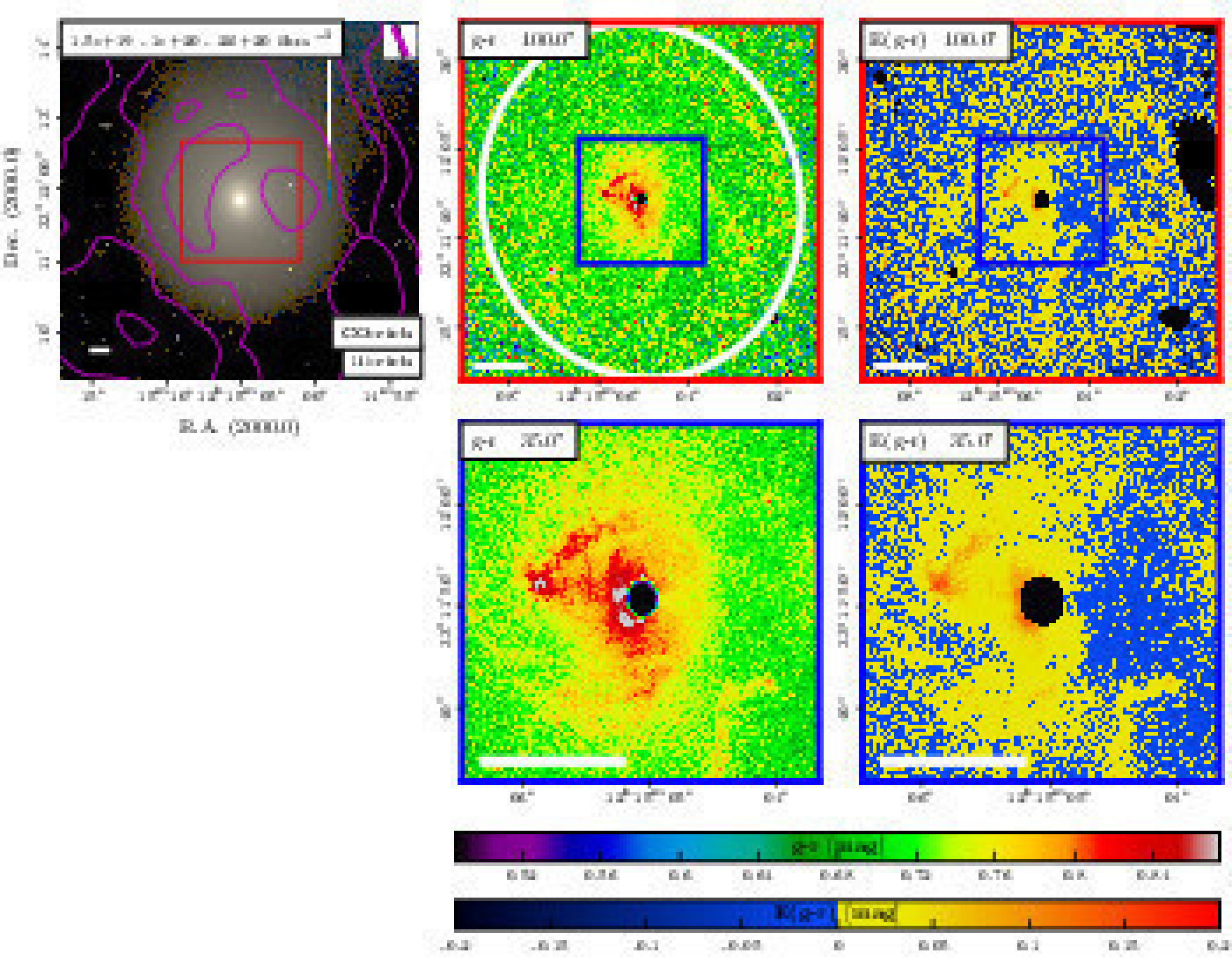}
\end{minipage}}
\makebox[\textwidth][c]{\begin{minipage}[l][-0.7cm][b]{.85\linewidth}
      \includegraphics[scale=0.55, trim={0 0.7cm 0 0},clip]{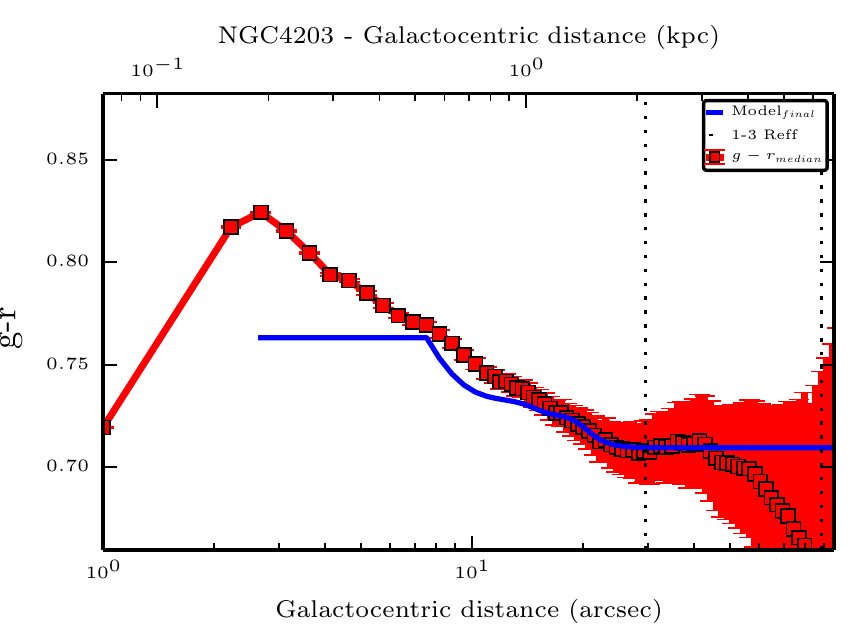}
      \includegraphics[scale=0.55]{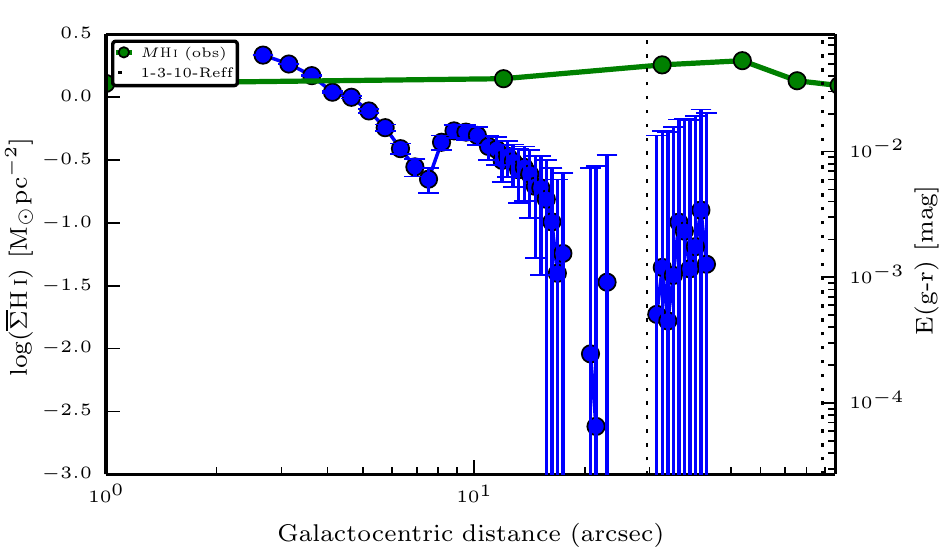}
\end{minipage}}
\caption{True image, colour map, colour excess map, and the radial profiles of NGC~4203.}
\label{fig:app_profiles}
\end{figure*}

%Page21
\clearpage
\begin{figure*}
\makebox[\textwidth][c]{\begin{minipage}[b][10.5cm]{.85\textwidth}
  \vspace*{\fill}
      \includegraphics[scale=0.85]{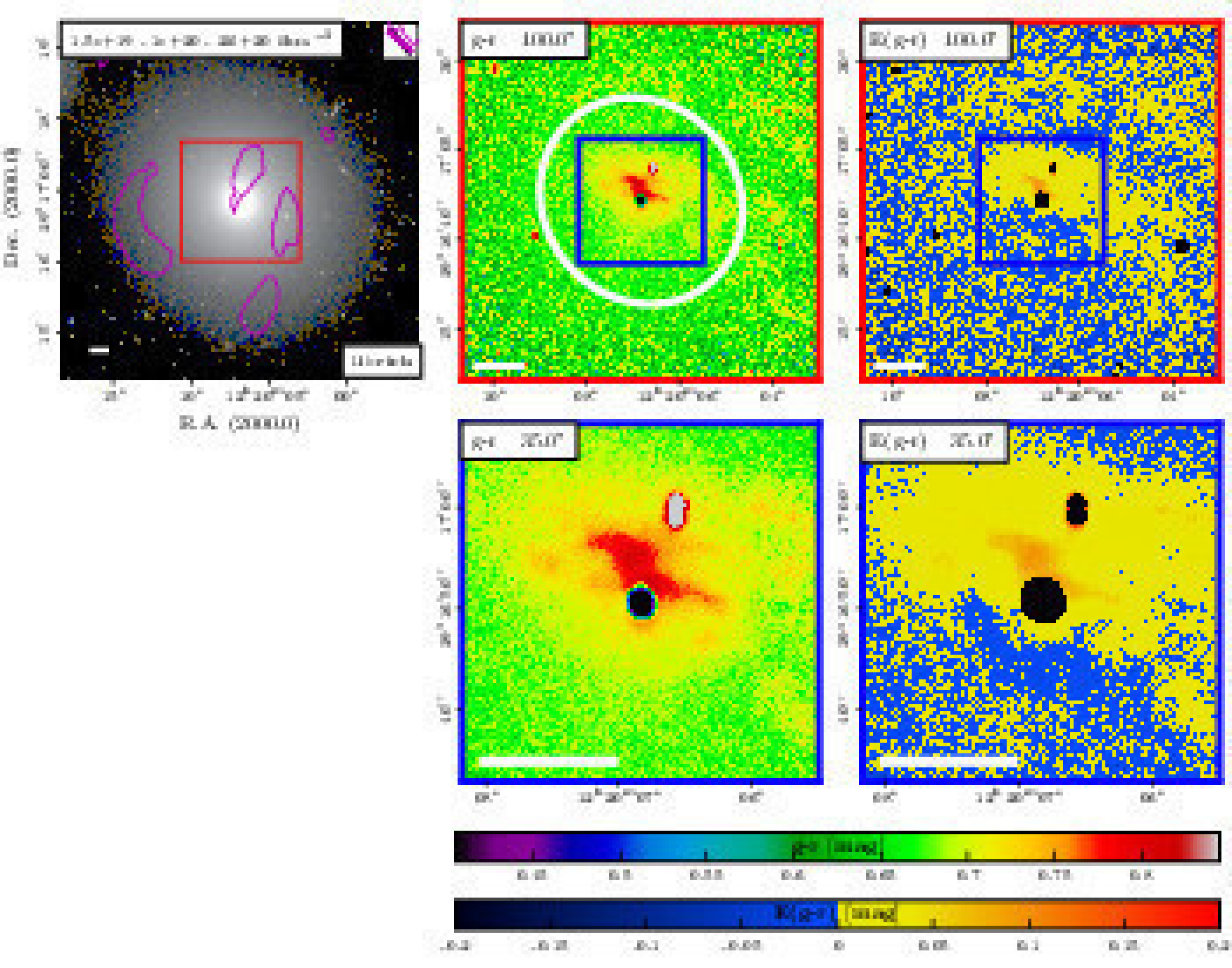}
\end{minipage}}
\makebox[\textwidth][c]{\begin{minipage}[l][-0.7cm][b]{.85\linewidth}
      \includegraphics[scale=0.55, trim={0 0.7cm 0 0},clip]{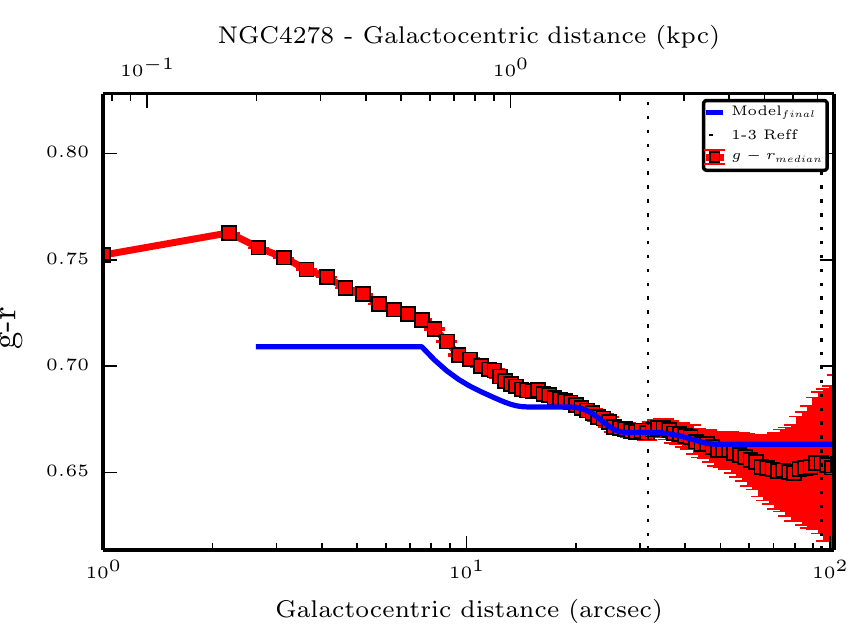}
      \includegraphics[scale=0.55]{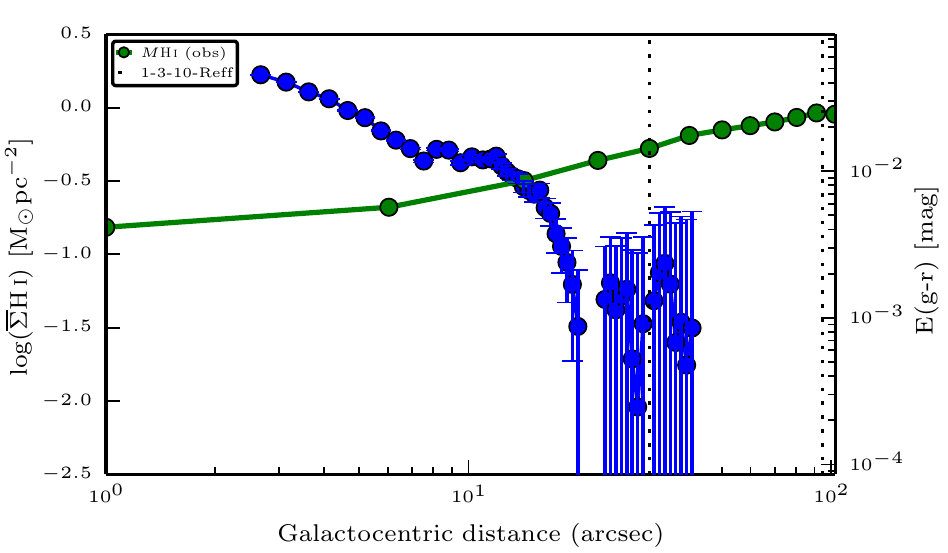}
\end{minipage}}
\caption{True image, colour map, colour excess map, and the radial profiles of NGC~4278.}
\label{fig:app_profiles}
\end{figure*}

\begin{figure*}
\makebox[\textwidth][c]{\begin{minipage}[b][11.6cm]{.85\textwidth}
  \vspace*{\fill}
      \includegraphics[scale=0.85]{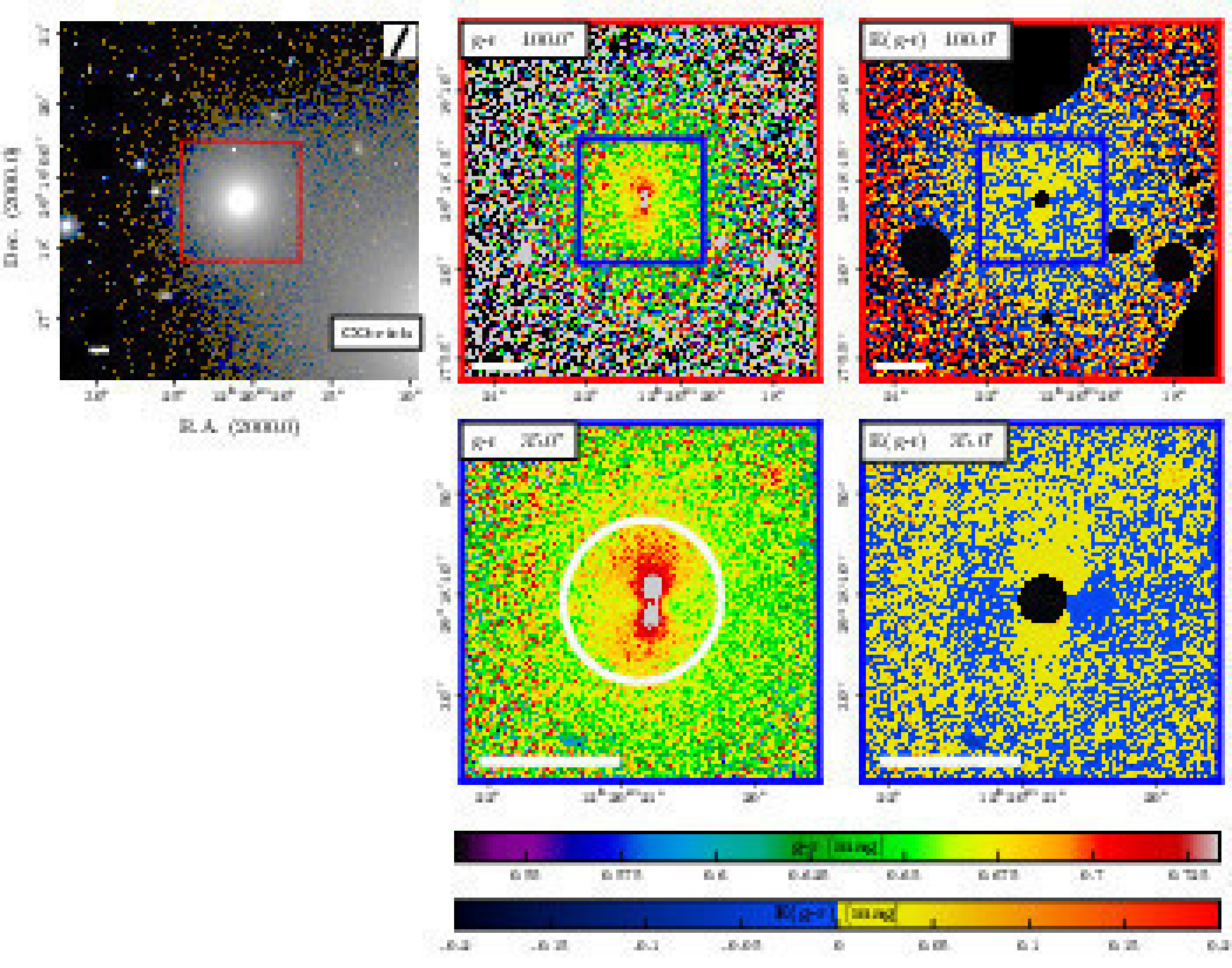}
\end{minipage}}
\makebox[\textwidth][c]{\begin{minipage}[l][-0.7cm][b]{.85\linewidth}
      \includegraphics[scale=0.55, trim={0 0.7cm 0 0},clip]{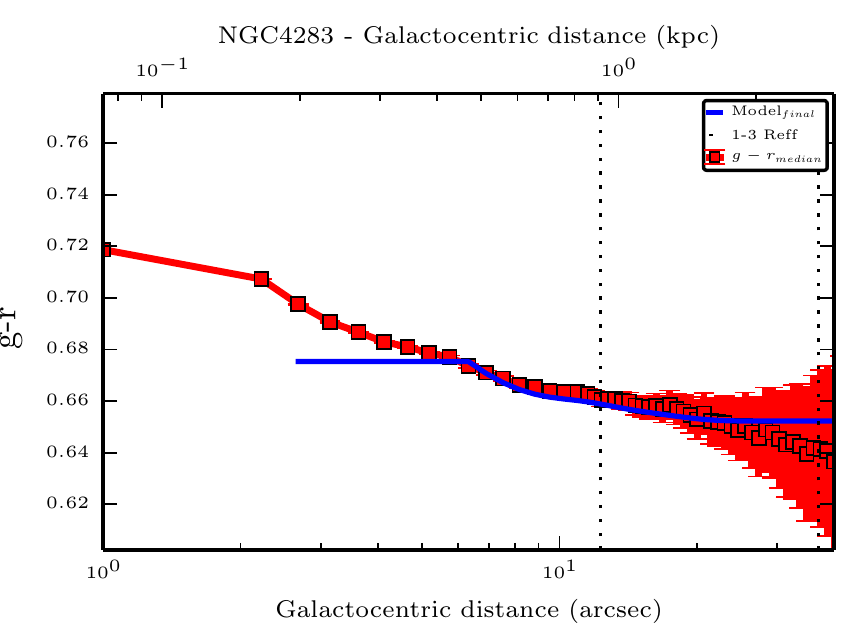}
      \includegraphics[scale=0.55]{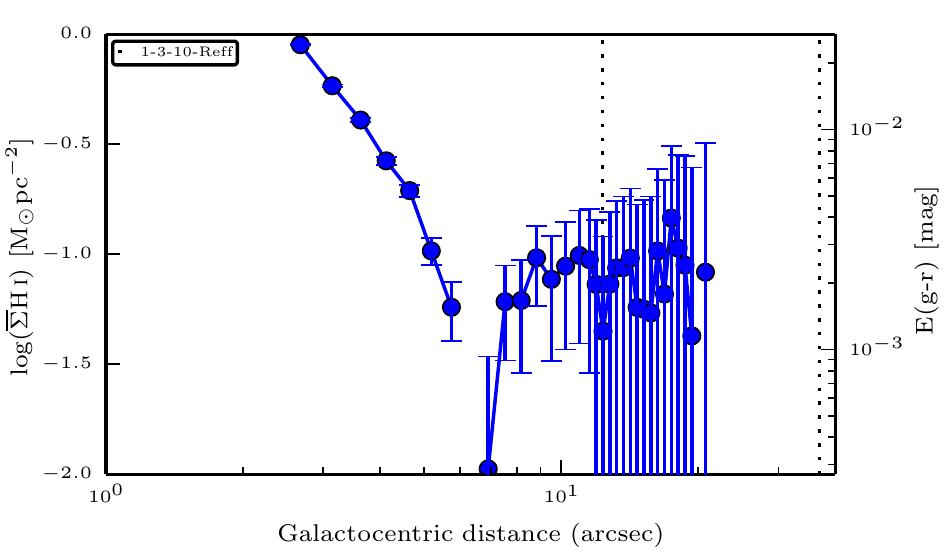}
\end{minipage}}
\caption{True image, colour map, colour excess map, and the radial profiles of NGC~4283.}
\label{fig:app_profiles}
\end{figure*}

%Page22
\clearpage
\begin{figure*}
\makebox[\textwidth][c]{\begin{minipage}[b][10.5cm]{.85\textwidth}
  \vspace*{\fill}
      \includegraphics[scale=0.85]{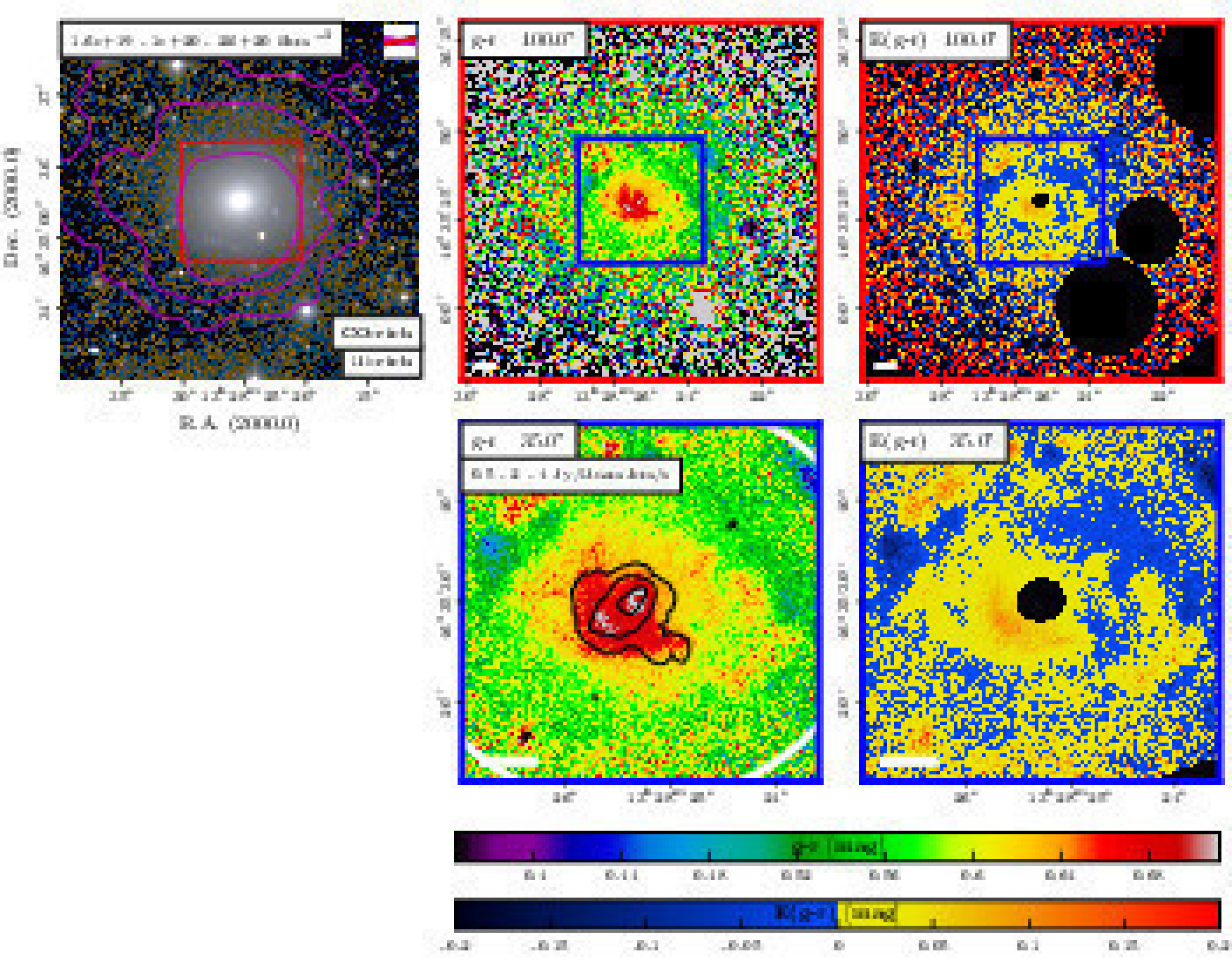}
\end{minipage}}
\makebox[\textwidth][c]{\begin{minipage}[l][-0.7cm][b]{.85\linewidth}
      \includegraphics[scale=0.55, trim={0 0.7cm 0 0},clip]{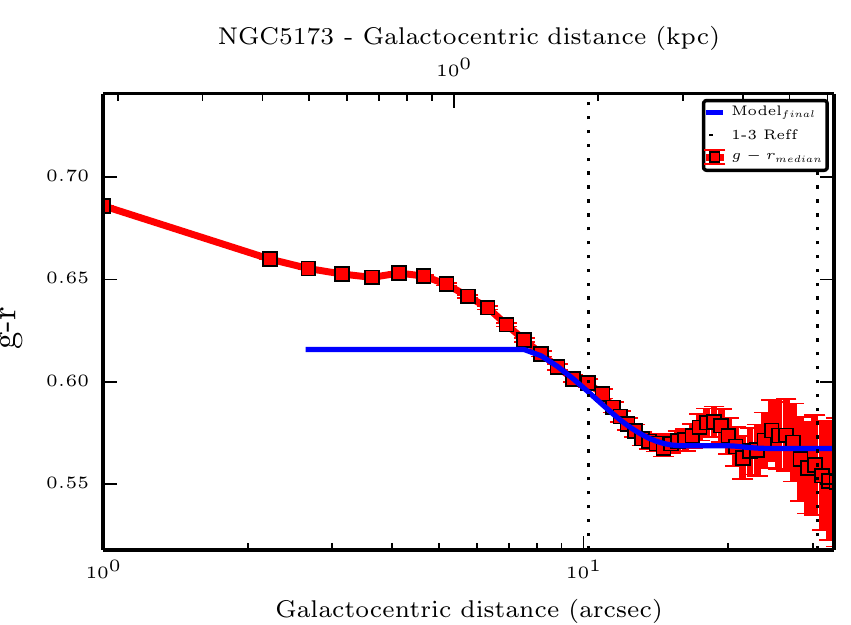}
      \includegraphics[scale=0.55]{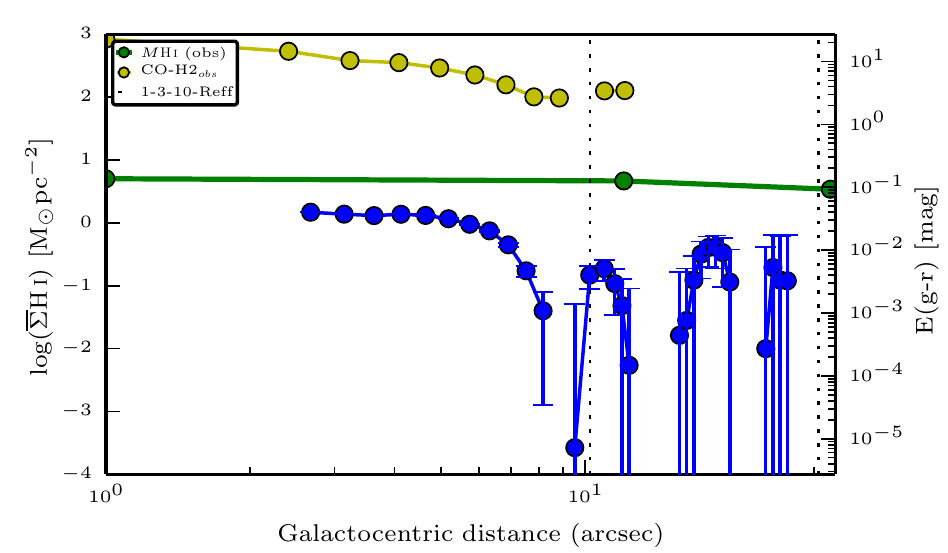}
\end{minipage}}
\caption{True image, colour map, colour excess map, and the radial profiles of NGC~5173.}
\label{fig:app_profiles}
\end{figure*}

\begin{figure*}
\makebox[\textwidth][c]{\begin{minipage}[b][11.6cm]{.85\textwidth}
  \vspace*{\fill}
      \includegraphics[scale=0.85]{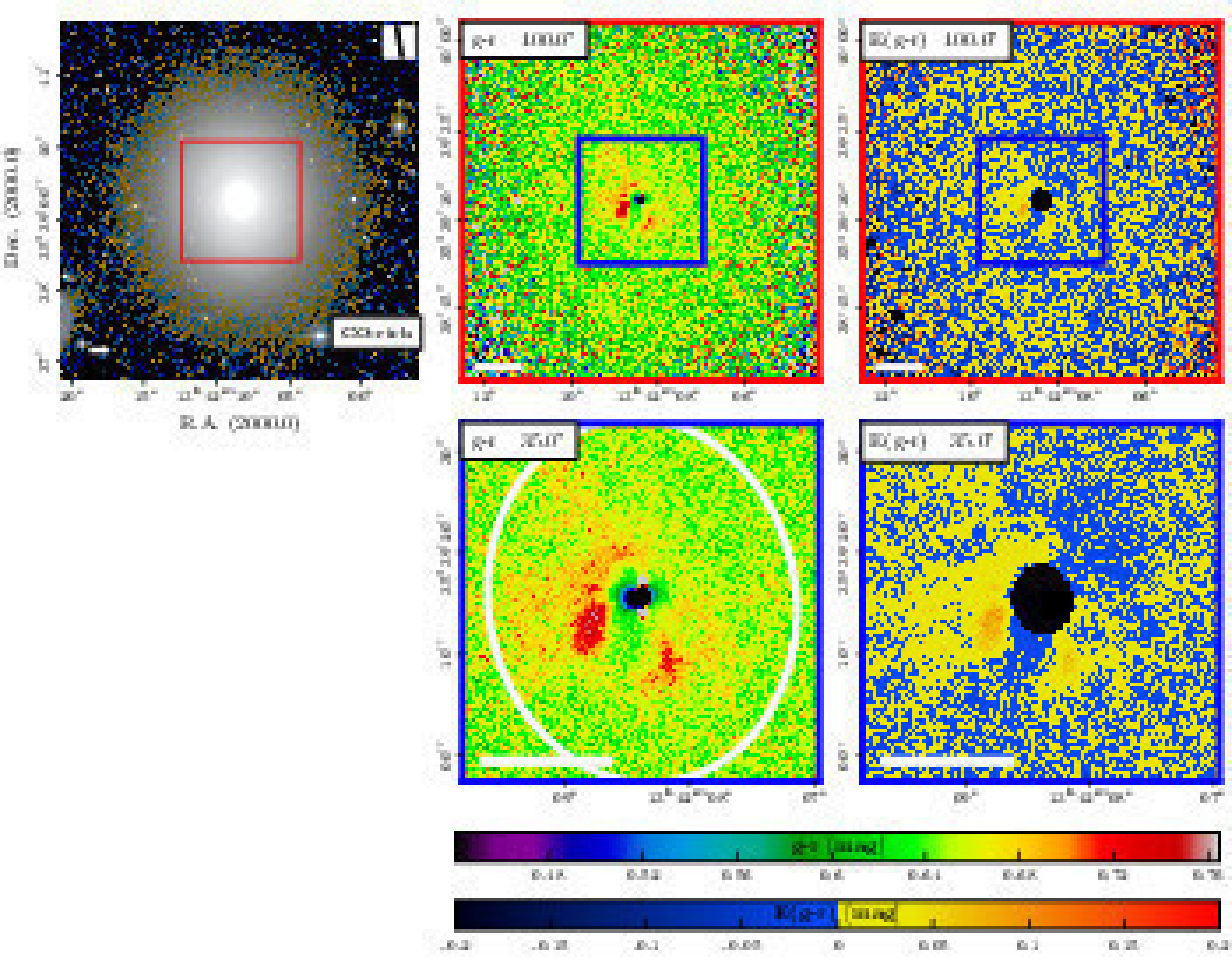}
\end{minipage}}
\makebox[\textwidth][c]{\begin{minipage}[l][-0.7cm][b]{.85\linewidth}
      \includegraphics[scale=0.55, trim={0 0.7cm 0 0},clip]{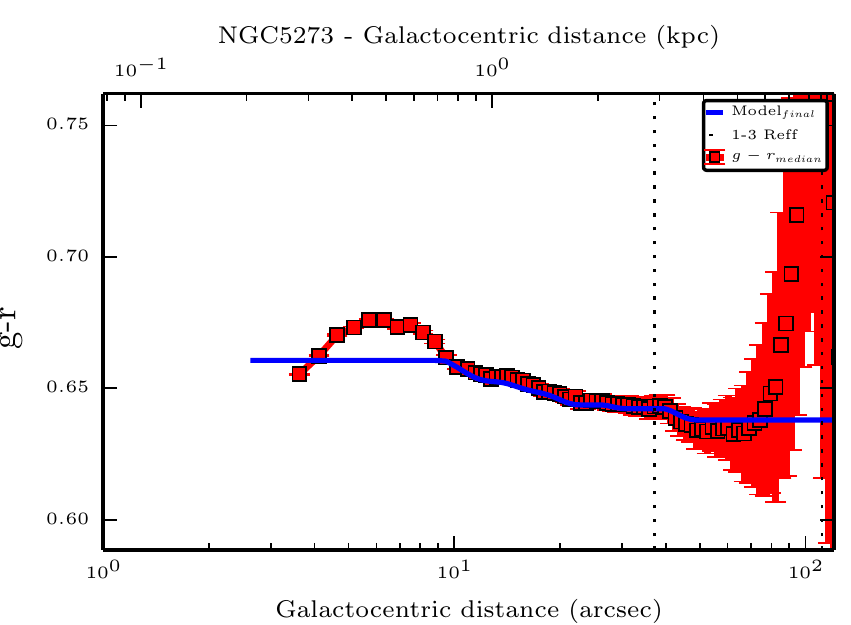}
      \includegraphics[scale=0.55]{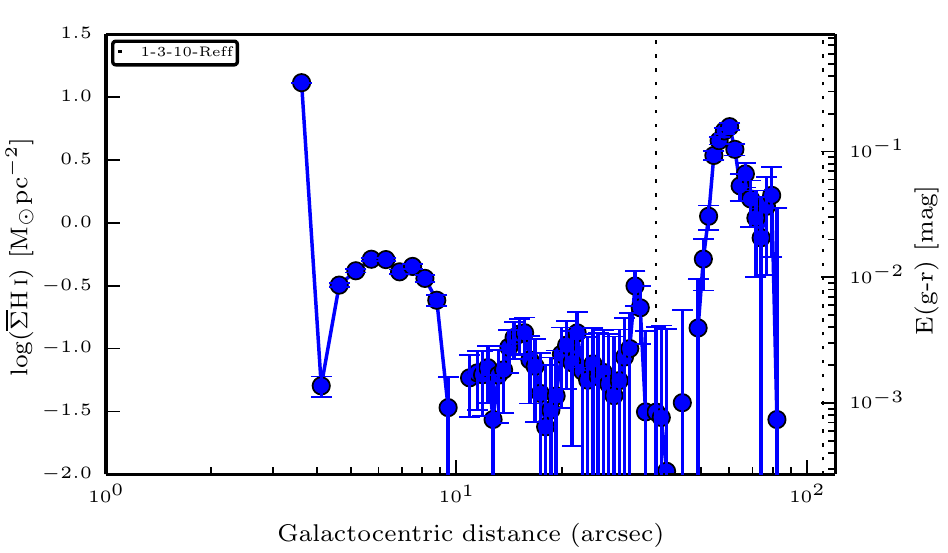}
\end{minipage}}
\caption{True image, colour map, colour excess map, and the radial profiles of NGC~5273.}
\label{fig:app_profiles}
\end{figure*}

%Page23
\clearpage
\begin{figure*}
\makebox[\textwidth][c]{\begin{minipage}[b][10.5cm]{.85\textwidth}
  \vspace*{\fill}
      \includegraphics[scale=0.85]{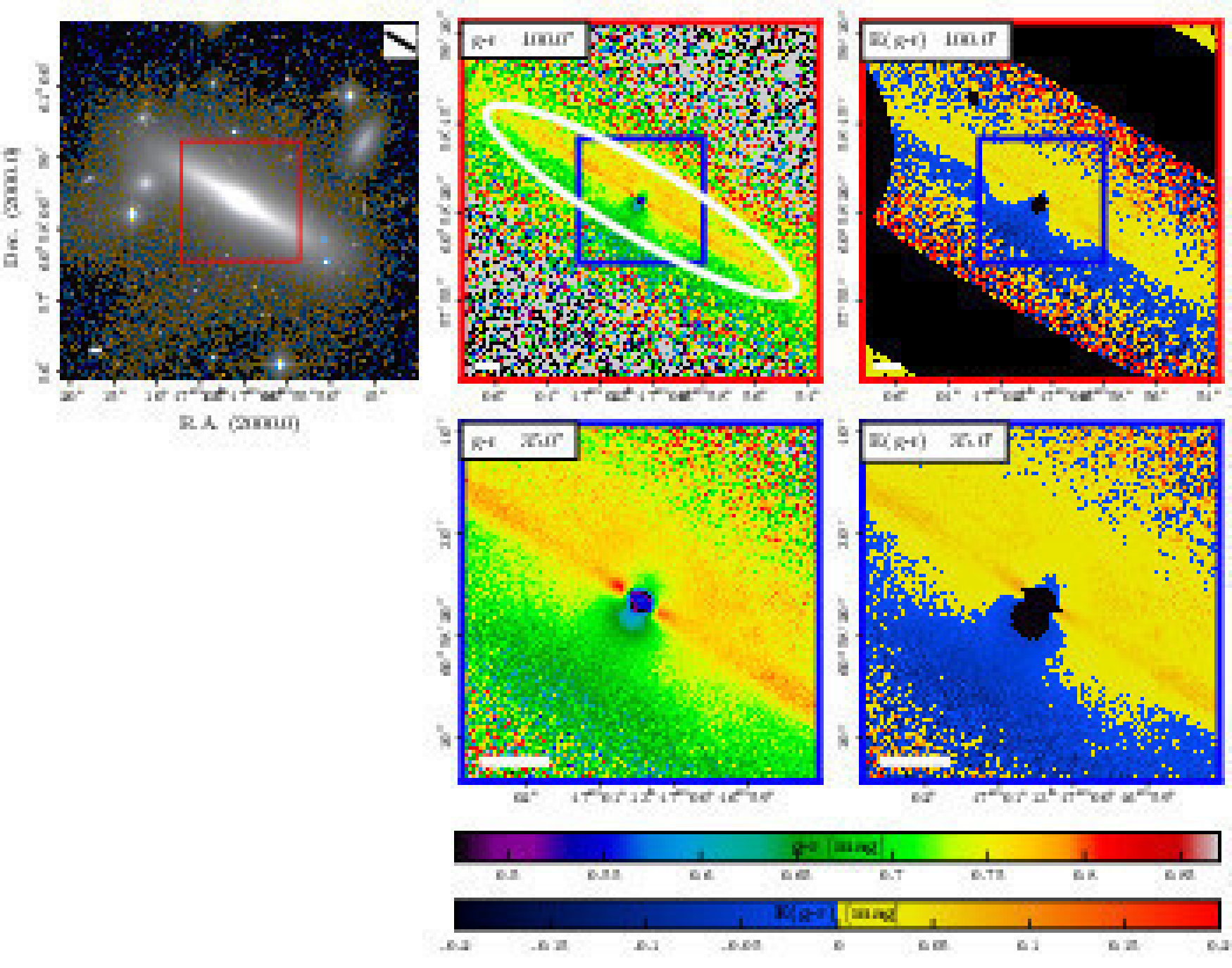}
\end{minipage}}
\makebox[\textwidth][c]{\begin{minipage}[l][-0.7cm][b]{.85\linewidth}
      \includegraphics[scale=0.55, trim={0 0.7cm 0 0},clip]{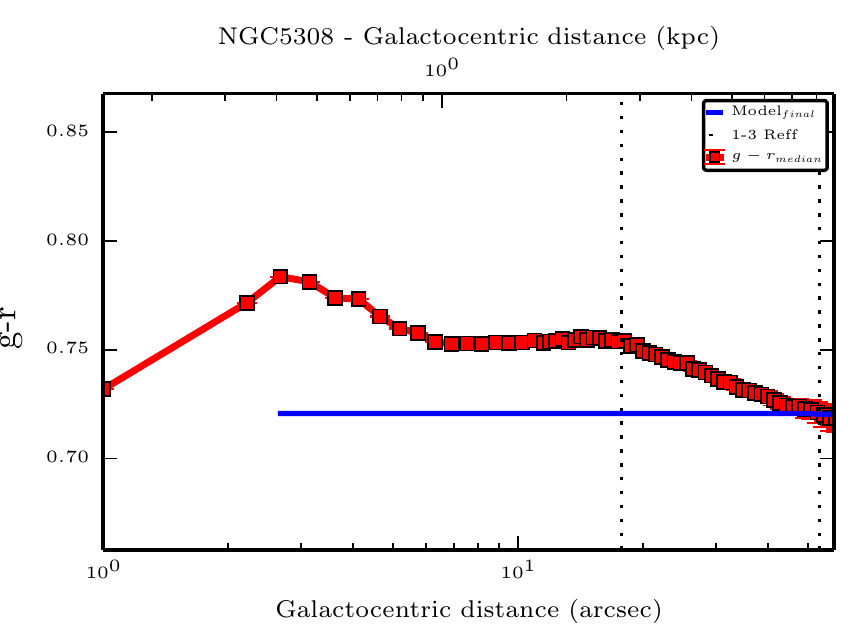}
      \includegraphics[scale=0.55]{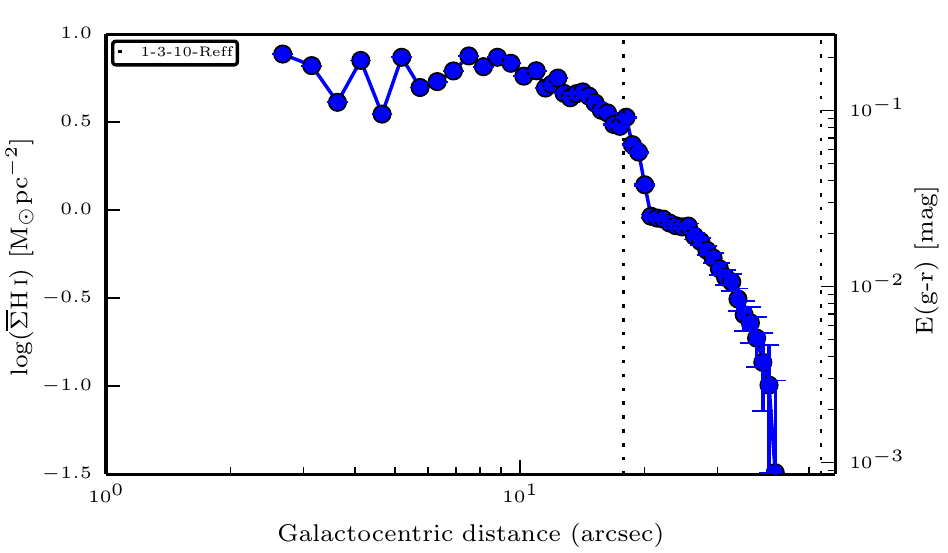}
\end{minipage}}
\caption{True image, colour map, colour excess map, and the radial profiles of NGC~5308.}
\label{fig:app_profiles}
\end{figure*}

\begin{figure*}
\makebox[\textwidth][c]{\begin{minipage}[b][11.6cm]{.85\textwidth}
  \vspace*{\fill}
      \includegraphics[scale=0.85]{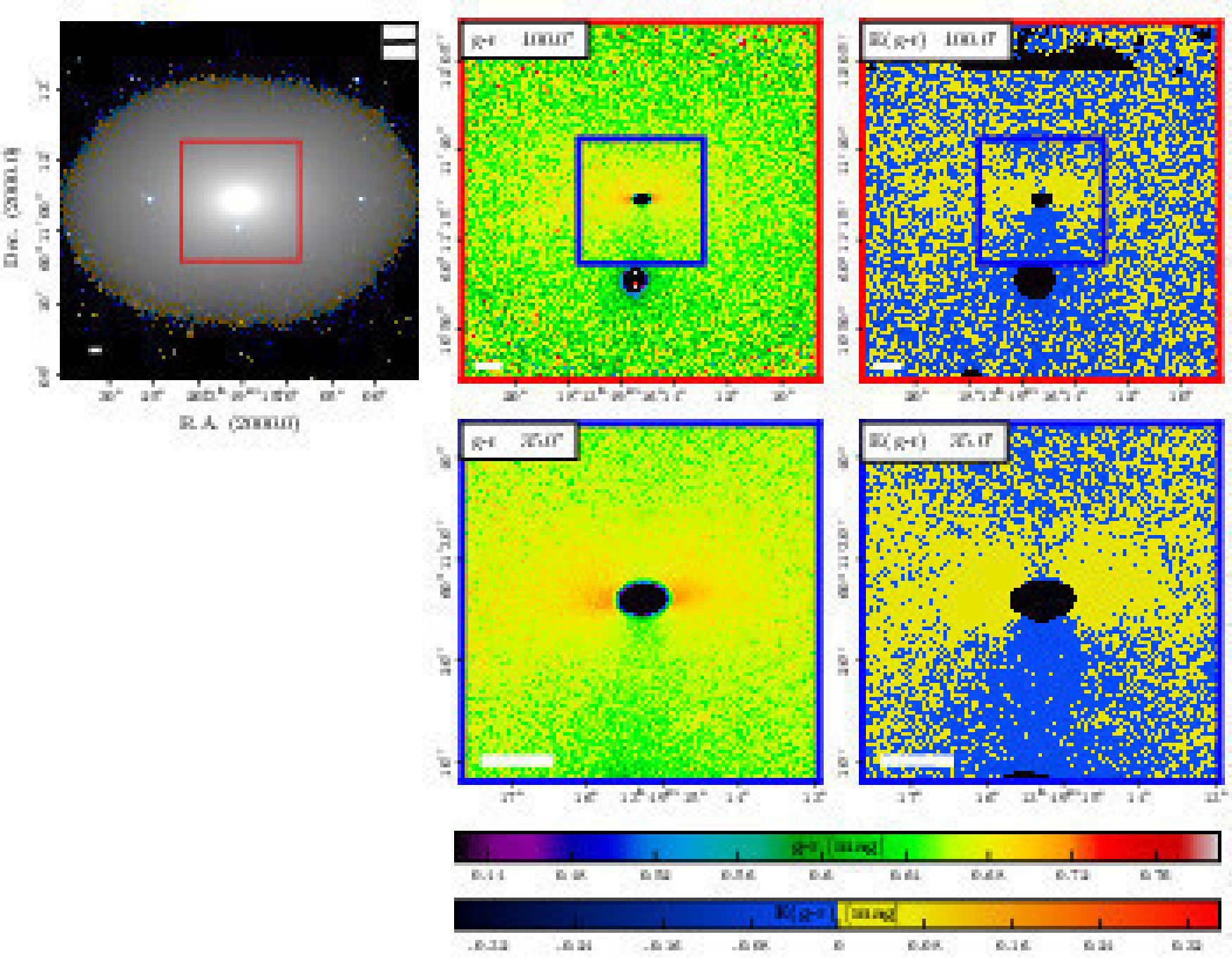}
\end{minipage}}
\makebox[\textwidth][c]{\begin{minipage}[l][-0.7cm][b]{.85\linewidth}
      \includegraphics[scale=0.55, trim={0 0.7cm 0 0},clip]{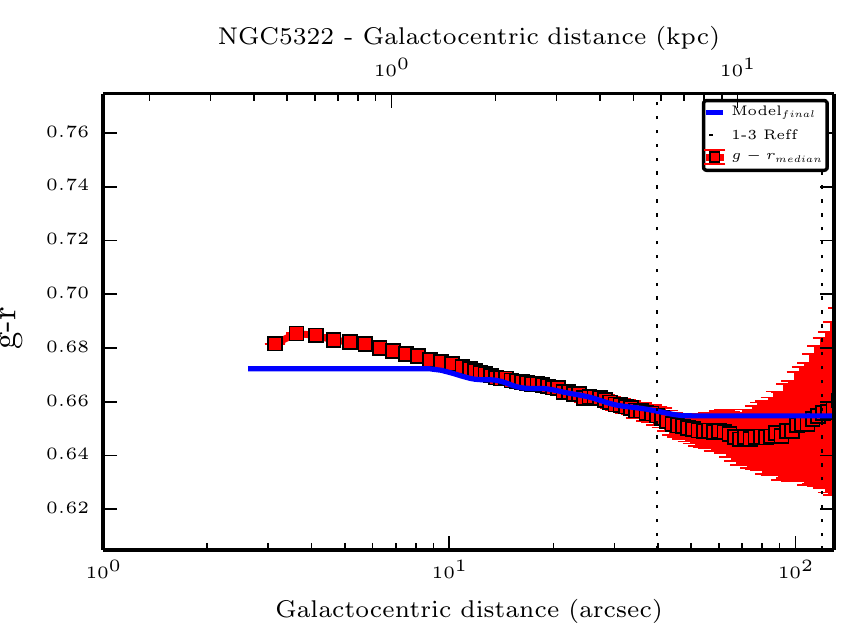}
      \includegraphics[scale=0.55]{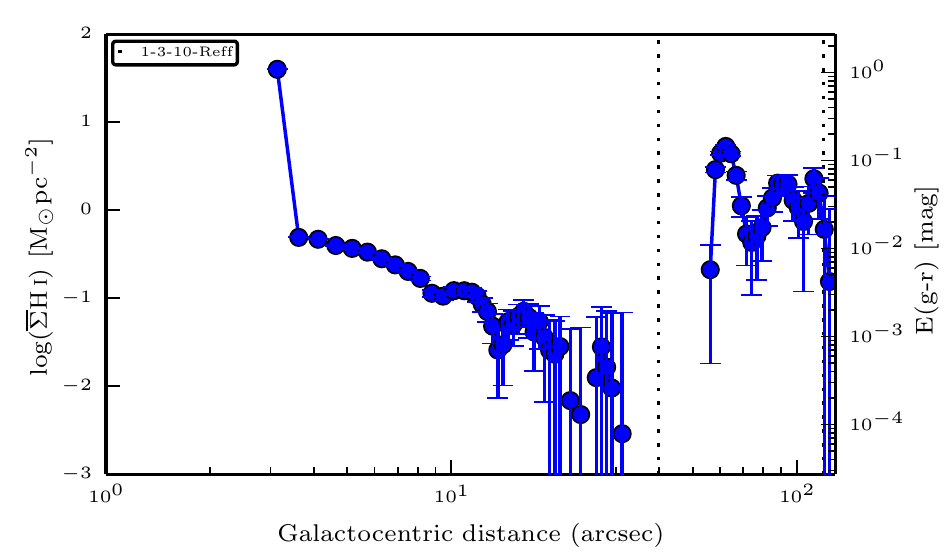}
\end{minipage}}
\caption{True image, colour map, colour excess map, and the radial profiles of NGC~5322.}
\label{fig:app_profiles}
\end{figure*}

%Page24
\clearpage
\begin{figure*}
\makebox[\textwidth][c]{\begin{minipage}[b][10.5cm]{.85\textwidth}
  \vspace*{\fill}
      \includegraphics[scale=0.85]{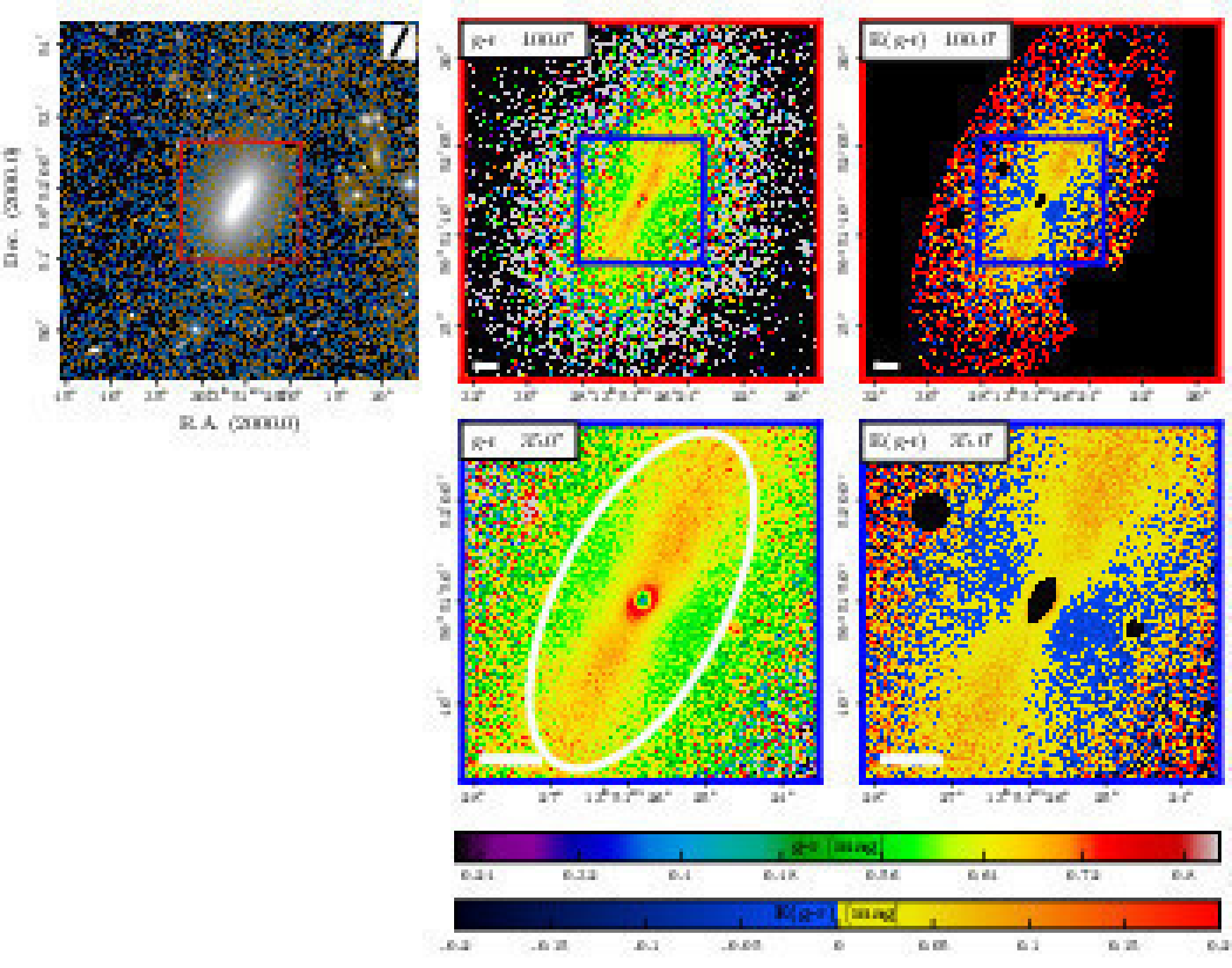}
\end{minipage}}
\makebox[\textwidth][c]{\begin{minipage}[l][-0.7cm][b]{.85\linewidth}
      \includegraphics[scale=0.55, trim={0 0.7cm 0 0},clip]{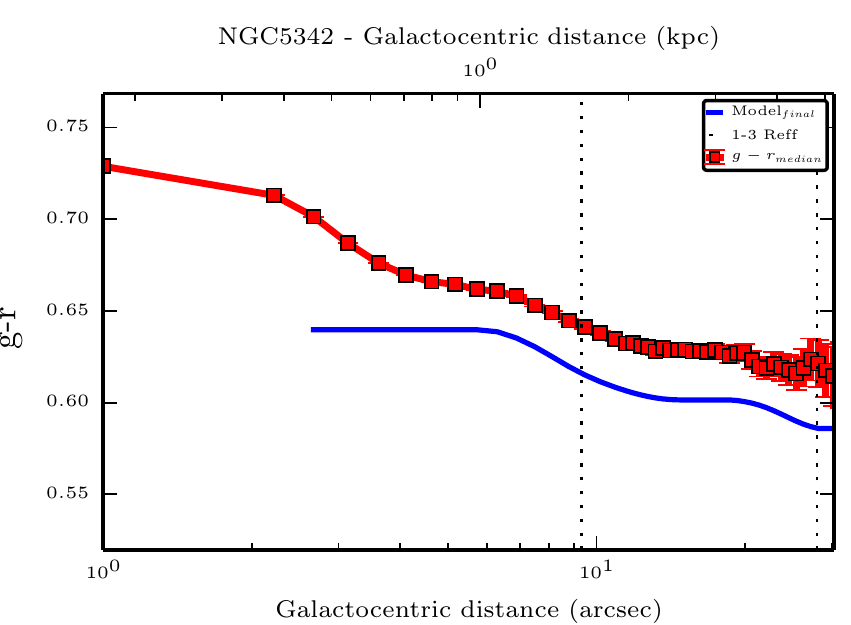}
      \includegraphics[scale=0.55]{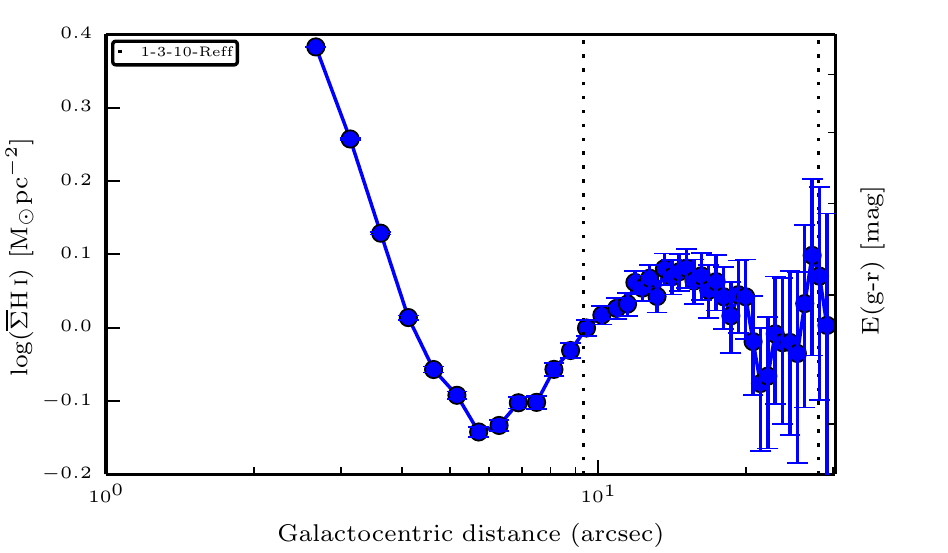}
\end{minipage}}
\caption{True image, colour map, colour excess map, and the radial profiles of NGC~5342.}
\label{fig:app_profiles}
\end{figure*}

\begin{figure*}
\makebox[\textwidth][c]{\begin{minipage}[b][11.6cm]{.85\textwidth}
  \vspace*{\fill}
      \includegraphics[scale=0.85]{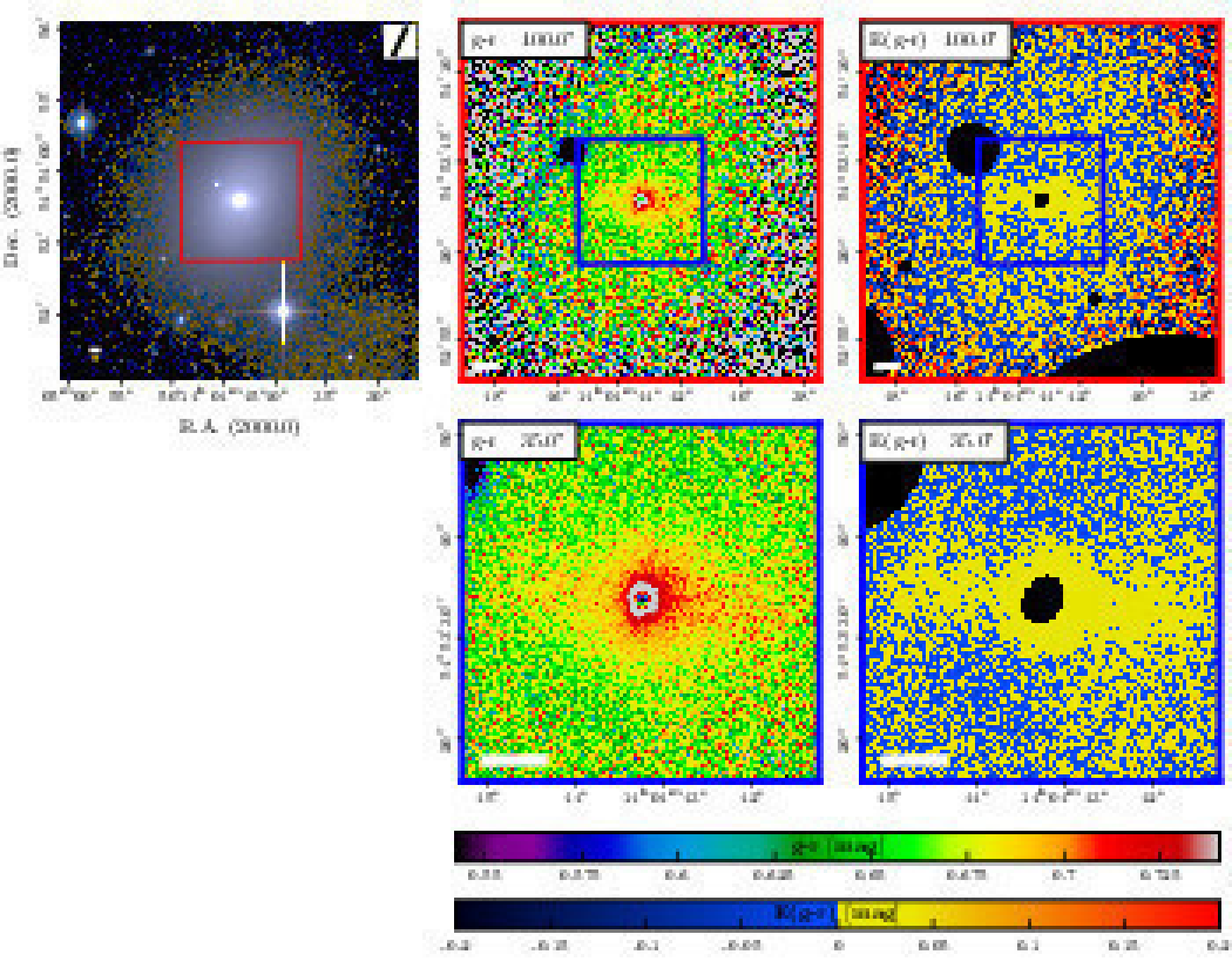}
\end{minipage}}
\makebox[\textwidth][c]{\begin{minipage}[l][-0.7cm][b]{.85\linewidth}
      \includegraphics[scale=0.55, trim={0 0.7cm 0 0},clip]{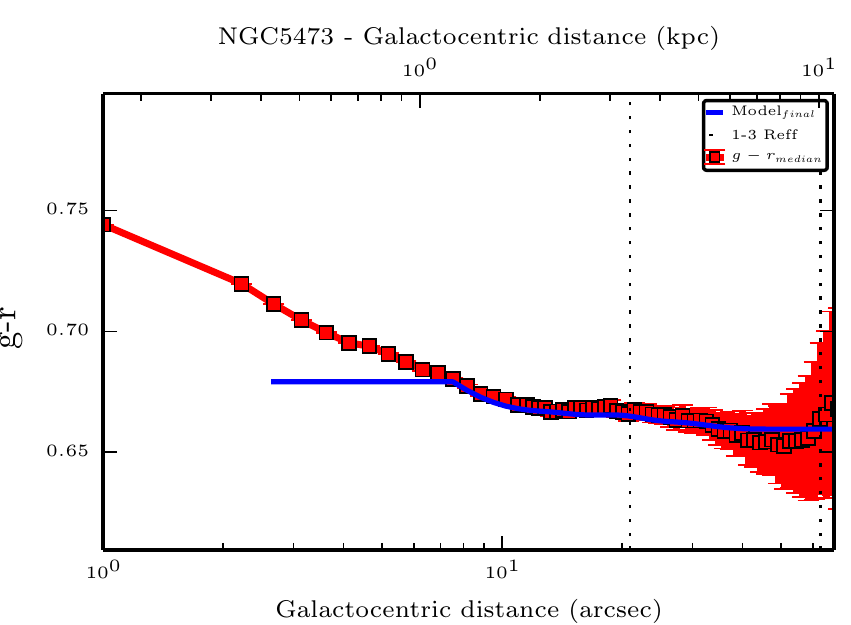}
      \includegraphics[scale=0.55]{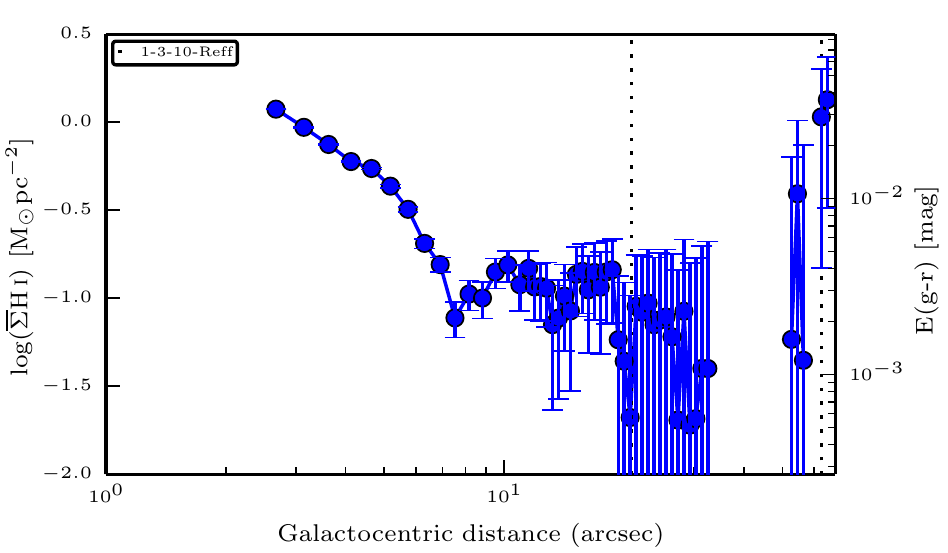}
\end{minipage}}
\caption{True image, colour map, colour excess map, and the radial profiles of NGC~5473.}
\label{fig:app_profiles}
\end{figure*}

%Page25
\clearpage
\begin{figure*}
\makebox[\textwidth][c]{\begin{minipage}[b][10.5cm]{.85\textwidth}
  \vspace*{\fill}
      \includegraphics[scale=0.85]{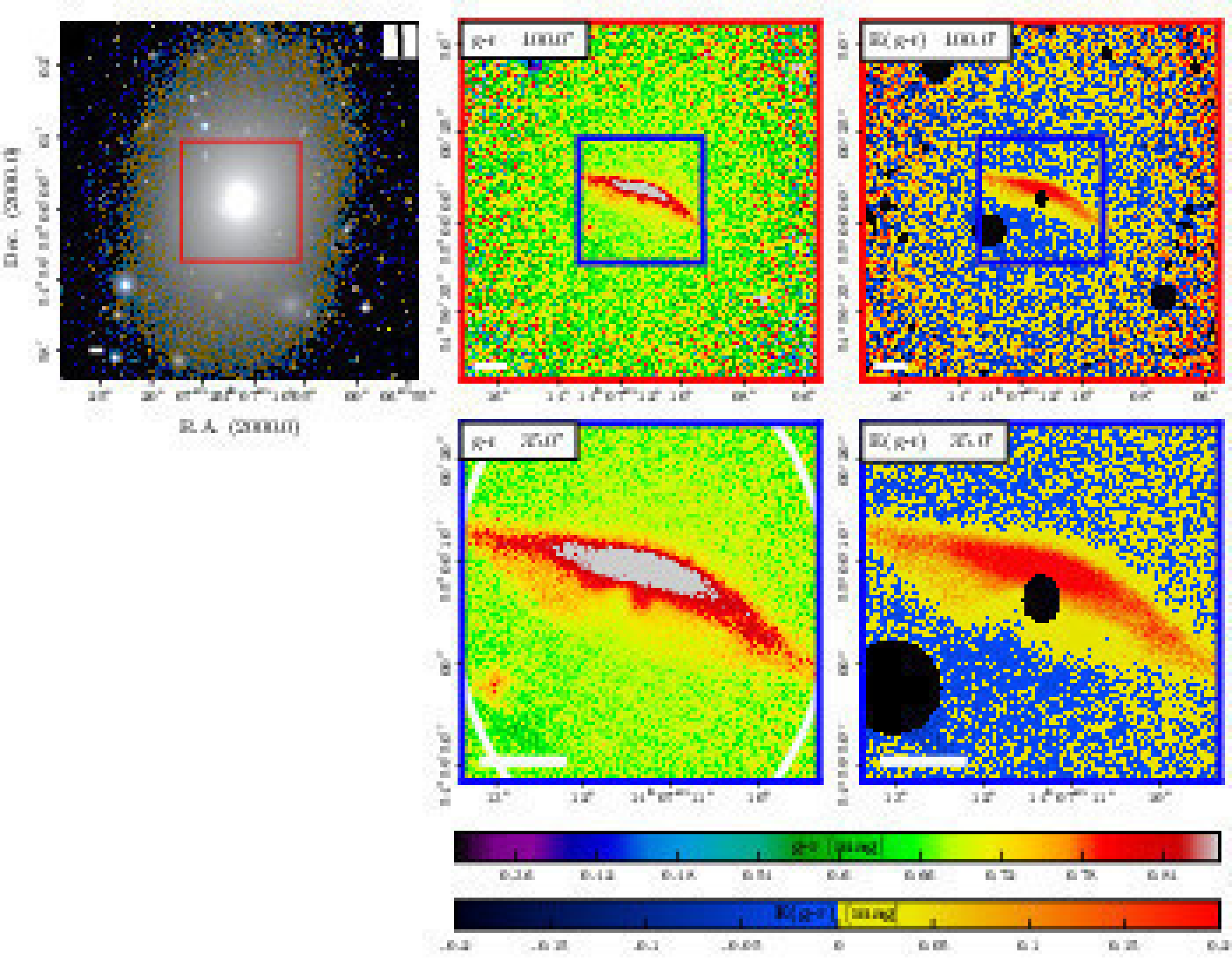}
\end{minipage}}
\makebox[\textwidth][c]{\begin{minipage}[l][-0.7cm][b]{.85\linewidth}
      \includegraphics[scale=0.55, trim={0 0.7cm 0 0},clip]{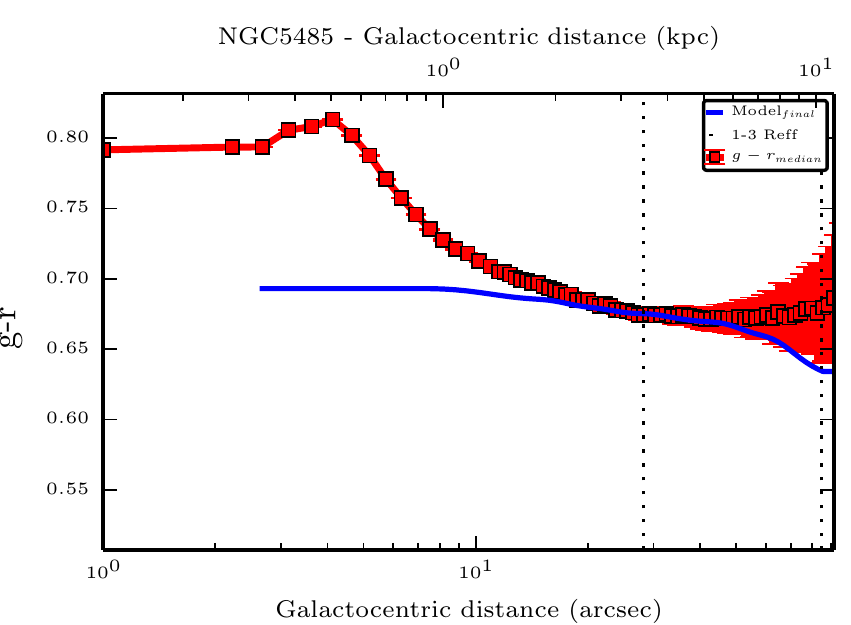}
      \includegraphics[scale=0.55]{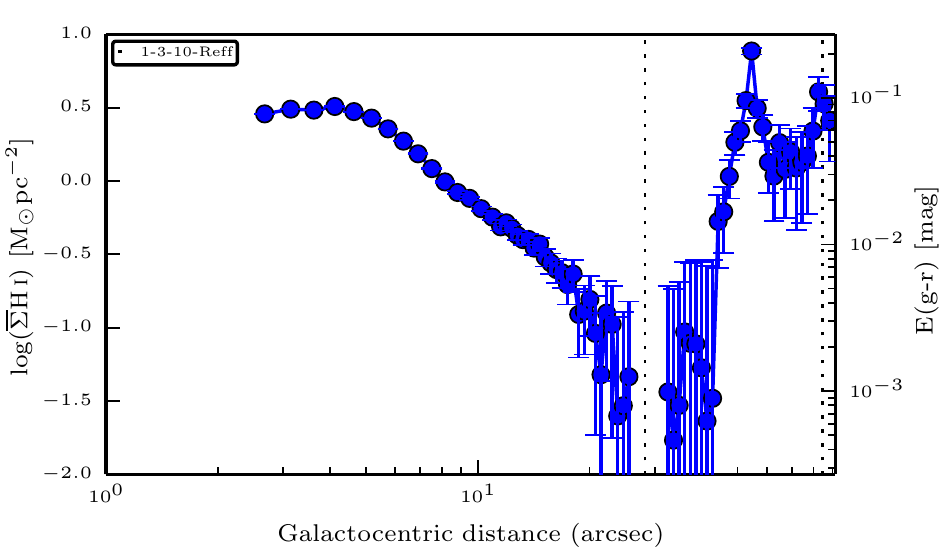}
\end{minipage}}
\caption{True image, colour map, colour excess map, and the radial profiles of NGC~5485.}
\label{fig:app_profiles}
\end{figure*}

\begin{figure*}
\makebox[\textwidth][c]{\begin{minipage}[b][11.6cm]{.85\textwidth}
  \vspace*{\fill}
      \includegraphics[scale=0.85]{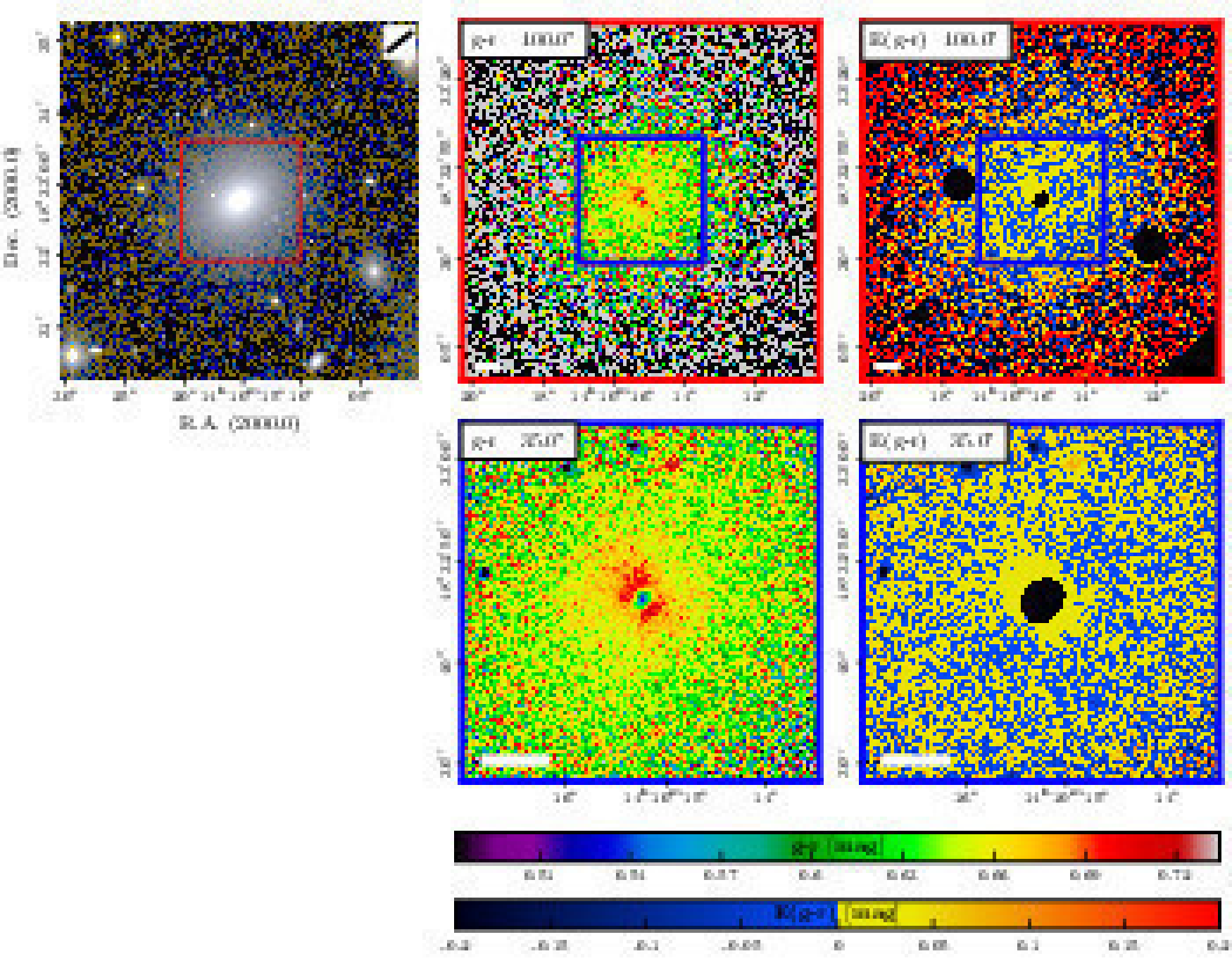}
\end{minipage}}
\makebox[\textwidth][c]{\begin{minipage}[l][-0.7cm][b]{.85\linewidth}
      \includegraphics[scale=0.55, trim={0 0.7cm 0 0},clip]{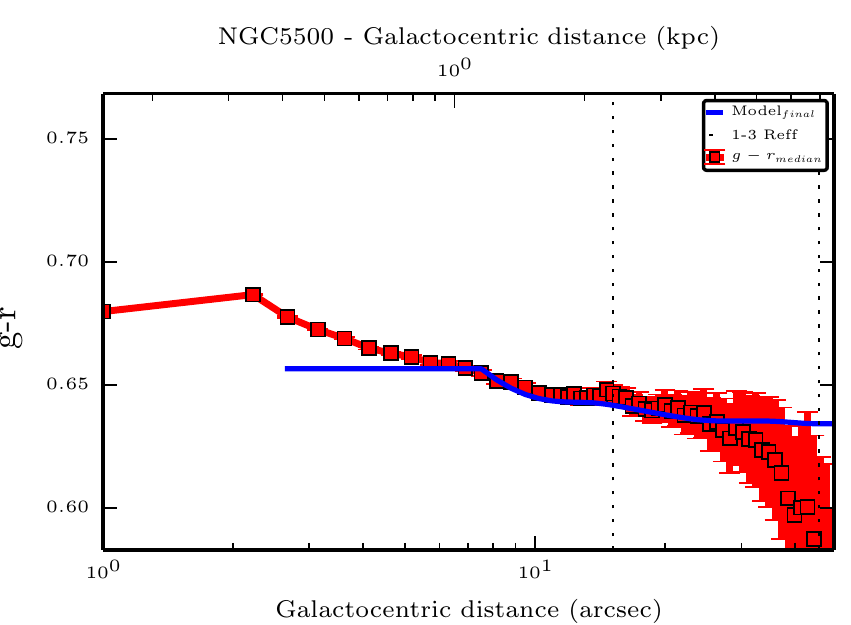}
      \includegraphics[scale=0.55]{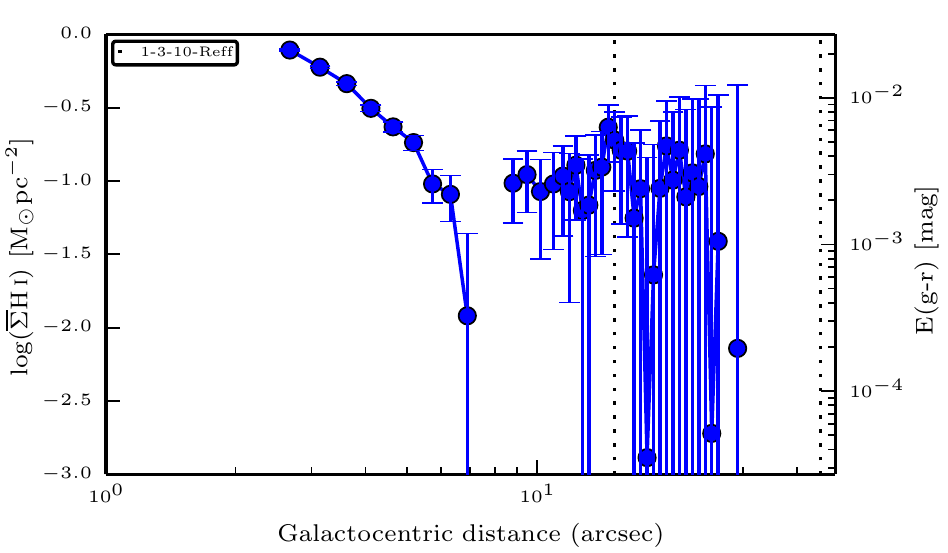}
\end{minipage}}
\caption{True image, colour map, colour excess map, and the radial profiles of NGC~5500.}
\label{fig:app_profiles}
\end{figure*}

%Page26
\clearpage
\begin{figure*}
\makebox[\textwidth][c]{\begin{minipage}[b][10.5cm]{.85\textwidth}
  \vspace*{\fill}
      \includegraphics[scale=0.85]{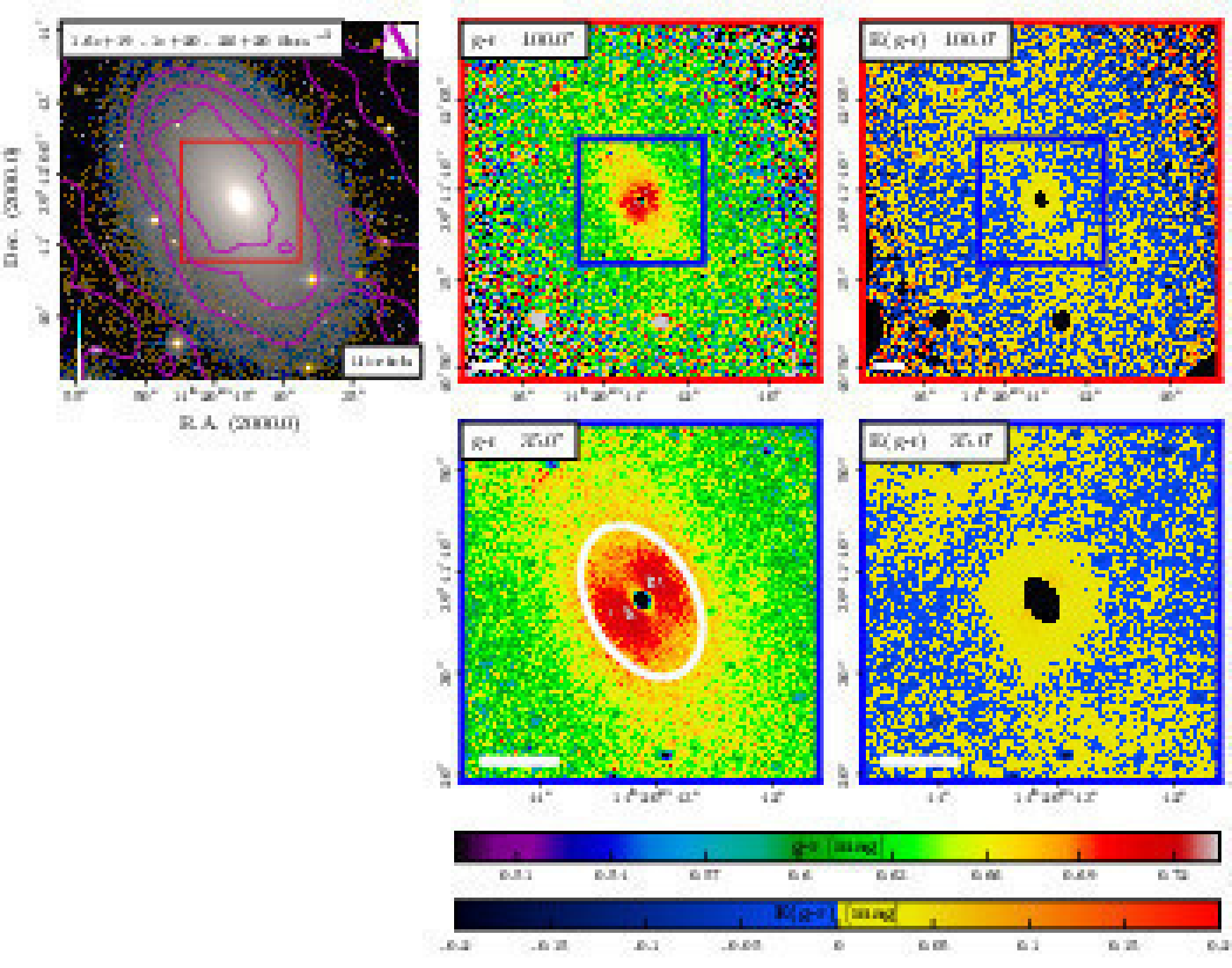}
\end{minipage}}
\makebox[\textwidth][c]{\begin{minipage}[l][-0.7cm][b]{.85\linewidth}
      \includegraphics[scale=0.55, trim={0 0.7cm 0 0},clip]{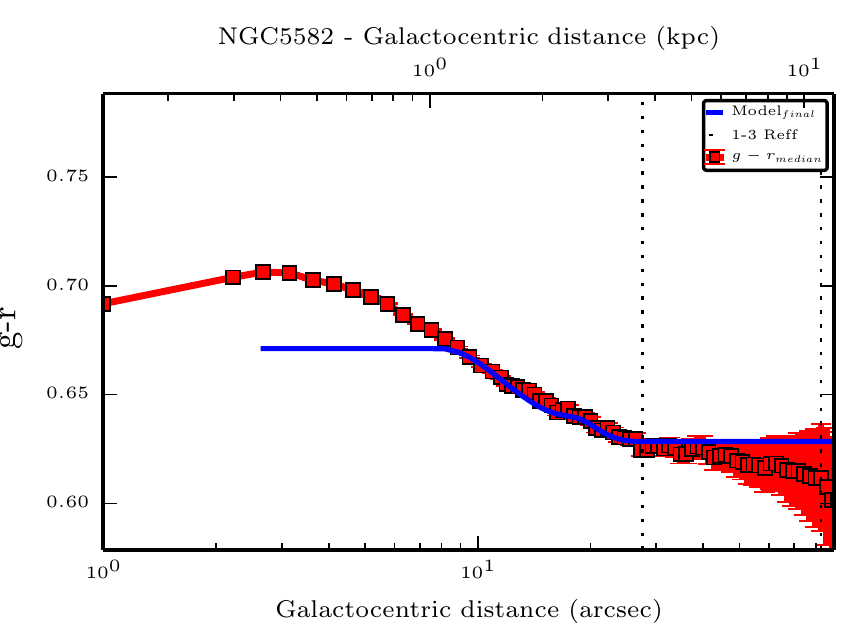}
      \includegraphics[scale=0.55]{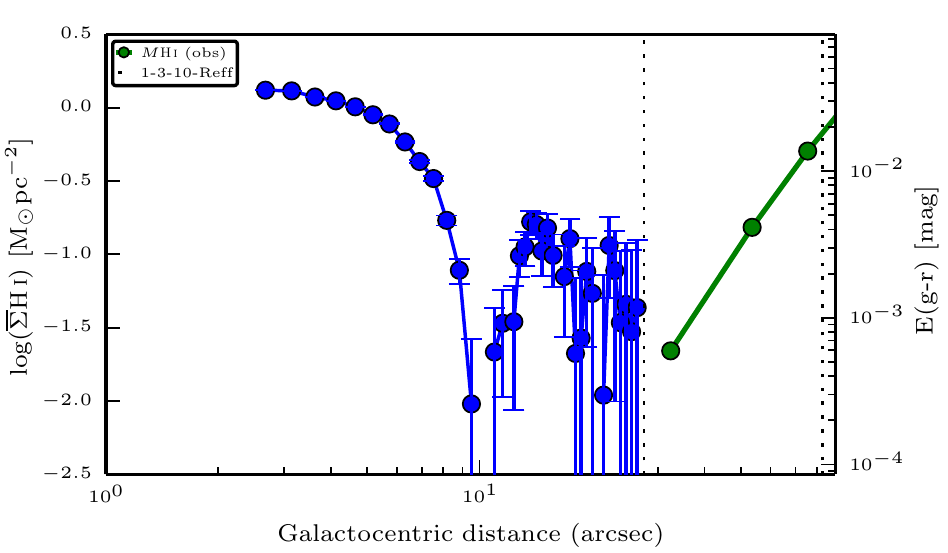}
\end{minipage}}
\caption{True image, colour map, colour excess map, and the radial profiles of NGC~5582.}
\label{fig:app_profiles}
\end{figure*}

\begin{figure*}
\makebox[\textwidth][c]{\begin{minipage}[b][11.6cm]{.85\textwidth}
  \vspace*{\fill}
      \includegraphics[scale=0.85]{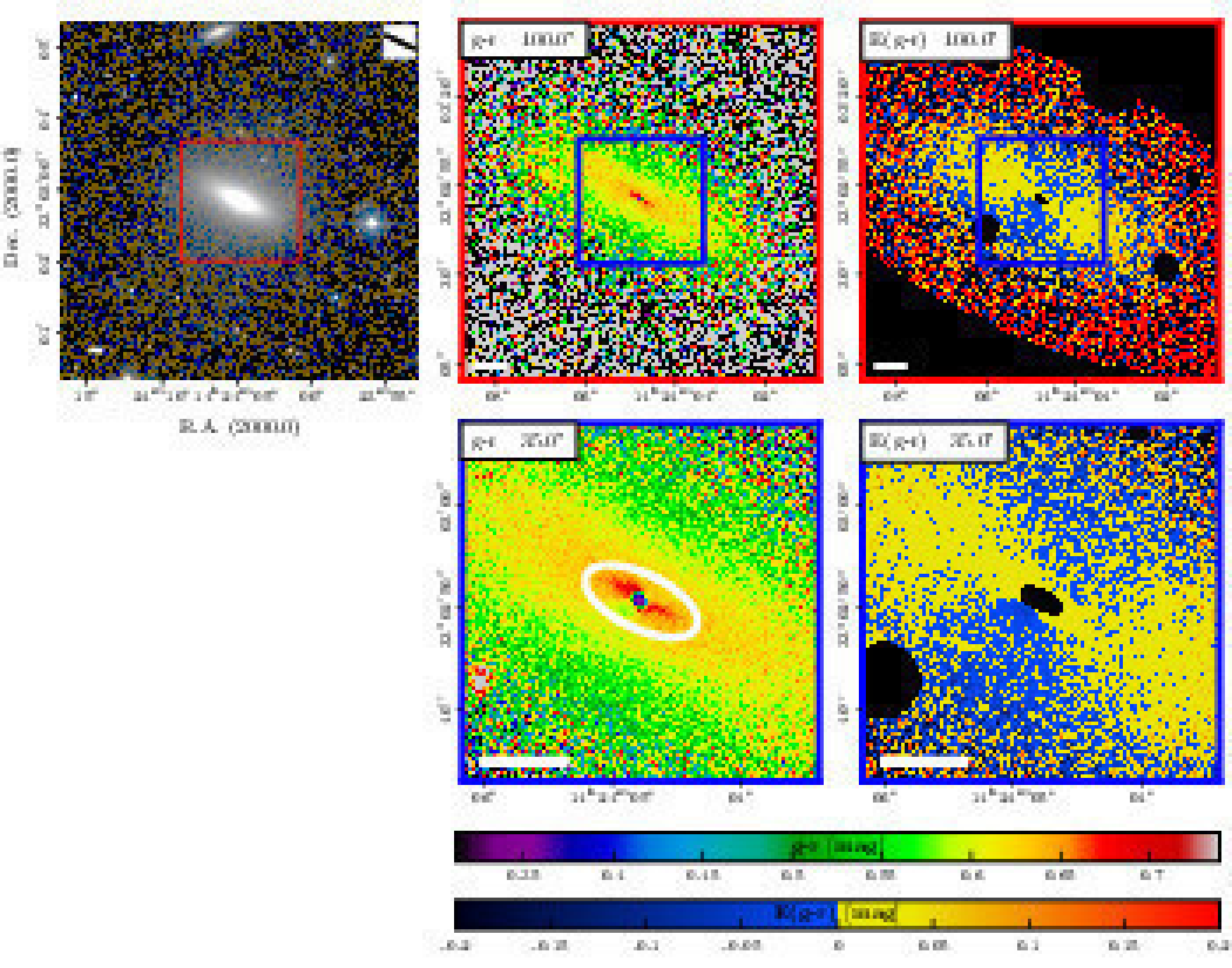}
\end{minipage}}
\makebox[\textwidth][c]{\begin{minipage}[l][-0.7cm][b]{.85\linewidth}
      \includegraphics[scale=0.55, trim={0 0.7cm 0 0},clip]{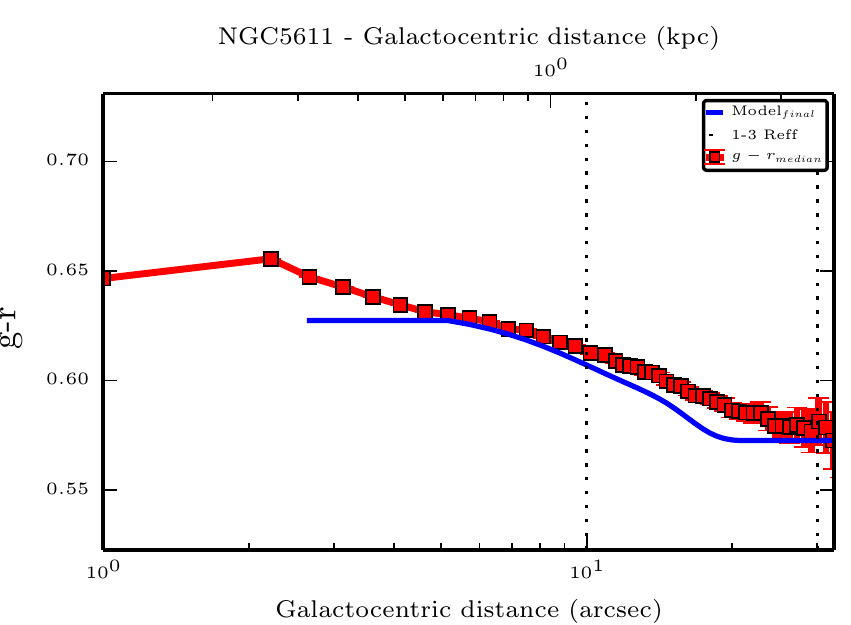}
      \includegraphics[scale=0.55]{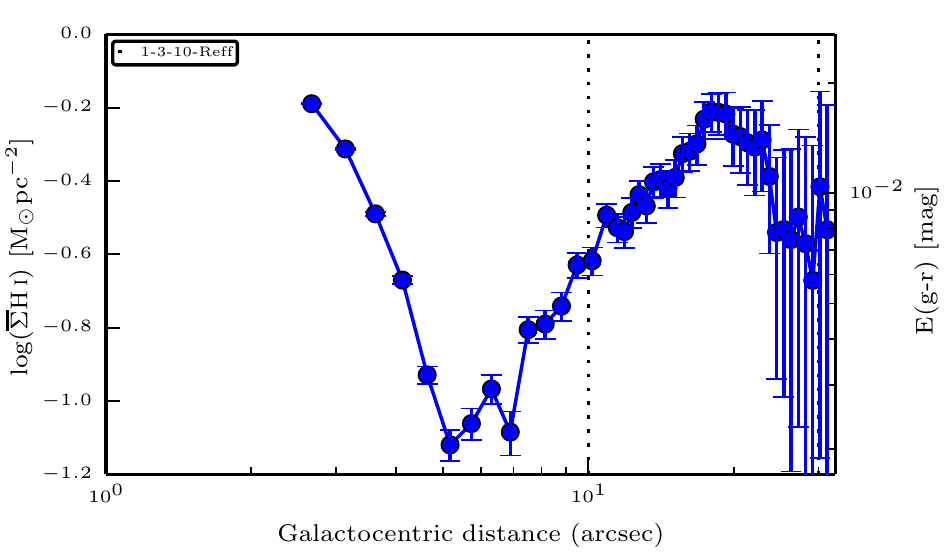}
\end{minipage}}
\caption{True image, colour map, colour excess map, and the radial profiles of NGC~5611.}
\label{fig:app_profiles}
\end{figure*}

%Page27
\clearpage
\begin{figure*}
\makebox[\textwidth][c]{\begin{minipage}[b][10.5cm]{.85\textwidth}
  \vspace*{\fill}
      \includegraphics[scale=0.85]{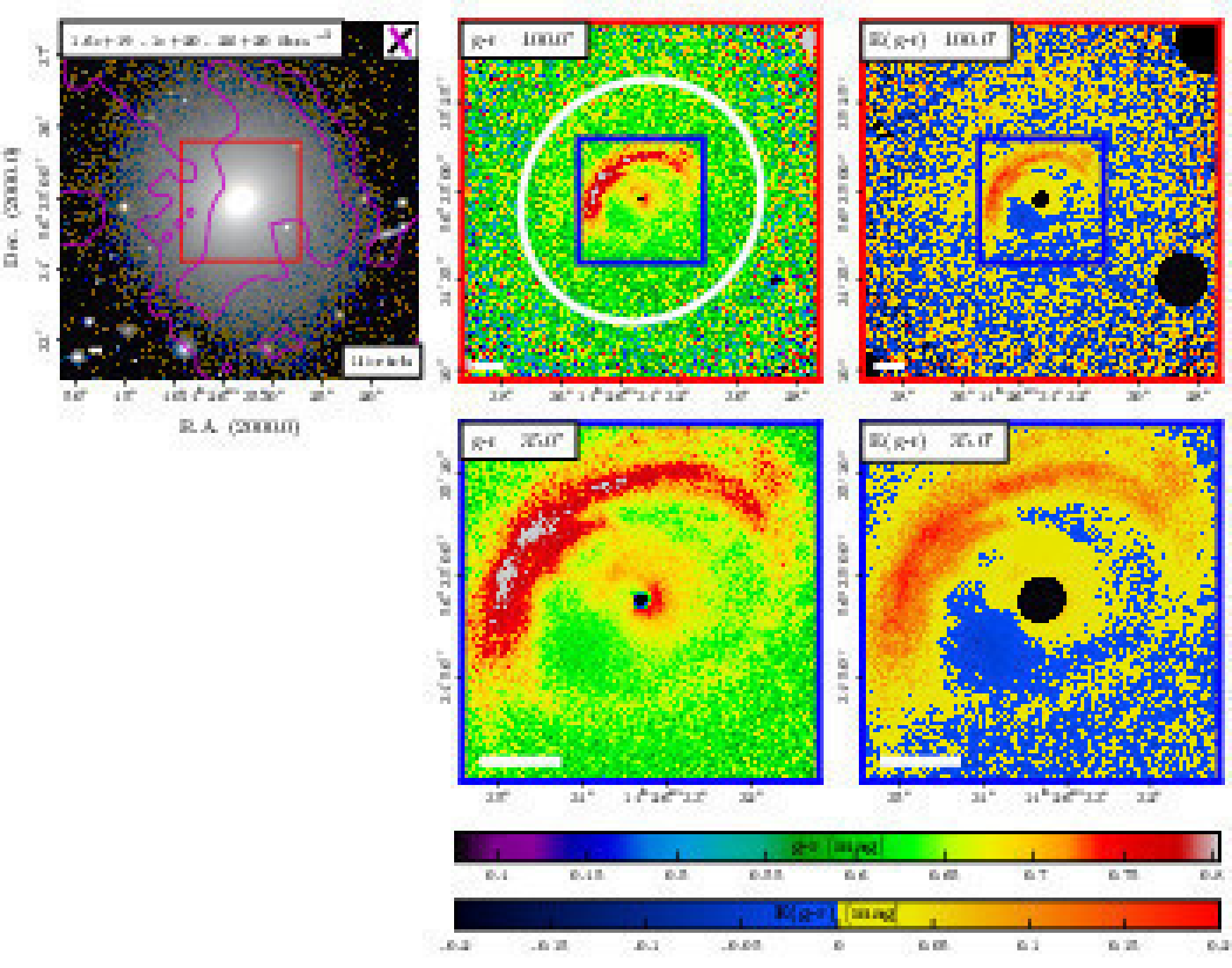}
\end{minipage}}
\makebox[\textwidth][c]{\begin{minipage}[l][-0.7cm][b]{.85\linewidth}
      \includegraphics[scale=0.55, trim={0 0.7cm 0 0},clip]{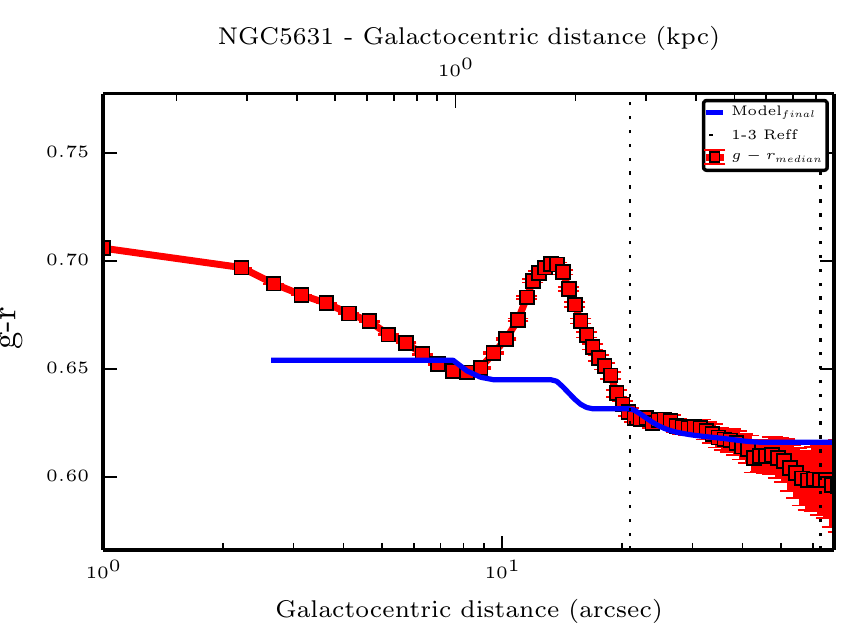}
      \includegraphics[scale=0.55]{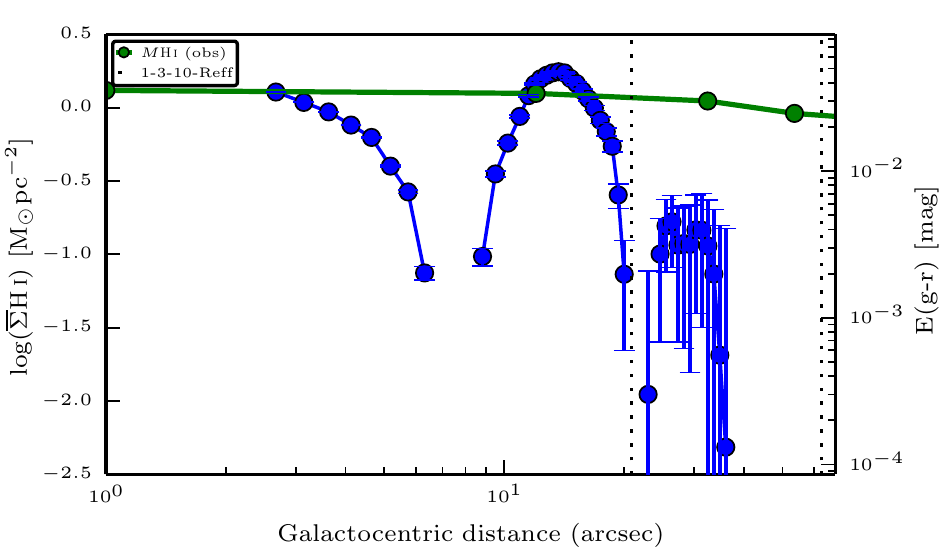}
\end{minipage}}
\caption{True image, colour map, colour excess map, and the radial profiles of NGC~5631.}
\label{fig:app_profiles}
\end{figure*}

\begin{figure*}
\makebox[\textwidth][c]{\begin{minipage}[b][11.6cm]{.85\textwidth}
  \vspace*{\fill}
      \includegraphics[scale=0.85]{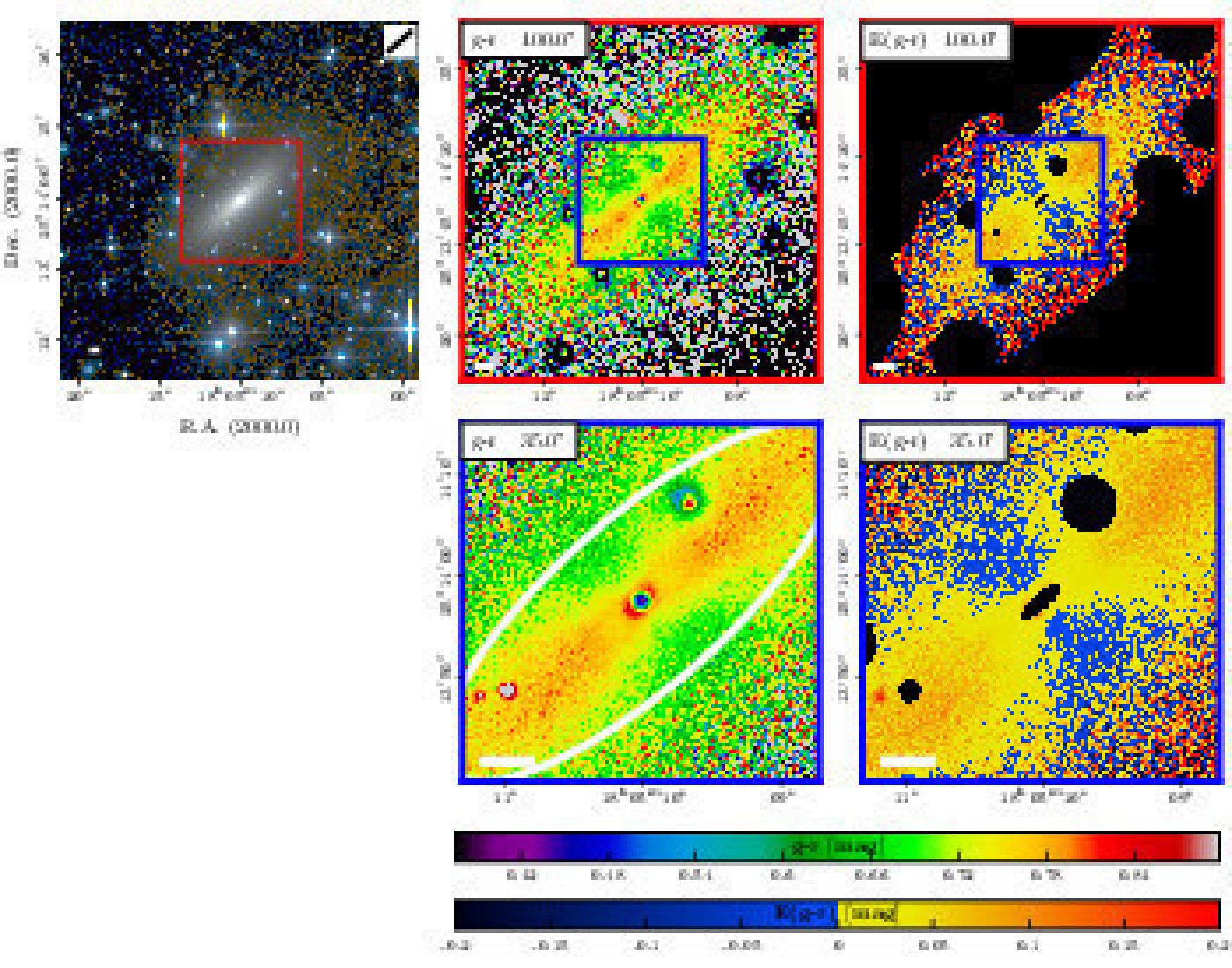}
\end{minipage}}
\makebox[\textwidth][c]{\begin{minipage}[l][-0.7cm][b]{.85\linewidth}
      \includegraphics[scale=0.55, trim={0 0.7cm 0 0},clip]{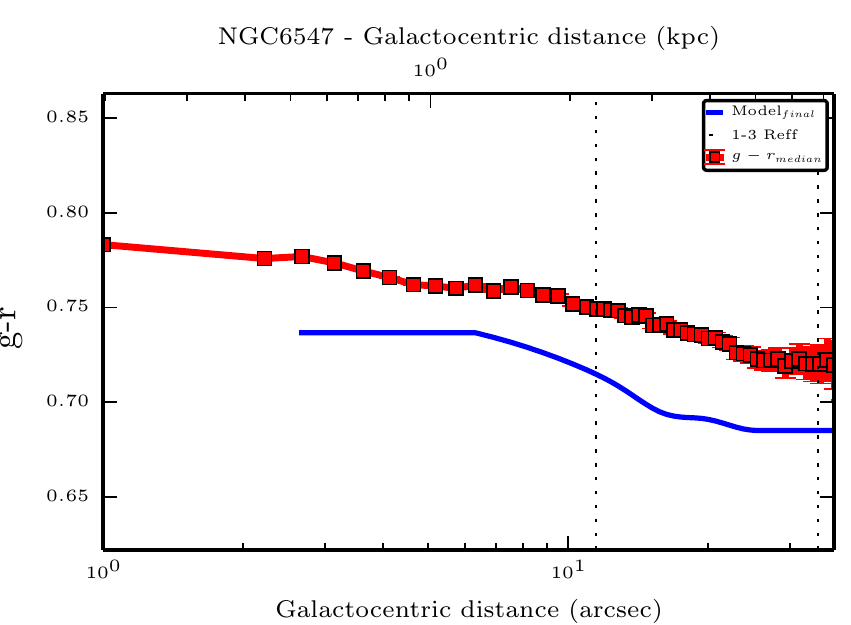}
      \includegraphics[scale=0.55]{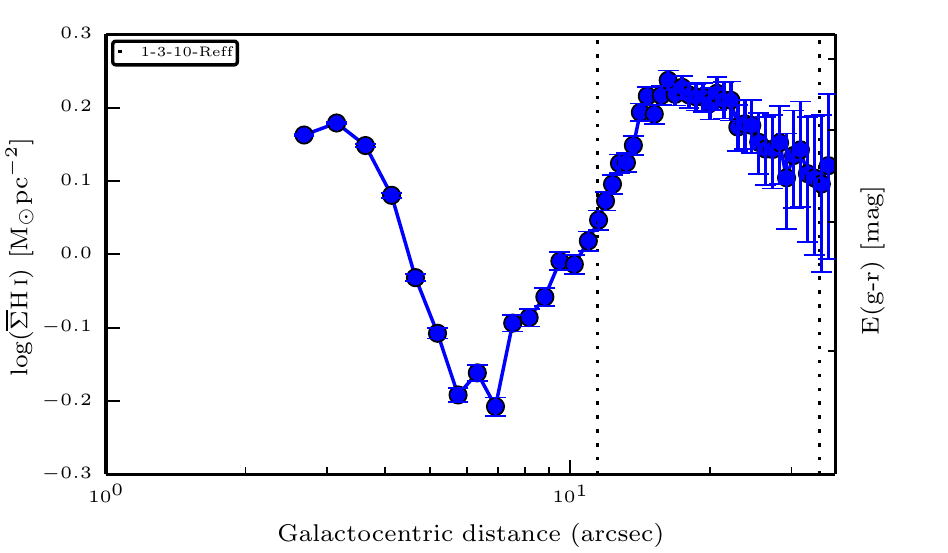}
\end{minipage}}
\caption{True image, colour map, colour excess map, and the radial profiles of NGC~6547.}
\label{fig:app_profiles}
\end{figure*}

%Page28
\clearpage
\begin{figure*}
\makebox[\textwidth][c]{\begin{minipage}[b][10.5cm]{.85\textwidth}
  \vspace*{\fill}
      \includegraphics[scale=0.85]{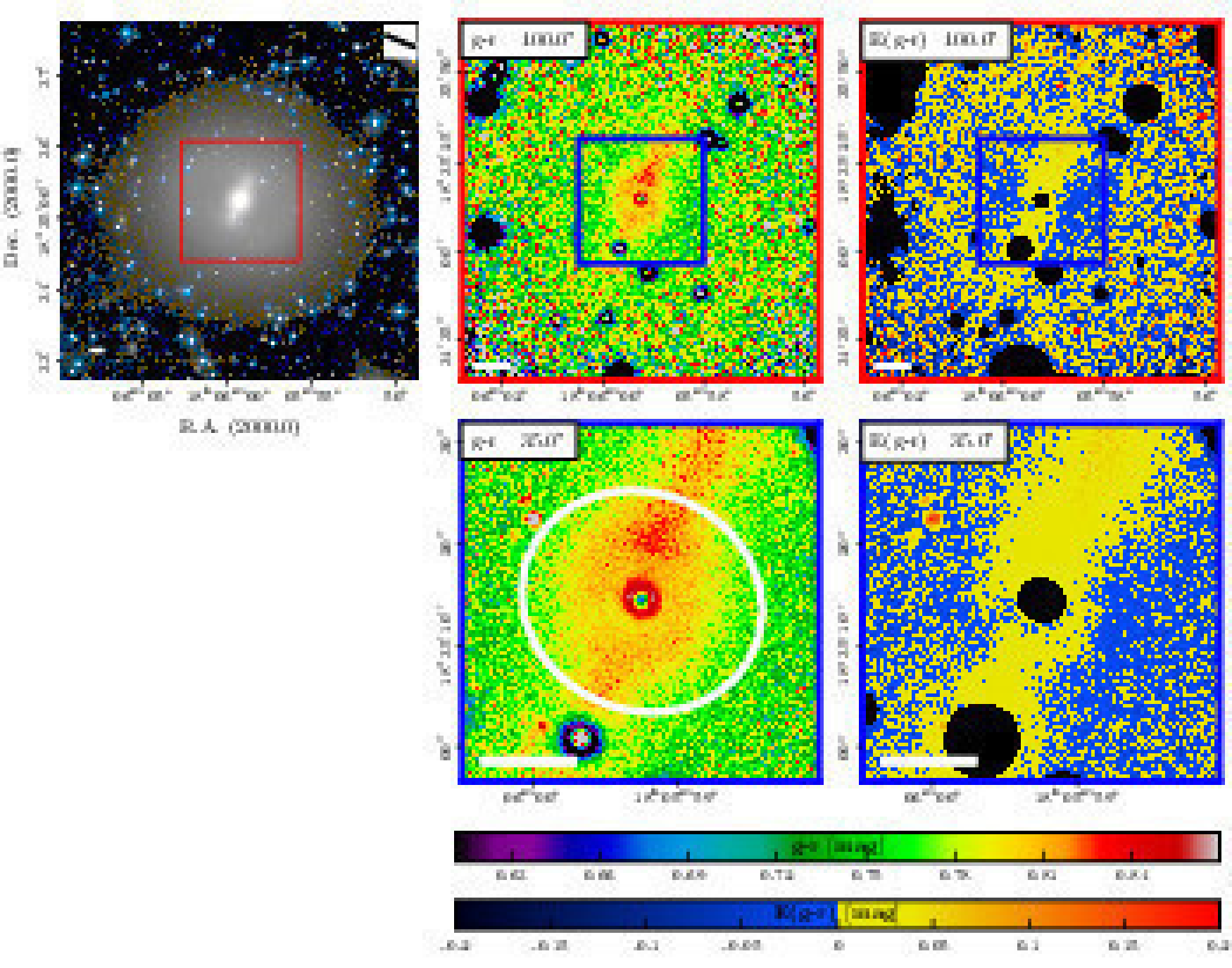}
\end{minipage}}
\makebox[\textwidth][c]{\begin{minipage}[l][-0.7cm][b]{.85\linewidth}
      \includegraphics[scale=0.55, trim={0 0.7cm 0 0},clip]{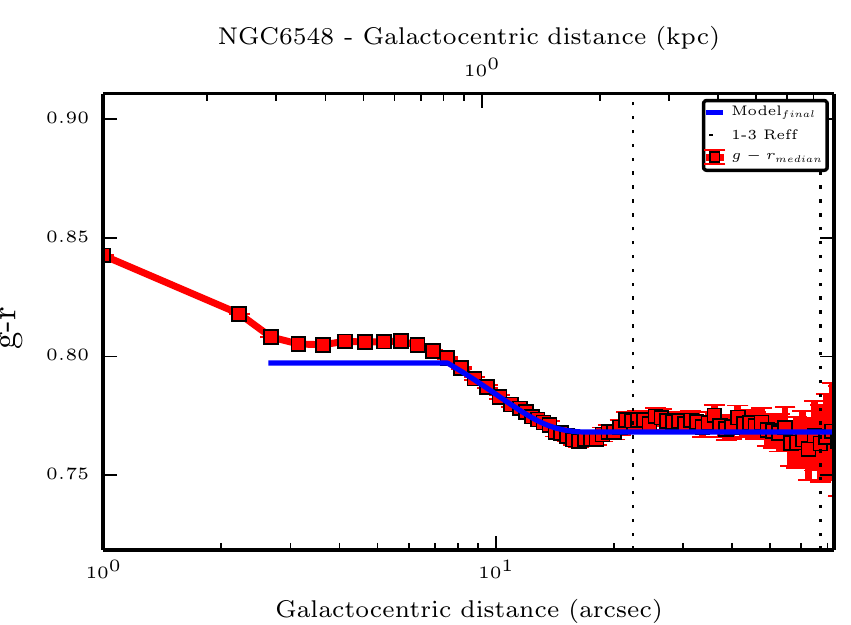}
      \includegraphics[scale=0.55]{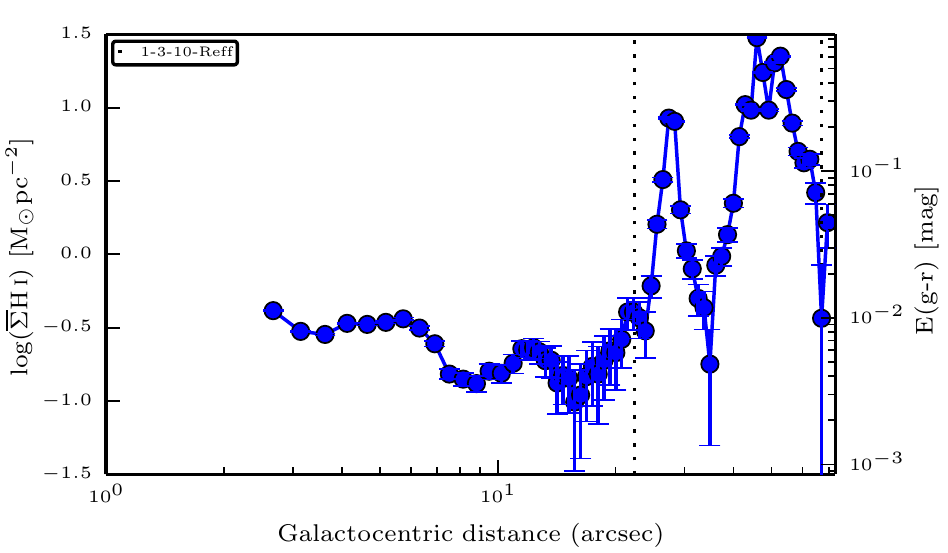}
\end{minipage}}
\caption{True image, colour map, colour excess map, and the radial profiles of NGC~6548.}
\label{fig:app_profiles}
\end{figure*}

\begin{figure*}
\makebox[\textwidth][c]{\begin{minipage}[b][11.6cm]{.85\textwidth}
  \vspace*{\fill}
      \includegraphics[scale=0.85]{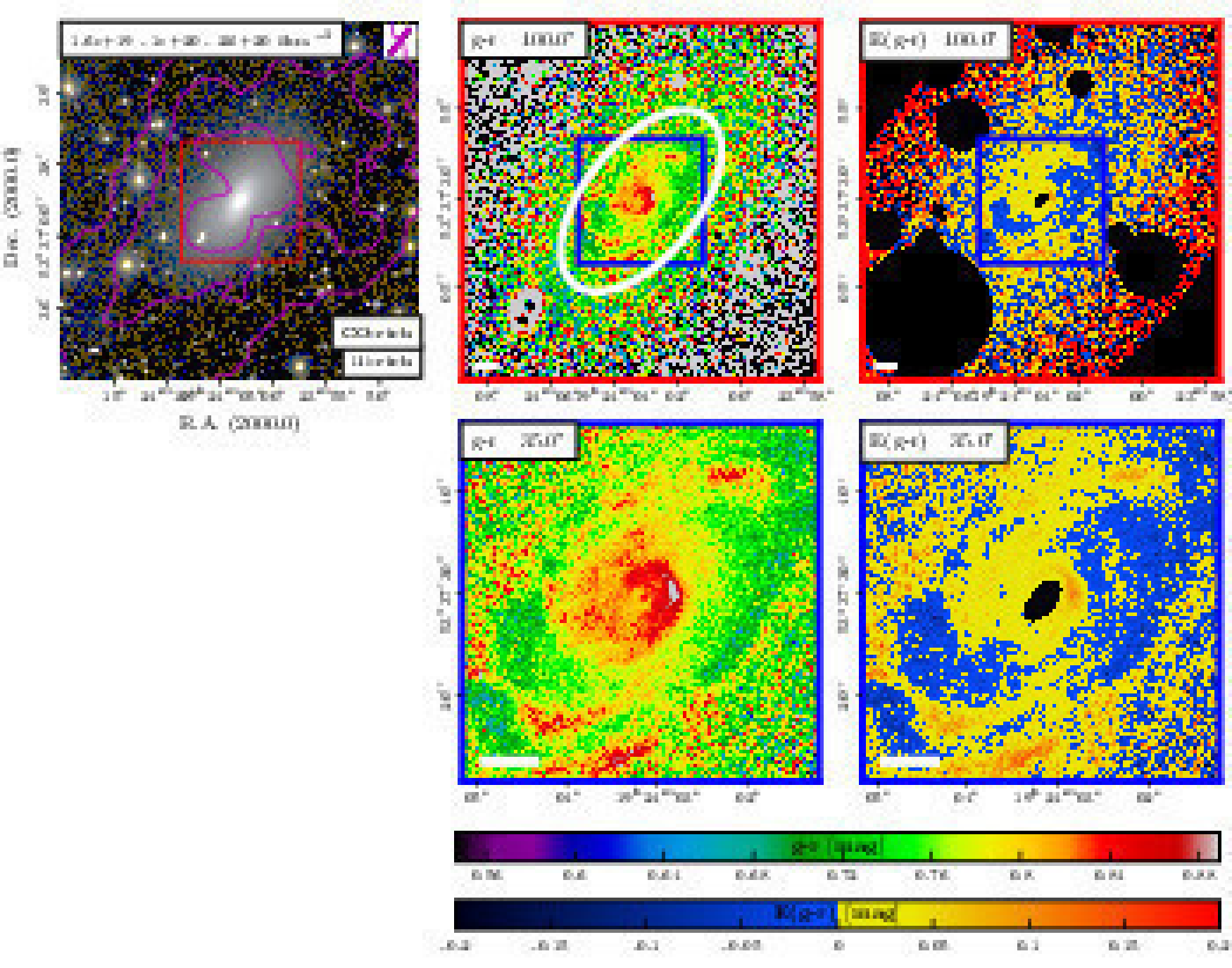}
\end{minipage}}
\makebox[\textwidth][c]{\begin{minipage}[l][-0.7cm][b]{.85\linewidth}
      \includegraphics[scale=0.55, trim={0 0.7cm 0 0},clip]{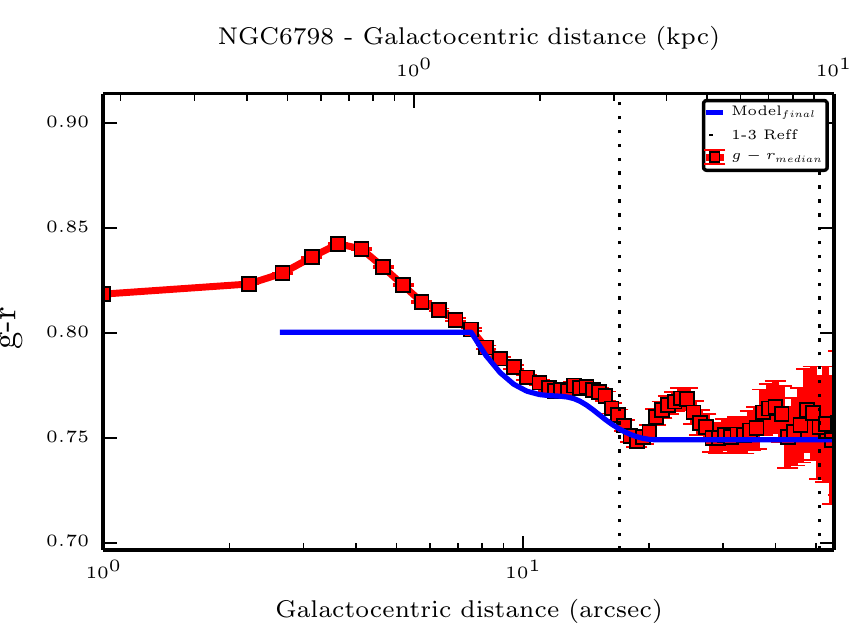}
      \includegraphics[scale=0.55]{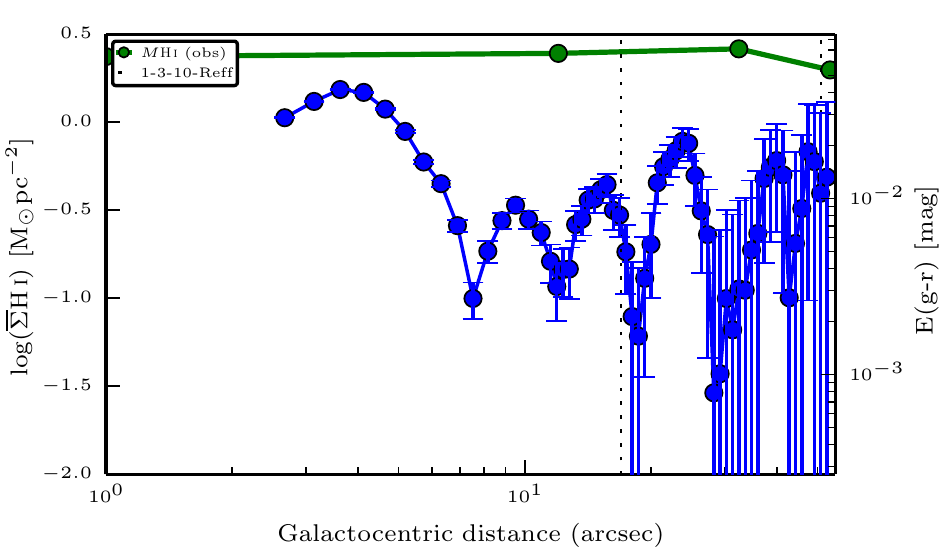}
\end{minipage}}
\caption{True image, colour map, colour excess map, and the radial profiles of NGC~6798.}
\label{fig:app_profiles}
\end{figure*}

%Page29
\clearpage
\begin{figure*}
\makebox[\textwidth][c]{\begin{minipage}[b][10.5cm]{.85\textwidth}
  \vspace*{\fill}
      \includegraphics[scale=0.85]{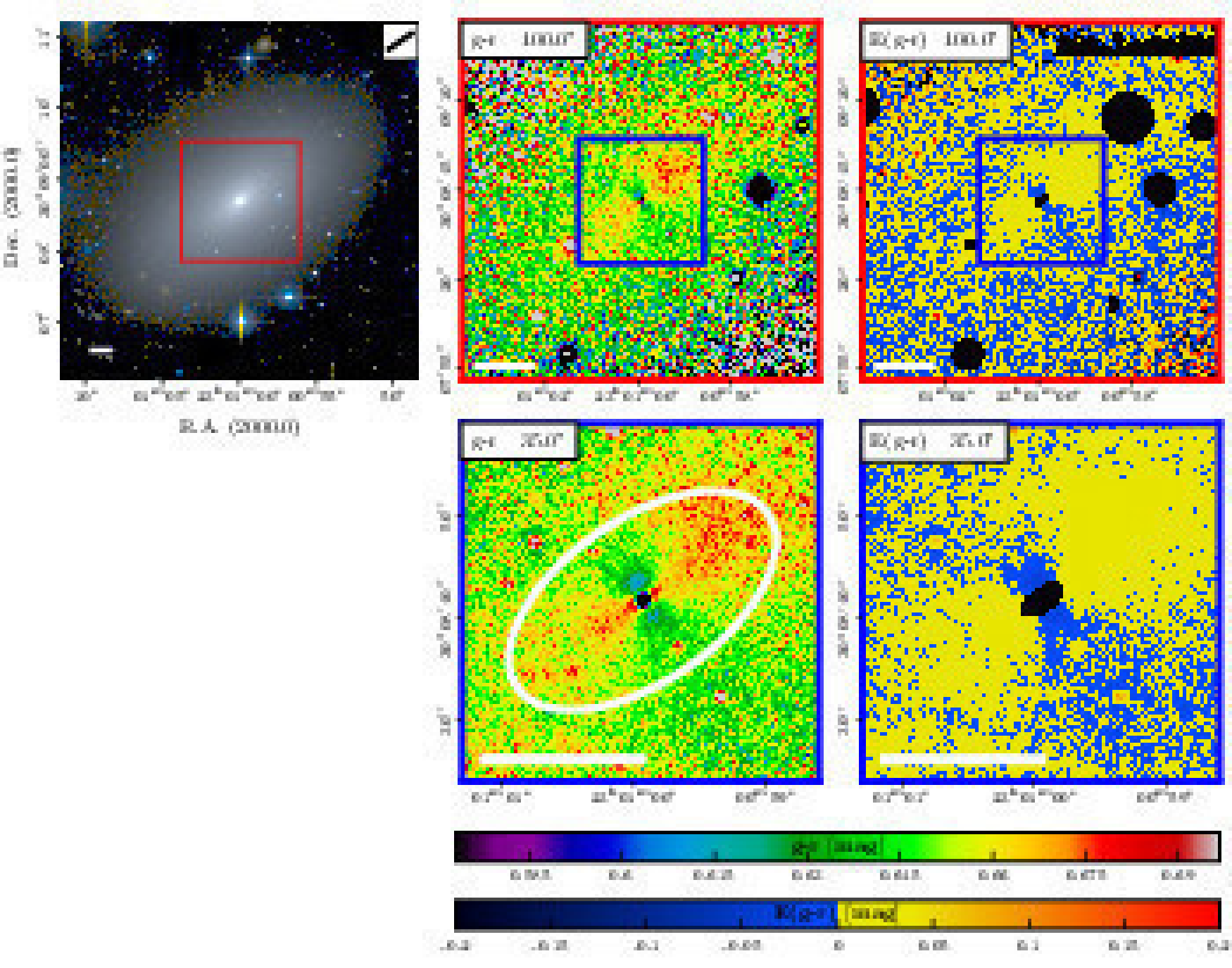}
\end{minipage}}
\makebox[\textwidth][c]{\begin{minipage}[l][-0.7cm][b]{.85\linewidth}
      \includegraphics[scale=0.55, trim={0 0.7cm 0 0},clip]{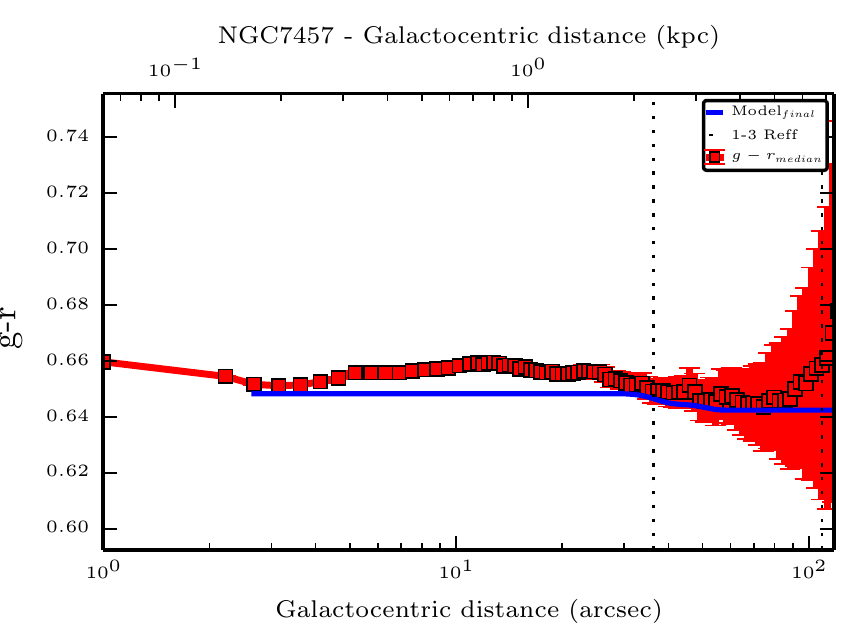}
      \includegraphics[scale=0.55]{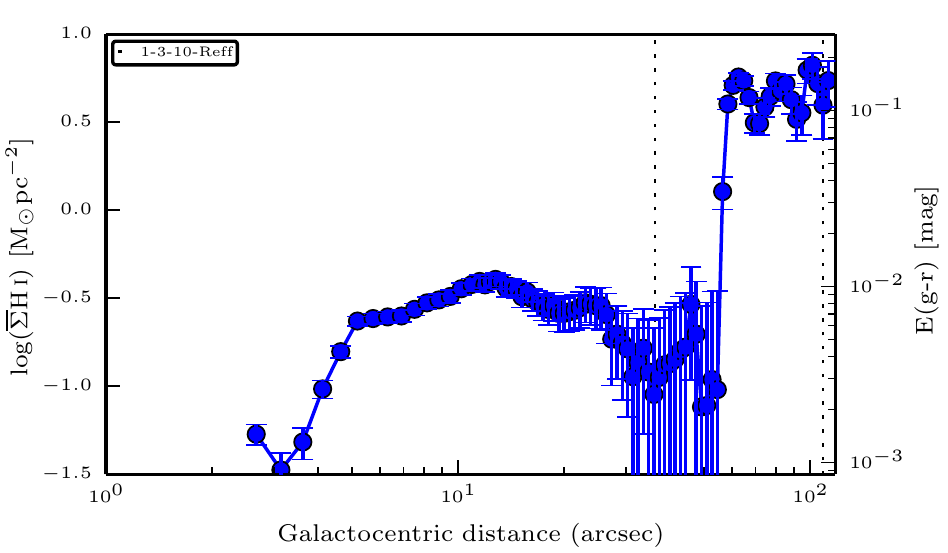}
\end{minipage}}
\caption{True image, colour map, colour excess map, and the radial profiles of NGC~7457.}
\label{fig:app_profiles}
\end{figure*}

\begin{figure*}
\makebox[\textwidth][c]{\begin{minipage}[b][11.6cm]{.85\textwidth}
  \vspace*{\fill}
      \includegraphics[scale=0.85]{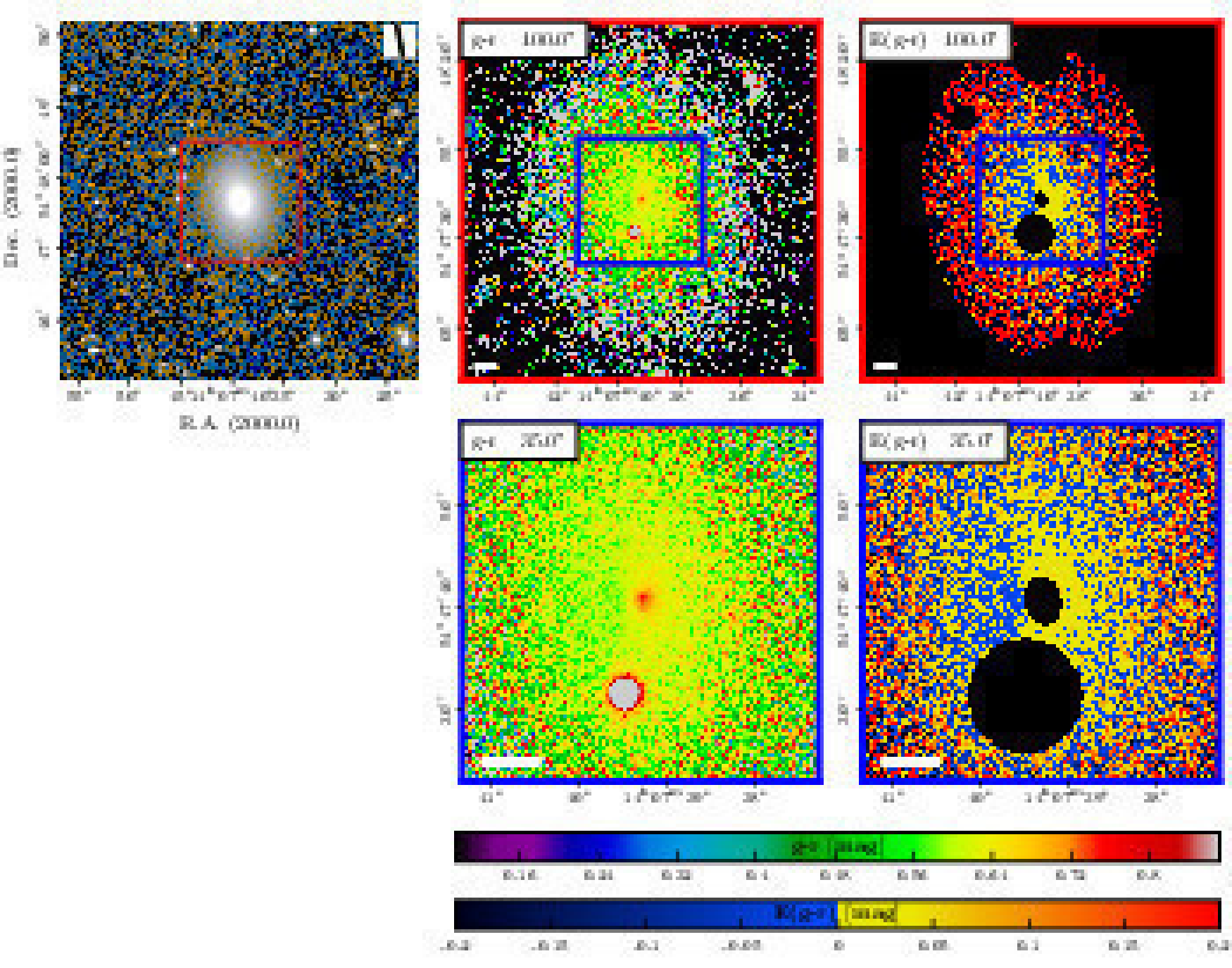}
\end{minipage}}
\makebox[\textwidth][c]{\begin{minipage}[l][-0.7cm][b]{.85\linewidth}
      \includegraphics[scale=0.55, trim={0 0.7cm 0 0},clip]{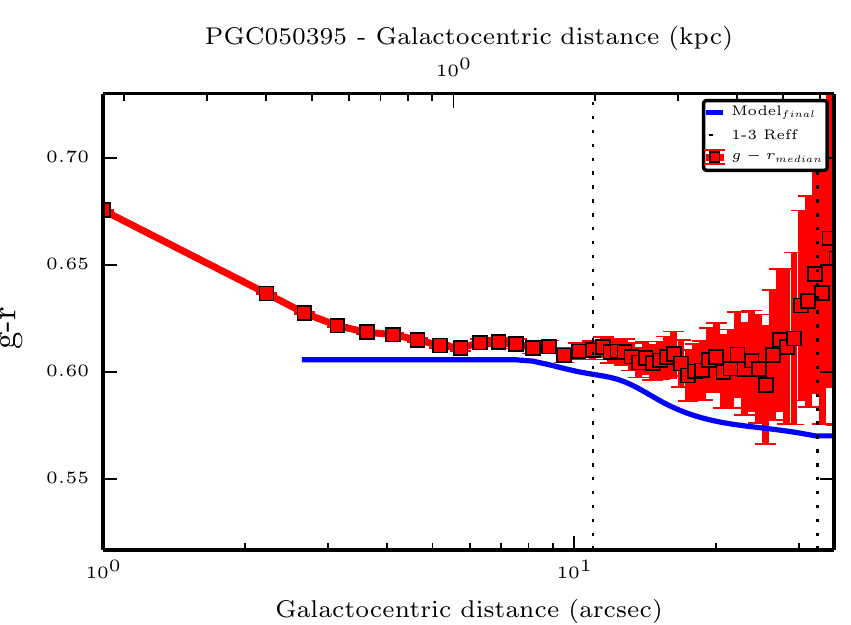}
      \includegraphics[scale=0.55]{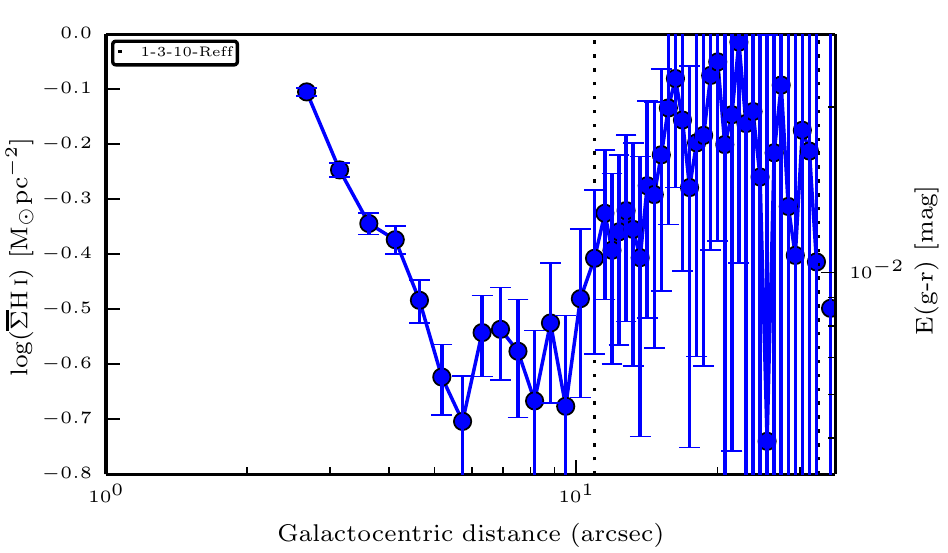}
\end{minipage}}
\caption{True image, colour map, colour excess map, and the radial profiles of PGC~050395.}
\label{fig:app_profiles}
\end{figure*}

%Page30
\clearpage
\begin{figure*}
\makebox[\textwidth][c]{\begin{minipage}[b][10.5cm]{.85\textwidth}
  \vspace*{\fill}
      \includegraphics[scale=0.85]{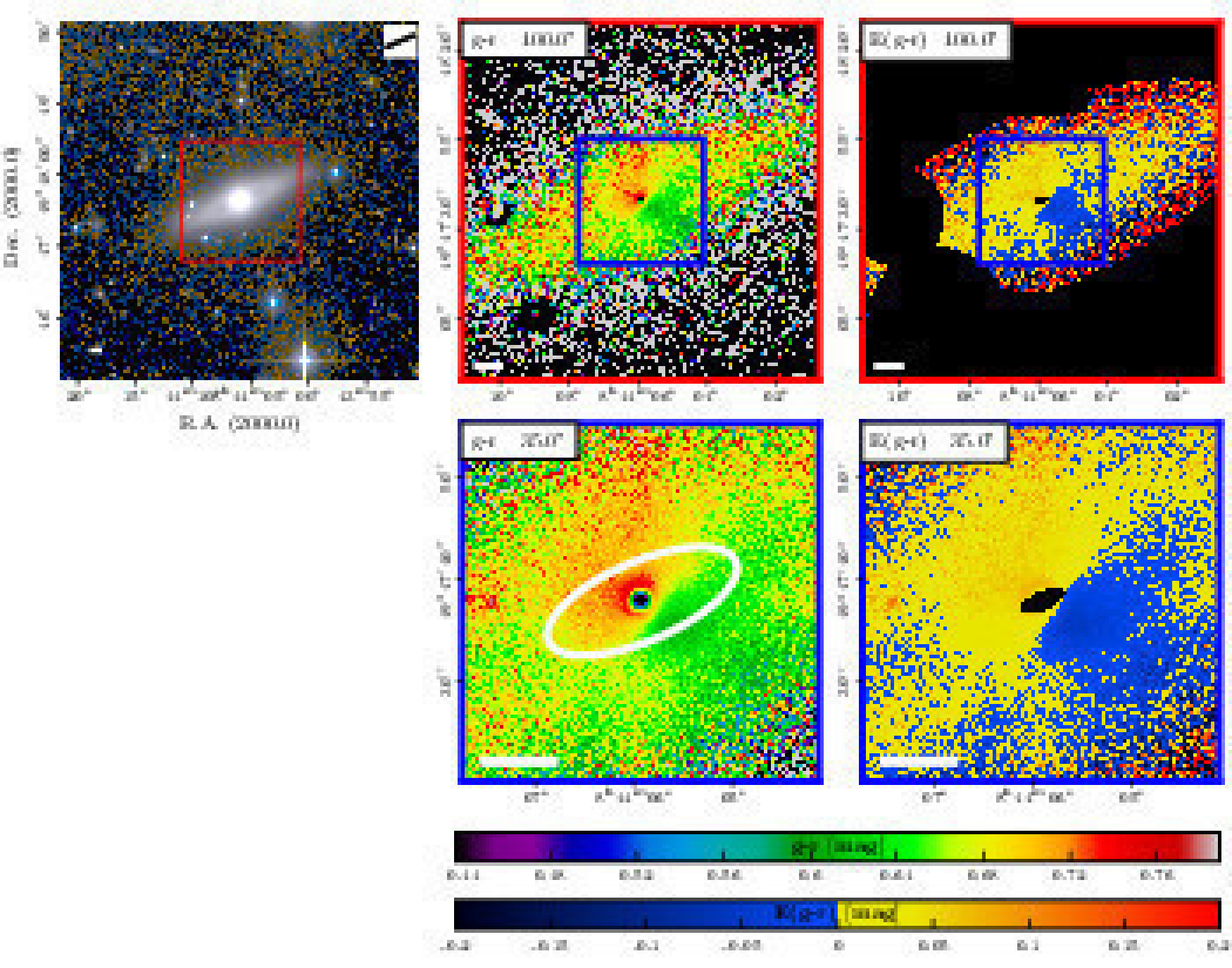}
\end{minipage}}
\makebox[\textwidth][c]{\begin{minipage}[l][-0.7cm][b]{.85\linewidth}
      \includegraphics[scale=0.55, trim={0 0.7cm 0 0},clip]{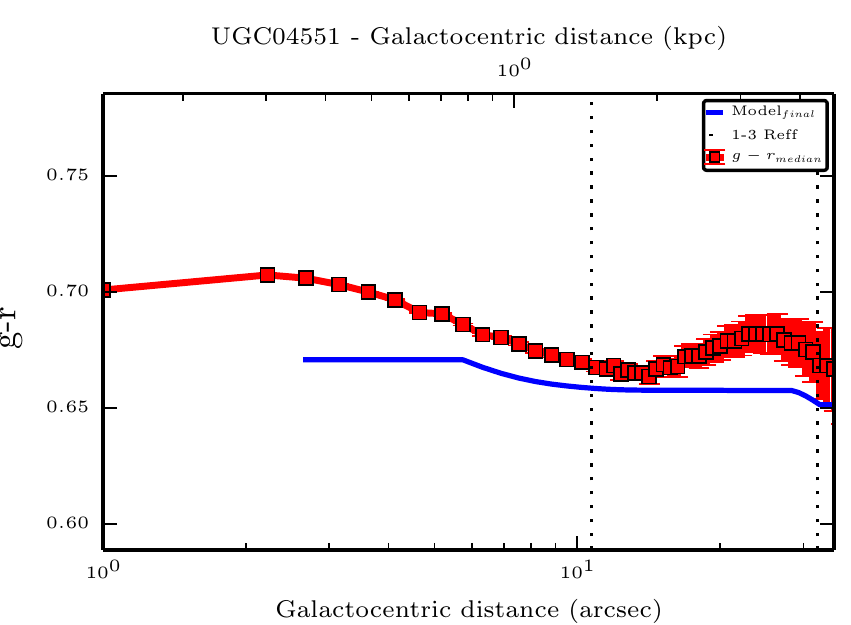}
      \includegraphics[scale=0.55]{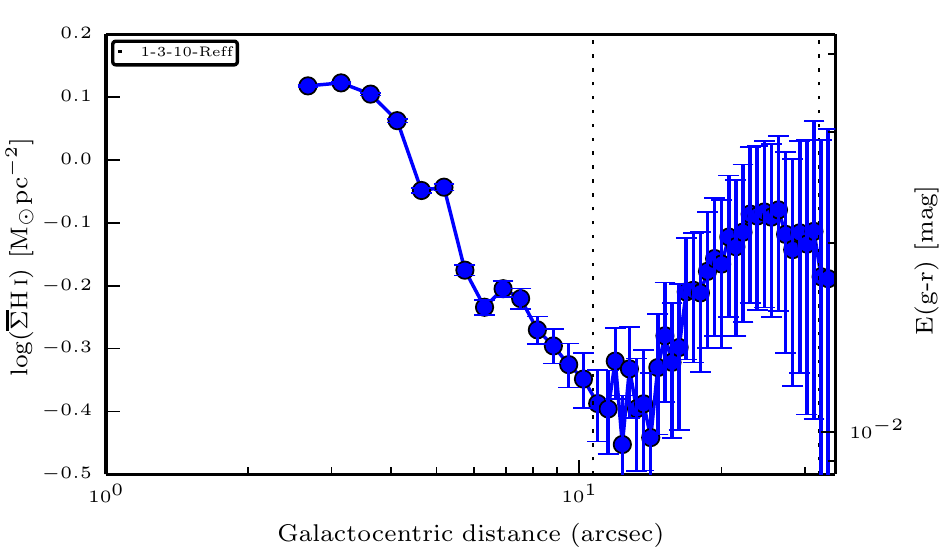}
\end{minipage}}
\caption{True image, colour map, colour excess map, and the radial profiles of UGC~04551.}
\label{fig:app_profiles}
\end{figure*}

\begin{figure*}
\makebox[\textwidth][c]{\begin{minipage}[b][11.6cm]{.85\textwidth}
  \vspace*{\fill}
      \includegraphics[scale=0.85]{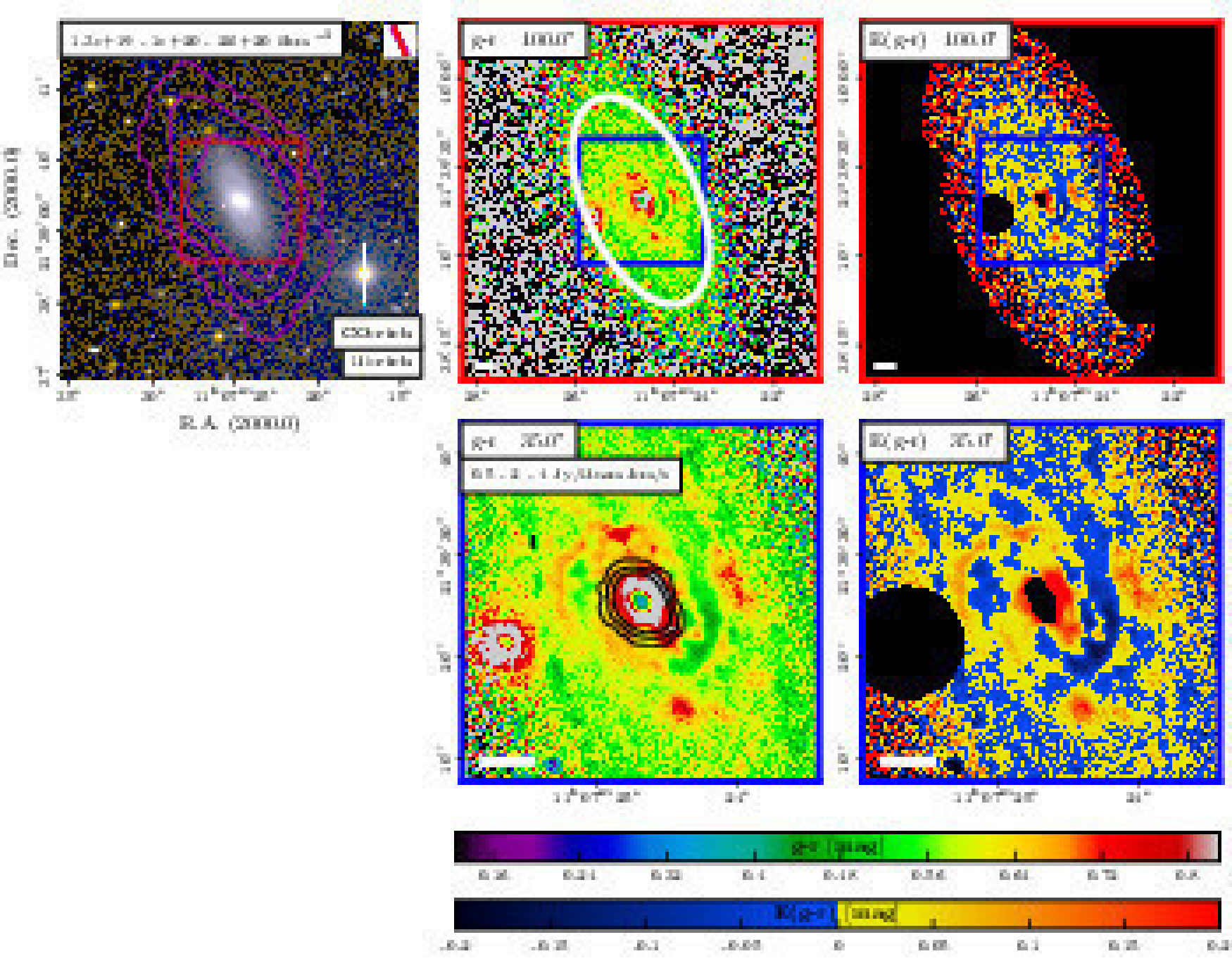}
\end{minipage}}
\makebox[\textwidth][c]{\begin{minipage}[l][-0.7cm][b]{.85\linewidth}
      \includegraphics[scale=0.55, trim={0 0.7cm 0 0},clip]{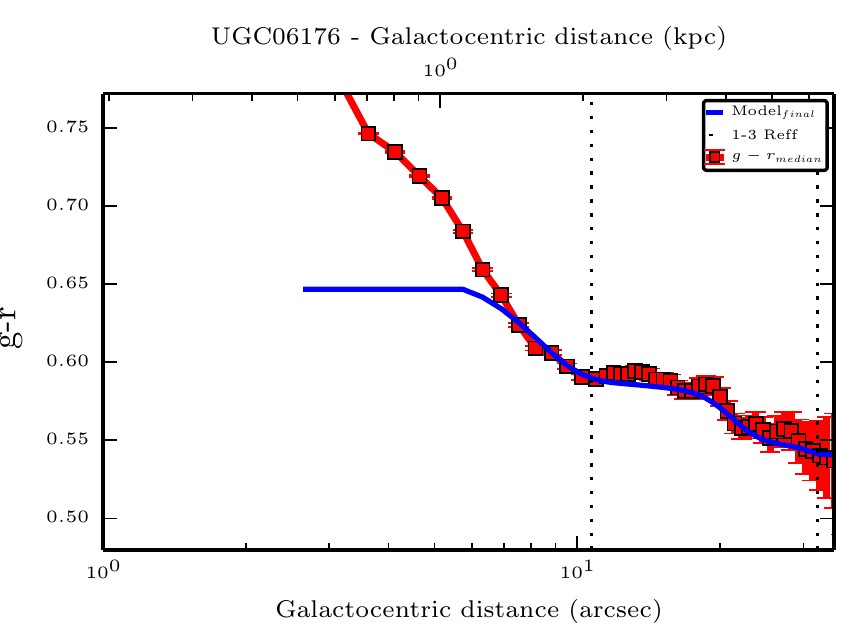}
      \includegraphics[scale=0.55]{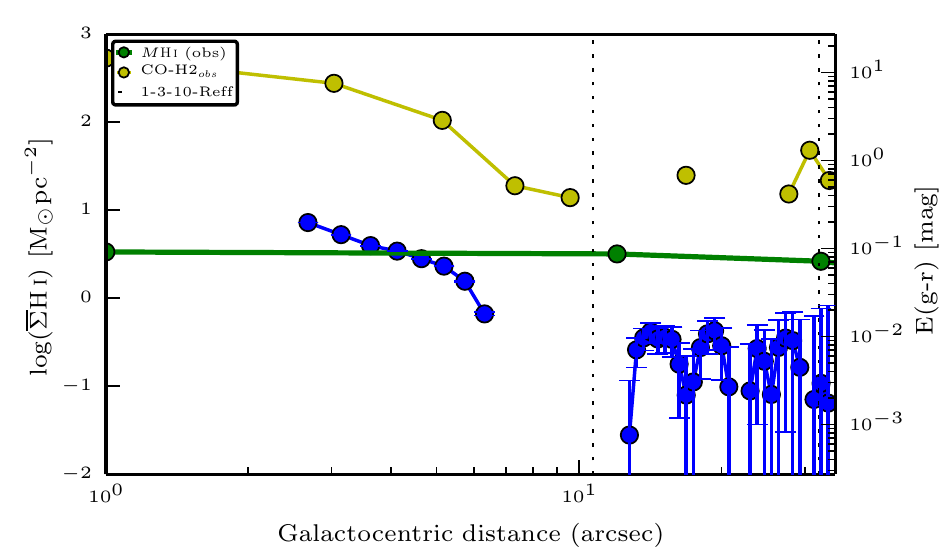}
\end{minipage}}
\caption{True image, colour map, colour excess map, and the radial profiles of UGC~06167.}
\label{fig:app_profiles}
\end{figure*}

%Page31
\clearpage
\begin{figure*}
\makebox[\textwidth][c]{\begin{minipage}[b][10.5cm]{.85\textwidth}
  \vspace*{\fill}
      \includegraphics[scale=0.85]{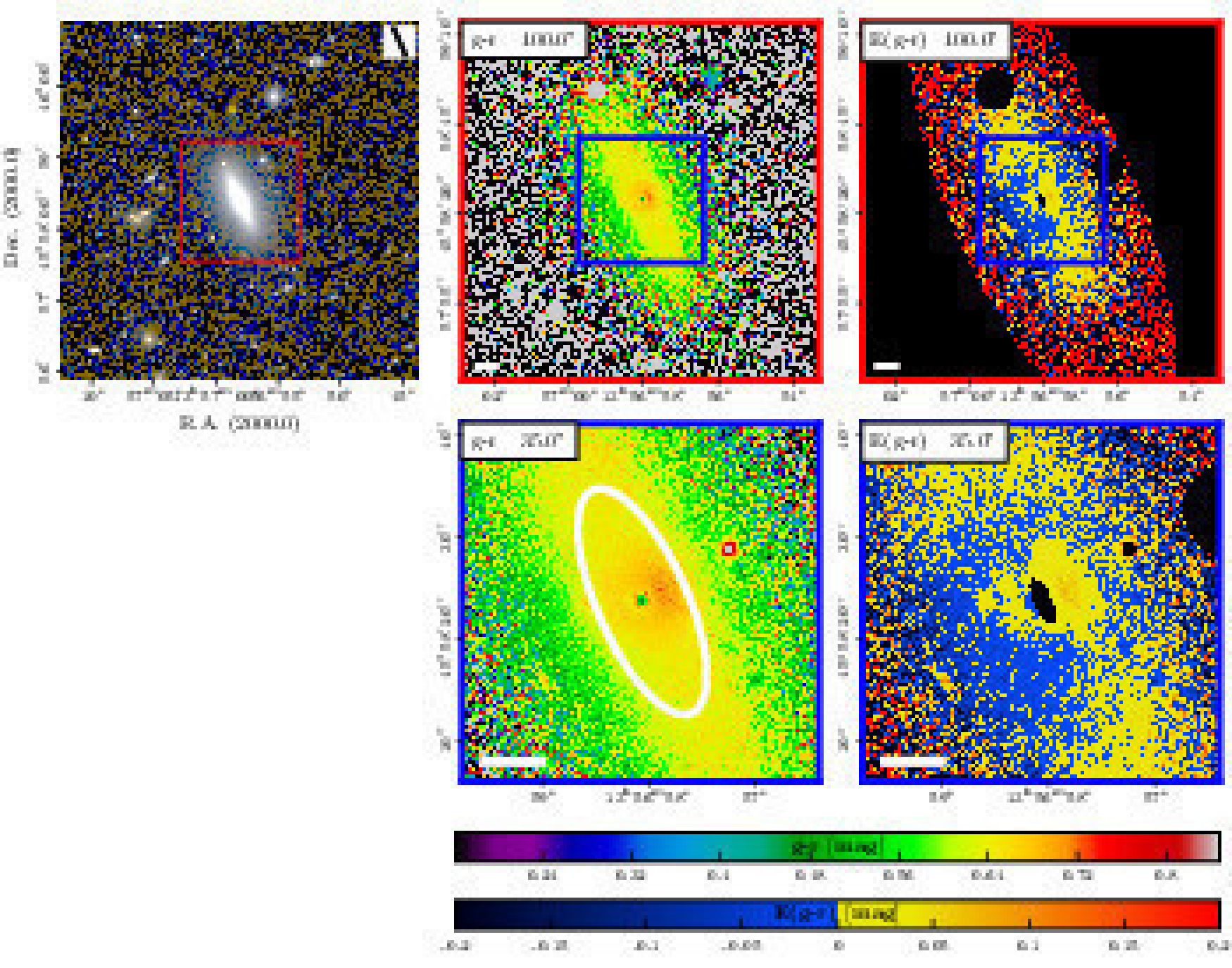}
\end{minipage}}
\makebox[\textwidth][c]{\begin{minipage}[l][-0.7cm][b]{.85\linewidth}
      \includegraphics[scale=0.55, trim={0 0.7cm 0 0},clip]{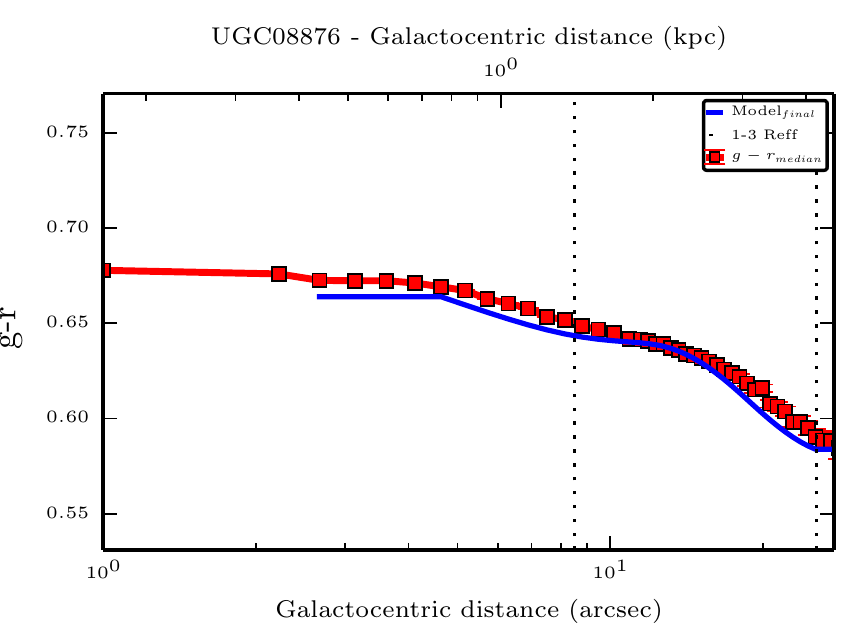}
      \includegraphics[scale=0.55]{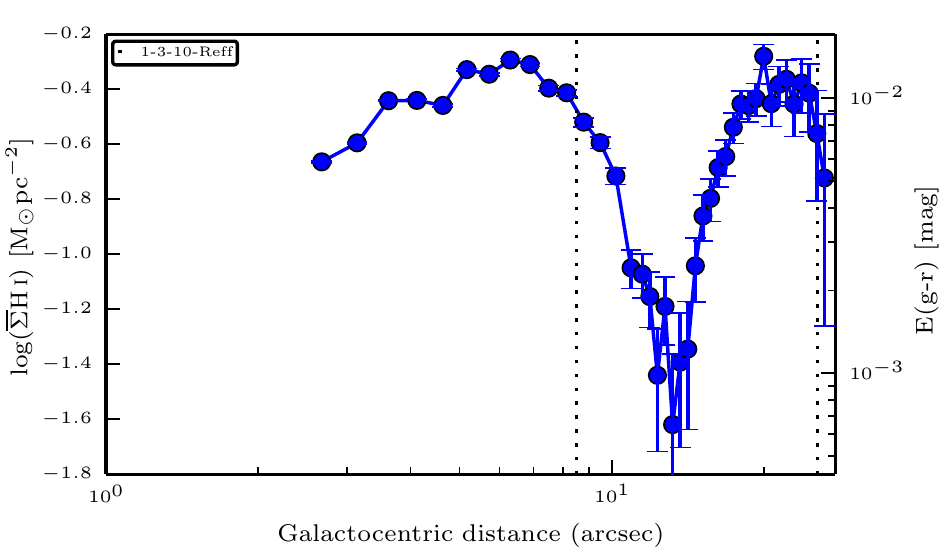}
\end{minipage}}
\caption{True image, colour map, colour excess map, and the radial profiles of UGC~08867.}
\label{fig:app_profiles}
\end{figure*}

\begin{figure*}
\makebox[\textwidth][c]{\begin{minipage}[b][11.6cm]{.85\textwidth}
  \vspace*{\fill}
      \includegraphics[scale=0.85]{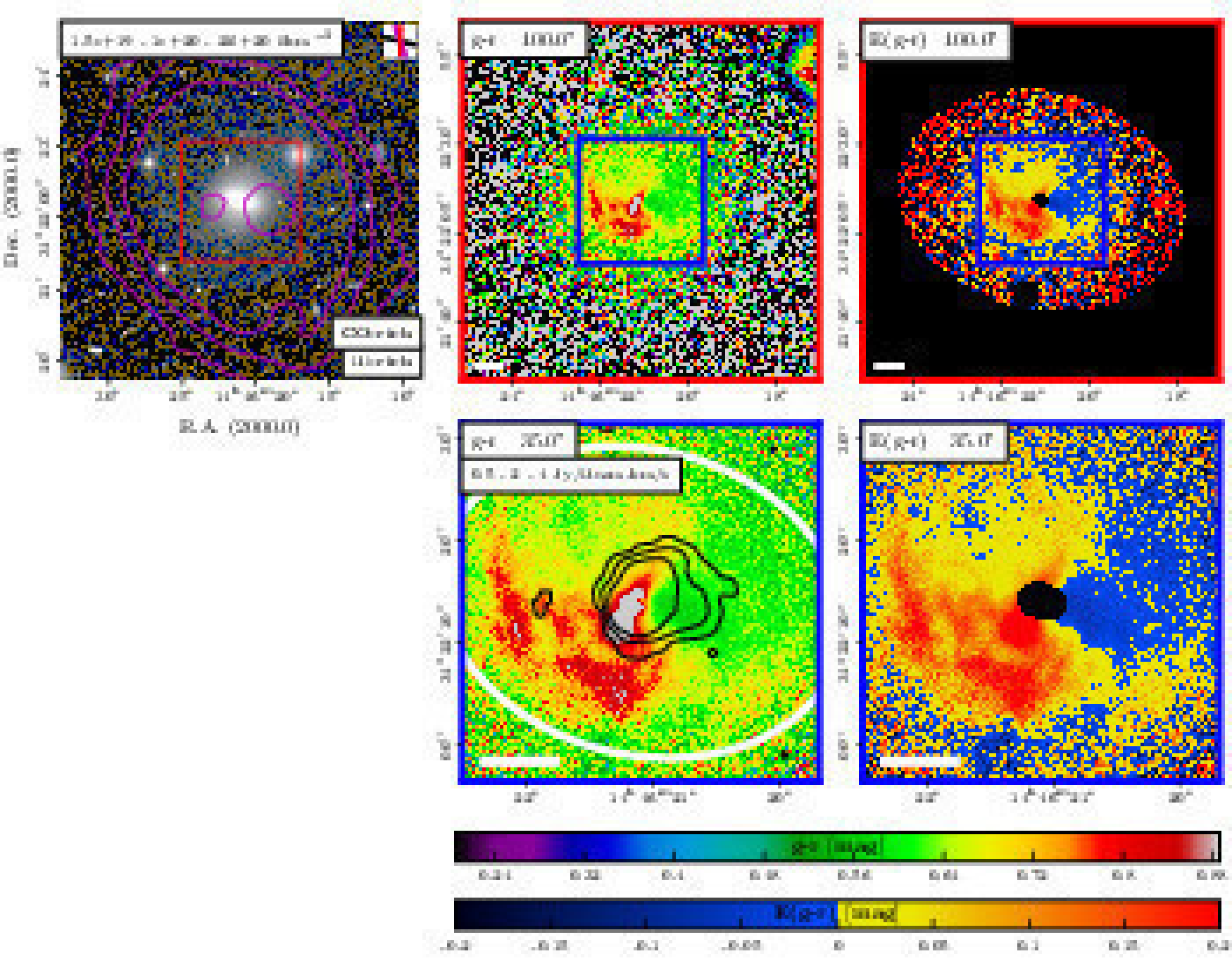}
\end{minipage}}
\makebox[\textwidth][c]{\begin{minipage}[l][-0.7cm][b]{.85\linewidth}
      \includegraphics[scale=0.55, trim={0 0.7cm 0 0},clip]{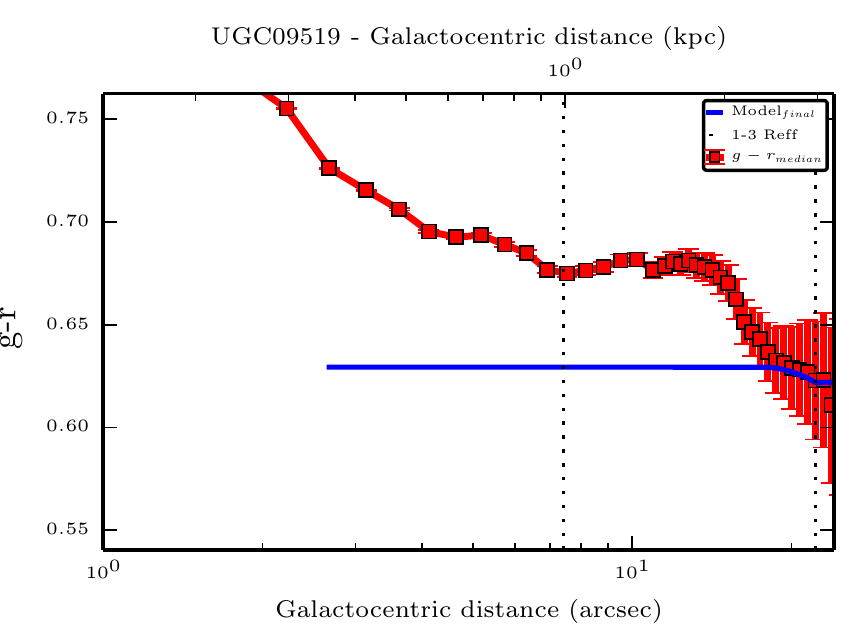}
      \includegraphics[scale=0.55]{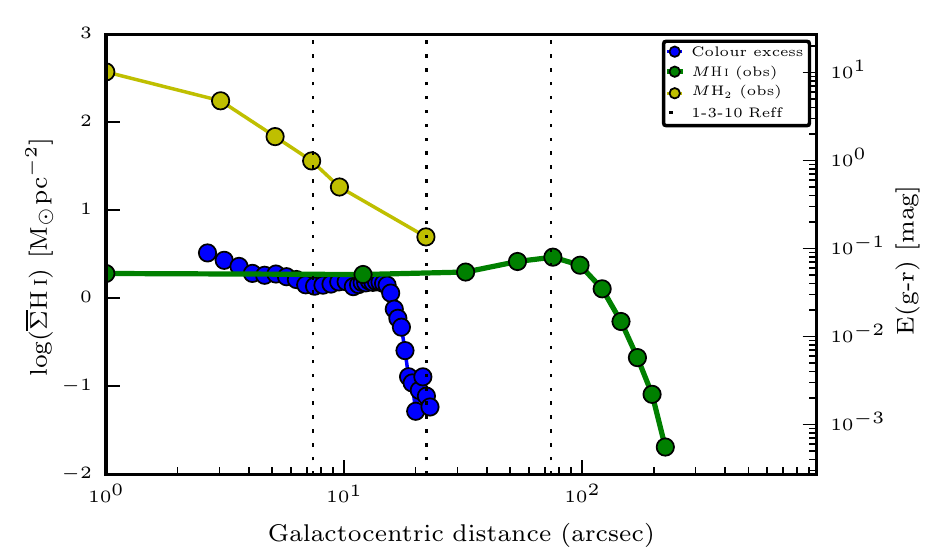}
\end{minipage}}
\caption{True image, colour map, colour excess map, and the radial profiles of UGC~09519.}
\label{fig:app_profiles}
\end{figure*}

\end{document}